\documentclass[PAPER, coverpage=false, atlasdraft=false, cernpreprint, texlive=2018, UKenglish, paper=a4, texmf]{atlasdoc}
 
\usepackage{atlaspackage}
\usepackage{atlasbiblatex}
 
\usepackage[jetetmiss=true,hepprocess=true]{atlasphysics}

\usepackage{multirow}
 
\graphicspath{{logos/}{figures/}}
 
\usepackage{ANA-TOPQ-2019-23-PAPER-defs}

\AtlasTitle{Measurements of differential cross-sections in top-quark pair events with a high transverse momentum top quark and limits on beyond the Standard Model contributions
to top-quark pair production with the ATLAS detector at $\sqrt{s}=13$ TeV}

\AtlasAbstract{
Cross-section measurements of top-quark pair production where the hadronically decaying top quark has transverse momentum greater than $355$ GeV and the other top quark decays into $\ell \nu b$ are presented using 139 fb$^{-1}$ of data collected by the ATLAS experiment during proton--proton collisions at the LHC. The fiducial cross-section at $\sqrt{s}=13$ TeV is measured to be $\sigma = 1.267 \pm 0.005 \pm 0.053$ pb, where the uncertainties reflect the limited number of data events and the systematic uncertainties, giving a total uncertainty of $4.2\%$. The cross-section is measured differentially as a function of variables characterising the $t\bar{t}$ system and additional radiation in the events. The results are compared with various Monte Carlo generators, including comparisons where the generators are reweighted to match a parton-level calculation at next-to-next-to-leading order. The reweighting improves the agreement between data and theory. The measured distribution of the top-quark transverse momentum is used to search for new physics in the context of the effective field theory framework. No significant deviation from the Standard Model is observed and limits are set on the Wilson coefficients of the dimension-six operators
$O_{tG}$ and $O_{tq}^{(8)}$, where the limits on the latter are the most stringent to date.
}
 
\author{The ATLAS Collaboration}
 
\AtlasRefCode{TOPQ-2019-23}
 
\PreprintIdNumber{CERN-EP-2022-003}

\AtlasJournalRef{\JHEP 06 (2022) 063}
\AtlasDOI{10.1007/JHEP06(2022)063}

\hypersetup{pdftitle={ATLAS document},pdfauthor={The ATLAS Collaboration}}
 
\begin{document}
 
\maketitle
 
\tableofcontents

\section{Introduction}
\label{sec-intro}
 
The high mass of the top quark means that it is often intimately involved in models of physics beyond the Standard
Model (SM). In effective models of new physics~\cite{WEINBERG1979327,BUCHMULLER1986621,BURGES1983464,Leung1986},
non-resonant deviations from the SM often appear
at high transverse momentum ($\pt$) of the top quark and high invariant mass of the top-quark--top-antiquark ($\ttbar$) system~\cite{Degrande_2011,Buckley:2015lku,Rosello:2015sck,Englert_2016}.
The high rate of $\ttbar$\ production at the Large Hadron Collider (LHC) provides a unique opportunity to
test for deviations from the SM predictions.
This motivates precise differential $\ttbar$~cross-section measurements of high-$\pt$\ (boosted) top quarks.
 
In this paper, $\ttbar$~events containing a boosted top quark that has decayed hadronically are selected
from data collected by the ATLAS detector in proton--proton~collisions at $\sqrt{s}=13$~\TeV\ from 2015 to 2018.
The selected events correspond to the semileptonic $\ttbar$~decay channel ($\ttbar \to WbWb \to \ell \nu b qq'b$) and the decay products are used to fully reconstruct
the kinematics of the $\ttbar$~system. The measured event properties are corrected for detector effects to obtain differential cross-sections as a function of various observables. These differential cross-section measurements characterise both the $\ttbar$ system and the additional radiation in the events. The analysis introduces a novel use of the
reconstructed top-quark mass to reduce the impact of uncertainties from the jet energy scale compared to the previous
ATLAS publication~\cite{TOPQ-2018-15} and significantly improves the precision of the measurements.
ATLAS has also measured the cross-section of $\ttbar$~events with boosted top quarks in the all-hadronic channel~\cite{TOPQ-2016-09} and
CMS has performed a measurement in the semileptonic channel~\cite{CMS-TOP-18-013}.
These three measurements used smaller datasets, corresponding to an integrated luminosity of $36$~\ifb, than the one used in this article, which corresponds to $139$~\ifb.
The measurements follow
on from analyses performed at $\sqrt{s}=8$~\TeV~\cite{TOPQ-2014-15, CMS-TOP-14-012}.
 
The measured cross-sections are compared with available SM predictions.
Precise predictions for SM $\ttbar$\ production are available at next-to-next-to-leading order (NNLO)
in quantum chromodynamics (QCD)~\cite{Czakon:2015owf,Czakon:2017wor,Catani:2019hip}.
The predictions are fully differential but are only available for stable top quarks or in the dilepton channel  ($\ttbar \to WbWb \to \ell \nu b \ell \nu b$)~\cite{Behring:2019iiv}.
Predictions at NLO in QCD are available for the semileptonic decay channel including the full decay of the top quarks
and parton shower and hadronisation effects in various Monte Carlo (MC) generators.
The measured cross-sections are compared with the NLO MC predictions and the impact of the higher-order corrections is tested
by reweighting the MC predictions at parton level to match predictions at NNLO in QCD.
 
The measurements presented in this article are also used to test for the presence of new physics beyond the SM.
The absence of direct evidence for the production of new particles beyond the SM at the LHC suggests
that any new physics is separated in mass from the SM fields. In this situation the new physics
can be parameterised in a model-independent way through the framework of effective field theory (EFT),
in which the Lagrangian of the SM is modified by adding an infinite series of higher-dimensional effective
operators~\cite{WEINBERG1979327,BUCHMULLER1986621,BURGES1983464,Leung1986} that are suppressed by the new-physics scale.
Assuming that the energy scale is sufficiently
large, the impact of new physics can be approximated by dimension-six operators.
The ability of the presented measurements to constrain new physics is illustrated by using the measured top-quark $\pt$\
distribution to set limits on the effective operators \Otg\ and \Otq, where the Warsaw basis~\cite{Grzadkowski_2010} is used to define the operators.
 
This paper is structured as follows. Section~\ref{sec-datamc} presents the data and simulated event samples.
Section~\ref{sec-evsel} discusses the object and event selection and the reconstruction of the $\ttbar$~system.
The methodology of the cross-section measurement, including the new method to use the reconstructed top-quark mass
to reduce the impact of the jet energy scale uncertainties, is described in Section~\ref{sec-xs}.
The systematic uncertainties are discussed in Section~\ref{sec-syst} and the results of the measurement are
compared with the theoretical predictions in Section~\ref{sec-results}. Section~\ref{sec-eft} presents the interpretation of the measured
top-quark $\pt$~distribution in the EFT framework. Finally, conclusions are presented in Section~\ref{sec-conclusions}.
 
\newcommand{\AtlasCoordFootnote}{
ATLAS uses a right-handed coordinate system with its origin at the nominal
interaction point (IP) in the centre of the detector and the \(z\)-axis along
the beam pipe. The \(x\)-axis points from the IP to the centre of the LHC ring,
and the \(y\)-axis points upwards. Cylindrical coordinates \((r,\phi)\) are used
in the transverse plane, \(\phi\) being the azimuthal angle around the
\(z\)-axis. The pseudorapidity is defined in terms of the polar angle \(\theta\)
as \(\eta = -\ln \tan(\theta/2)\). The rapidity is defined as \(y = (1/2)[(E+p_z)/(E-p_z)]\).
Angular distance is measured in units of \(\Delta R \equiv \sqrt{(\Delta\eta)^{2} + (\Delta\phi)^{2}}\).}

\section{Data and simulated event samples}
\label{sec-datamc}
 
The ATLAS detector~\cite{PERF-2007-01,PIX-2018-001,ATLAS-TDR-19} surrounds one of the collision points at the LHC.\footnote{\AtlasCoordFootnote}
The detector consists of
an inner tracking system surrounded by a superconducting solenoid producing a 2~T axial
magnetic field, electromagnetic and hadronic calorimeters and an external muon spectrometer
incorporating three toroidal magnet assemblies.
An extensive software suite~\cite{ATL-SOFT-PUB-2021-001} is used in the reconstruction and analysis of
real and simulated data, in detector operations, and in the trigger and data acquisition systems of the experiment.
 
The analysis was performed on data collected from
$pp$~collisions at $\sqrt{s}=13$~\TeV\ during 2015--2018. The dataset must fulfil standard data
quality requirements~\cite{DAPR-2018-01} and corresponds to an integrated luminosity of $139$~\ifb.
The uncertainty in the integrated luminosity is $1.7\%$~\cite{ATLAS-CONF-2019-021},
obtained using the LUCID-2 detector \cite{LUCID2} for the primary luminosity measurements.
Events are required to pass a single-electron or single-muon trigger~\cite{TRIG-2018-05,TRIG-2018-01}.

MC simulated event samples are used to determine background contributions, derive corrections
for detector effects, simulate potential new-physics contributions, and to compare with data.
Samples were processed using either the full ATLAS detector simulation~\cite{SOFT-2010-01} based on \GEANT~\cite{Agostinelli:2002hh},
or with a faster simulation making use of parameterised showers in the calorimeters~\cite{ATL-PHYS-PUB-2010-013}.
The effects of multiple collisions during a single bunch crossing (pile-up) were simulated by overlaying
additional inelastic $pp$~collisions generated with \PYTHIA[8]~\cite{Sjostrand:2014zea} and the A3~\cite{ATL-PHYS-PUB-2016-017}
set of tuned parameters (tune) onto the primary simulated events. These events were then processed with the same reconstruction software as the data.
The top-quark mass ($m_t$)\ is set to $172.5$~\GeV\ in all samples aside from those used to study the impact of the uncertainty in $m_t$.
 
The nominal simulated \ttbar\ sample was generated using \POWHEGBOX[v2]~\cite{Frixione:2007nw,Nason:2004rx, Frixione:2007vw, Alioli:2010xd}
(hereafter referred to as \POWHEG), which provides matrix elements at NLO in the strong coupling constant,
with the \NNPDF[3.0nlo]~\cite{Ball:2014uwa} parton distribution functions (PDFs). The $h_{\text{damp}}$~parameter, which
controls the matching of the matrix element to the parton shower and effectively regulates the high-$\pt$\ radiation against which the \ttbar\ system recoils,
was set to $1.5m_t$~\cite{ATL-PHYS-PUB-2016-020}. The functional form of the renormalisation (\muR) and factorisation (\muF) scales was
set to $\sqrt{m_{\textrm{t}}^2 + p^2_{\textrm{T}}}$, where $p_{\textrm{T}}$\ is the transverse momentum of the top quark.
\PYTHIA[8.230] was used to model the parton shower, hadronisation
and underlying event, using the A14 tune~\cite{ATL-PHYS-PUB-2014-021} and the \NNPDF[2.3lo]~\cite{Ball:2012cx} set of PDFs.
In the figures, this sample is referred to as PWG+PY8.
 
Additional \ttbar\ samples are used to assess the uncertainty in the modelling of \ttbar\ events and to compare
with the data measurements.
The dependence of the analysis on the \hdamp\ parameter is tested using a sample where the
parameter is varied as described in Ref.~\cite{ATL-PHYS-PUB-2017-007}.
The impact of using a different parton shower and hadronisation model is evaluated
using a sample produced with
\POWHEG interfaced to \HERWIG[7.04]~\cite{Bahr:2008pv,Bellm:2015jjp}.
The settings in \POWHEG~are the same as for the nominal sample, and the H7UE tune~\cite{Bellm:2015jjp}
and the \MMHT[lo] PDF set~\cite{Harland-Lang:2014zoa} are used for \HERWIG.
To assess the uncertainty due to the choice of generator,
events were generated with \MGNLO[2.6.0]~\cite{Alwall:2014hca} and the \NNPDF[3.0nlo]\ PDF set.
The choice of \mur\ and \muf\ is the same as for the \POWHEG set-up.
The events were interfaced with \PYTHIA[8.230].
The uncertainty due to the top-quark mass is evaluated by using samples generated in the same way as the nominal
\ttbar\ sample, but with the top-quark mass changed to $172, 173, 169$~and $176$~\GeV.
These samples were simulated using the fast simulation of the calorimeter, while all the previously described $\ttbar$ samples used the full simulation of the calorimeter.
A version of the nominal sample was also produced using the fast simulation to ensure samples with consistent simulation settings were compared when evaluating the impact of the top-quark mass.
 
All \ttbar\ samples are normalised to the cross-section prediction at NNLO
in QCD including the resummation of next-to-next-to-leading logarithmic (NNLL) soft-gluon terms calculated using
\TOPpp[2.0]~\cite{Beneke:2011mq,Cacciari:2011hy,Baernreuther:2012ws,Czakon:2012zr,Czakon:2012pz,Czakon:2013goa,Czakon:2011xx}.
For proton--proton collisions at a centre-of-mass energy of \rts~=~\SI{13}{\TeV}, this cross-section corresponds to
$\sigma(\ttbar)_{\textrm{NNLO+NNLL}} = \mathrm{832\pm51~pb}$ using a top-quark mass of $\mtop = 172.5$~\GeV.
The uncertainties in the cross-section due to the PDF and \alphas are calculated using the PDF4LHC prescription~\cite{Butterworth:2015oua}
with the \MSTW[nnlo] 68\% CL~\cite{Martin:2009iq,Martin:2009bu}, \CT[10nnlo]~\cite{Lai:2010vv,Gao:2013xoa} and \NNPDF[2.3] 5f FFN~\cite{Ball:2012cx} PDF sets,
and are added in quadrature to the effect of the scale uncertainty.
Predictions for $\ttbar$~production at NNLO matched to the parton shower to produce particle-level predictions are not yet available for the semileptonic final state.
In order to evaluate the impact of NNLO QCD corrections, the MC events are reweighted at parton level to match higher-order predictions.
The reweighting is performed on the three variables $\pt(t)$, $m(\ttbar)$\ and $\pt(\ttbar)$, using the kinematics of the top quarks in the MC samples after initial- and final-state radiation.
The predictions for $\pt(t)$\ and $m(\ttbar)$\ are calculated at NNLO in QCD with NLO electroweak (EW) corrections~\cite{Czakon:2017wor} with the \NNPDF[3.0qed] PDF set using the dynamic renormalisation
and factorisation scales $m_\text{T}(t)/2$\ for $\pt(t)$ and $H_\text{T}/4$ for $m(\ttbar)$ as proposed in Ref.~\cite{Czakon:2017wor}. The prediction for $\pt(\ttbar)$\ is calculated at NNLO in
QCD~\cite{Catani:2019hip,Grazzini:2017mhc} with the NNPDF3.0 PDF set and with \muR and \muF set to $H_\text{T}/4$. All the predictions use $m_t=173.3$~\GeV\footnote{It
was verified that the changes in the distributions from the reweighting procedure are much larger than those expected from changing $m_t$~by $0.8$~\GeV.}.
The reweighting was performed iteratively~\cite{Serkin:2021bbn}, such that at the end of the procedure the reweighted MC sample agrees well with the higher-order
prediction for each of the three variables. These samples are referred to as being reweighted to the NNLO prediction in the remainder of the document.
The reweighted predictions themselves are not equivalent to complete NNLO plus parton shower calculations and are used to estimate the effect of the NNLO contributions on the measured observables.

To examine the predictions provided by generators that use higher-order calculations for the additional jets in $\ttbar$\ events, a sample
was generated using \SHERPA[2.2.10]~\cite{Bothmann:2019yzt}. The sample uses NLO-accurate
matrix elements for up to one additional parton, and LO-accurate
matrix elements for up to four additional partons, calculated with the
\AtlasMC{Comix}~\cite{Gleisberg:2008fv} and
\OPENLOOPS~\cite{Buccioni:2019sur,Cascioli:2011va,Denner:2016kdg} libraries. They are
matched with the \SHERPA parton shower~\cite{Schumann:2007mg} using
the \MEPSatNLO
prescription~\cite{Hoeche:2011fd,Hoeche:2012yf,Catani:2001cc,Hoeche:2009rj}
and the set of tuned parameters developed by the \SHERPA authors
to match the \NNPDF[3.0nnlo] set of PDFs~\cite{Ball:2014uwa}.
The central scale has the functional
form $\mu^{2} = m_{t}^{2} + \frac{1}{2}(p_{\mathrm{T},t}^{2} +
p_{\mathrm{T},\bar{t}}^{2})$. The CKKW matching scale~\cite{Catani:2001cc} of the
additional emissions was set to 30\,\GeV. As this sample contains parts of the inclusive NNLO corrections to
$\ttbar$\ there are two possibilities for normalising the sample: first to use the prediction provided by the generator (referred to as NLO norm.\ in the following) and second
to normalise the sample to the NNLO+NNLL prediction in the same way as for \ttbar\ samples from the other MC generators.
The prediction of the inclusive $\ttbar$\ cross-section from the NNLO+NNLL calculation is $21\%$\ higher  than the prediction from the \SHERPA[] generator.
 
The possible contributions from high-energy-scale new physics are modelled using an EFT approach.
In this approach the SM Lagrangian is expanded with higher-dimensional operators according to:
\begin{equation}\nonumber
\mathcal{L}_\text{EFT} = \mathcal{L}_\text{SM}+\sum_{i,D}\frac{C_{i}^{D}}{\Lambda^{D-4}}O_{i}^{D},
\label{eq:lagr}
\end{equation}
where each operator $O_{i}^{D}$~of dimension $D$~has a corresponding Wilson coefficient $C_i$, and
$\Lambda$~is the energy scale associated with the new physics.
Assuming that the energy scale is high, the sum can be truncated at dimension six, which is the first term that gives non-zero contributions
when assuming lepton and baryon number conservation.
At dimension six, assuming baryon number conservation and minimal flavour violation~\cite{D_Ambrosio_2002}, there are 59 independent operators.
The SM is recovered by setting all the Wilson coefficients to zero.
This analysis is restricted
to two operators in the Warsaw basis~\cite{Grzadkowski_2010} that
are expected to have a significant impact on $\ttbar$~production: $\Otg$\ and $\Otq$.
MC samples were generated with the \AtlasMC[1.0.0]{SMEFT@NLO} UFO model~\cite{degrande2020automated} at LO
to provide events with $\left(\ctg, \ctq\right) = (\pm1 \text{~or}\ 0,~\pm1 \text{~or}\ 0)$ and $\Lambda = 1$~\TeV.
The renormalisation and factorisation scales were set to $\muR=\muF=m_t$\ and the $\{m_W,m_Z,G_\mu\}$\ EW input scheme was used as outlined in Ref.~\cite{Brivio:2792440}.
Two sets of samples were produced: one set includes contributions proportional to $\Lambda^{-2}$, corresponding to the interference between the SM and dimension-six operators,
and the second set additionally includes terms proportional to $\Lambda^{-4}$, corresponding to the square of dimension-six operators.
An additional sample including the $\Lambda^{-2}$\ and  $\Lambda^{-4}$ contributions was generated
with $\left(\ctg, \ctq\right) = (0.2, 0.2)$\ and $\Lambda = 1$~\TeV\ in order to test for possible biases in the EFT fit.
The parton shower and hadronisation were performed using \PYTHIA[8.244].
 
Backgrounds from other processes that include the decay of at least one $W$~or $Z$~boson into leptons were simulated using MC generators.
Top-quark production in association with a $W$ boson ($tW$)
and the production of single top quarks in the $s$-channel were
modelled by the
\POWHEG~\cite{Re:2010bp,Alioli:2009je}
generator at NLO in QCD using the five-flavour scheme and the
\NNPDF[3.0nlo] set of PDFs interfaced to \PYTHIA[8.230].
The diagram removal scheme~\cite{Frixione:2008yi} was used to
remove interference and overlap between $tW$~and $\ttbar$\ production.
The production of single top quarks in the $t$-channel was simulated using \POWPY in the four-flavour scheme~\cite{Frederix:2012dh}
with the corresponding \NNPDF[3.0nlo] set of PDFs.
The production of $V+$\,jets ($V=W\text{~or~}Z$) was simulated with \SHERPA[2.2.1]
using NLO matrix elements for up to two partons, and LO matrix elements
for up to four partons, calculated with the \AtlasMC{Comix}
and \OPENLOOPS libraries. They
were matched with the \SHERPA parton shower using the MEPS@NLO
prescription. 
The \NNPDF[3.0nnlo] set of PDFs was used and the samples
were normalised to an NNLO prediction~\cite{Anastasiou:2003ds}.
Events with diboson final states ($VV$) were also simulated with the
\SHERPA generator using
matrix elements at NLO accuracy in QCD for up to one additional parton
and at LO accuracy for up to three additional parton
emissions. The showering and hadronisation were performed in the same way as for the $V+$\,jets samples.
Production of $\ttV$\ events
forms a small background in the analysis and was modelled using the
\MGNLO[2.3.3]~\cite{Alwall:2014hca} generator at NLO with the
\NNPDF[3.0nlo] parton distribution functions.
Finally, $t\bar{t}H$~events were modelled using the
\POWHEG~generator at NLO with the \NNPDF[3.0nlo]~PDF set.
The events were interfaced to \PYTHIA[8.230] in a similar way to the nominal $t\bar{t}$~sample.
 
\section{Event selection and reconstruction}
\label{sec-evsel}
 
All events must contain a primary vertex with at least two associated tracks with $\pT>0.5$~\GeV.
The vertex with the highest $\sum \pT^2$\ of the associated tracks is taken as the primary vertex.
Electrons are reconstructed from energy deposits in the electromagnetic (EM) calorimeter matched to a track
in the inner detector. They must have transverse energy $\eT > 27$~\GeV\ and pass the `Tight' likelihood-based
requirement~\cite{EGAM-2018-01}. They must have pseudorapidity $|\eta|<2.47$~and be outside the transition region
between the barrel and endcap EM calorimeters ($1.37 < |\eta| < 1.52$).
Electrons are required to be isolated
by applying the `Tight' requirements on the sum of nearby energy in the calorimeter and the sum of the momenta
of nearby tracks~\cite{EGAM-2018-01}.
The track associated with the electron must satisfy a requirement of $|d_0|/\sigma_{d_0} < 5$ on the transverse impact parameter significance calculated relative to the beam line,
and a requirement of $|\zzsth| < 0.5$~mm on the longitudinal impact parameter calculated relative to the event primary vertex,
where $\theta$\ is the polar angle of the track.
Muons are reconstructed by combining a track found in the inner detector with a matching track found in
the muon spectrometer. Selected muons must have $\pT > 27$~\GeV\ and $|\eta|<2.5$, and pass the `Medium' identification
requirements and `Tight' isolation
requirements~\cite{MUON-2018-03}.
The muons must have impact parameters satisfying $|d_0|/\sigma_{d_0} < 3$\ and $|\zzsth| < 0.5$~mm.
 
Jets are reconstructed using the \antikt\ clustering algorithm~\cite{Cacciari:2008gp,Fastjet} with radius parameter $R = 0.4$
starting from particle-flow objects that exploit both calorimeter and track measurements~\cite{PERF-2015-09}.
Jets are calibrated using measurements in both simulation and data~\cite{JETM-2018-05} and are required
to have $\pt > 26$~\GeV\ and $|\eta|<2.5$\ so that they are within the acceptance of the inner detector.
Jets with $\pt < 60$~\GeV\ must also pass a pile-up rejection threshold placed on the output of the multivariate jet-vertex tagger (JVT)~\cite{PERF-2014-03}.
Jets that are close to an electron, $\Delta R(e, j) < 0.2$, are removed to avoid double counting the energy of the electron.
Jets that have less than three tracks and are
either close to a muon ($\Delta R(\mu, j) < 0.2$) or have a track that is part of the muon are also removed. This avoids
counting the energy deposits of muons as jets.
An additional requirement of $\Delta R(\ell, j) > 0.4$ then ensures that electrons and muons are well separated from jets;
leptons failing this requirement are rejected.
Jets that contain $b$-hadrons ($b$-jets) are identified by the use
of the DL1r multivariate algorithm~\cite{FTAG-2018-01,ATL-PHYS-PUB-2017-013}. The selected working point results in an efficiency of $77\%$\ per
$b$-jet, as measured in simulated $\ttbar$\ events. These jets are hereafter referred to as $b$-tagged jets.

Highly boosted top quarks ($\pT \gtrsim 2\mtop$) that decay hadronically can produce decay products that are not resolved
as three separate jets. These boosted top quarks are therefore identified using large-radius jets so as to capture all the decay products in a single jet.
These \largeR\ jets are reconstructed by applying the \antikt\ clustering algorithm with radius parameter $R = 1.0$ to the selected $R=0.4$ (\smallR) jets~\cite{Nachman:2014kla}.
Any \smallR\ jets that have less than $5\%$ of the $\pT$\ of the corresponding \largeR\ jet are removed from that \largeR\ jet. This trimming procedure~\cite{Krohn:2009th,Nachman:2014kla} is designed
to remove jets that are more likely to originate from pile-up.
The \largeR\ jets must have $\pT > 355$~\GeV\ and $|\eta|<2.0$. In order to select jets consistent with a hadronically decaying top
quark, the \largeR\ jets must contain at least one $b$-tagged jet and have an invariant mass in the range $120 < m < 220$~\GeV.
If more than one \largeR\ jet passes these requirements, the one with
the highest $\pT$\ is assumed to be the one originating from the hadronically decaying top quark. This selected \largeR\ jet is referred to as the top-tagged jet.
The missing transverse momentum ($\MET$) is reconstructed from the negative vector sum of calibrated
leptons, \smallR\ jets and the soft term (calculated using other tracks associated with the primary vertex)~\cite{PERF-2016-07}.
 
Events are required to have exactly one selected lepton, at least one top-tagged jet and at least two $b$-tagged jets.
At least one of the $b$-tagged jets must not be a constituent of the top-tagged jet.
The selected lepton must match a corresponding electron or muon trigger object.
Consistency with the expected boosted topology is ensured by requiring the lepton to be close to a $b$-tagged jet,
$\Delta R(\ell, b) < 2.0$,~and the same $b$-tagged jet must not be a constituent of the top-tagged jet.
A requirement of $\Delta R(e, t) > 1.0$ prevents the selection of \largeR\ jets seeded by a high-$\pT$ electron.
To reduce the multijet background, events must have  $\MET > 20$~\GeV\ and
$\MET + \mT^W > 60$~\GeV, where $\mT^{W}=\sqrt{2p^{\ell}_\text{T}\MET\left( 1 - \cos{\Delta\phi}\left(p^{\ell}_\text{T},\MET\right)\right)}$
is the transverse mass of the $W$ boson.
The invariant mass of the lepton
and the nearest $b$-tagged jet, $m_{\ell b}$, must be less than $180$~\GeV.
This selection requirement retains signal events where the lepton and $b$-jet originate from an on-shell top-quark decay, while rejecting events from $tW$\ single
top-quark production~\cite{TOPQ-2017-05}.
 
The kinematics of the top quarks are obtained from the selected objects. The selected top-tagged jet is used as
the estimate of the hadronically decaying top quark. The leptonically decaying top quark is reconstructed from the four-vector sum
of the lepton, the closest $b$-tagged jet and the neutrino. The $x$~and $y$~components of the missing transverse momentum provide estimates
of the corresponding components of the neutrino four-vector.
The $z$-component is calculated using the constraint
that the lepton--neutrino system has invariant mass equal to the $W$\ boson mass~\cite{PDG}.
If there are two real solutions to the corresponding quadratic equation, the solution that
gives the smallest value for the mass of the leptonically decaying top quark is used. In the case of
complex solutions, only the real part is used.
Any jets other than the constituents of the top-tagged
jet and the $b$-tagged jet used in reconstructing the leptonically decaying top quark are referred to as additional jets.

 
\section{Cross-section measurements}
\label{sec-xs}
 
The strategy used to measure differential cross-sections
is to correct (unfold) the data for detector effects and can be summarised in
the following equations:
 
\begin{gather}
\frac{\dif\sigma}{\dif X} = \frac{N^u_i}{\mathcal{L}\Delta X_i} \label{eqn-unf}\\
N^u_i = \frac{1}{f_{\mathrm{eff}}^i} \sum_j M^{-1}_{ij} f_{\mathrm{acc}}^j \left(N^j_\mathrm{d}(\mathrm{\JSF}) - N^j_\mathrm{b} \right), \nonumber
\end{gather}
 
where $X$~represents the variable being measured.
The data are first corrected by applying a jet energy scale factor (JSF) that ensures that the mean of the reconstructed top-quark mass agrees with the simulation, as discussed in Section~\ref{sec-xs-jsf}.
The number of events in each bin after that correction, $N^j_\mathrm{d}(\mathrm{\JSF})$, then have the background contributions ($N^j_\mathrm{b}$) subtracted.
The yields are then corrected with the factor $f^j_{\mathrm{acc}}$ to account for $\ttbar$~events that do not pass the fiducial requirements.
An iterative Bayesian unfolding~\cite{DAgostini:1995} (denoted by $M^{-1}_{ij}$) implemented in RooUnfold~\cite{Adye:2011gm} is used to correct for the limited resolution of the detector.
The factor $f_{\mathrm{eff}}^i$ corrects for events that pass the fiducial requirements but do not pass the detector-level event selection.
Finally, the events are corrected for the integrated luminosity ($\mathcal{L}$) and the bin width ($\Delta X_i$).
The same methodology is used for double-differential distributions.
The inclusive cross-section is determined by using all the selected events in a single bin as input to Eq.~(\ref{eqn-unf}).
All the correction factors are determined
using the nominal $\ttbar$~simulation.
In addition to the absolute differential cross-section measurements, normalised differential cross-section measurements
are produced by dividing by the measured inclusive cross-section. The normalised measurements provide a way to evaluate
how well a distribution's shape agrees between data and theory without considering the overall normalisation.
This is relevant, for example, in new-physics searches where the normalisation of $\ttbar$\ production can be fitted from the data, while the
shape relies on the simulation~\cite{EXOT-2016-17}.
As the normalised measurements contain less information than the absolute measurements,
they are presented in Appendix~\ref{app:normalisedResults}.
The different components of the measurement methodology are discussed in the following subsections and the section
is concluded by comparing the observed data with the expectation from the $\ttbar$\ simulation and the background estimates.
 
\subsection{Background estimate}
\label{sec-xs-blg}
 
The event selection is designed to reject the vast majority of background events such that the modelling of the background processes has a minor impact on the measurements.
The main backgrounds to the signal process are expected to be from $tW$~single top-quark production and $W+$\,jets production.
These backgrounds, along with smaller backgrounds from $\ttbar V$, $\ttbar H$, $Z\,+$\,jets and diboson production, are estimated
from the simulated samples described in Section~\ref{sec-datamc}. Events can also pass the selection if the lepton originates
from the decay of a light- or heavy-flavour hadron, or due to the misreconstruction of a jet as a lepton. These events are referred to as the multijet background in the remainder of the paper.
The contribution from this background is estimated using the matrix method~\cite{ATLAS-CONF-2014-058}, which uses a selection of events in data
where the lepton requirements are less stringent than the nominal requirements.

\subsection{Correction of the jet energy}
\label{sec-xs-jsf}
 
Uncertainties in the jet energy scale are significant in many $\ttbar$~cross-section measurements~\cite{TOPQ-2020-02,TOPQ-2018-18,TOPQ-2018-15,CMS-TOP-18-002,CMS-TOP-17-002}.
In order to reduce the impact of such uncertainties in this analysis,
the reconstructed invariant mass of the selected hadronically decaying
top quark ($\mtopjet$) is used in conjunction with the precisely known value of the top-quark mass~\cite{CMS-TOP-14-022,TOPQ-2017-03}.
The overall jet energy scale difference between data and simulation for \smallR\ jets is parameterised with a jet energy scale factor, \JSF.
Assuming that the top-quark mass is known, the $\mtopjet$~distribution can be used to measure the \JSF\ in the data.
In this way, any overall jet energy scale difference between data and simulation is absorbed into the JSF and does not impact the extracted cross-sections,
thereby reducing the impact of the jet energy scale uncertainties on the measured cross-sections.
The jet energy scale uncertainties, which are included in the measurement as described in Section~\ref{sec-syst},
are not completely suppressed by the method because they have kinematic dependencies that cannot be absorbed by the \JSF.
The uncertainty in the top-quark mass effectively limits the precision to which the overall jet energy scale difference between data and simulation can be absorbed
by the method, but the high precision of the top-quark mass measurements means the corresponding systematic uncertainty is not significant in the measurements (see Section~\ref{sec-syst-ttbarmodel}).
Figure~\ref{fig:JSFscalingExample:mtopjsf} shows the $\mtopjet$~distribution for three different \JSF\ values.
The distributions are obtained by applying the JSF to the energy of every jet in the nominal $\ttbar$\ sample and
the simulation samples used for the background estimates.
The JSF is applied on top of the standard ATLAS jet calibration discussed in Section~\ref{sec-evsel}.
Figure~\ref{fig:JSFscalingExample:jsfmap} shows that the mean of the $\mtopjet$~distribution\footnote{The mean is calculated from events that have $135 < \mtopjet < 205$~\GeV.}
($\meanmtopjet$) depends linearly on the \JSF, as expected from the fact the top-tagged jet four-vector is a sum of the four-vectors of the constituent \smallR\ jets.
The $\meanmtopjet$~value in data is used to determine the JSF in data from this linear relationship.
The distribution of $\mtopjet$~in data is shown in Figure~\ref{fig:JSFscalingExample:mtopdata} with $\mathrm{JSF}=1$.
In order to visualise possible differences in the shape of the distribution, the expectation is normalised to the data yield.
Good agreement is seen between the data and the expectation, and the \JSF\ is measured to be $\JSFval \pm \JSFunc$, where the uncertainty
is due to the number of observed data events.
The energy of each reconstructed jet in the data is then corrected with $1/ \mathrm{\JSF}$
and the event selection is reapplied, yielding the corrected distributions that are used as input for the cross-section calculation.
The statistical uncertainty of the JSF is included in the measurement as discussed in Section~\ref{sec-syst-jsfstat}.
 
\begin{figure}[htbp]
\centering
\subfloat[]{\includegraphics[width=0.33\linewidth]{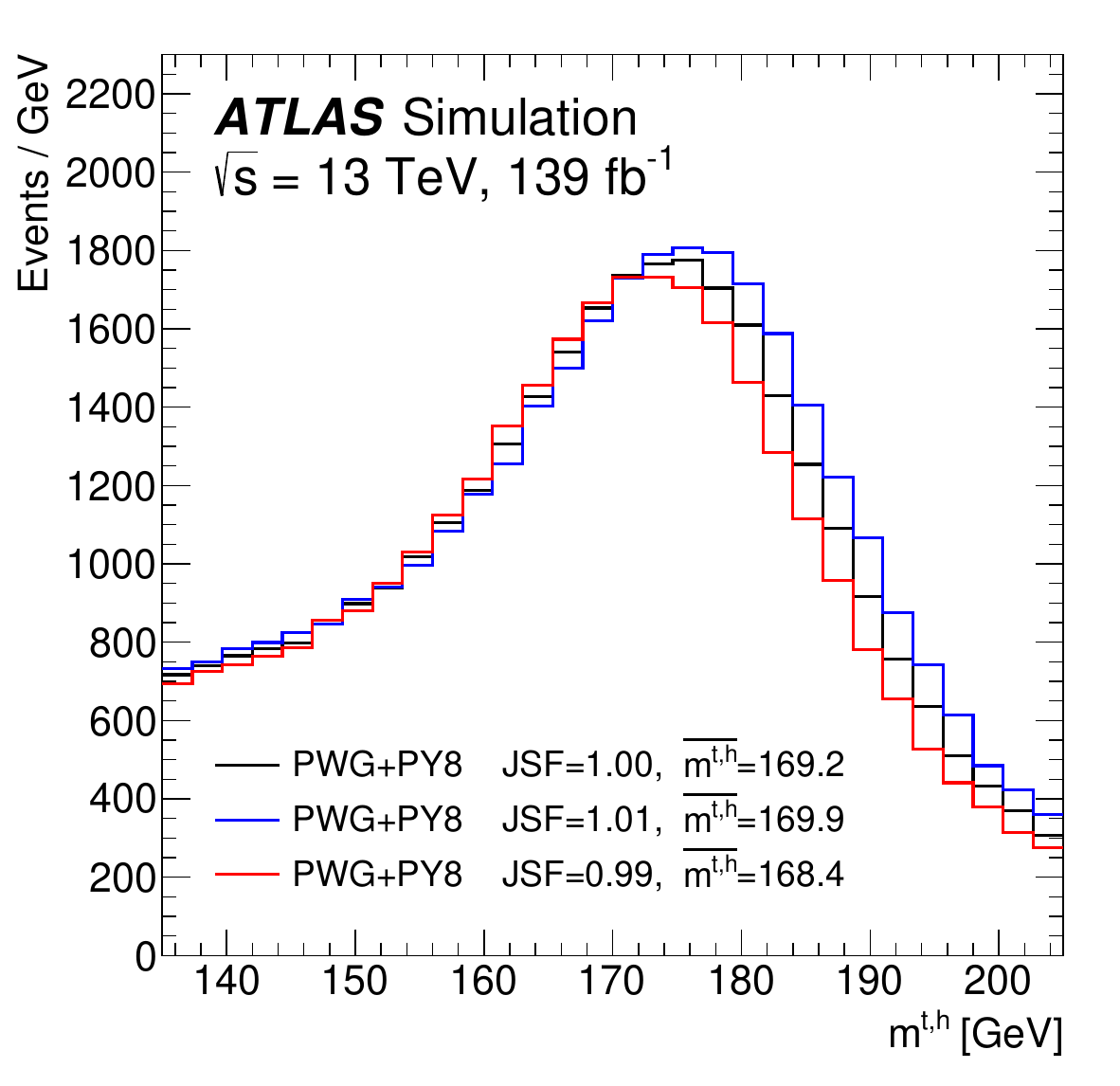}\label{fig:JSFscalingExample:mtopjsf}}
\subfloat[]{\includegraphics[width=0.33\linewidth]{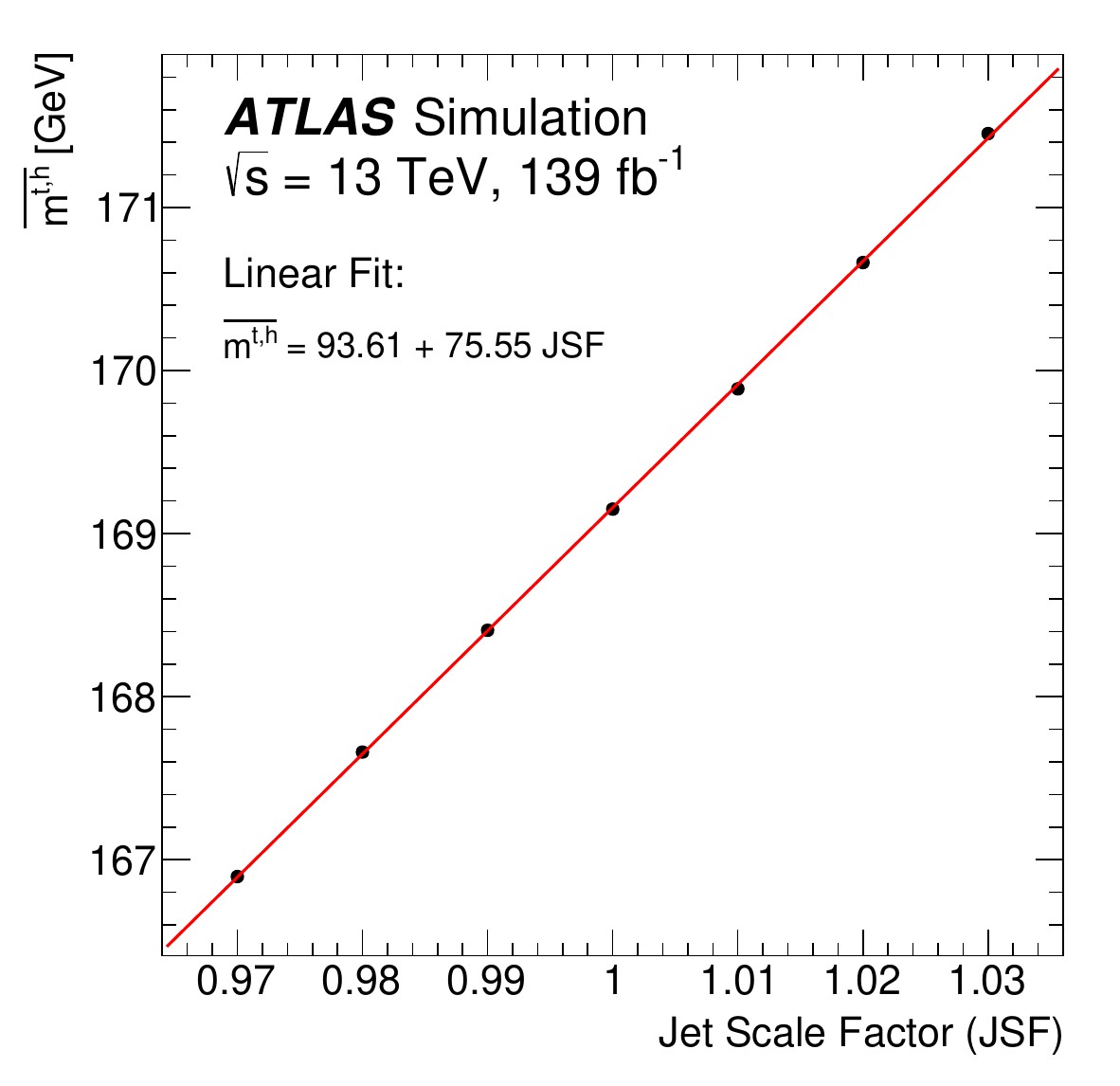}\label{fig:JSFscalingExample:jsfmap}}
\subfloat[]{\includegraphics[width=0.325\linewidth]{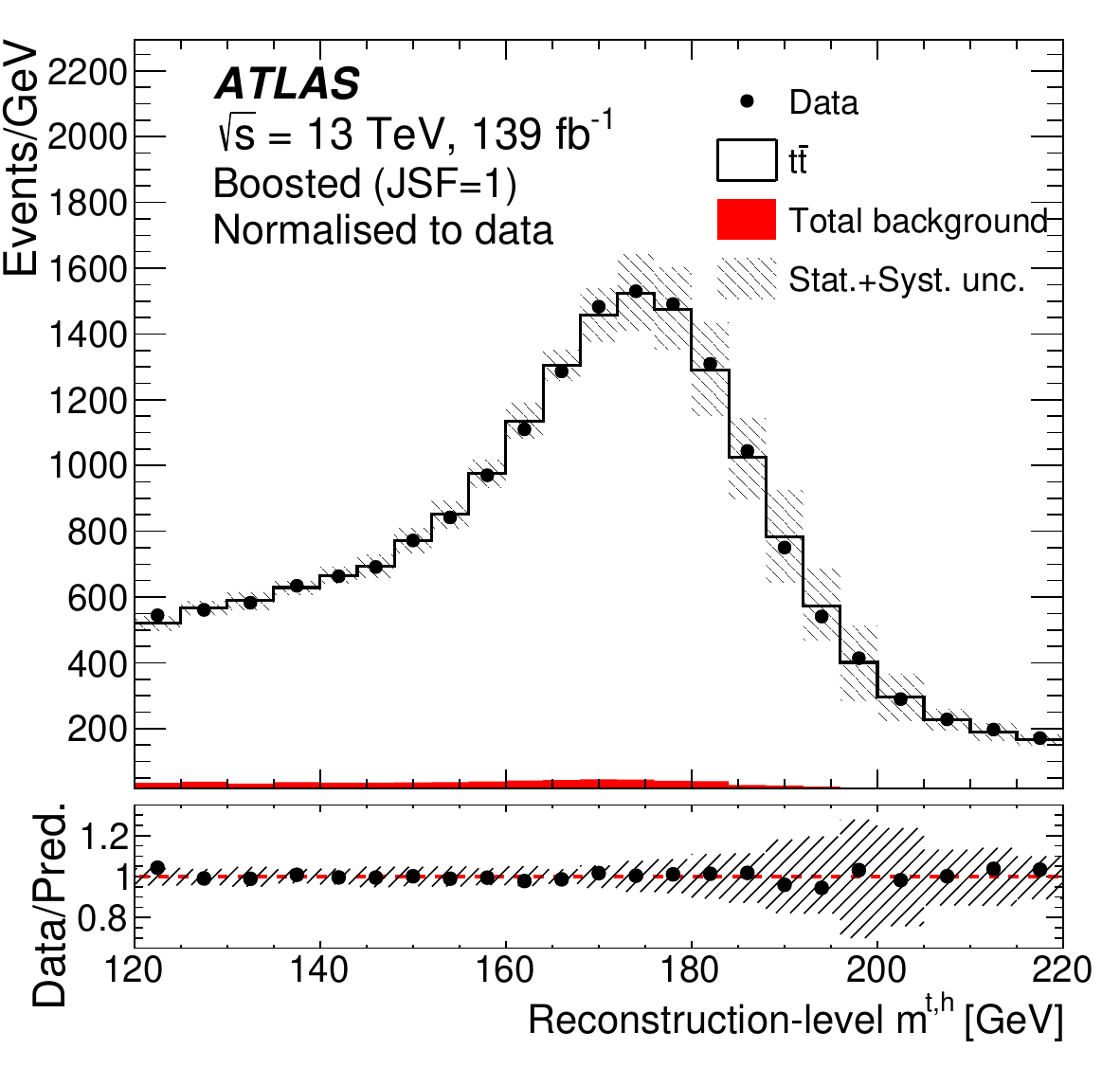}\label{fig:JSFscalingExample:mtopdata}}
\caption{(a) The distribution of the invariant mass of the top-tagged jet ($\mtopjet$) for the expected signal plus background for three example values of the JSF. PWG+PY8 corresponds to the \POWPY sample. (b) The mean of the same distribution ($\meanmtopjet$) as a function of JSF and a linear fit to the simulated samples. Uncertainties in the mean due to limited MC sample size are included for each point, but are too small to be visible.
(c) The $\mtopjet$~distribution observed in the data compared with the expected signal and background processes for $\mathrm{JSF}=1$, where the expectation is normalised to the number
of events observed in data.}
\label{fig:JSFscalingExample}
\end{figure}

\subsection{Fiducial requirements}
\label{sec-xs-fid}
 
The cross-sections are measured at particle level with a set of requirements designed to keep the measured phase-space close
to that selected by the event selection.
Particle-level objects are defined using only stable particles, defined to have a lifetime $> 30$ ps, taken from the MC generator record.
Electrons, muons and neutrinos not originating, either directly or through a $\tau$-lepton decay, from a hadron are referred to as prompt in the following.
The four-momenta of prompt photons within $\Delta R<0.1$~are added to the leptons (`dressing') and the corresponding photons are removed from the list of stable particles.
The dressed leptons, considered in the particle-level selection, are then required to have $\pt > 27$~\GeV\ and $|\eta| < 2.5$.
 
Particle-level \smallR\ jets are obtained by clustering all stable particles with $|\eta| < 4.5$ with the \antikt\ algorithm ($R=0.4$), excluding dressed leptons and prompt neutrinos.
These jets are required to have $\pt > 26$~\GeV\ and $|\eta| < 2.5$. A \smallR\ jet is considered a $b$-jet if it is ghost-matched~\cite{Cacciari:2008gn} to a $b$-hadron with $\pt > 5$~\GeV.
Dressed muons and electrons separated by $\Delta R<0.4$\ from a jet are excluded.
The \largeR\ jets at particle level are built by clustering the selected \smallR\ jets with the \antikt\ algorithm ($R=1.0$), using the same approach as for the data.
To remain consistent with the detector-level selection, the \largeR\ jets are required to have $\pt > 355$~\GeV, $|\eta|<2.0$, an invariant mass between 120~\GeV\ and 220~\GeV,
and one of the \smallR\ jets used to build the \largeR\ jet must be a $b$-jet. If more than one \largeR\ jet passes the selection the one with higher $\pt$ is considered the hadronically decaying top-quark candidate.
The particle-level $\MET$\ is the magnitude of the vector sum of the transverse momenta of all the neutrinos.
The reconstruction of the leptonically decaying top-quark candidate is analogous to that in data, using the lepton, $b$-jet and neutrino (see Section~\ref{sec-evsel}).
The constraint on the \Wboson\ boson mass is used to calculate the neutrino $p_z$\ in the same way as in data.
The particle-level events must pass selection requirements which are the same as in the data selection, but applied to the particle-level objects.
 
\subsection{Corrections for detector effects}
\label{sec-xs-corr}
 
Top-quark pair events that fail the fiducial selection requirements detailed in Section~\ref{sec-xs-fid}~can still pass the event selection requirements.
For example, the top-quark \largeR\ jet can fail the $\pt$~requirement at particle level, but due to the limited detector resolution can be measured
above the requirement at detector level. These effects are corrected for with the factor $f_{\mathrm{acc}}^j$. An example of the distribution
of this factor for $\pThad$~is shown in Figure~\ref{fig:corrections:acc}. The corrections are largest at low $\pt$~due to the resolution effects.
The iterative unfolding uses as input the migration matrix $M$~which encodes the relationship between the particle-level and detector-level distributions.
For the measurements performed as a function of two observables, the matrix spans the full two-dimensional space, such that all migrations are accounted for.
The binning of the migration matrix is chosen such that the majority of the events lie on the diagonal and there are sufficient simulated events in each bin
to achieve stable estimates of the systematic uncertainties.
The migration matrix for $\pThad$~is shown in Figure~\ref{fig:corrections:mig}~and good correspondence between particle level
and detector level is observed.
The number of iterations used in the unfolding is determined by minimising the average correlation factor as defined in Ref.~\cite{Schmitt:2016orm}.
For a limited number of variables, the number of iterations is adjusted to ensure good closure of the method as discussed in the following subsection,
resulting in either two, three or four iterations being used.
The final correction, $f_{\mathrm{eff}}^i$, is for events that pass the particle-level selection but fail the detector-level selection.
This efficiency, shown in Figure~\ref{fig:corrections:eff}~for $\pThad$, is around $0.3$, with the largest effect being due to the efficiency of the $b$-tagging, which is
approximately $77\%$~per $b$-jet. The shape of $f_{\mathrm{eff}}^i$\ originates from the fact that the selection efficiency depends on the kinematics of the events.
For example, the efficiency of the $b$-tagging is highest for jets with $\pT \sim 100$~\GeV\ and is lower at higher and lower $\pT$\ values.
 
\begin{figure}[htbp]
\centering
\subfloat[]{\includegraphics[width=0.33\linewidth]{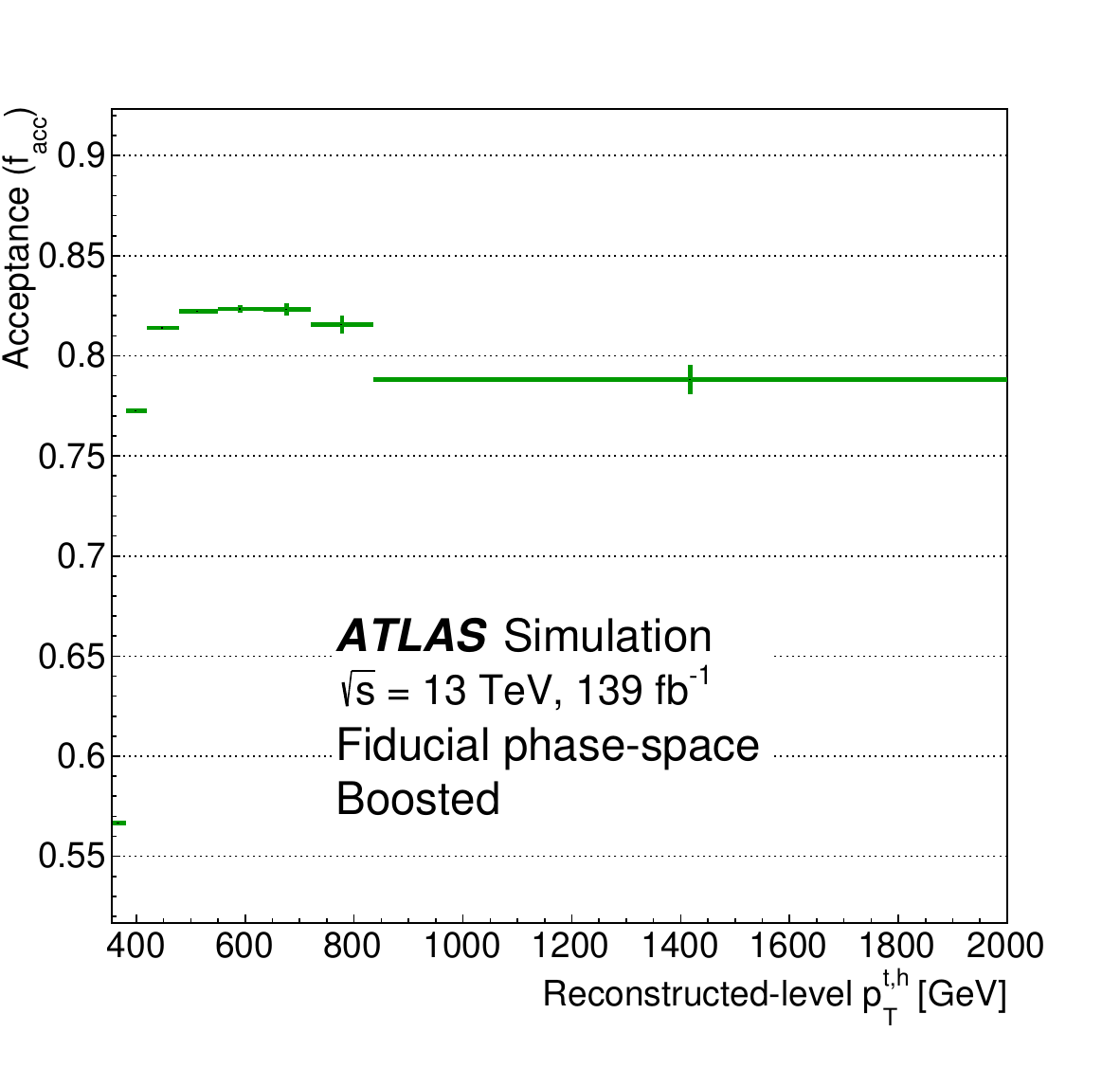}\label{fig:corrections:acc}}
\subfloat[]{\includegraphics[width=0.33\linewidth]{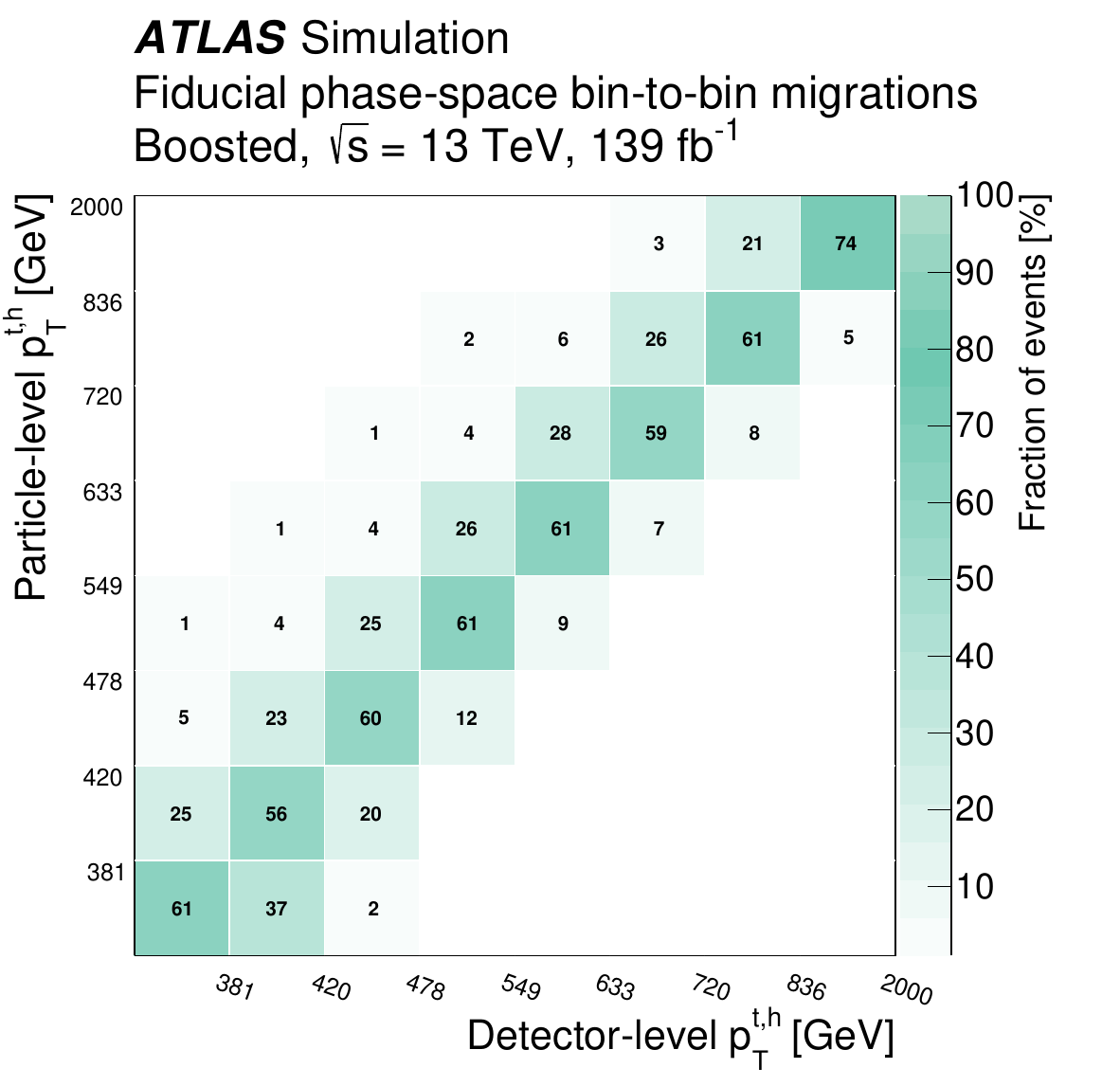}\label{fig:corrections:mig}}
\subfloat[]{\includegraphics[width=0.33\linewidth]{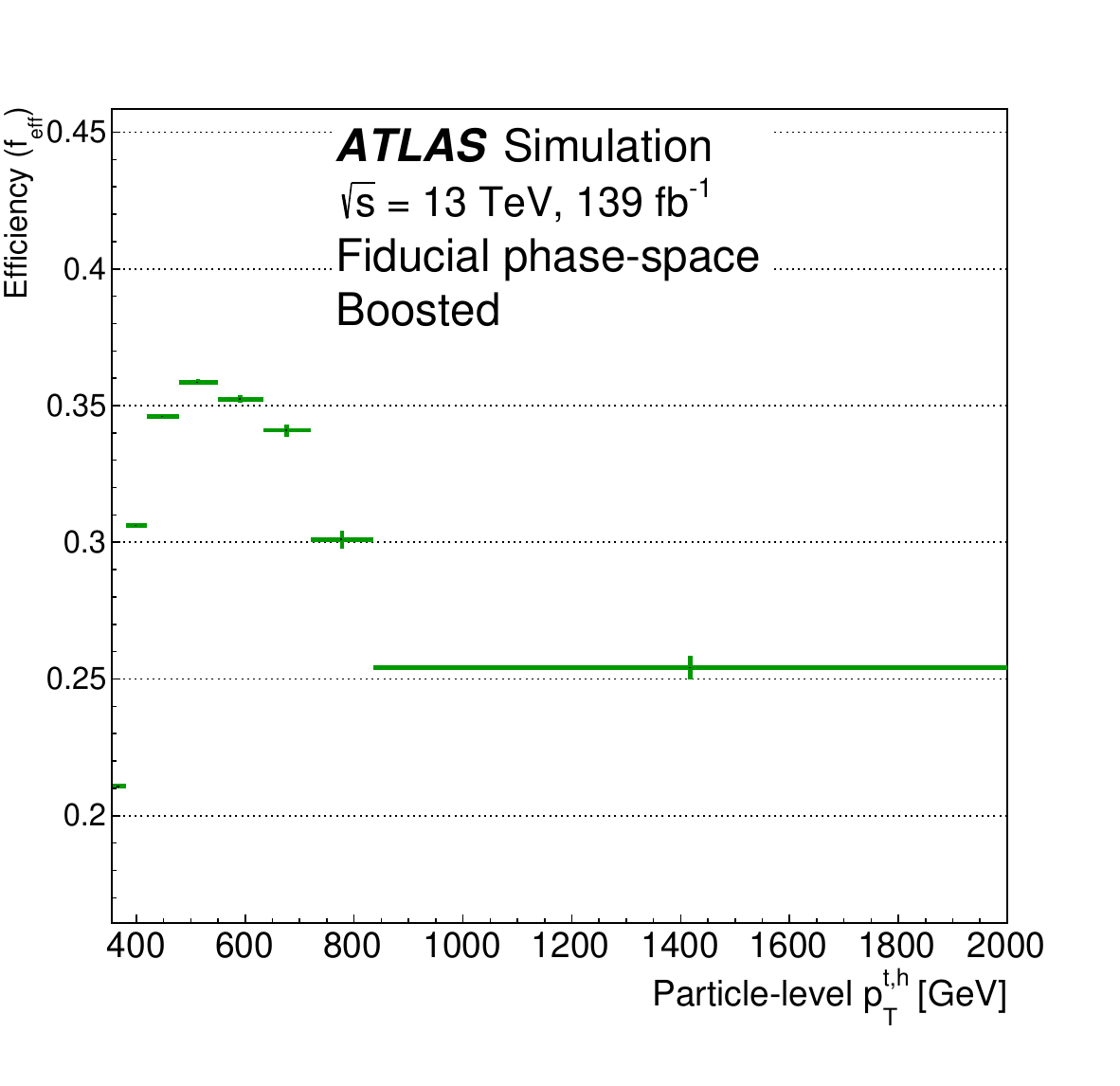}\label{fig:corrections:eff}}
\caption{The acceptance and efficiency corrections for $\pThad$~are shown in (a) and (c). The migration matrix for $\pThad$~is shown in (b).
For the migration matrix, the numerical value of the bin is displayed if it is $>1\%$.}
\label{fig:corrections}
\end{figure}
 
\subsection{Validation of the measurement technique}
\label{sec-xs-val}
 
Tests are performed to ensure the measurement is unbiased within the expected statistical uncertainty.
In order to check for biases due to the iterative procedure, the simulated samples are divided into two equally sized samples. One sample
is used as pseudo-data and the other sample is used to derive the correction factors and migration matrix. Good closure is observed
between the unfolded pseudo-data and the corresponding particle-level distributions.
In order to check that there is no significant bias originating from the choice of model used to derive the corrections applied to the data,
pseudo-data are created with distorted differential distributions and injected into the measurement procedure.
To test that the unfolding is not sensitive to the differences seen between the data and the nominal expectation,
modified distributions are derived by reweighting the nominal $\ttbar$\ MC sample as a function of particle-level observables such that the reweighted detector-level distributions are similar to the observed data.
To test that the unfolding would not be biased if new physics is present in the data, modified distributions are also created by adding EFT contributions
to the nominal $\ttbar$\ MC sample.
Three separate tests are performed. In the first two tests, the MC samples that include terms up to $\Lambda^{-2}$\ are rescaled to produce pseudo-data equivalent
to $(\ctg, \ctq)=(0.25, 0)$~and $(\ctg, \ctq)=(0, 0.25)$.
The third test uses pseudo-data derived from the MC sample including terms up to $\Lambda^{-4}$ with $(\ctg, \ctq)=(0.2, 0.2)$.
Figure~\ref{fig:biastest} shows the result of these tests for $\pThad$.
The figures display the ratio of the unfolded pseudo-data to the corresponding particle-level prediction. In each case, the ratio is found to be compatible
with unity, demonstrating that there is no significant bias from the measurement technique.
The size of the difference between the nominal MC sample and the pseudo-data at particle level is also shown in the figure by the dotted lines.
The sample with $(\ctg, \ctq)=(0.2, 0.2)$\ is also used to verify that the extracted \JSF\ does not change significantly
from the value extracted when using the nominal SM MC sample.
The possible biases originating from systematic uncertainties are discussed in Section~\ref{sec-syst}.
 
\begin{figure}[htbp]
\centering
\subfloat[]{\includegraphics[width=0.44\linewidth]{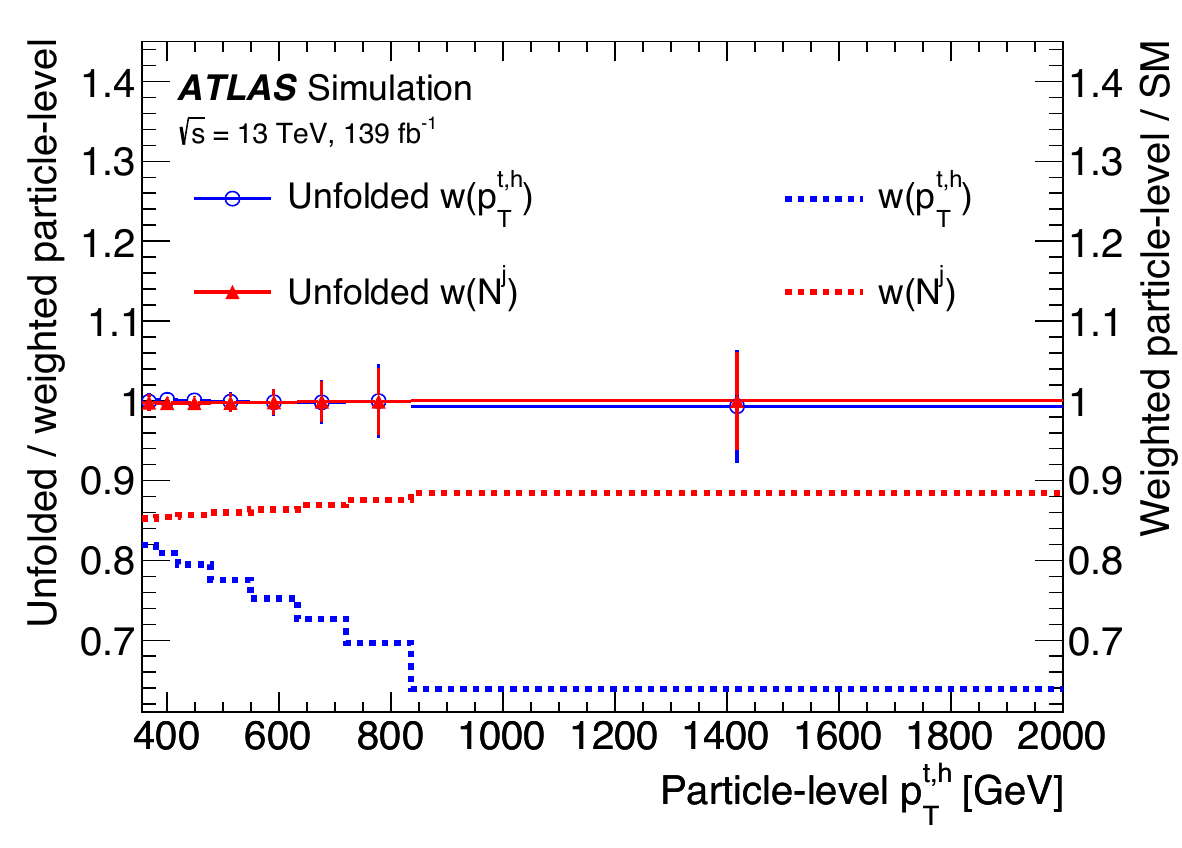}}
\hspace{0.2cm}
\subfloat[]{\includegraphics[width=0.44\linewidth]{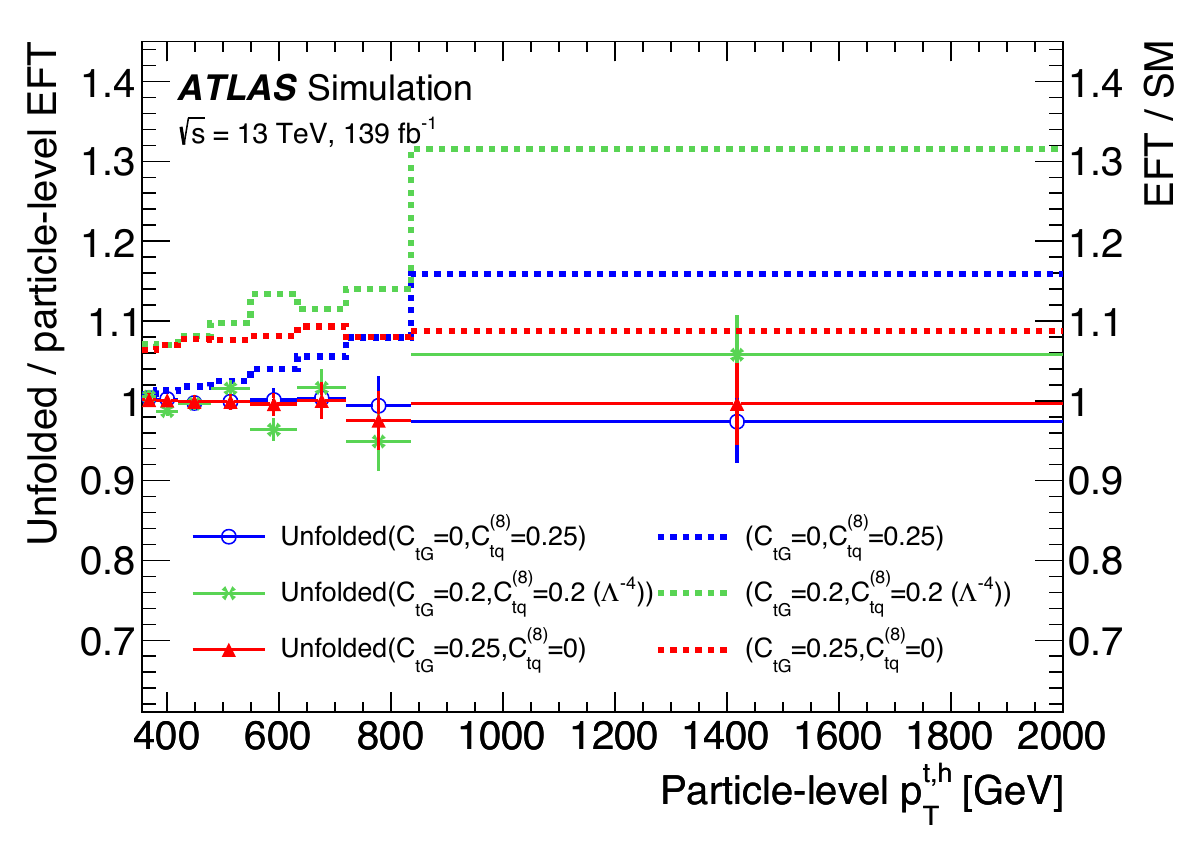}}
\caption{
The ratio of unfolded pseudo-data to the corresponding particle-level prediction as a function of $\pThad$~is shown in (a)
for pseudo-data built by reweighting the MC sample to match the data in the $\pThad$ (solid blue) and $\Ne$ (solid red) distributions.
The equivalent ratio is shown in (b) for pseudo-data built from the nominal
MC sample and adding contributions corresponding to $\ctg=0.25$~(blue), $\ctq=0.25$~(red, including only $\Lambda^{-2}$ contributions) and $(\ctg, \ctq=0.2, 0.2)$\ (green, including $\Lambda^{-2}$\ and  $\Lambda^{-4}$\ contributions).
The dotted lines show in (a) the ratio of the weighted particle-level distribution to the particle-level distribution from the nominal $\ttbar$\ MC sample and in
(b) the ratio of the particle-level distribution of the EFT to the nominal $\ttbar$\ sample. The $y$-axis on the right of each plot refers to these dotted lines.
}
\label{fig:biastest}
\end{figure}
 
\FloatBarrier
 
\subsection{Choice of observables}
\label{sec-meas-obs}
 
The measured observables can be split into two categories, those characterising the kinematics of the
top quarks and those probing additional radiation in the events. In the first category are:
\begin{itemize}
\item $\pThad$, $\pTlep$:
the transverse momenta
of the hadronically and leptonically decaying top quarks,
\item $\yhad$, $\ylep$:
the rapidity of the top quarks,
\item $\yttbar$: the rapidity of the $\ttbar$~system,
\item $\mttbar$: the invariant mass of the $\ttbar$ system,
\item $\dPhibt$: the azimuthal separation between the hadronically decaying top quark and the $b$-tagged jet associated with the leptonically decaying top quark,
\item $\HTttbar$: the scalar sum of the transverse momenta of the top quarks.
\end{itemize}
The $\HTttbar$, $m^{\ttbar}$~and top-quark $\pt$\ distributions are expected to be sensitive to new-physics effects.
The rapidity distributions can potentially probe the parton distribution functions~\cite{Czakon_2017}.
These variables were measured in previous publications~\cite{TOPQ-2018-15,CMS-TOP-18-013} but
less precisely than in this analysis.
In the second category are a set of variables
that seek to probe the additional radiation in detail:
\begin{itemize}
\item $\pt^{\ttbar}$: the transverse momentum of the $\ttbar$~system,
\item $\Ne$: the number of additional jets,
\item $\pTe{1}$, $\pTe{2}$: transverse momenta of the two leading additional jets,
\item $\dPhieth{1}$, $\dPhieth{2}$: azimuthal angles between the two leading additional jets and the hadronically decaying top quark,
\item $\dPhiee$: azimuthal angle between the two leading additional jets,
\item $\dPhitt$: azimuthal angle between the two top quarks,
\item $\meth$: invariant mass of the leading additional jet and the hadronically decaying top quark,
\item $\HTall$: scalar sum of the transverse momenta of the top quarks and the additional jets.
\end{itemize}
The variables involving one or two additional jets use the subsets of selected events that contain at least one or two additional jets.
These eight variables were not measured previously for highly boosted $\ttbar$~events. Finally, a set of two-dimensional cross-section
measurements that characterise the additional radiation as a function of $\pThad$~and $\Ne$ are presented:
\begin{itemize}
\item $\pTe{1}$ in bins of $\pThad$,
\item $\pTe{1}$ in bins of $\Ne$,
\item $\dPhieth{1}$ in bins of $\pThad$,
\item $\dPhieth{1}$ in bins of $\Ne$.
\end{itemize}
 
\subsection{Observed data distributions}
 
A total of 75\,743 events are observed after the \JSF\ correction and data selection. The
data are compared with the expectation from the simulation and background estimates
in Table~\ref{table:Background_yields}.
The uncertainties shown in the table are the combined effect of the statistical, detector systematic, and background modelling uncertainties (described in Section~\ref{sec-syst}). These uncertainties are shown because they do not change the predicted cross-section and hence directly impact the measured results.
The selected sample is expected to be dominated by $\ttbar$~events
and have a purity $>95\%$. The largest background contribution is from $tW$~single top-quark production.
The data yield is observed to be $14\%$~lower than the expectation from the nominal signal and background predictions, which is consistent
with the trend observed in the previous ATLAS measurement,
where the cross-section at high top-quark $\pT$~was found to be smaller than the expectation from \POWPY~\cite{TOPQ-2018-15}.
Figures~\ref{fig:dataMC1}--\ref{fig:dataMC3} show the data distributions for the selected observables at reconstruction level.
In order to visualise the level of agreement in the shape of the distributions, the sum of the $\ttbar$\ and expected background events
is normalised to the data yield in these figures.
The simulation is found to give a reasonable
description of the data; however, in some distributions trends can be observed in the ratio between data and expectation.
In particular, the $\pThad$~and $\pTlep$~distributions are seen to be softer
in data than predicted, and the shape of $\dPhitt$ differs between data and simulation. As discussed in Section~\ref{sec-xs-val},
these differences are used to check that the unfolding procedure is robust against potential mismodelling of the expected $\ttbar$~distributions.
No significant biases are observed in these tests.
 
\tableVSpace{1.1}
\begin{table}[htbp]
\center
\caption{Event yields for measured data, simulated \ttbar signal and background events. The uncertainty values are symmetrised and indicate the combined effect of statistical, detector and background modelling uncertainties.}
\label{table:Background_yields}
\begin{tabular}[t]{@{}llllll@{}} \toprule
Process & Expected events \\ \midrule
\ttbar & \blank{0}84\,200 $\pm$ 2600 \\
Single top quark & \blank{00\,}1710 $\pm$ \blank{0}280 \\
\ttV\ (\ttW+\,\ttZ\,+\,\ttH) & \blank{00\,0}850 $\pm$ \blank{0}110 \\
Multijet & \blank{00\,0}560 $\pm$ \blank{0}370 \\
$W+$\,jets & \blank{00\,0}420 $\pm$ \blank{0}160 \\
$Z\,+$\,jets & \blank{00\,00}84 $\pm$ \blank{00}43 \\
Diboson & \blank{00\,00}41 $\pm$ \blank{00}21 \\ \midrule
Total prediction & \blank{0}87\,900 $\pm$ 2700 \\ \midrule
Data & \blank{0}75\,743 \\ \bottomrule
\end{tabular}
\end{table}
 
\FloatBarrier
 
\begin{figure}[htbp]
\centering
\subfloat[]{
\includegraphics[width=0.32\textwidth]{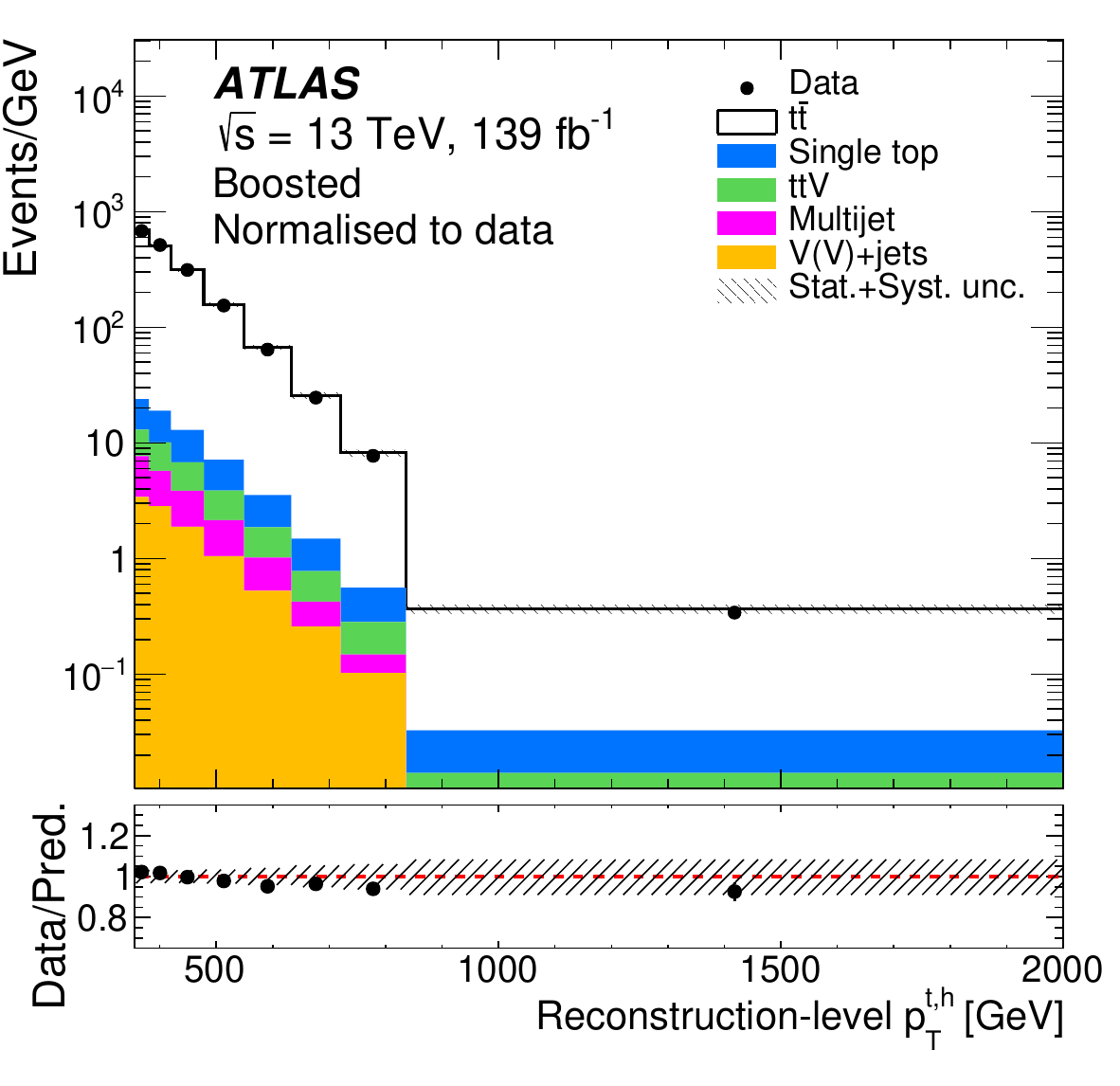}}
\subfloat[]{
\includegraphics[width=0.32\textwidth]{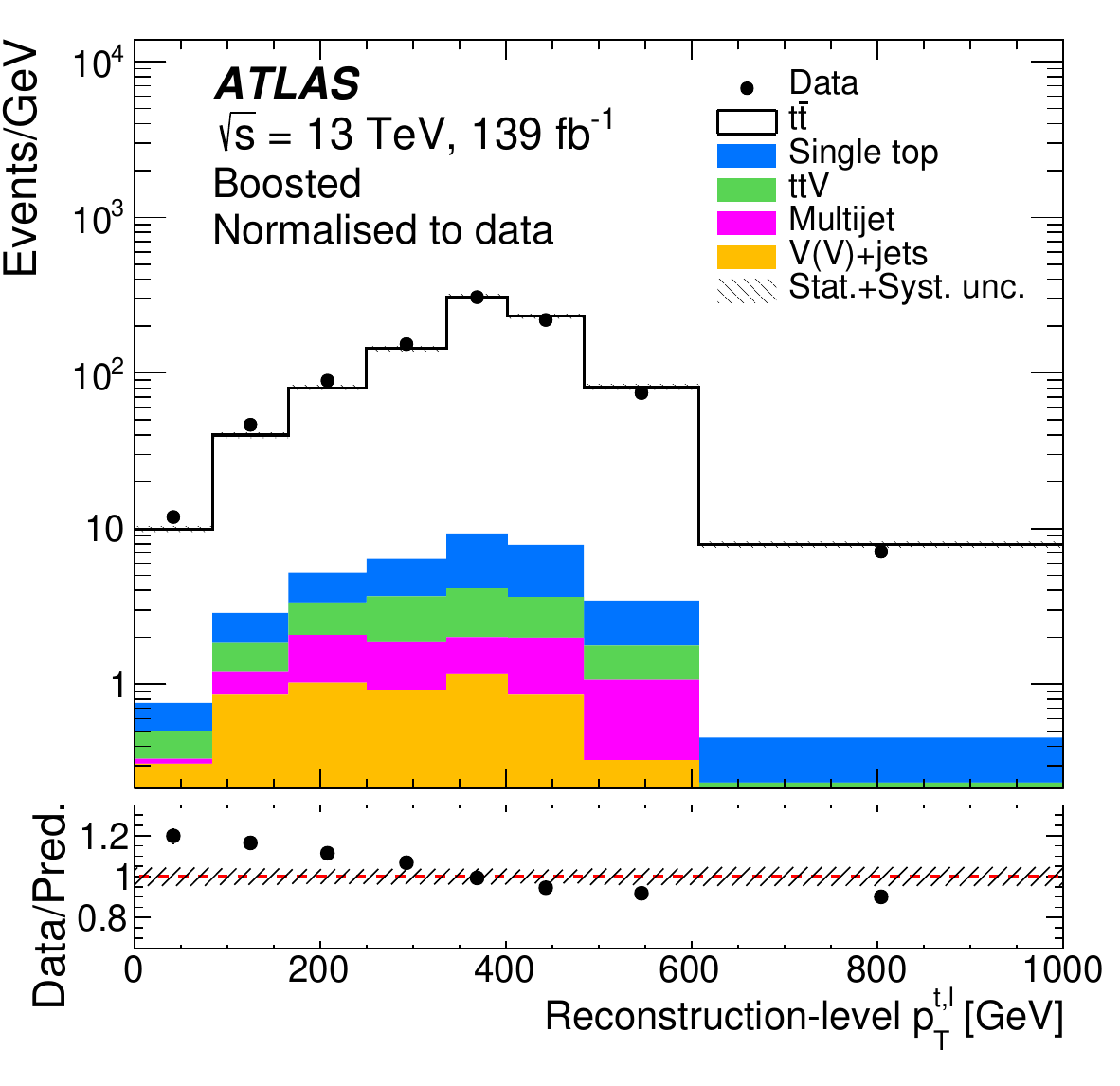}}
\subfloat[]{
\includegraphics[width=0.32\textwidth]{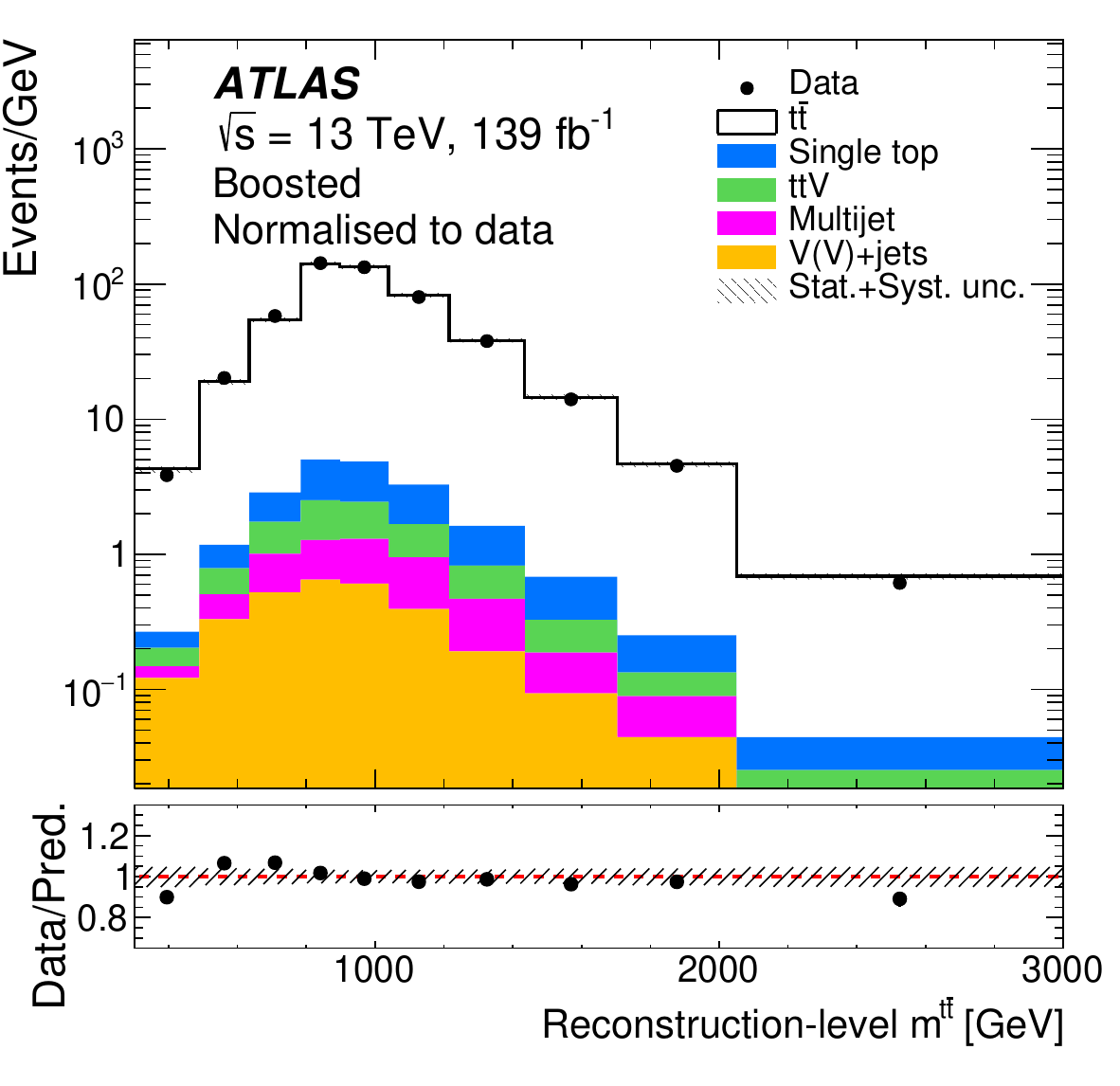}}
 
\subfloat[]{
\includegraphics[width=0.32\textwidth]{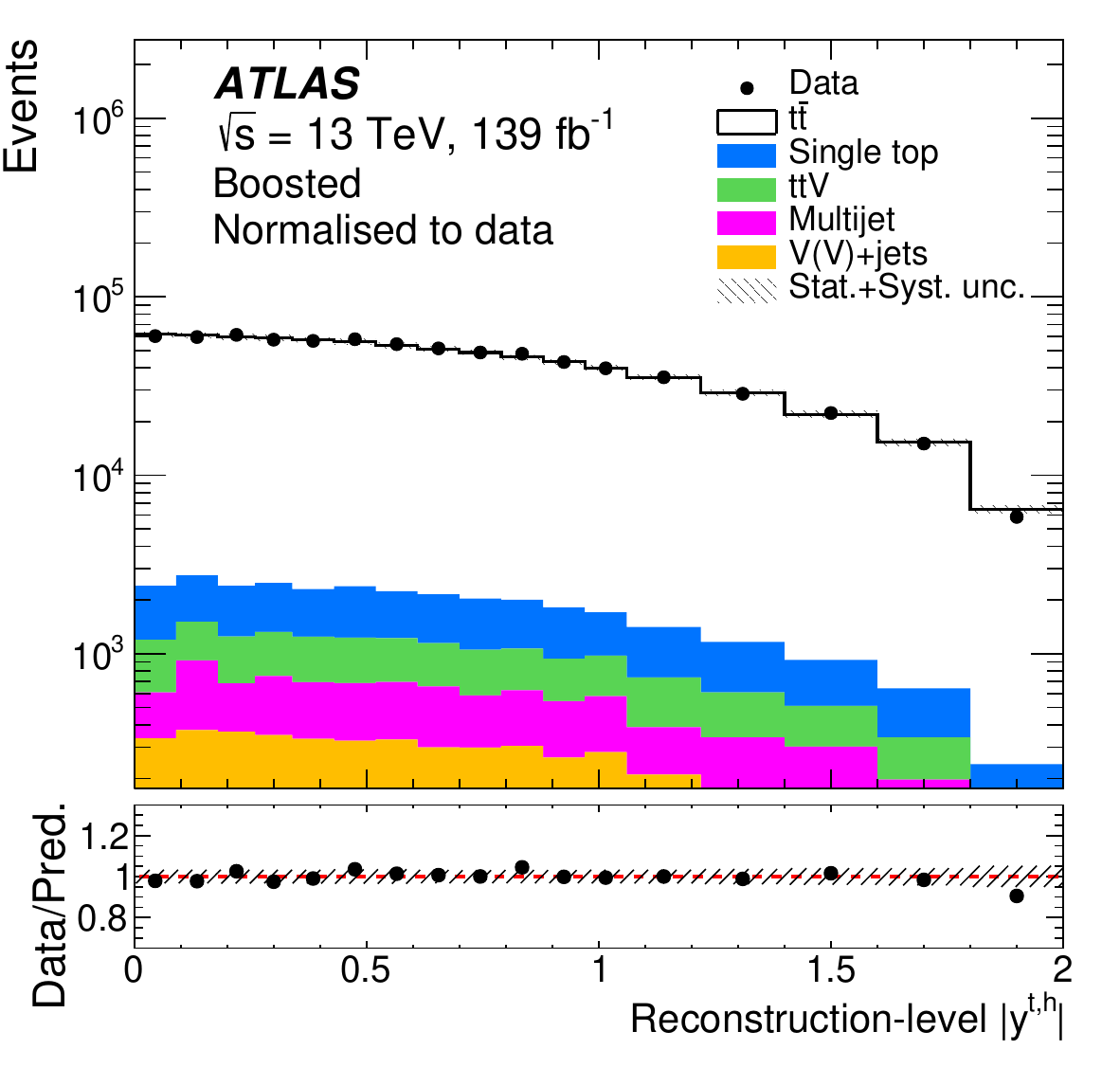}}
\subfloat[]{
\includegraphics[width=0.32\textwidth]{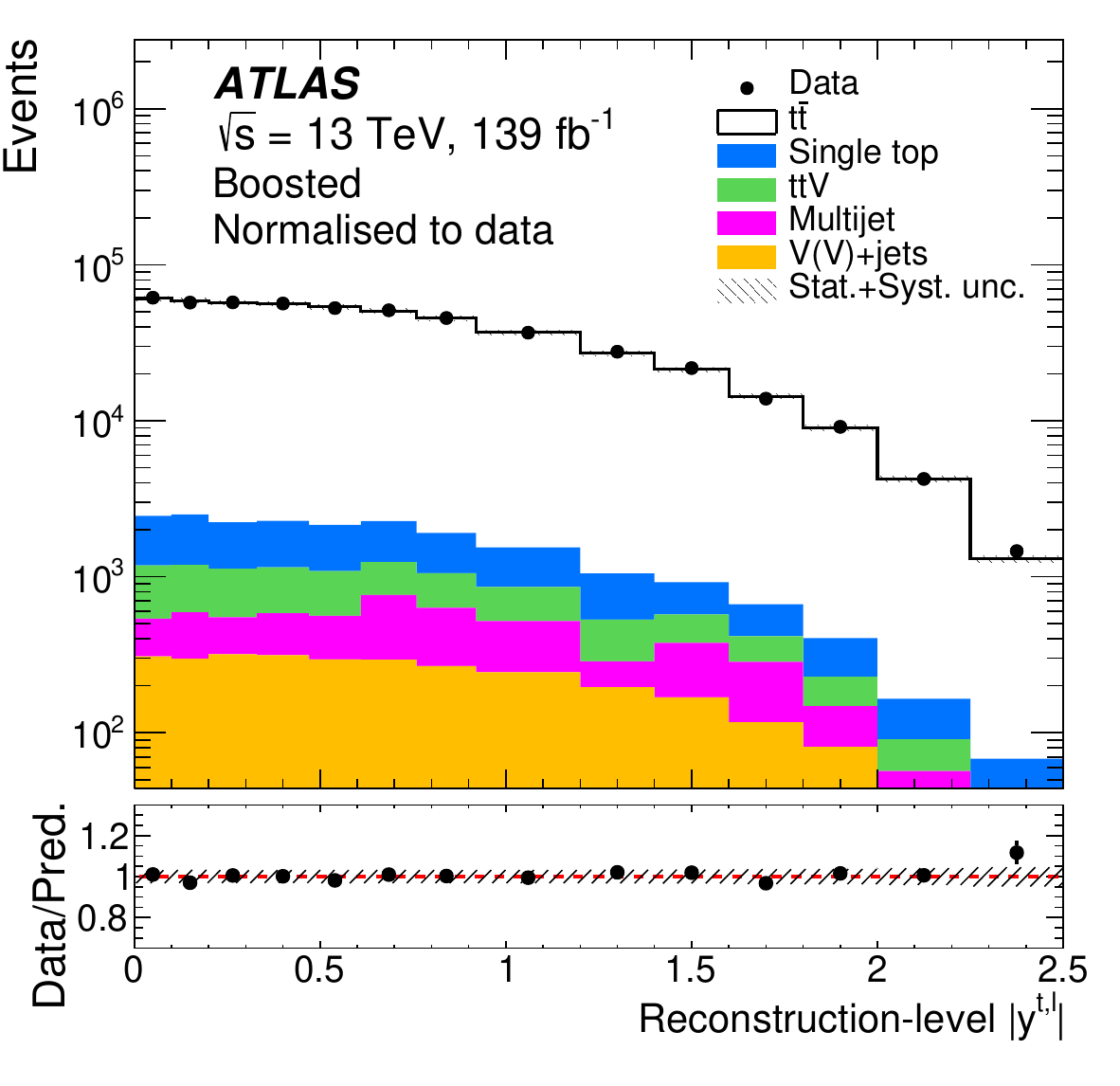}}
\subfloat[]{
\includegraphics[width=0.32\textwidth]{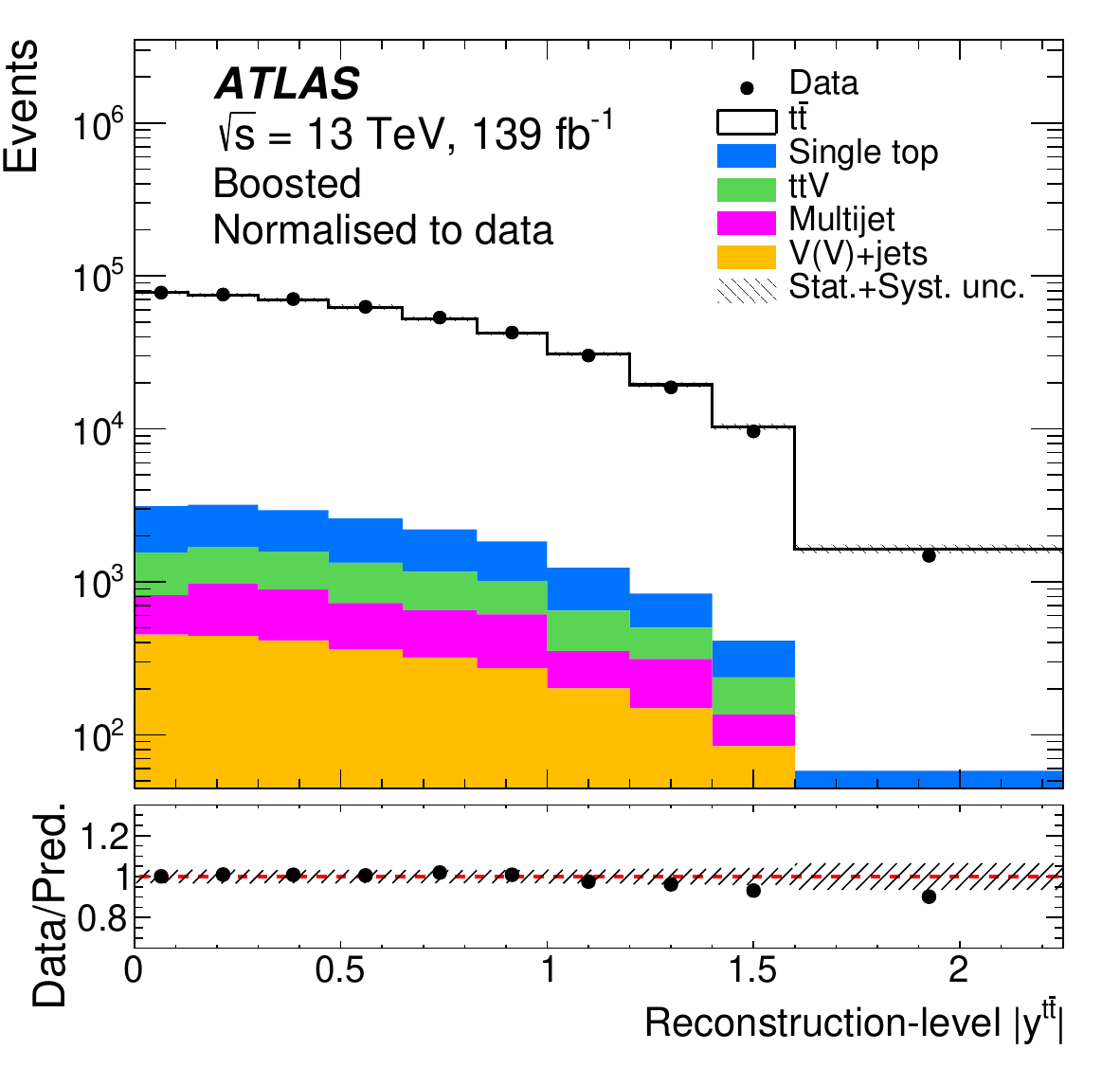}}
 
\subfloat[]{
\includegraphics[width=0.32\textwidth]{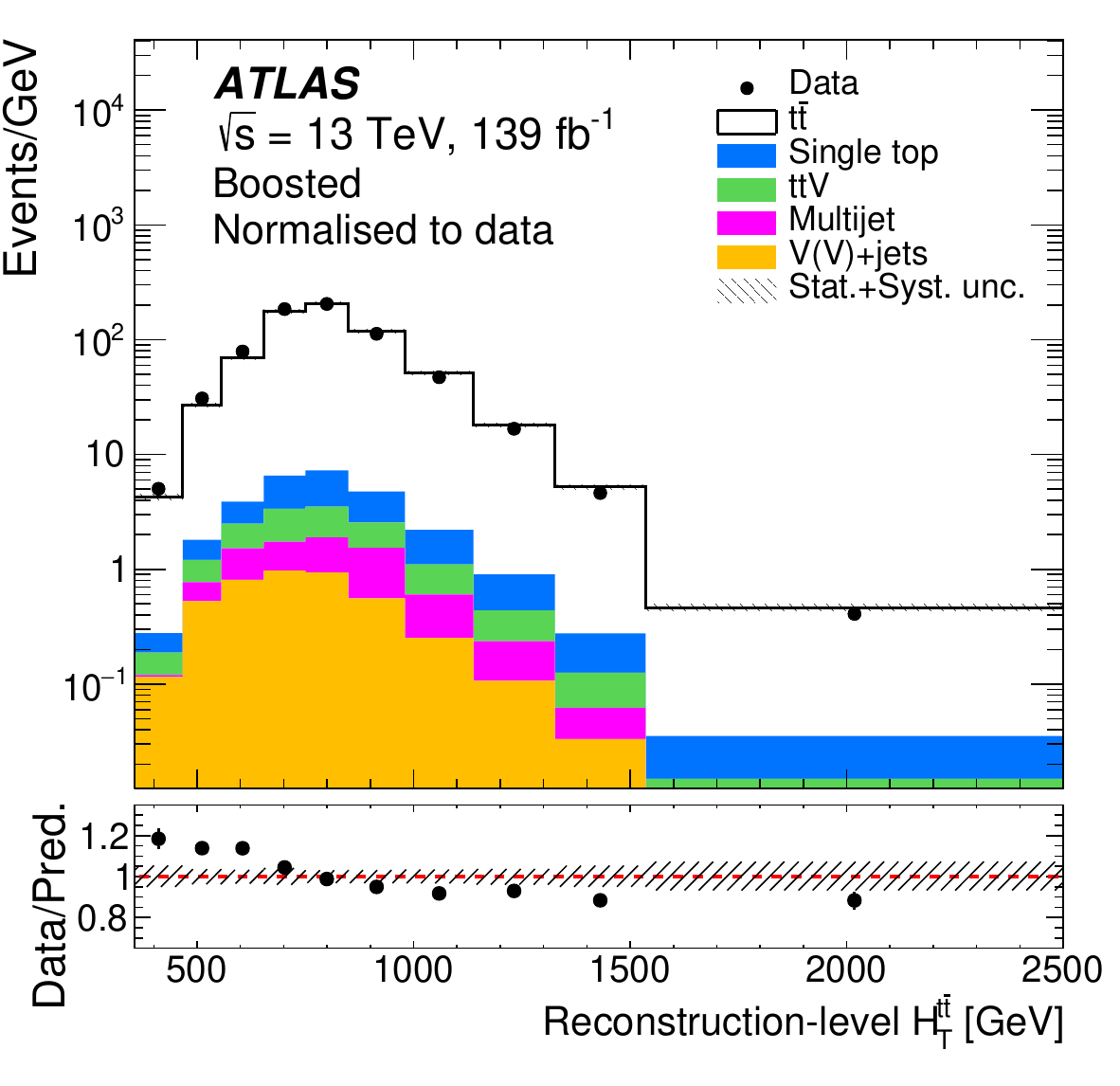}}
\subfloat[]{
\includegraphics[width=0.32\textwidth]{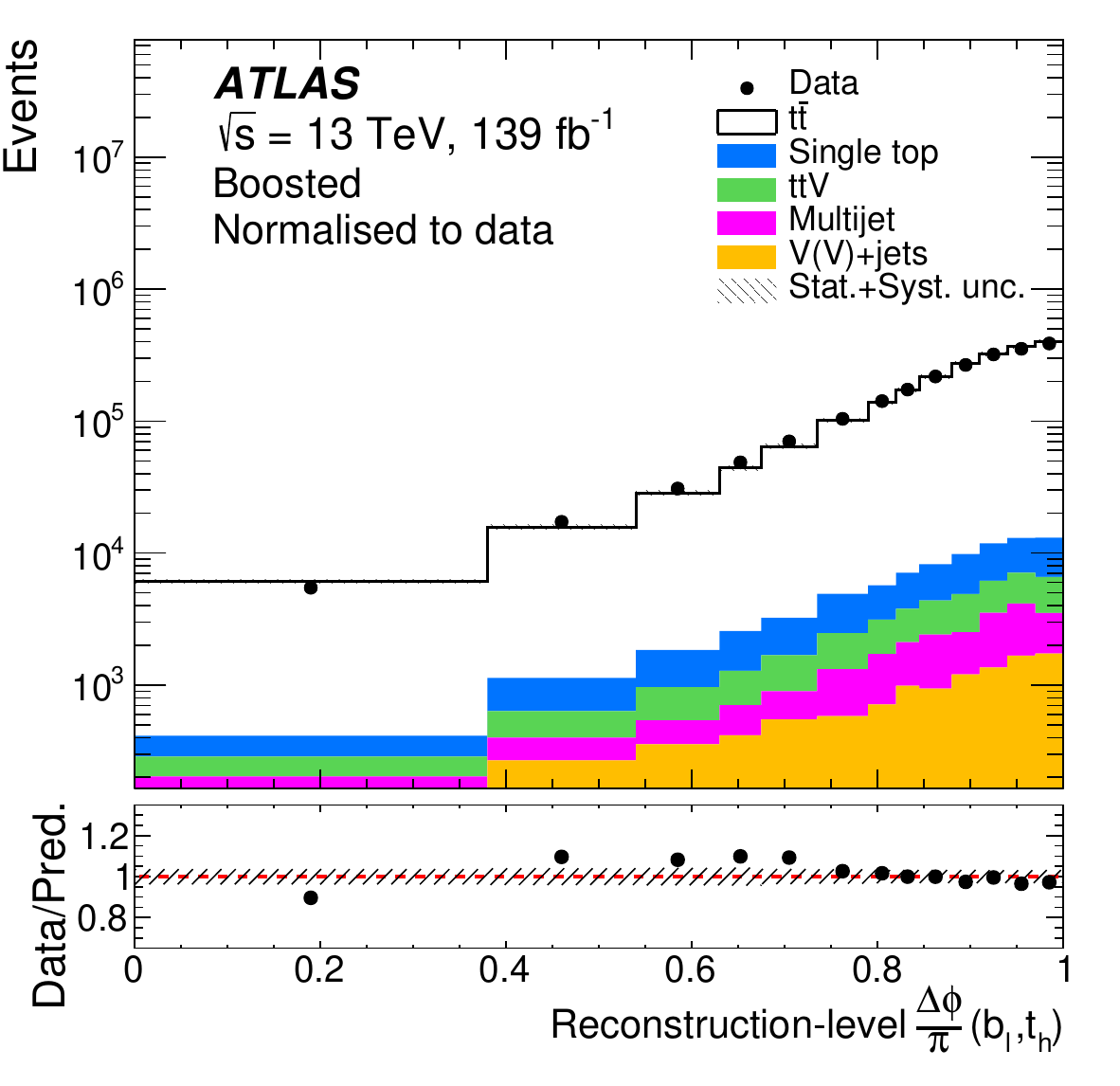}}
\caption{Distributions of the observables sensitive to the kinematics of the top quarks: (a) $\pThad$, (b) $\pTlep$, (c) $\mttbar$, (d) $\yhad$, (e) $\ylep$,
(f) $\yttbar$, (g) $\HTttbar$~and (h) $\dPhibt$.
The data are compared with the expectation from the simulation and
background estimates. The total prediction is normalised to the same number of entries as the data.
The lower panel in each subfigure shows the ratio of the data to the normalised expectation.
The shaded band represents the uncertainties originating from the limited data sample size and systematic uncertainties ($t\bar{t}$\ modelling uncertainties
are not included).}
\label{fig:dataMC1}
\end{figure}
 
\begin{figure}[htbp]
\centering
\subfloat[]{
\includegraphics[width=0.32\textwidth]{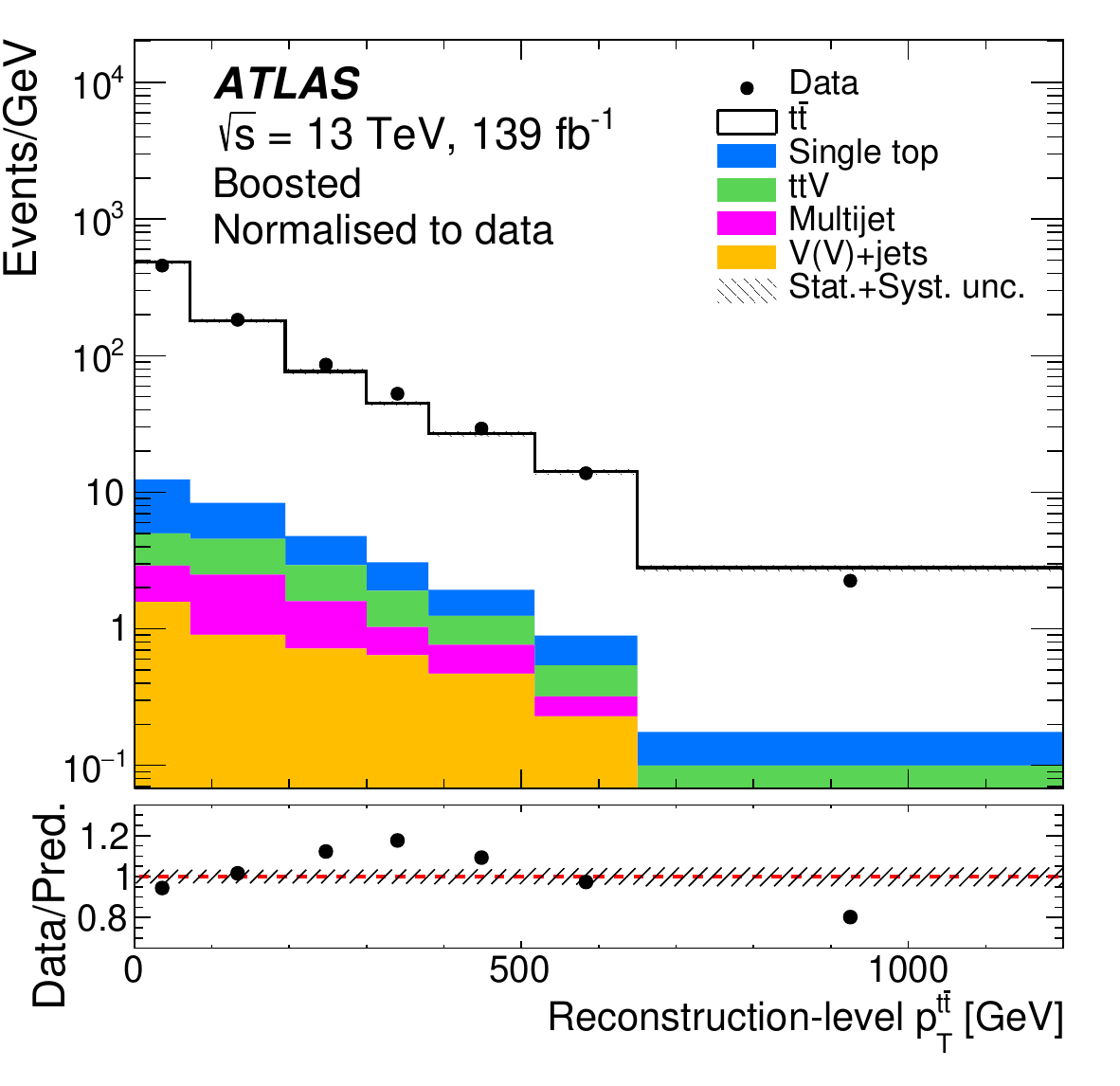}}
\subfloat[]{
\includegraphics[width=0.32\textwidth]{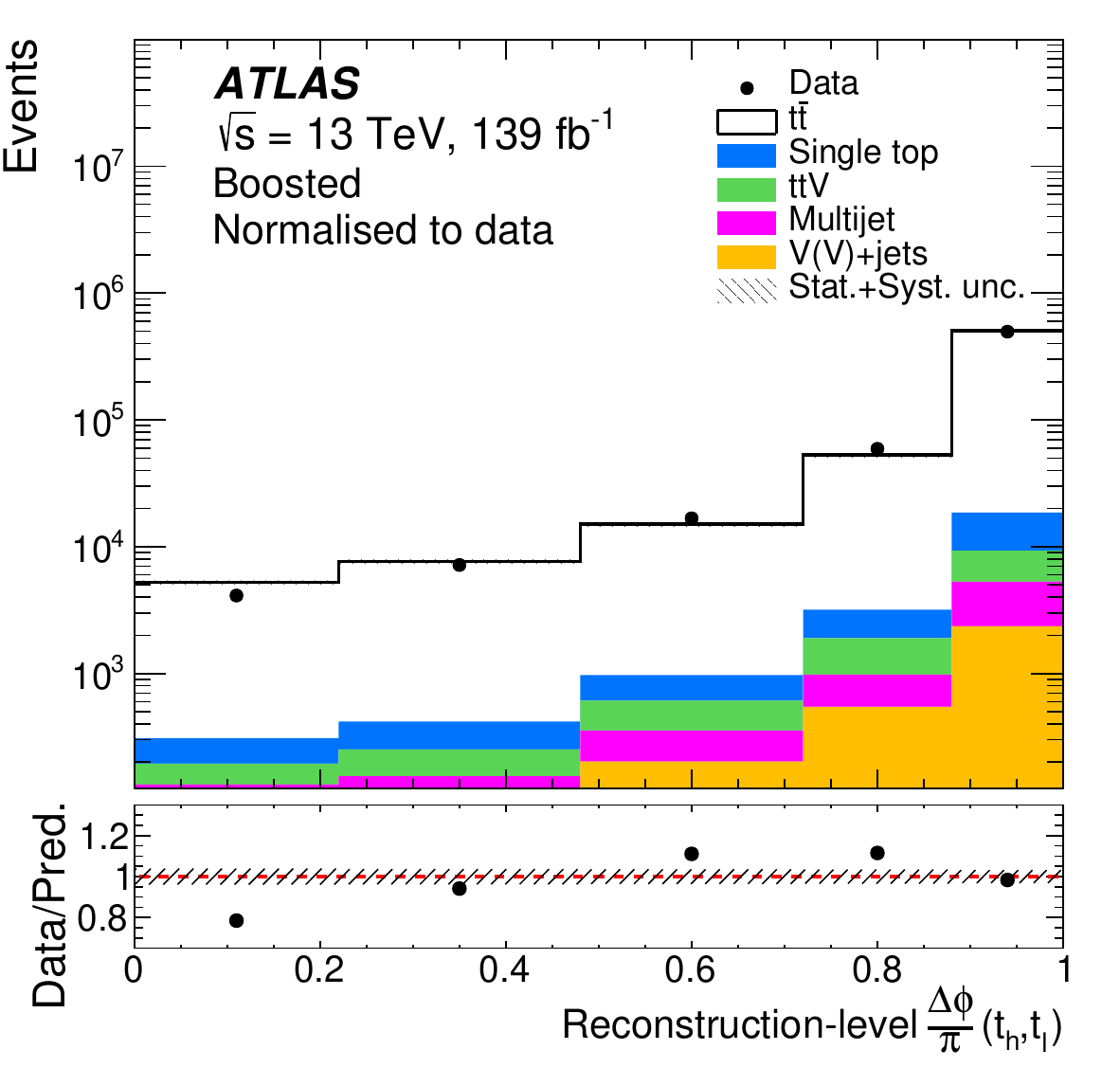}}
\subfloat[]{
\includegraphics[width=0.32\textwidth]{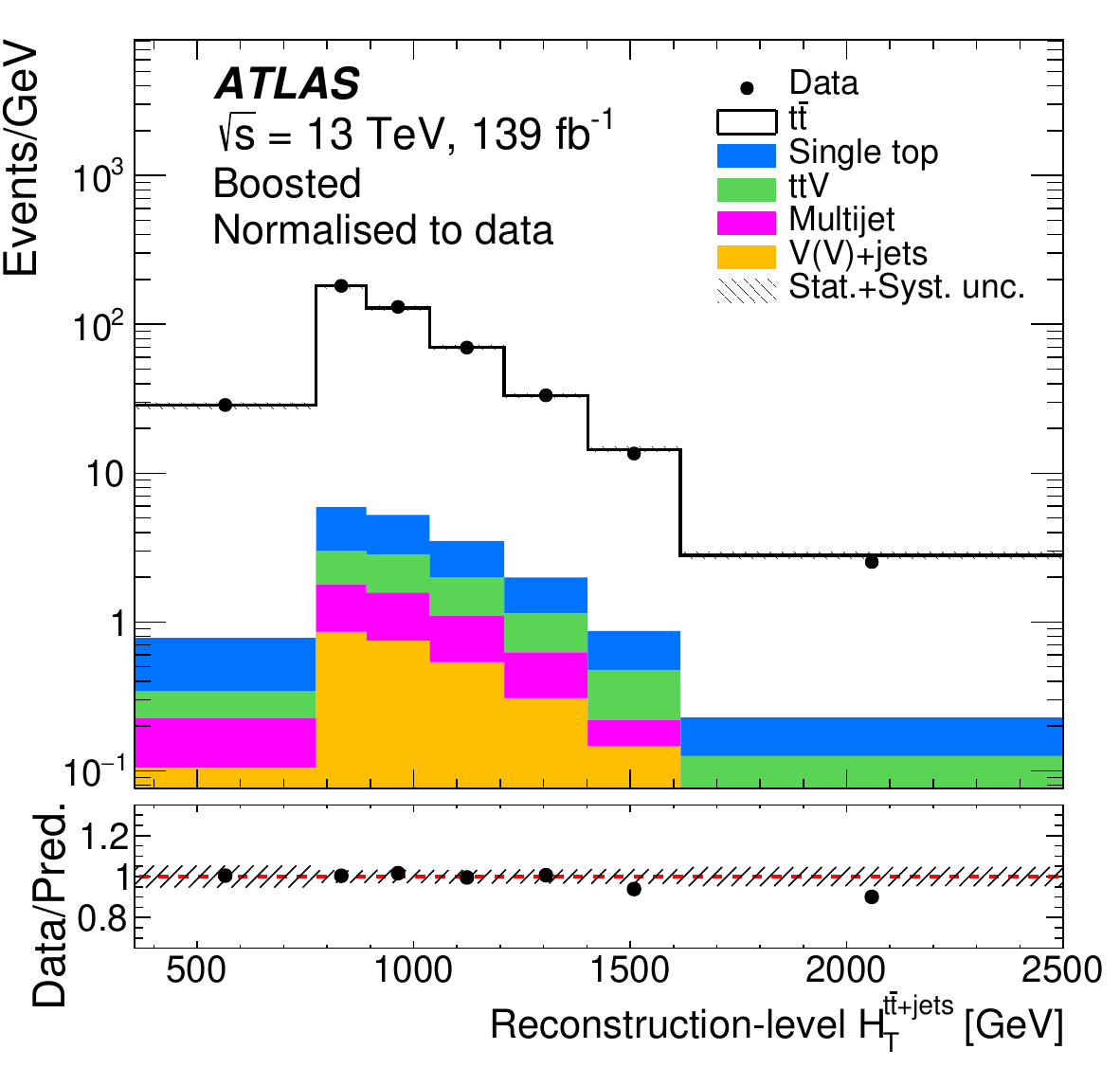}}
 
\subfloat[]{
\includegraphics[width=0.32\textwidth]{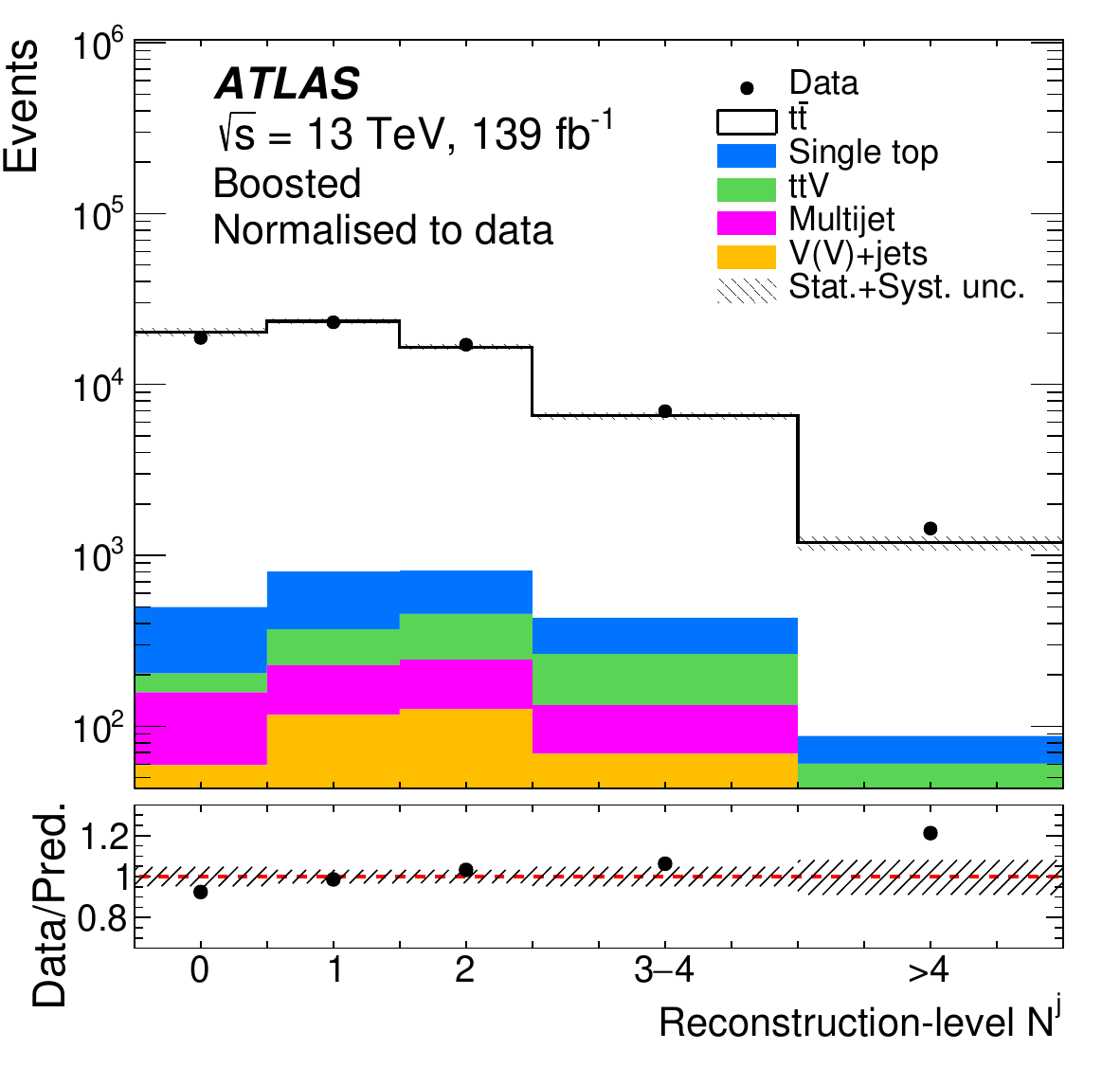}}
\subfloat[]{
\includegraphics[width=0.32\textwidth]{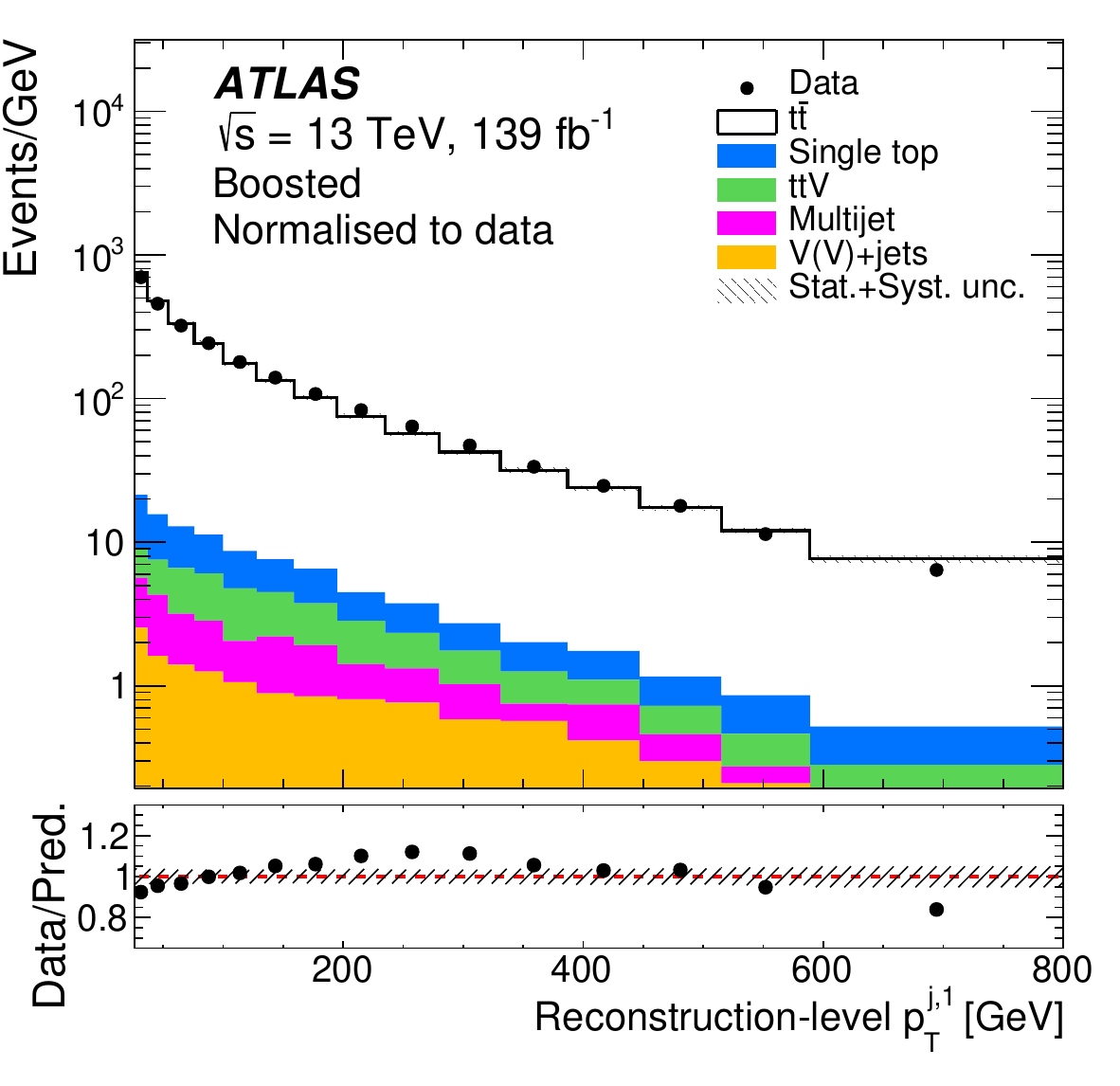}}
\subfloat[]{
\includegraphics[width=0.32\textwidth]{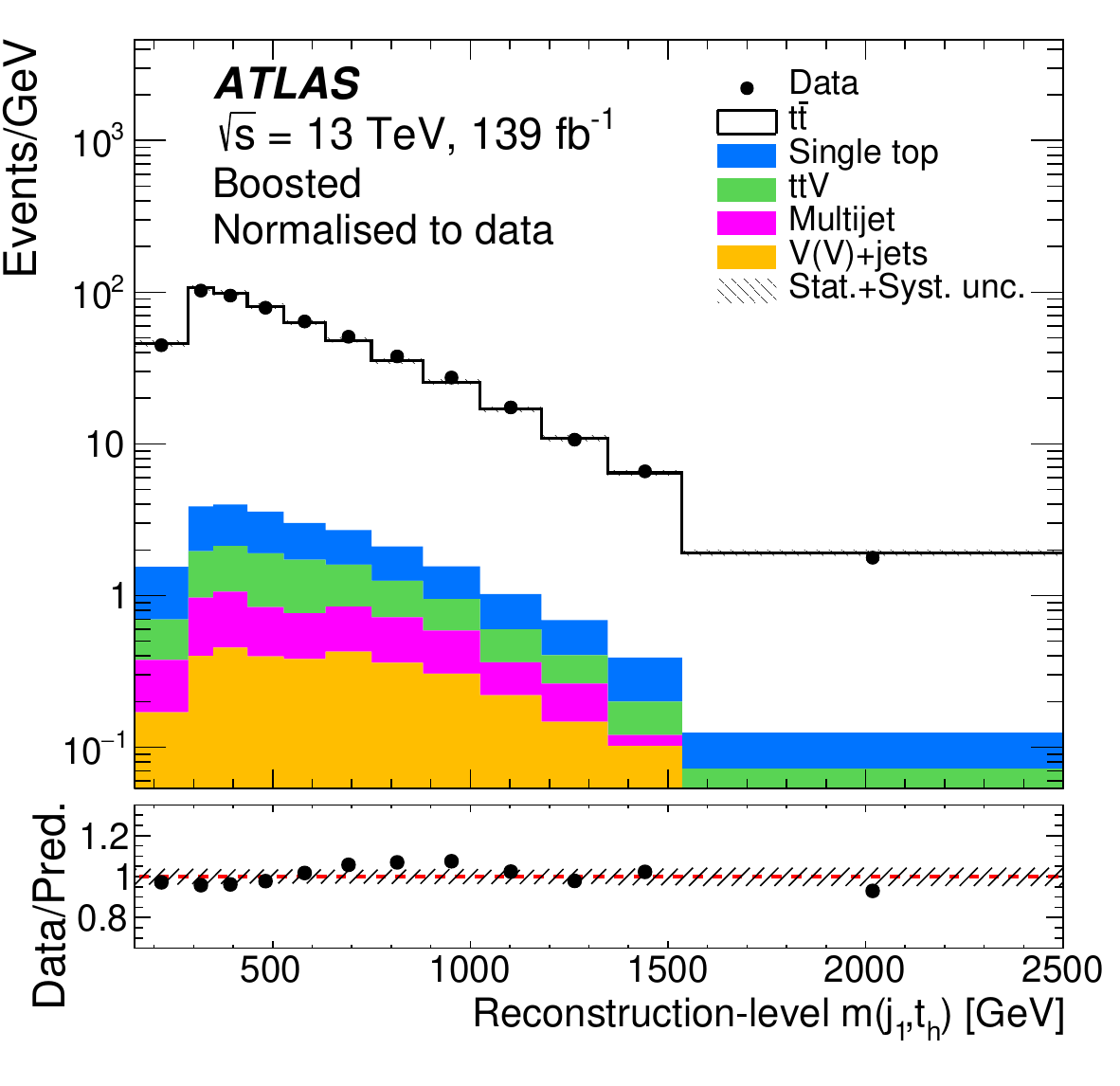}}
 
\subfloat[]{
\includegraphics[width=0.32\textwidth]{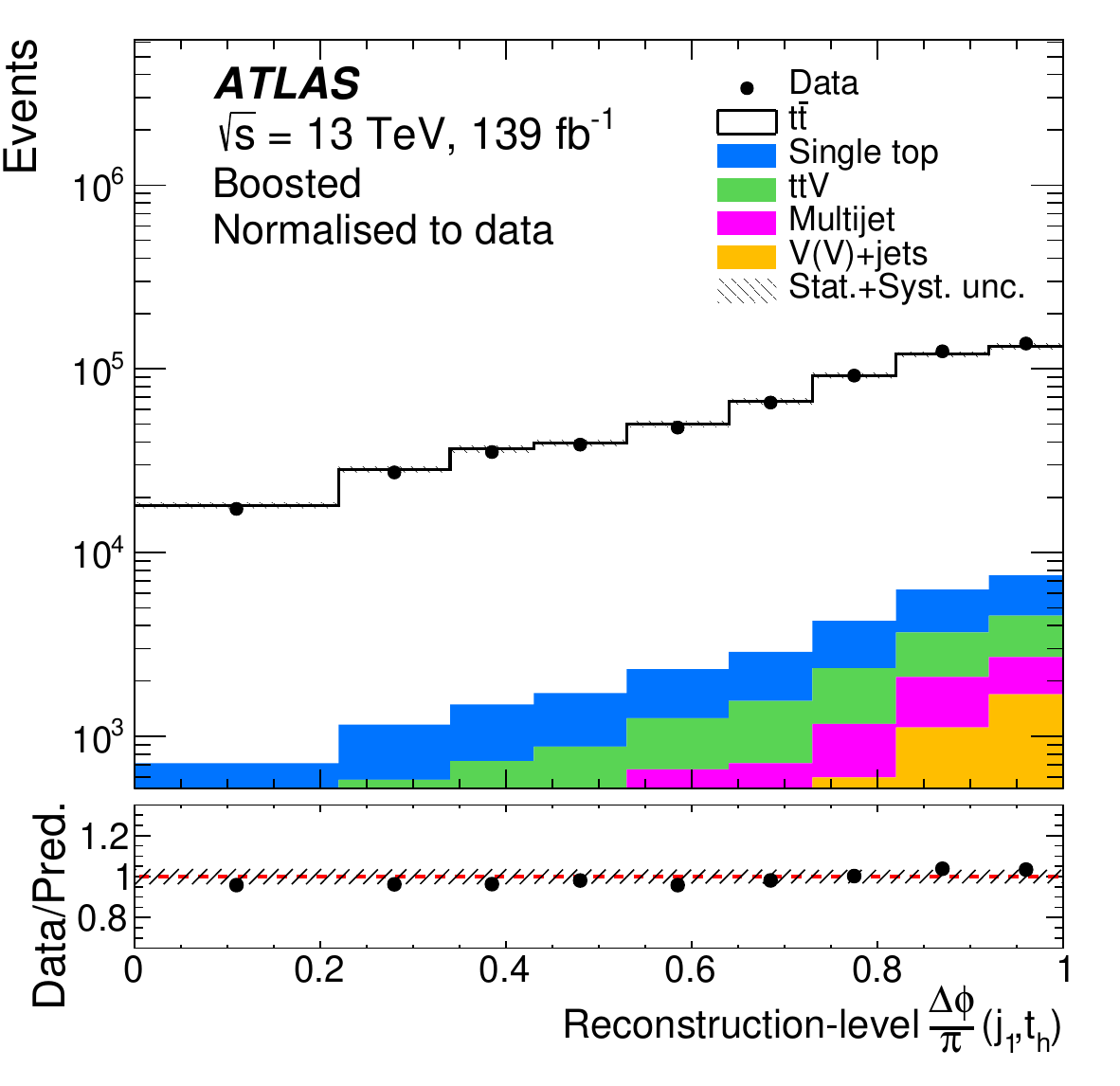}}
\caption{Distributions of observables sensitive to jets produced in association with the $\ttbar$\ system:
(a) $\pTttbar$, (b) $\dPhitt$, (c) $\HTall$, (d) $\Ne$,
(e) $\pTe{1}$, (f) $\meth$ and (g) $\dPhieth{1}$.
The data are compared with the expectation from the simulation and
background estimates. The total prediction is normalised to the same number of entries as the data.
The lower panel in each subfigure shows the ratio of the data to the normalised expectation.
The shaded band represents the uncertainties originating from the limited data sample size and systematic uncertainties ($t\bar{t}$\ modelling uncertainties
are not included).
}
\label{fig:dataMC2}
\end{figure}
 
\begin{figure}[htbp]
\centering
\subfloat[]{
\includegraphics[width=0.32\textwidth]{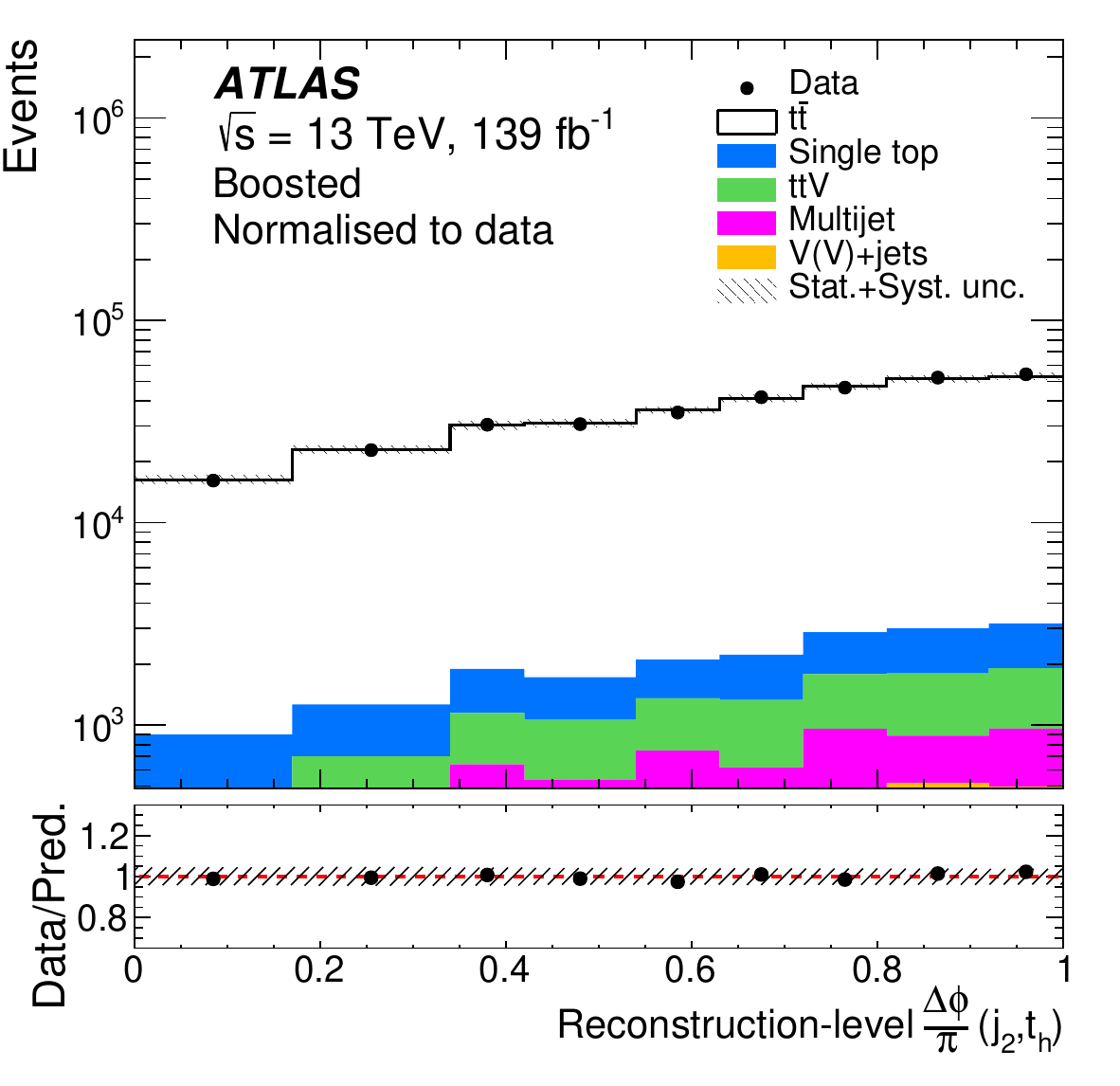}}
\subfloat[]{
\includegraphics[width=0.32\textwidth]{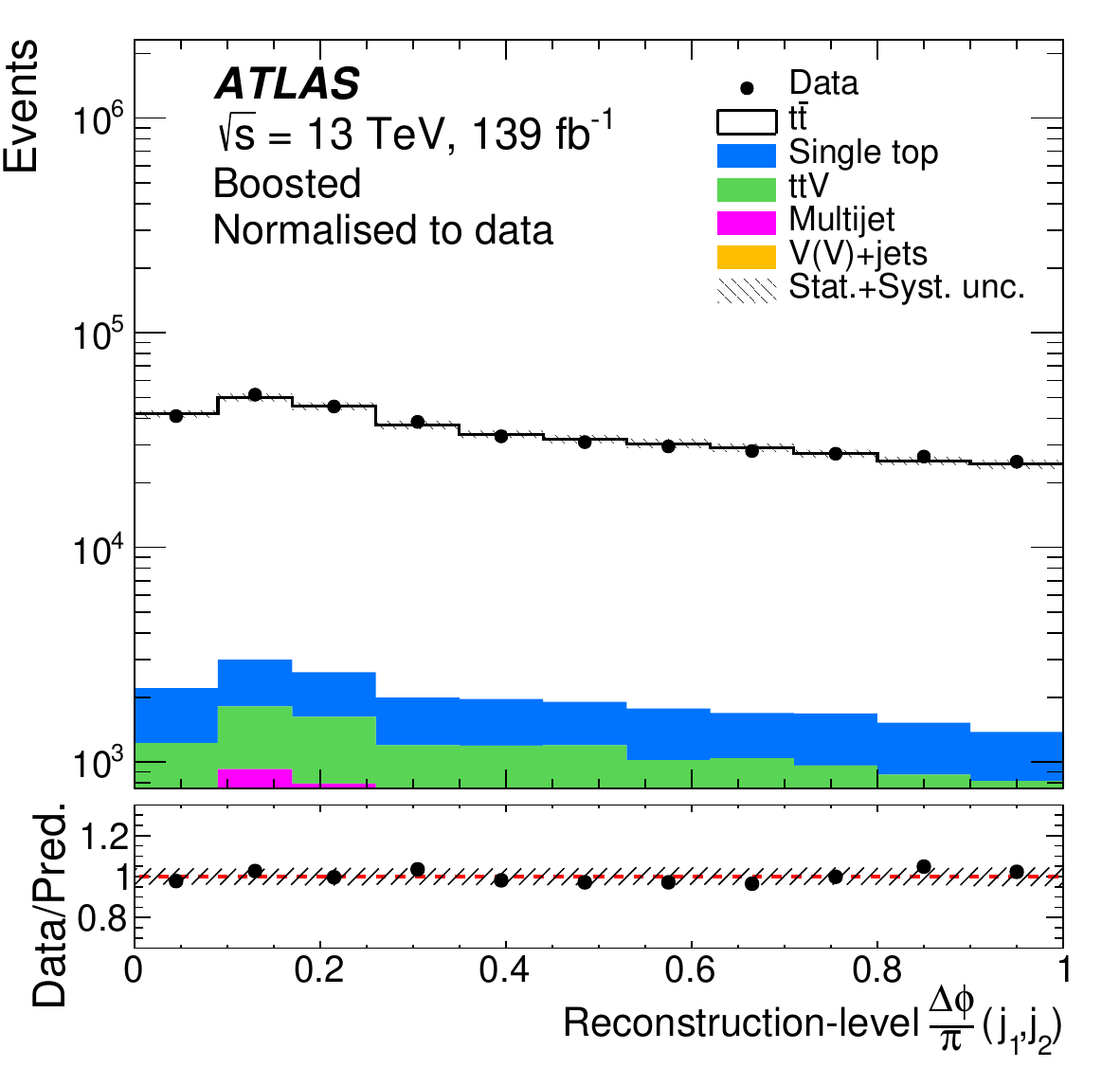}}
\subfloat[]{
\includegraphics[width=0.32\textwidth]{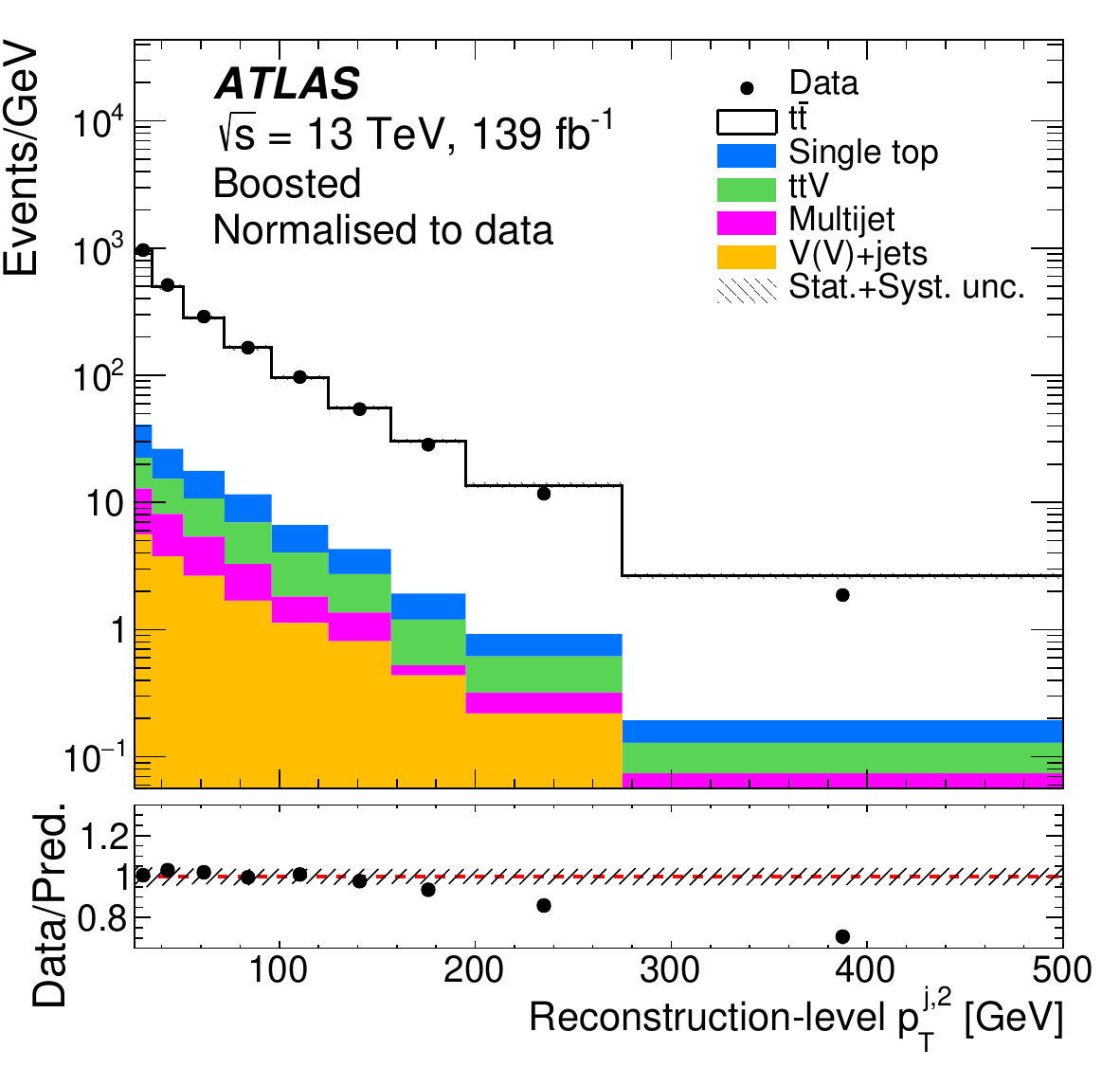}}
\caption{Distributions of observables sensitive to the second leading additional jet:
(a) $\dPhieth{2}$, (b) $\dPhiee$ and (c) $\pTe{2}$.
The data are compared with the expectation from the simulation and
background estimates. The total prediction is normalised to the same number of entries as the data. The lower panel in each subfigure shows the ratio of the data
to the normalised expectation. The shaded band represents the uncertainties originating from the limited data sample size and systematic uncertainties ($t\bar{t}$\ modelling uncertainties
are not included).
}
\label{fig:dataMC3}
\end{figure}
 
\section{Systematic uncertainties}
\label{sec-syst}

Systematic uncertainties affect the measured cross-section through the unfolding corrections discussed in Section~\ref{sec-xs}.
The systematic uncertainty from each source is evaluated by creating pseudo-data where
the source of uncertainty was varied in the simulation and background model.
The modified pseudo-data sample is treated as if it is data and differential cross-sections are extracted using the analysis procedure described in Section~\ref{sec-xs}.
Of particular relevance is the JSF, which is extracted for each pseudo-data sample as described in Section~\ref{sec-xs-jsf}.
In this way, correlations between the different aspects of the analysis methodology are fully accounted for.
The difference
between the cross-sections extracted from the varied pseudo-data and the cross-sections extracted from the nominal
simulation is used as an estimate of the impact of that uncertainty source. For the signal modelling components, where alternative \ttbar\ simulations are employed, the cross-sections extracted from the pseudo-data are compared with the corresponding particle-level distributions to assess the impact of the uncertainty.
The total uncertainty is then calculated assuming
all the systematic uncertainties are uncorrelated.
A description of the different uncertainty components is given in the following subsections.
Table~\ref{table:Results_inclusive_xsection_uncertainties_JSFcomp} summarises the impact of the systematic uncertainties on the inclusive cross-section measurement.
The benefit of the \JSF\ correction procedure is evaluated
by comparing the impact of each source for the case where no \JSF\ correction is applied with the nominal analysis set-up.
The precision of the measurement is seen to improve significantly due to the \JSF\ correction.
When examining the impact of the individual uncertainty sources, some distinct patterns are apparent. Uncertainty sources such as the $b$-tagging and lepton uncertainties
are not affected by the \JSF\ correction because they do not influence either the jet energy measurement or the $\mtopjet$\ distribution.
The impact of the jet energy scale uncertainty is largely reduced because the \JSF\ is able to absorb any overall jet energy scale differences between data and simulation.
The top-quark mass uncertainty increases with the introduction of the \JSF\ procedure because the mean of the $\mtopjet$\ distribution
is directly related to the top-quark mass.
The data and MC statistical uncertainties are also increased by the use of $\meanmtopjet$; the increased statistical
uncertainty is traded for the reduced systematic uncertainties.
The reduction of the uncertainties is also shown in Figure~\ref{fig:jes_jsf}, which displays the effect of the \JSF\ correction on
the uncertainty for three differential distributions, both for the total uncertainty and for the jet energy scale (JES) uncertainty.
In most bins the JES uncertainty is reduced due to the introduction of the \JSF. There are a small number of bins where the JES uncertainty increases due to the \JSF;
this occurs when the kinematics of the jets in these bins are very different from the average jet kinematics in the sample and hence the average correction provided
by the \JSF\ causes the systematic uncertainty from the JES to increase.
Figure~\ref{fig:uncbreakdown} summarises the impact of the systematic uncertainties on three example observables.
The modelling uncertainties generally have the largest impact on the measurement, while the JES and $b$-tagging uncertainties are important in particular
phase-space regions.
 
\begin{table}[pthb]
\center
\caption{Fractional uncertainty breakdown for the inclusive \ttbar cross-section both with and without the JSF method applied.}
\label{table:Results_inclusive_xsection_uncertainties_JSFcomp}
\begin{tabular}[t]{@{}lcclll@{}} \toprule
Source & Uncertainty [\%] & Uncertainty [\%] (no JSF) \\ \midrule
Statistical (data) &  $\pm0.4$ & $\pm0.4$ \\
JSF statistical (data) &  $\pm0.4$ & --- \\ \midrule
Statistical (MC) & $\pm0.2$ & $\pm0.1$ \\
Hard scatter & $\pm0.5$ & $\pm0.8$ \\
Hadronisation & $\pm2.0$ & $\pm1.8$ \\
\multirow{2}{*}{\raisebox{0.5ex}{Radiation (ISR/FSR + \hdamp)}} & $+1.0$ & $+1.4$ \\
& \raisebox{1ex}{$-1.6$} & \raisebox{1ex}{\,$-2.3$} \vspace{-1ex}\\
PDF & $\pm0.1$ & $\pm0.1$ \\
\multirow{2}{*}{\raisebox{0.5ex}{Top-quark mass}} & $+0.8$ & \multirow{2}{*}{\raisebox{0.5ex}{$\pm0.1$}}\\
& \raisebox{1ex}{$-1.1$} & \vspace{-1ex}\\
Jets & $\pm0.7$ & $\pm4.2$ \\
\btag & $\pm2.4$ & $\pm2.4$ \\
Leptons & $\pm0.8$ & $\pm0.8$ \\
\etmiss & $\pm0.1$ & $\pm0.1$ \\
Pile-up & $\pm0.4$ & $\pm0.0$ \\
Luminosity & $\pm1.8$ & $\pm1.8$ \\
Background modelling & $\pm0.6$ & $\pm0.6$ \\ \midrule
\multirow{2}{*}{\raisebox{0.5ex}{Total systematic uncertainty}} & $+4.1$ & $+5.8$ \\
& \raisebox{1ex}{$-4.3$} & \raisebox{1ex}{\,$-$6.0} \vspace{-1ex}\\ \midrule
\multirow{2}{*}{\raisebox{0.5ex}{Total}} & $+4.1$ & $+5.8$ \\
& \raisebox{1ex}{$-4.3$} & \raisebox{1ex}{\,$-6.0$} \vspace{-1ex}\\ \bottomrule
\end{tabular}
\end{table}
 
\FloatBarrier
 
\begin{figure}[htbp]
\centering
\subfloat[]{
\includegraphics[width=0.33\textwidth]{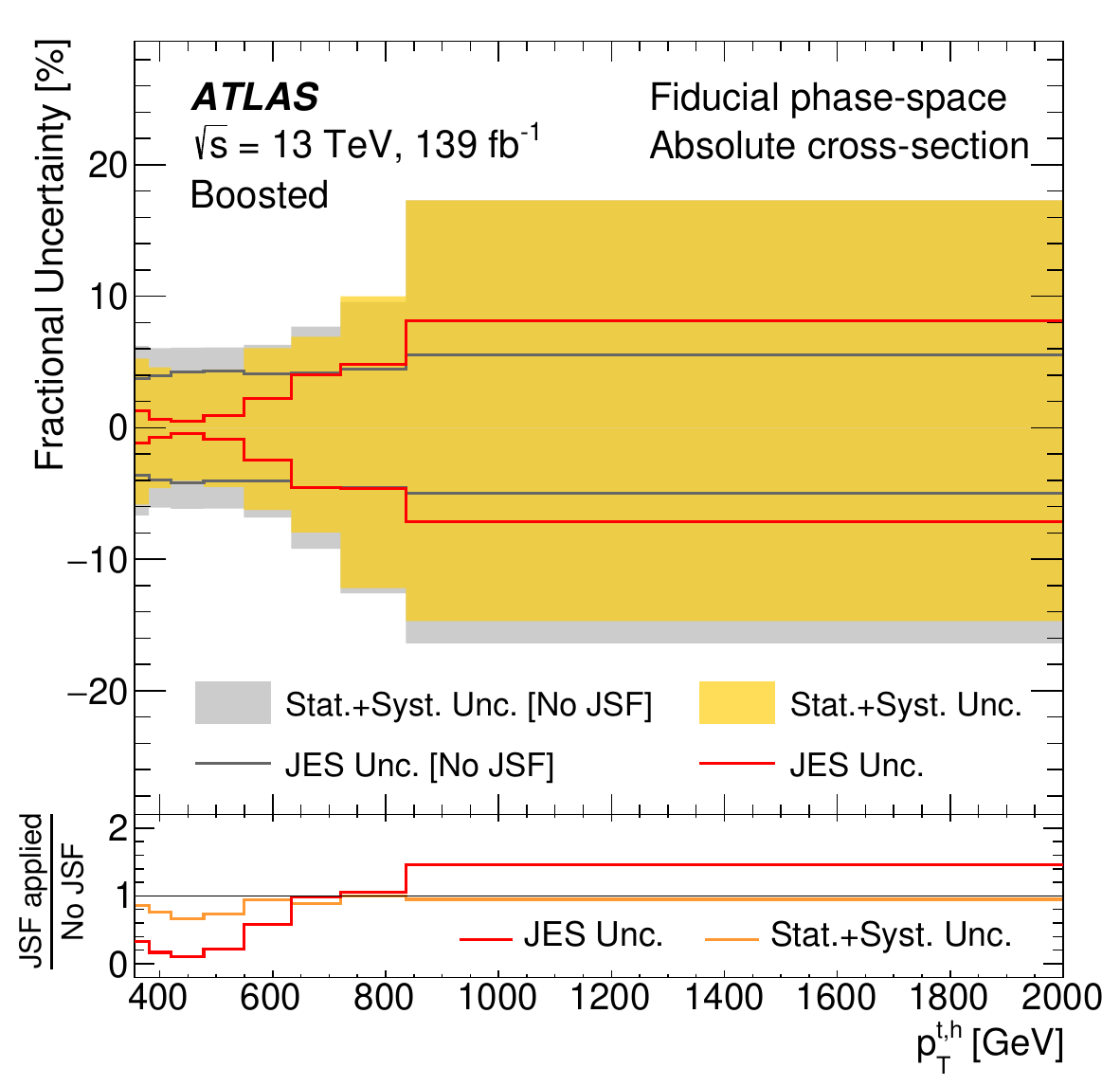}}
\subfloat[]{
\includegraphics[width=0.33\textwidth]{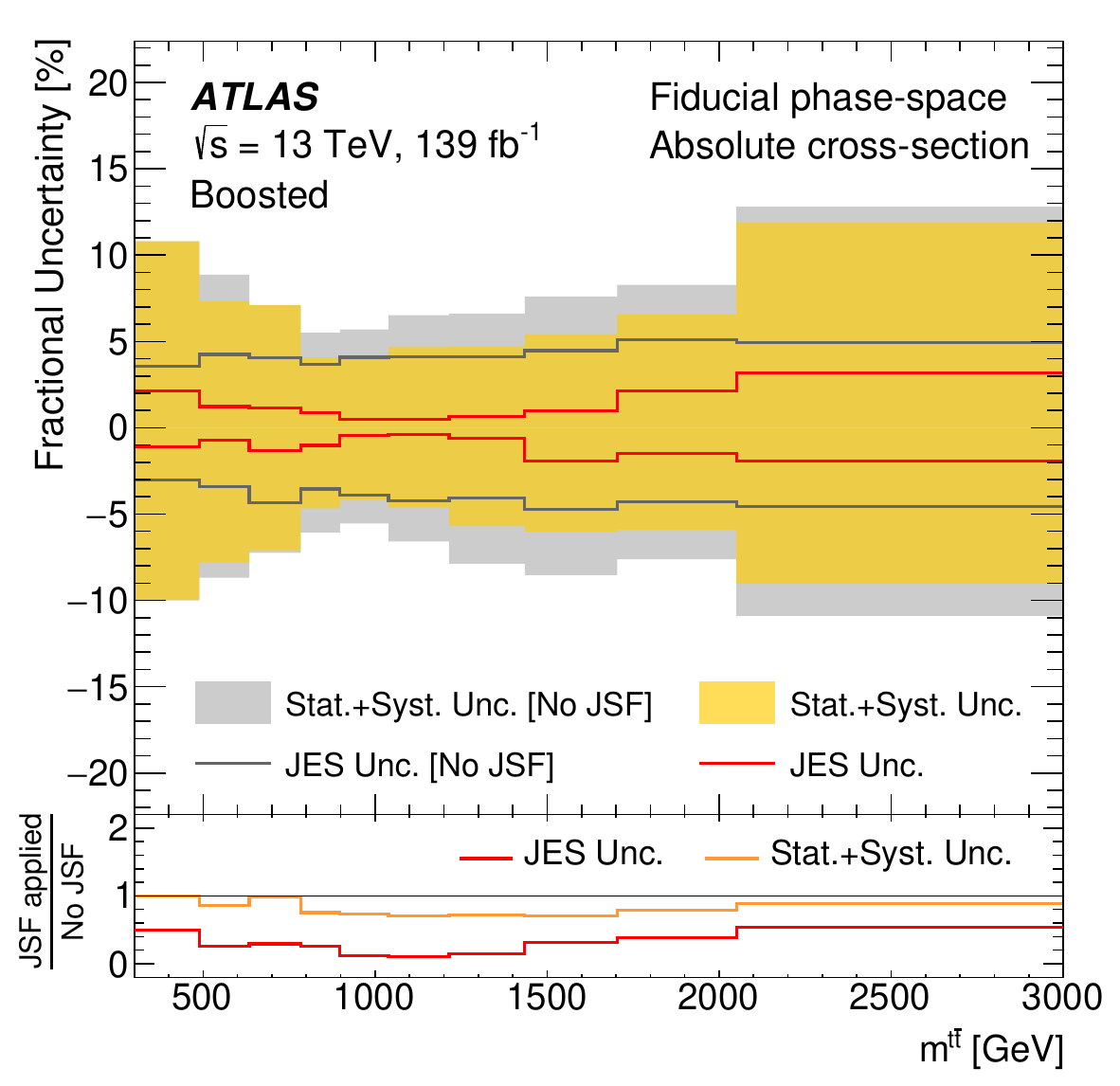}}
\subfloat[]{
\includegraphics[width=0.33\textwidth]{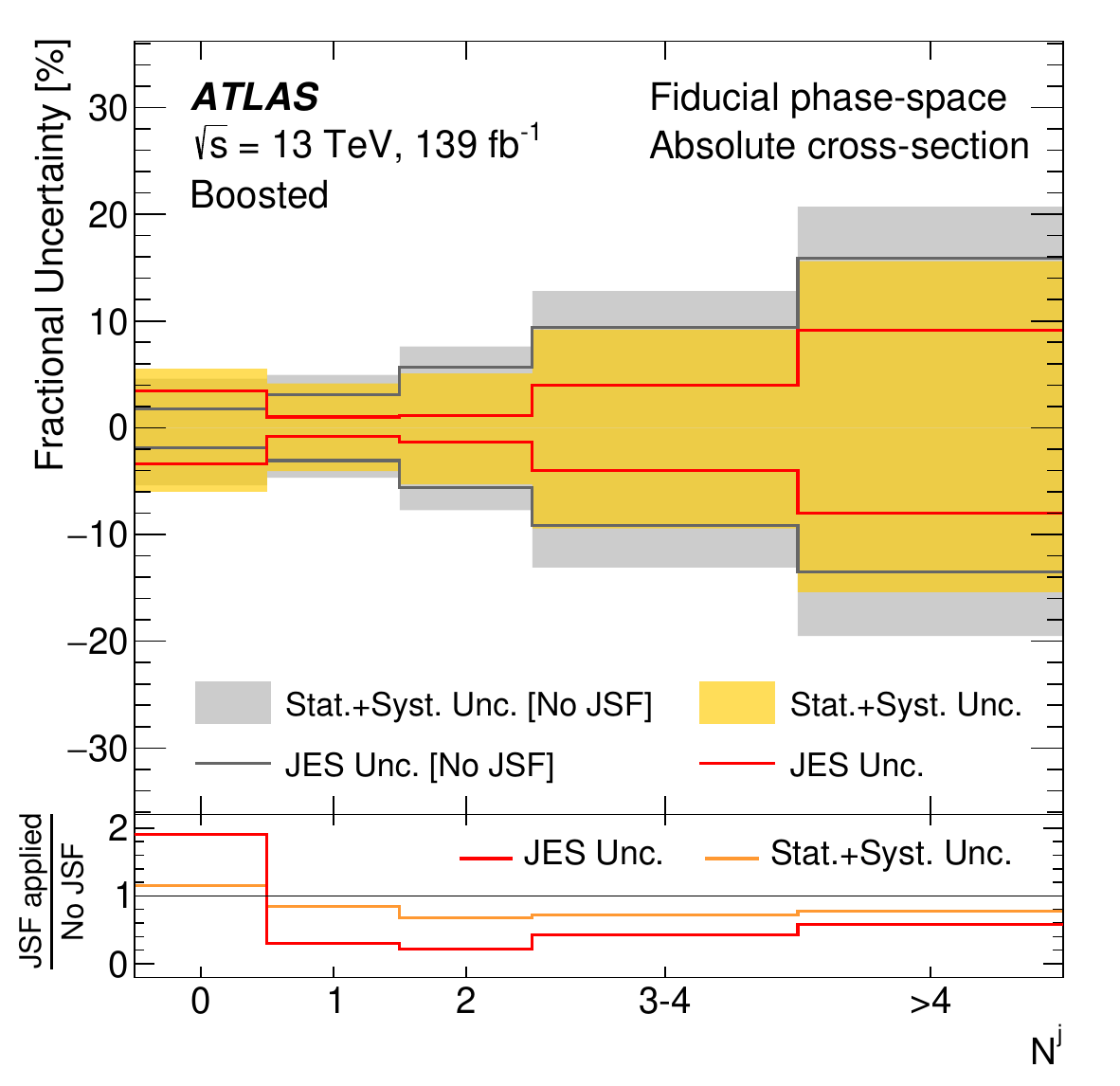}}
\caption{Effect of the \JSF\ correction on the total uncertainty of the cross-section measurements as a function of (a) $\pT$~of the hadronically decaying top quark ($\pThad$), (b) invariant mass of the $\ttbar$~system ($\mttbar$)
and (c) the number of additional jets in the event ($\Ne$).
The yellow (grey) bands represent the total uncertainty with (without) the \JSF\ correction applied.
The red (grey) line in the upper pad shows the JES uncertainty with (without) the JSF method.
The bottom pad shows the ratios of the absolute size of the uncertainty with and without the \JSF\ correction applied, in red for the JES uncertainty and in yellow for the total uncertainty.}
\label{fig:jes_jsf}
\end{figure}
 
\begin{figure}[htbp]
\centering
\subfloat[]{
\includegraphics[width=0.33\textwidth]{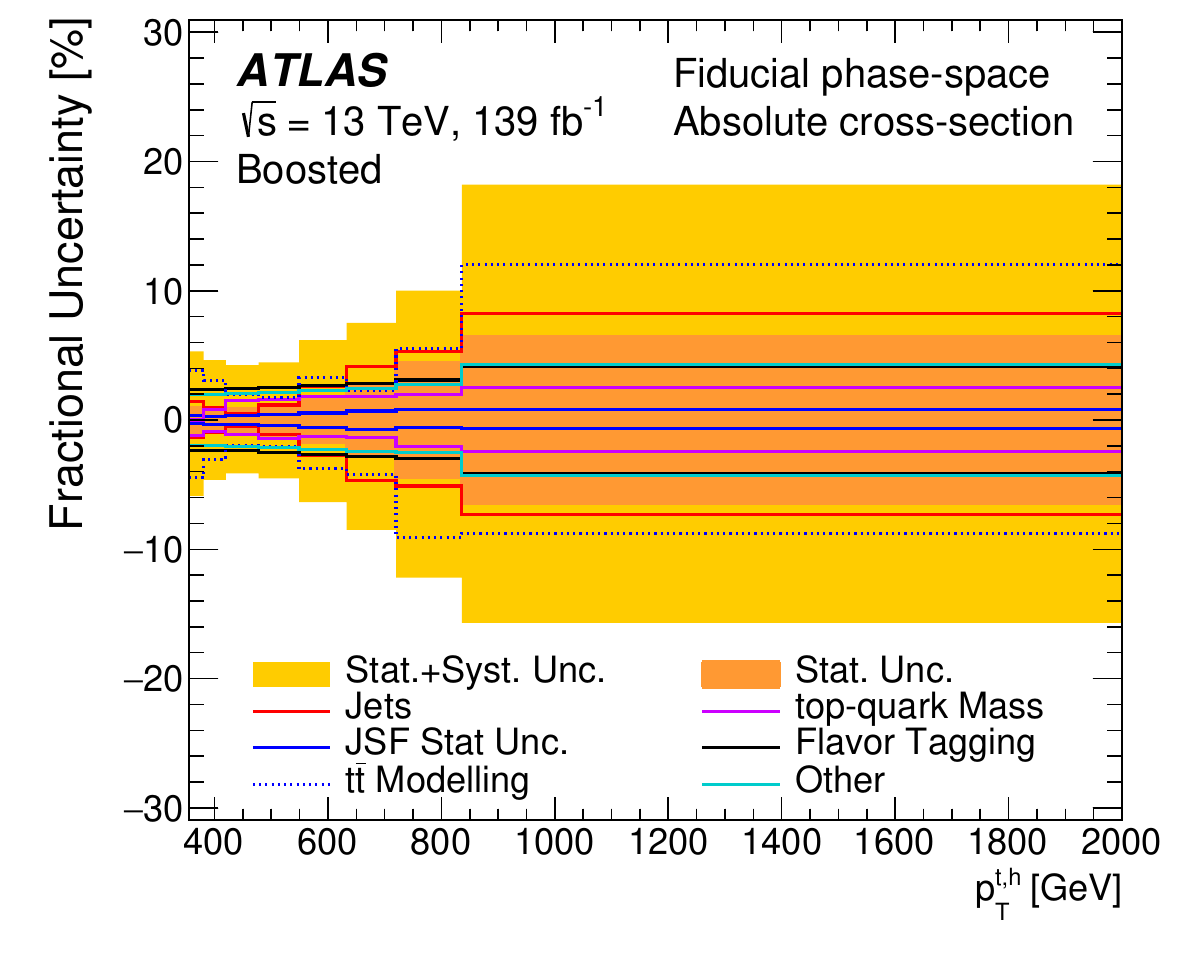}\label{fig:uncbreakdown:toppt}}
\subfloat[]{
\includegraphics[width=0.33\textwidth]{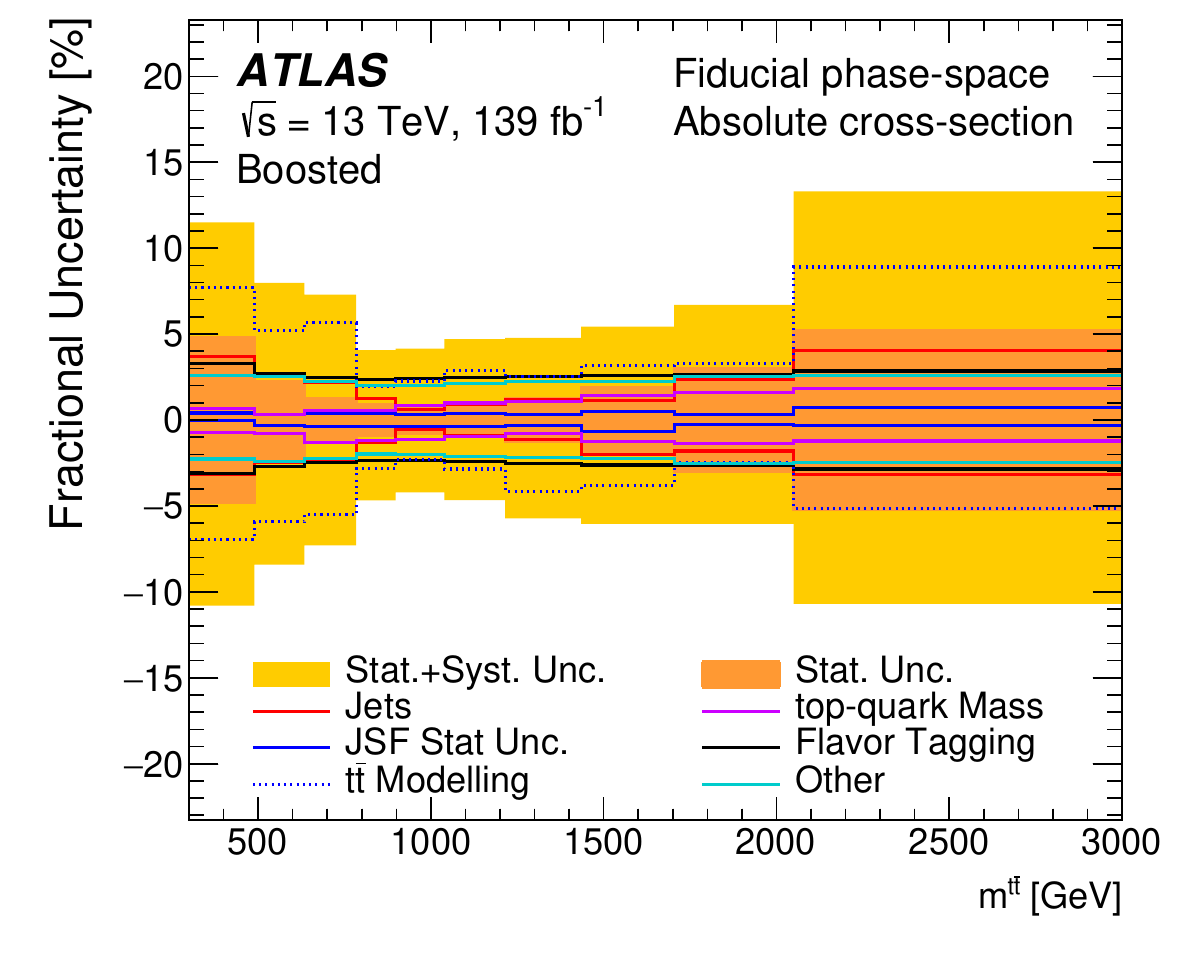}}
\subfloat[]{
\includegraphics[width=0.33\textwidth]{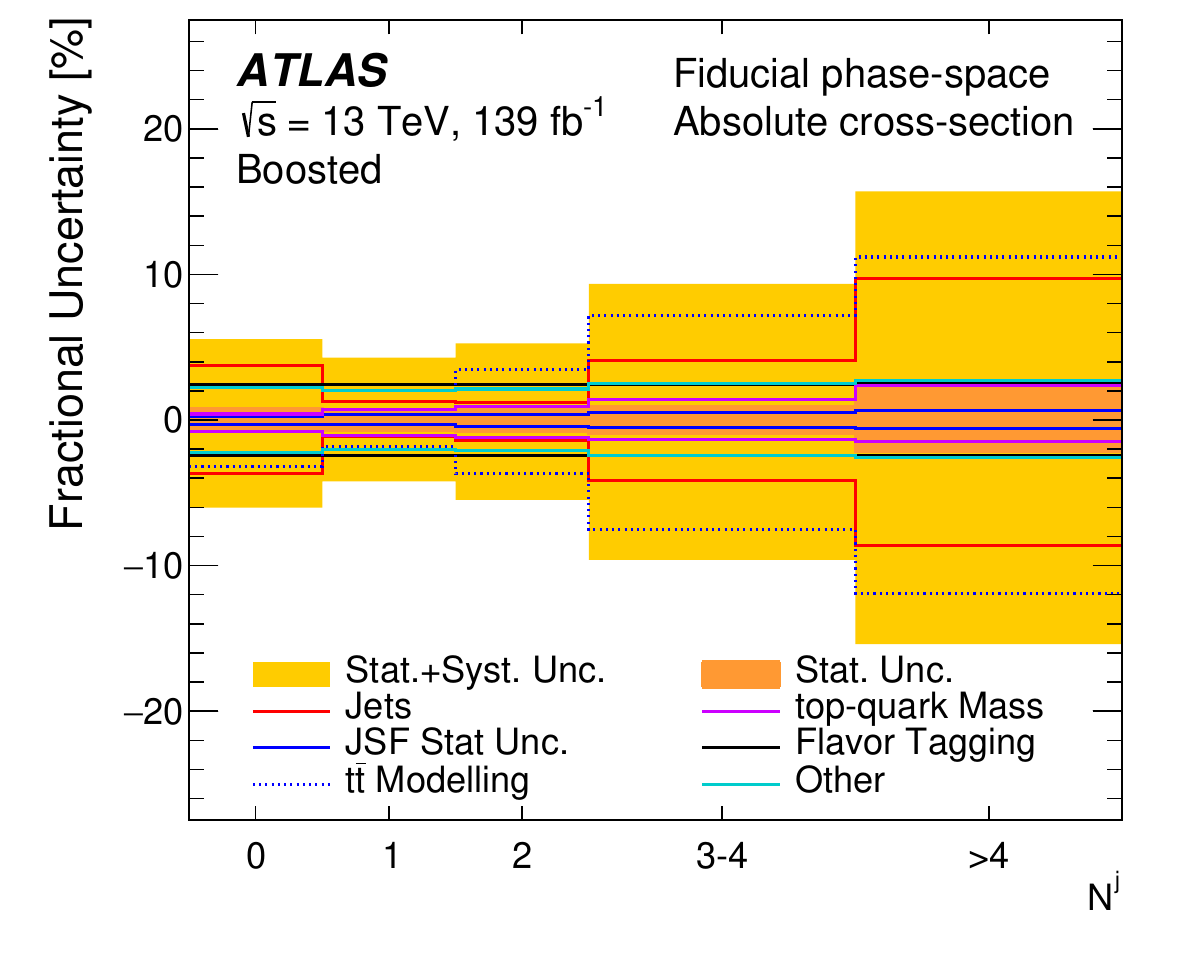}}
\caption{Fractional uncertainties of the absolute cross-section measurement as a function of
(a) $\pT$~of the hadronically decaying top quark ($\pThad$),
(b) invariant mass of the $\ttbar$~system ($\mttbar$)
and
(c) the number of additional jets in the event ($\Ne$). The line labelled `Jets' includes the uncertainties from the JES, JER and JVT requirements. The line labelled `$\ttbar$\ Modelling' includes
all the uncertainties discussed in Section~\ref{sec-syst-ttbarmodel}, with the exception of the uncertainty in the top-quark mass, which is shown separately.
}
\label{fig:uncbreakdown}
\end{figure}
 
\subsection{Lepton reconstruction and identification}
 
The uncertainty in the efficiency to reconstruct and identify electrons and muons was obtained by studying
$Z \to ee / \mu\mu$~events as discussed in Refs.~\cite{EGAM-2018-01, MUON-2018-03}.
Similar studies were performed to determine the uncertainty in the trigger efficiencies for electrons and muons~\cite{TRIG-2018-05, TRIG-2018-01}.
The impact of these uncertainties on the analysis is small, but they are the largest part of the `Leptons' entry in Table~\ref{table:Results_inclusive_xsection_uncertainties_JSFcomp}.
The uncertainties in the electron and muon energy / momentum scales and resolutions were determined
using resonance decays~\cite{EGAM-2018-01,PERF-2015-10} and are found to have a negligible impact on the analysis.
 
\subsection{Jet reconstruction and $b$-tagging}
 
The JES and jet energy resolution of \smallR\ jets were determined using a combination of simulation,
test beam and in situ measurements~\cite{JETM-2018-05}.
The corresponding uncertainties are evaluated using a model with 30 independent components for the jet energy scale
and 8 independent components for the jet energy resolution.
The uncertainties are propagated to the \largeR\ jets such that the correlations between the energies of \smallR\ and \largeR jets
are maintained.
The impact of the JES uncertainties on the measurement
is significantly reduced by the \JSF\ procedure; for example, the uncertainty in the inclusive cross-section due to the JES  is reduced from $4.0\%$\ to
$0.4\%$.
The JES uncertainties have their largest impact in regions with high jet multiplicity.
The impact of the jet energy resolution uncertainty is generally smaller than that due to the jet energy scale uncertainty.
The uncertainty in the efficiency of the JVT requirement for pile-up suppression is also considered~\cite{PERF-2014-03}.
 
The performance of the $b$-tagging algorithm has been calibrated in the data~\cite{FTAG-2018-01}. The corresponding uncertainties
are propagated to the analysis by using an uncertainty model containing $9 / 4 / 4$~independent variations for the $b$-/$c$-/light-jet
calibrations and two components for the MC-based extrapolation to jets with very high $\pt$.
These uncertainties have a moderate impact on the fiducial cross-section measurement as seen in Table~\ref{table:Results_inclusive_xsection_uncertainties_JSFcomp}
and do not significantly vary in size as a function of the measured observables, as seen in Figure~\ref{fig:uncbreakdown}.
 
\subsection{\JSF\ statistical uncertainty}
\label{sec-syst-jsfstat}
 
The statistical uncertainty due to the limited number of data events in the determination of the \JSF\ is evaluated by
performing pseudo-experiments in which the data yields in the
$\mtopjet$~distribution are varied according to a Poisson distribution. As the $\mtopjet$~distribution is found to be largely uncorrelated with
all measured observables, this uncertainty is assumed to be uncorrelated with the statistical uncertainty of the cross-section measurement.
A similar procedure is used to assess the statistical uncertainty of the \JSF\ due to the limited size of the simulated event samples.

\subsection{\ttbar modelling}
\label{sec-syst-ttbarmodel}
 
Uncertainties in the modelling of $\ttbar$~production affect the unfolding corrections as well as the $\mtopjet$~distribution
that is used to determine the \JSF. Several separate variations of the $\ttbar$~model are considered in the analysis.
 
The uncertainty due to the choice of parton shower and hadronisation models is assessed by
using the events generated by \POWHER to build pseudo-data, repeating the analysis, and comparing the unfolded
distributions with the particle-level prediction of \POWHER.
The impact of this uncertainty is quite important
and largest at low $\pThad$.
The uncertainty originating from the choice of generator
is assessed by 
using pseudo-data created from the \MGNLOPY sample.
The impact of this uncertainty
is relatively small.
The uncertainty originating from the scales used in the matrix elements and the parton shower is evaluated
by using pseudo-data built from the samples with variations of the \hdamp\ parameter,
and by using pseudo-data created by reweighting the nominal sample to correspond to different values of the \mur\ and
\muf\ scales in the matrix elements, the parameters in the showering tune~\cite{ATL-PHYS-PUB-2017-007}, and the \mur\ scale in the final-state parton shower.
The reweighted samples using the changes in the scales in the matrix elements and the changes in the parameter values in the showering tune are referred to as the ISR variations.
The reweighted samples using the varied scale in the final-state parton shower are referred to as the FSR variations.
These modelling uncertainties are found to be particularly important at high $\pThad$ (Figure~\ref{fig:uncbreakdown:toppt}).
 
The extraction of the \JSF\ relies on the measured value of the top-quark mass, which is known to a precision
of around $0.5$~\GeV~\cite{TOPQ-2017-03,CMS-TOP-14-022}.
This uncertainty is evaluated by using pseudo-data built from MC samples where the top-quark mass is varied from its nominal value of $172.5$~\GeV.
Thanks to the high precision of the top-quark mass measurements, the impact on the analysis is small, although the use of the \JSF\ method increases the impact of this uncertainty,
as seen in Table~\ref{table:Results_inclusive_xsection_uncertainties_JSFcomp}.
 
The uncertainty in the parton distribution functions is evaluated using the 30 eigenvectors of PDF4LHC30~\cite{Butterworth:2015oua}. The impact of the uncertainty is found to be very small.
 
\subsection{Background modelling}
 
The uncertainty in the modelling of single top-quark production is assessed by using samples where the scales are varied in a similar way to the $\ttbar$~samples
described above.
The uncertainty in the subtraction of the $\ttbar$~events from the $tW$~sample is assessed by using an alternative sample
that uses the diagram subtraction scheme~\cite{Frixione:2008yi} instead of the diagram removal scheme.
Uncertainties in the cross-sections of the single-top-quark processes are also included.
 
The uncertainties in the $W$+\,jets background are evaluated by reweighting the MC samples to correspond to different values of
the scales in the matrix elements and the parton shower~\cite{ATL-PHYS-PUB-2017-006}. The small $Z$\,+\,jets and diboson backgrounds
are assigned uncertainties of $50\%$ to cover potential mismodelling of these backgrounds. The $\ttbar V$~and $\ttbar H$~processes
are assigned an uncertainty of $13\%$~\cite{deFlorian:2016spz}.
 
The uncertainty in the multijet background is assessed by comparing the estimate from the matrix method
with the estimate from an alternative method based on fitting MC templates to the $\MET$~and $\MET + \mT^W$~distributions.
These comparisons result in an uncertainty of $65\%$~in the multijet background estimate.
 
The impact of the background uncertainties on the measurements is generally small, and less than in the previous
ATLAS measurement thanks to the tighter selection requirements (particularly the one on $m_{\ell b}$).
 
\subsection{Luminosity and other uncertainties}
 
The calibration of the integrated luminosity has an uncertainty of $1.7\%$~\cite{ATLAS-CONF-2019-021}. The uncertainty
is important for the inclusive cross-section measurement but less so for the differential measurements.
The uncertainty in the pile-up modelling is evaluated by varying the mean number of interactions in the simulation and this uncertainty
is found to be small.
The uncertainty in the $\MET$ originates from the possible miscalibration of the tracks in the soft term and it was derived from the $\pT$ imbalance between the soft and hard components in data--simulation comparisons~\cite{PERF-2016-07}.
The uncertainty due to the limited number of simulated events is evaluated by varying the number of events according
to the statistical uncertainties of the simulated samples and is found to be smaller than the statistical uncertainty of the data.
 
\section{Results}
\label{sec-results}
 
The fiducial cross-section (as defined in Section~\ref{sec-xs-fid}) is measured to be $1.267 \pm 0.005 \pm 0.053$~pb, where
the first uncertainty is due to the size of the data sample and the second originates from the systematic uncertainties.
This cross-section is measured to a relative precision of $4.2\%$. This is smaller than the calculated uncertainty of $6.1\%$ in the
inclusive $\ttbar$\ cross-section at NNLO+NNLL, which is composed
of $4.2\%$\ from PDF uncertainties, $3\%$\ from scale variations and $2.8\%$\ from the uncertainty in the top-quark mass.
A comparison with the SM predictions obtained with different MC set-ups (each normalised to the inclusive NNLO+NNLL $\ttbar$~cross-section) is shown in Figure~\ref{fig:inclusive_cross_section}.
All MC set-ups give predictions that are higher than the data, with the \POWPY and
\MGNLOPY predictions being around two standard deviations above the data.\footnote{
The estimate of the uncertainty of the ratio between data and expectation includes the experimental uncertainty and the $6.1\%$~uncertainty on the $\ttbar$\ cross-section, assuming the two are uncorrelated.}
Significantly better agreement is seen after reweighting the MC simulations to the differential NNLO predictions, indicating
the corrections are relevant given the precision of the measurement.
 
\begin{figure}[b]
\centering
\includegraphics[width=0.45\linewidth]{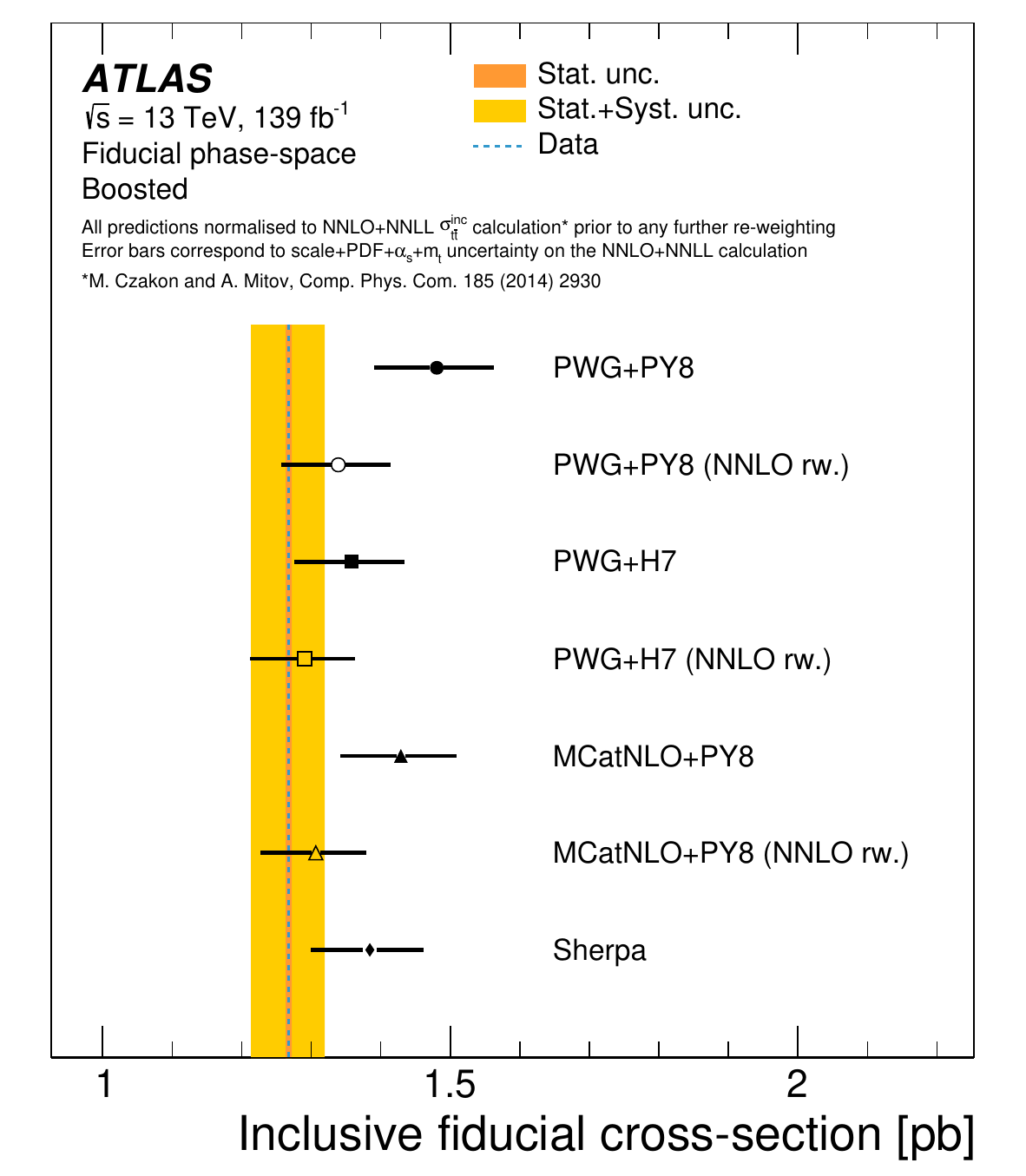}
\caption{The fiducial cross-section at particle level for boosted \ttbar production measured in data (dashed line) is compared with several NLO predictions (closed markers).
All the MC samples are normalised to the NNLO+NNLL prediction for the total $\ttbar$\ cross-section ($\sigma^{\textrm{inc}}_{\ttbar}$).
The open markers show the predictions from the MC generators after they were reweighted at parton level to match NNLO predictions for $\pt(t)$, $m(\ttbar)$\ and $\pt(\ttbar)$.
The yellow band represents the total uncertainty of the measured cross-section, while the orange band shows the statistical component.
The uncertainties
in the predictions are evaluated as the quadrature sum of the $\alphas$, PDF, $m_t$ and scale uncertainties present in the NNLO+NNLL prediction.
PWG+PY8 corresponds to the \POWPY sample, PWG+H7 to the \POWHER sample, and MCatNLO+PY8 to the \MGNLOPY sample.}
\label{fig:inclusive_cross_section}
\end{figure}
 
The level of agreement for the differential cross-section measurements is quantified by
calculating $\chi^2$\ values according to:
\begin{equation}\nonumber
\chi^2 = V^\mathrm{T} C^{-1} V
\end{equation}
where $V$~is the vector of residuals between the measured and predicted cross-sections and $C$~is the covariance
matrix of the measured data (including both the statistical and systematic uncertainties). No uncertainties in the theoretical
predictions are included in the $\chi^2$~calculation. The observed $\chi^2$~and the
number of degrees of freedom are used to calculate $p$-values.
Table~\ref{table:chi2_table_NNLO_absolute} shows the $\chi^2$~values for the different NLO generators
discussed in Section~\ref{sec-datamc}. The effect of the NNLO reweighting is quantified in this table
by also showing the $\chi^2$\ values for the different generators with the NNLO reweighting applied.
Table~\ref{table:chi2_table_absolute} shows the observed $\chi^2$\ values
for the different \POWPY set-ups discussed in Section~\ref{sec-datamc}.
 
\tableVSpace{1.1}
\begin{table}[h]
\centering
\caption{$\chi^2$ and $p$-values quantifying the level of agreement between the absolute unfolded spectra, several NLO+PS predictions and the respective NNLO reweighted spectrum. PWG+PY8 corresponds to the \POWPY sample, PWG+H7 to the \POWHER sample and MC@NLO+PY8 to the \MGNLOPY sample.}
\resizebox{\textwidth}{!}{
\begin{tabular}[t]{|c| r @{/} l r  || r @{/} l r  || r @{/} l r  || r @{/} l r  || r @{/} l r  || r @{/} l r  |} \toprule
Observable& \multicolumn{3}{c|}{\textsc{PWG+PY8}}& \multicolumn{3}{c|}{\textsc{PWG+PY8(NNLO weight)}}& \multicolumn{3}{c|}{\textsc{MC@NLO+PY8}}& \multicolumn{3}{c|}{\textsc{MC@NLO+PY8(NNLO weight)}}& \multicolumn{3}{c|}{\textsc{PWG+H7}}& \multicolumn{3}{c|}{\textsc{PWG+H7(NNLO weight)}}\\ \midrule
& \multicolumn{2}{c}{$\chi^{2}$/NDF} &  ~$p$-value& \multicolumn{2}{c}{$\chi^{2}$/NDF} &  ~$p$-value& \multicolumn{2}{c}{$\chi^{2}$/NDF} &  ~$p$-value& \multicolumn{2}{c}{$\chi^{2}$/NDF} &  ~$p$-value& \multicolumn{2}{c}{$\chi^{2}$/NDF} &  ~$p$-value& \multicolumn{2}{c}{$\chi^{2}$/NDF} &  ~$p$-value\\
\hline
\hline
$\pThad$ &{\ } 26 & 8 &<0.01 &{\ } 5 & 8 &0.79 &{\ } 18 & 8 &0.03 &{\ } 4 & 8 &0.85 &{\ } 7 & 8 &0.56 &{\ } 3 & 8 &0.94\\
$\pTlep$ &{\ } 78 & 8 &<0.01 &{\ } 28 & 8 &<0.01 &{\ } 144 & 8 &<0.01 &{\ } 10 & 8 &0.27 &{\ } 43 & 8 &<0.01 &{\ } 18 & 8 &0.02\\
$\pTttbar$ &{\ } 162 & 7 &<0.01 &{\ } 46 & 7 &<0.01 &{\ } 171 & 7 &<0.01 &{\ } 22 & 7 &<0.01 &{\ } 122 & 7 &<0.01 &{\ } 39 & 7 &<0.01\\
$\HTall$ &{\ } 36 & 7 &<0.01 &{\ } 7 & 7 &0.42 &{\ } 17 & 7 &0.02 &{\ } 23 & 7 &<0.01 &{\ } 21 & 7 &<0.01 &{\ } 12 & 7 &0.10\\
$\HTttbar$ &{\ } 86 & 10 &<0.01 &{\ } 37 & 10 &<0.01 &{\ } 110 & 10 &<0.01 &{\ } 16 & 10 &0.10 &{\ } 47 & 10 &<0.01 &{\ } 28 & 10 &<0.01\\
$\yhad$ &{\ } 47 & 17 &<0.01 &{\ } 27 & 17 &0.06 &{\ } 37 & 17 &<0.01 &{\ } 23 & 17 &0.15 &{\ } 30 & 17 &0.03 &{\ } 26 & 17 &0.07\\
$\ylep$ &{\ } 40 & 14 &<0.01 &{\ } 17 & 14 &0.26 &{\ } 29 & 14 &0.01 &{\ } 12 & 14 &0.58 &{\ } 28 & 14 &0.01 &{\ } 19 & 14 &0.16\\
$\yttbar$ &{\ } 30 & 10 &<0.01 &{\ } 8 & 10 &0.58 &{\ } 23 & 10 &0.01 &{\ } 6 & 10 &0.81 &{\ } 14 & 10 &0.19 &{\ } 7 & 10 &0.74\\
$\mttbar$ &{\ } 52 & 10 &<0.01 &{\ } 24 & 10 &<0.01 &{\ } 81 & 10 &<0.01 &{\ } 7 & 10 &0.74 &{\ } 29 & 10 &<0.01 &{\ } 22 & 10 &0.02\\
$\pTe{1}$ &{\ } 115 & 15 &<0.01 &{\ } 38 & 15 &<0.01 &{\ } 413 & 15 &<0.01 &{\ } 194 & 15 &<0.01 &{\ } 143 & 15 &<0.01 &{\ } 69 & 15 &<0.01\\
$\pTe{2}$ &{\ } 46 & 9 &<0.01 &{\ } 19 & 9 &0.02 &{\ } 25 & 9 &<0.01 &{\ } 74 & 9 &<0.01 &{\ } 42 & 9 &<0.01 &{\ } 29 & 9 &<0.01\\
$\Ne$ &{\ } 32 & 5 &<0.01 &{\ } 12 & 5 &0.03 &{\ } 76 & 5 &<0.01 &{\ } 78 & 5 &<0.01 &{\ } 57 & 5 &<0.01 &{\ } 62 & 5 &<0.01\\
$\dPhieth{1}$ &{\ } 17 & 9 &0.05 &{\ } 8 & 9 &0.53 &{\ } 150 & 9 &<0.01 &{\ } 80 & 9 &<0.01 &{\ } 42 & 9 &<0.01 &{\ } 30 & 9 &<0.01\\
$\dPhieth{2}$ &{\ } 8 & 9 &0.56 &{\ } 5 & 9 &0.84 &{\ } 8 & 9 &0.57 &{\ } 25 & 9 &<0.01 &{\ } 85 & 9 &<0.01 &{\ } 76 & 9 &<0.01\\
$\dPhibt$ &{\ } 95 & 13 &<0.01 &{\ } 34 & 13 &<0.01 &{\ } 145 & 13 &<0.01 &{\ } 16 & 13 &0.23 &{\ } 52 & 13 &<0.01 &{\ } 25 & 13 &0.02\\
$\dPhitt$ &{\ } 111 & 5 &<0.01 &{\ } 36 & 5 &<0.01 &{\ } 134 & 5 &<0.01 &{\ } 82 & 5 &<0.01 &{\ } 90 & 5 &<0.01 &{\ } 36 & 5 &<0.01\\
$\dPhiee$ &{\ } 24 & 11 &0.01 &{\ } 16 & 11 &0.13 &{\ } 31 & 11 &<0.01 &{\ } 69 & 11 &<0.01 &{\ } 237 & 11 &<0.01 &{\ } 215 & 11 &<0.01\\
$\meth$ &{\ } 50 & 12 &<0.01 &{\ } 20 & 12 &0.06 &{\ } 221 & 12 &<0.01 &{\ } 48 & 12 &<0.01 &{\ } 41 & 12 &<0.01 &{\ } 19 & 12 &0.08\\
$\pTe{1}$ vs $\Ne$ &{\ } 355 & 21 &<0.01 &{\ } 205 & 21 &<0.01 &{\ } 633 & 21 &<0.01 &{\ } 316 & 21 &<0.01 &{\ } 263 & 21 &<0.01 &{\ } 159 & 21 &<0.01\\
$\pTe{1}$ vs $\pThad$ &{\ } 115 & 17 &<0.01 &{\ } 53 & 17 &<0.01 &{\ } 383 & 17 &<0.01 &{\ } 152 & 17 &<0.01 &{\ } 121 & 17 &<0.01 &{\ } 74 & 17 &<0.01\\
$\dPhieth{1}$ vs $\pThad$ &{\ } 69 & 21 &<0.01 &{\ } 43 & 21 &<0.01 &{\ } 427 & 21 &<0.01 &{\ } 223 & 21 &<0.01 &{\ } 78 & 21 &<0.01 &{\ } 60 & 21 &<0.01\\
$\dPhieth{1}$ vs $\Ne$ &{\ } 109 & 19 &<0.01 &{\ } 64 & 19 &<0.01 &{\ } 545 & 19 &<0.01 &{\ } 250 & 19 &<0.01 &{\ } 85 & 19 &<0.01 &{\ } 60 & 19 &<0.01\\
\bottomrule
\end{tabular}
}
\label{table:chi2_table_NNLO_absolute}
\end{table}
\tableVSpace{1.1}
\begin{table}[!h]
\centering
\caption{$\chi^2$ and $p$-values quantifying the level of agreement between the absolute unfolded spectra and several NLO+PS predictions. PWG+PY8 corresponds to the \POWPY sample. \textsc{SHERPA} (NLO norm.) refers to the \SHERPA sample with its default normalisation. All other samples are normalised to the inclusive NNLO+NNLL $\ttbar$\ cross-section prediction.}
\resizebox{\textwidth}{!}{
\begin{tabular}[t]{|c| r @{/} l r  || r @{/} l r  || r @{/} l r  || r @{/} l r  || r @{/} l r  || r @{/} l r  |} \toprule
Observable& \multicolumn{3}{c|}{\textsc{PWG+PY8}}& \multicolumn{3}{c|}{\textsc{PWG+PY8(ISR Down)}}& \multicolumn{3}{c|}{\textsc{PWG+PY8(ISR Up)}}& \multicolumn{3}{c|}{\textsc{PWG+PY8($h_{\mathrm{damp}}=3m_{t}$)}}& \multicolumn{3}{c|}{\textsc{SHERPA}}& \multicolumn{3}{c|}{\textsc{SHERPA} (NLO norm.)}\\ \midrule
& \multicolumn{2}{c}{$\chi^{2}$/NDF} &  ~$p$-value& \multicolumn{2}{c}{$\chi^{2}$/NDF} &  ~$p$-value& \multicolumn{2}{c}{$\chi^{2}$/NDF} &  ~$p$-value& \multicolumn{2}{c}{$\chi^{2}$/NDF} &  ~$p$-value& \multicolumn{2}{c}{$\chi^{2}$/NDF} &  ~$p$-value& \multicolumn{2}{c}{$\chi^{2}$/NDF} &  ~$p$-value\\
\hline
\hline
$\pThad$ &{\ } 26 & 8 &<0.01 &{\ } 26 & 8 &<0.01 &{\ } 25 & 8 &<0.01 &{\ } 36 & 8 &<0.01 &{\ } 12 & 8 &0.15 &{\ } 11 & 8 &0.19\\
$\pTlep$ &{\ } 78 & 8 &<0.01 &{\ } 144 & 8 &<0.01 &{\ } 20 & 8 &0.01 &{\ } 50 & 8 &<0.01 &{\ } 12 & 8 &0.13 &{\ } 11 & 8 &0.22\\
$\pTttbar$ &{\ } 162 & 7 &<0.01 &{\ } 243 & 7 &<0.01 &{\ } 340 & 7 &<0.01 &{\ } 108 & 7 &<0.01 &{\ } 70 & 7 &<0.01 &{\ } 57 & 7 &<0.01\\
$\HTall$ &{\ } 36 & 7 &<0.01 &{\ } 38 & 7 &<0.01 &{\ } 96 & 7 &<0.01 &{\ } 52 & 7 &<0.01 &{\ } 39 & 7 &<0.01 &{\ } 34 & 7 &<0.01\\
$\HTttbar$ &{\ } 86 & 10 &<0.01 &{\ } 119 & 10 &<0.01 &{\ } 46 & 10 &<0.01 &{\ } 72 & 10 &<0.01 &{\ } 28 & 10 &<0.01 &{\ } 22 & 10 &0.01\\
$\yhad$ &{\ } 47 & 17 &<0.01 &{\ } 46 & 17 &<0.01 &{\ } 46 & 17 &<0.01 &{\ } 55 & 17 &<0.01 &{\ } 25 & 17 &0.10 &{\ } 20 & 17 &0.29\\
$\ylep$ &{\ } 40 & 14 &<0.01 &{\ } 45 & 14 &<0.01 &{\ } 34 & 14 &<0.01 &{\ } 45 & 14 &<0.01 &{\ } 24 & 14 &0.05 &{\ } 18 & 14 &0.19\\
$\yttbar$ &{\ } 30 & 10 &<0.01 &{\ } 32 & 10 &<0.01 &{\ } 23 & 10 &<0.01 &{\ } 35 & 10 &<0.01 &{\ } 22 & 10 &0.02 &{\ } 20 & 10 &0.03\\
$\mttbar$ &{\ } 52 & 10 &<0.01 &{\ } 78 & 10 &<0.01 &{\ } 75 & 10 &<0.01 &{\ } 53 & 10 &<0.01 &{\ } 31 & 10 &<0.01 &{\ } 25 & 10 &<0.01\\
$\pTe{1}$ &{\ } 115 & 15 &<0.01 &{\ } 136 & 15 &<0.01 &{\ } 272 & 15 &<0.01 &{\ } 74 & 15 &<0.01 &{\ } 140 & 15 &<0.01 &{\ } 98 & 15 &<0.01\\
$\pTe{2}$ &{\ } 46 & 9 &<0.01 &{\ } 12 & 9 &0.23 &{\ } 196 & 9 &<0.01 &{\ } 81 & 9 &<0.01 &{\ } 41 & 9 &<0.01 &{\ } 19 & 9 &0.02\\
$\Ne$ &{\ } 32 & 5 &<0.01 &{\ } 51 & 5 &<0.01 &{\ } 27 & 5 &<0.01 &{\ } 41 & 5 &<0.01 &{\ } 23 & 5 &<0.01 &{\ } 16 & 5 &<0.01\\
$\dPhieth{1}$ &{\ } 17 & 9 &0.05 &{\ } 34 & 9 &<0.01 &{\ } 22 & 9 &<0.01 &{\ } 23 & 9 &<0.01 &{\ } 10 & 9 &0.38 &{\ } 11 & 9 &0.25\\
$\dPhieth{2}$ &{\ } 8 & 9 &0.56 &{\ } 7 & 9 &0.67 &{\ } 22 & 9 &0.01 &{\ } 19 & 9 &0.03 &{\ } 6 & 9 &0.74 &{\ } 3 & 9 &0.96\\
$\dPhibt$ &{\ } 95 & 13 &<0.01 &{\ } 116 & 13 &<0.01 &{\ } 294 & 13 &<0.01 &{\ } 119 & 13 &<0.01 &{\ } 51 & 13 &<0.01 &{\ } 28 & 13 &0.01\\
$\dPhitt$ &{\ } 111 & 5 &<0.01 &{\ } 164 & 5 &<0.01 &{\ } 207 & 5 &<0.01 &{\ } 79 & 5 &<0.01 &{\ } 36 & 5 &<0.01 &{\ } 39 & 5 &<0.01\\
$\dPhiee$ &{\ } 24 & 11 &0.01 &{\ } 17 & 11 &0.12 &{\ } 41 & 11 &<0.01 &{\ } 38 & 11 &<0.01 &{\ } 26 & 11 &<0.01 &{\ } 20 & 11 &0.05\\
$\meth$ &{\ } 50 & 12 &<0.01 &{\ } 111 & 12 &<0.01 &{\ } 93 & 12 &<0.01 &{\ } 43 & 12 &<0.01 &{\ } 65 & 12 &<0.01 &{\ } 40 & 12 &<0.01\\
$\pTe{1}$ vs $\Ne$ &{\ } 355 & 21 &<0.01 &{\ } 495 & 21 &<0.01 &{\ } 488 & 21 &<0.01 &{\ } 254 & 21 &<0.01 &{\ } 193 & 21 &<0.01 &{\ } 137 & 21 &<0.01\\
$\pTe{1}$ vs $\pThad$ &{\ } 115 & 17 &<0.01 &{\ } 192 & 17 &<0.01 &{\ } 256 & 17 &<0.01 &{\ } 87 & 17 &<0.01 &{\ } 133 & 17 &<0.01 &{\ } 87 & 17 &<0.01\\
$\dPhieth{1}$ vs $\pThad$ &{\ } 69 & 21 &<0.01 &{\ } 104 & 21 &<0.01 &{\ } 56 & 21 &<0.01 &{\ } 73 & 21 &<0.01 &{\ } 42 & 21 &<0.01 &{\ } 32 & 21 &0.06\\
$\dPhieth{1}$ vs $\Ne$ &{\ } 109 & 19 &<0.01 &{\ } 201 & 19 &<0.01 &{\ } 66 & 19 &<0.01 &{\ } 91 & 19 &<0.01 &{\ } 35 & 19 &0.01 &{\ } 26 & 19 &0.14\\
\bottomrule
\end{tabular}
}
\label{table:chi2_table_absolute}
\end{table}
 
Figure~\ref{fig:diffxstop} shows the measured top-quark transverse momentum distributions, the invariant mass of the $\ttbar$~system,
the rapidities of the top quarks and the $\ttbar$~system, $\HTttbar$\ and $\dPhibt$. In this figure, the distributions are compared with those from the nominal MC sample (with and without NNLO reweighting),
the variations of the nominal MC sample used for the systematic uncertainties, and the alternative NLO generators.
Figure~\ref{fig:diffxstopNNLO} shows the same distributions, this time compared with the three NLO generator set-ups, in each case showing the impact
of the NNLO reweighting.
As observed in previous measurements, the $\pT$~distributions of the top quarks are seen to be softer in data than in simulation.
Reweighting the MC predictions to the NNLO parton-level prediction gives significantly better agreement between data
and predictions.
This is because the NNLO prediction has a softer top-quark
$\pT$\ distribution than the MC set-ups predict and hence results in a lower predicted cross-section in the fiducial region as seen in Figure~\ref{fig:inclusive_cross_section}.
It is worth noting that the $\pTlep$~distribution
is somewhat anti-correlated with $\pTttbar$~(correlation coefficient of $-0.55$) due to the selection
requirements and hence differences in this distribution and the corresponding large $\chi^2$\ values seen in Tables~\ref{table:chi2_table_NNLO_absolute} and~\ref{table:chi2_table_absolute}
may originate from mismodelling of the $\pTttbar$~distribution rather than
the $\pTlep$~distribution.
The measured $\mttbar$~distribution has a relatively poor $\chi^2$~for all the MC models and in this case the impact of the NNLO reweighting is largest for the
\MGNLOPY set-up, where it brings the prediction into agreement with the data.
The shapes of the top-quark rapidity distributions show good agreement between the data and the MC models.
There is a small slope in the ratio of MC simulation to data in $\yttbar$; however, this disagreement is not found to be significant.
The NNLO reweighting is found to have a very small impact on the level of agreement in the shapes of the rapidity distributions.
 
\begin{figure}
\centering
\subfloat[]{
\includegraphics[width=0.32\textwidth]{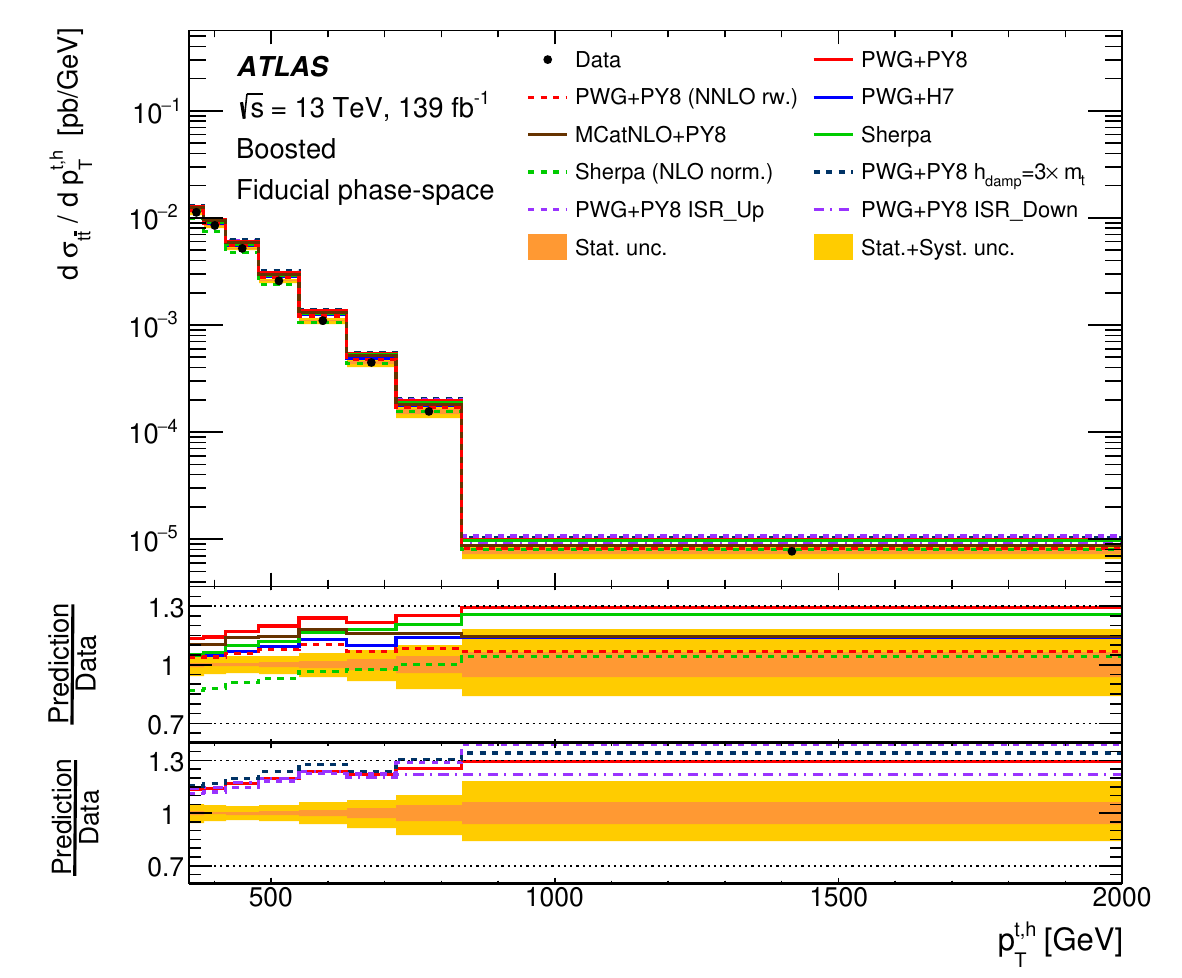}\label{fig:diffxstop:toppt}}
\subfloat[]{
\includegraphics[width=0.32\textwidth]{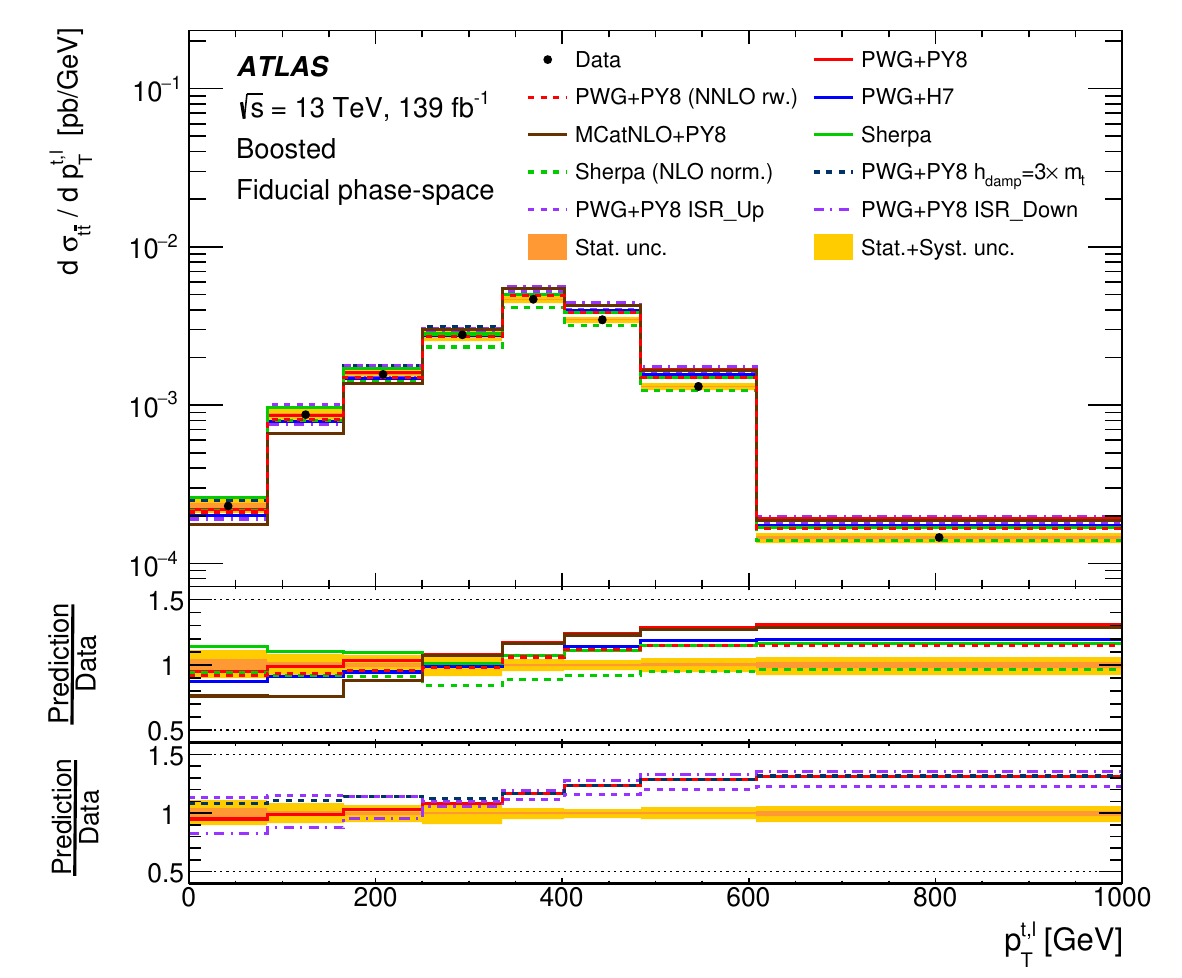}}
\subfloat[]{
\includegraphics[width=0.32\textwidth]{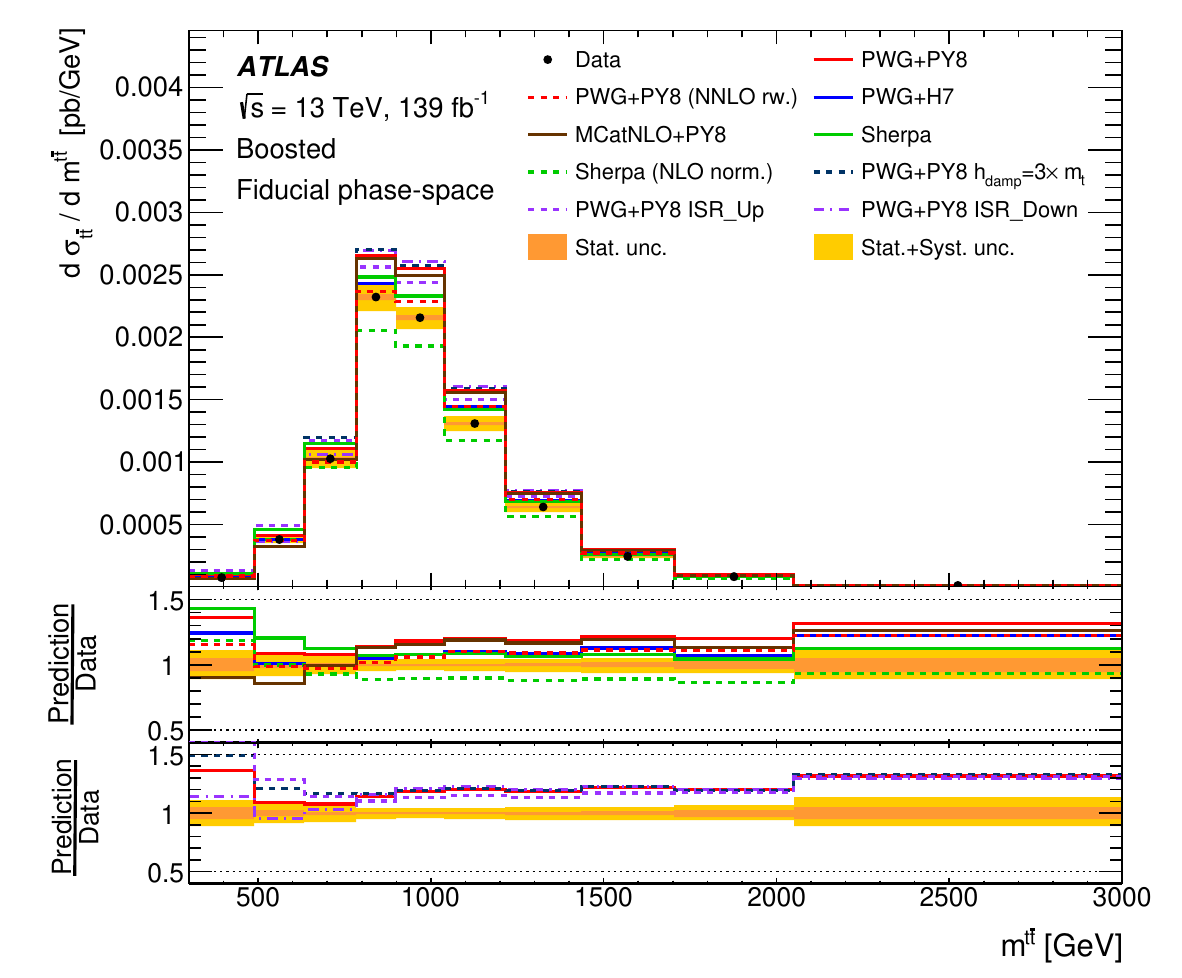}}
 
\subfloat[]{
\includegraphics[width=0.32\textwidth]{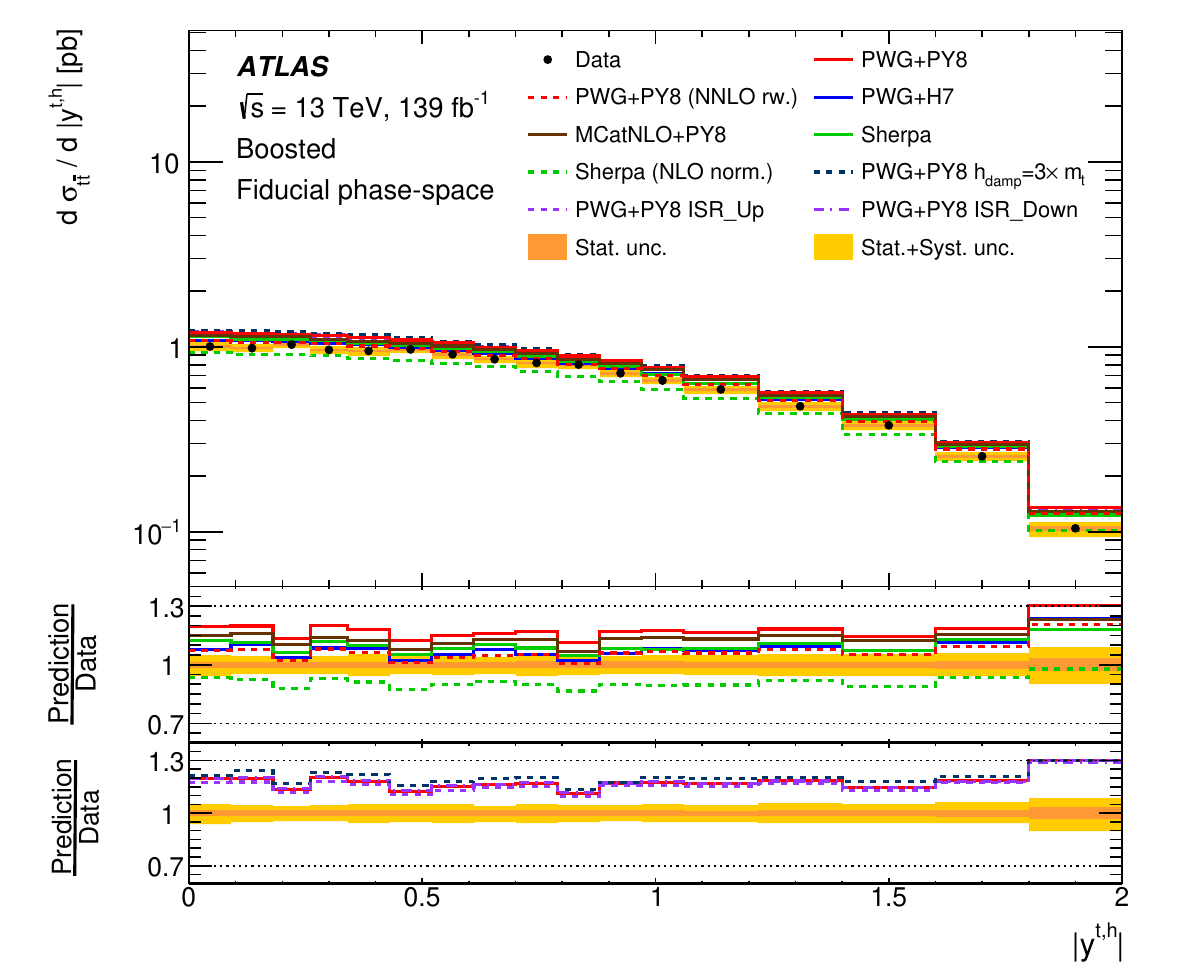}}
\subfloat[]{
\includegraphics[width=0.32\textwidth]{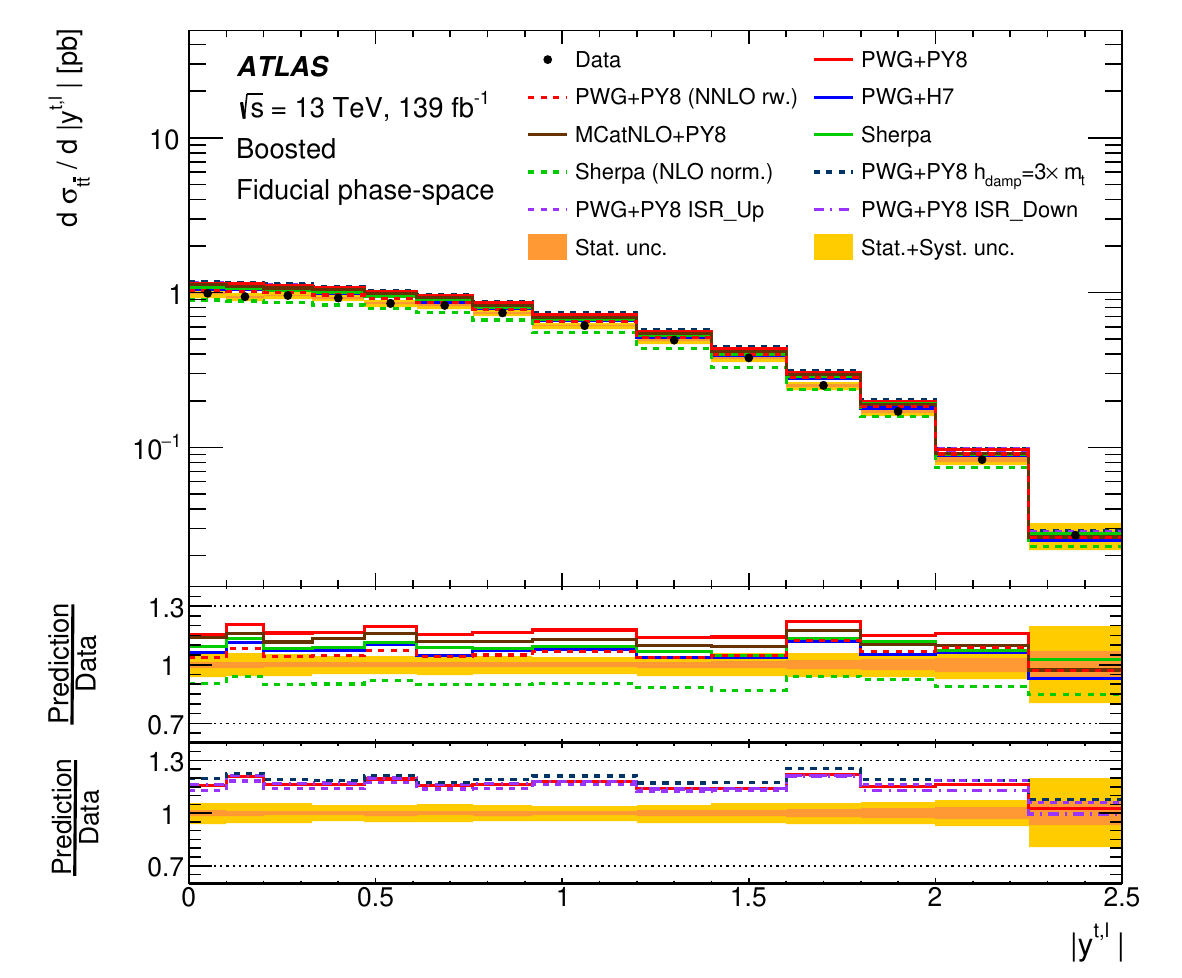}}
\subfloat[]{
\includegraphics[width=0.32\textwidth]{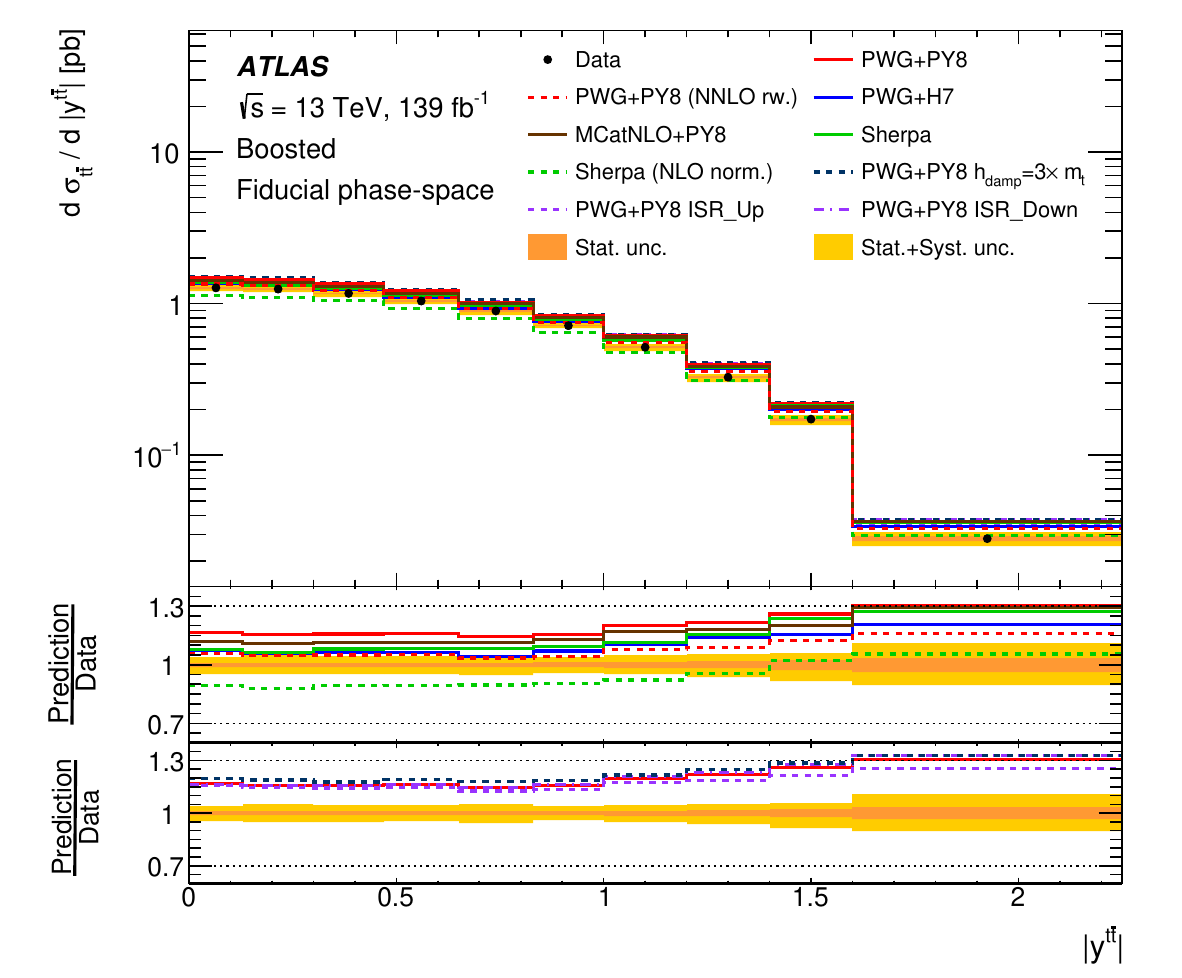}}
 
\subfloat[]{
\includegraphics[width=0.32\textwidth]{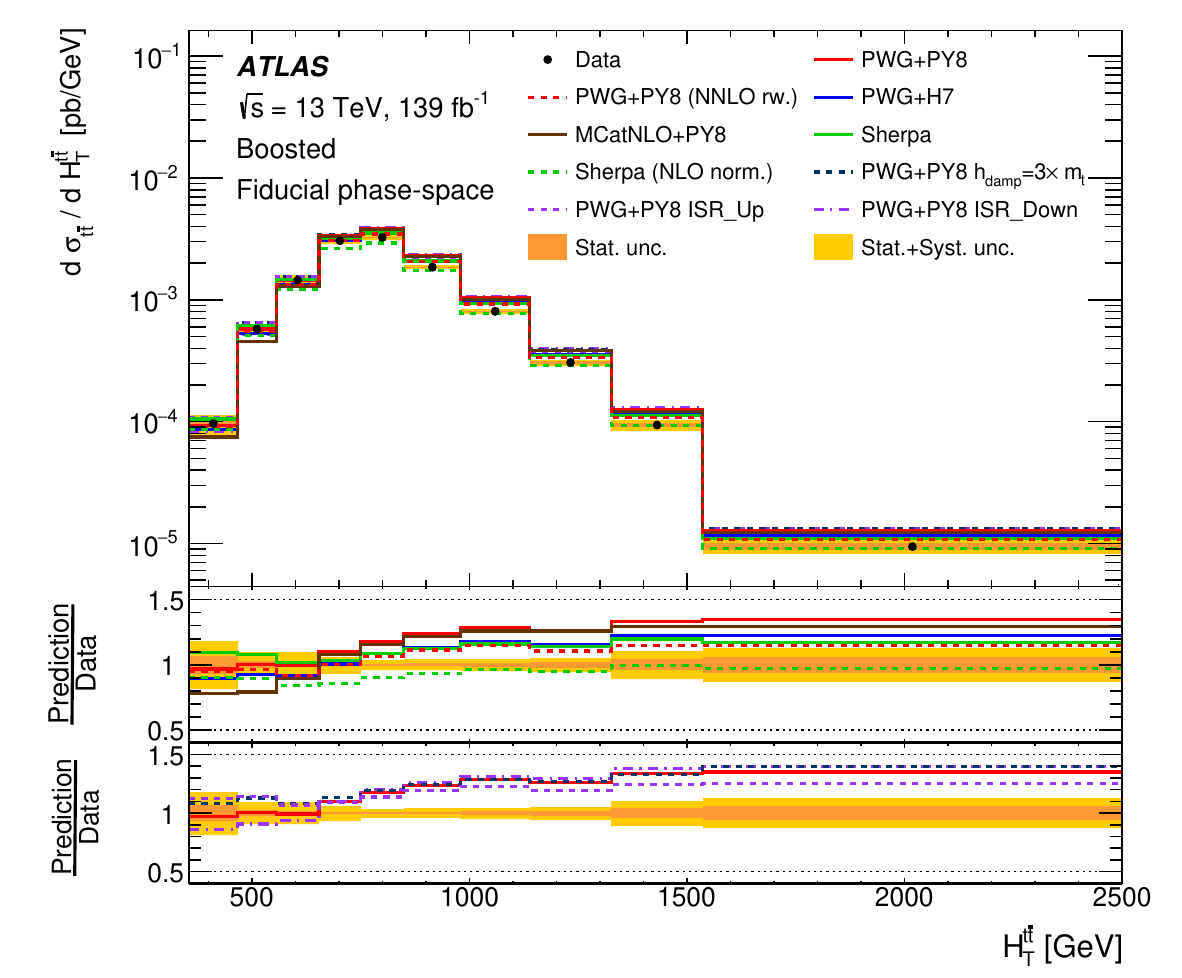}}
\subfloat[]{
\includegraphics[width=0.32\textwidth]{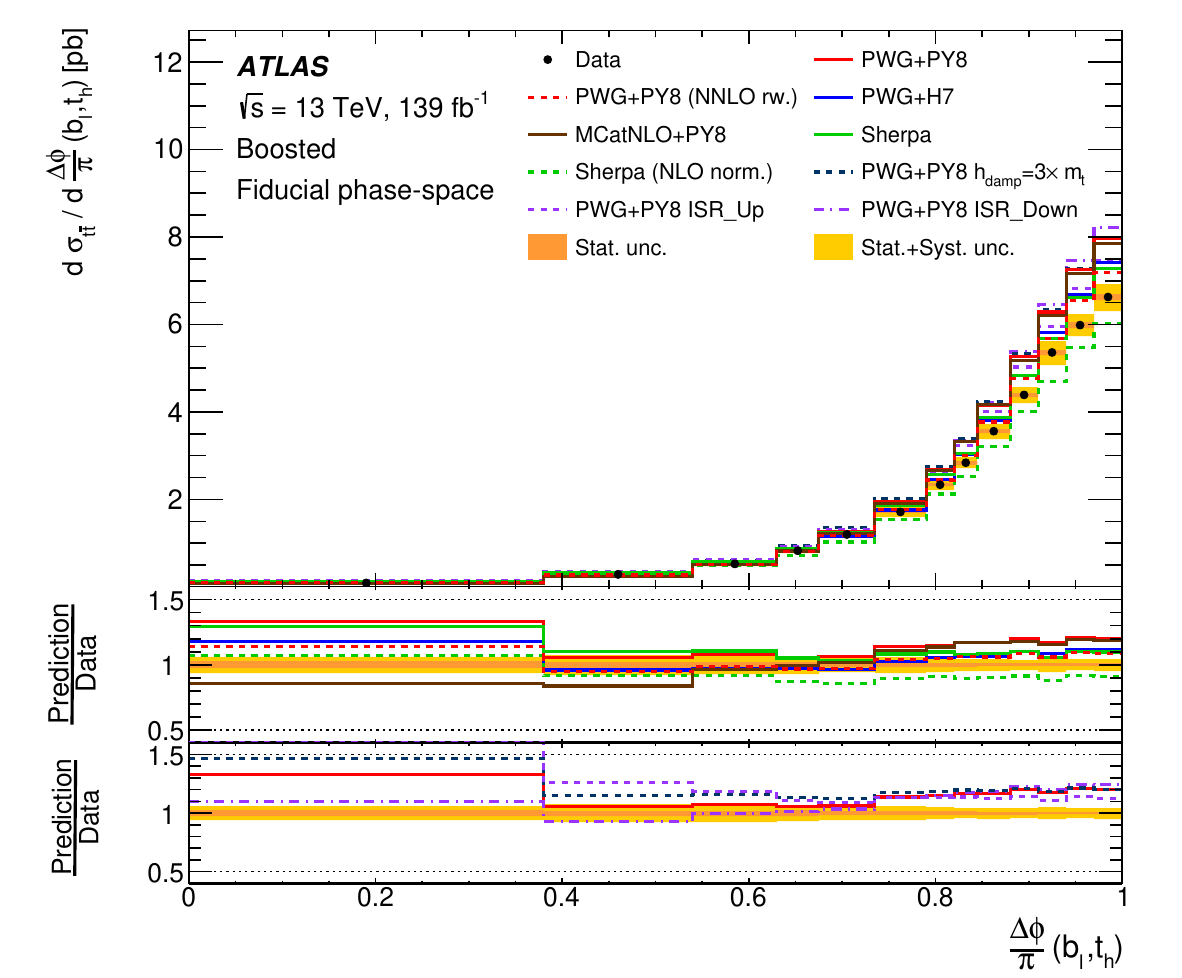}}
\caption{Differential cross-section measurements for the observables related to the kinematics of the top quarks. The cross-section is shown
as a function of (a) $\pThad$, (b) $\pTlep$, (c) $\mttbar$, (d) $\yhad$, (e) $\ylep$,
(f) $\yttbar$, (g) $\HTttbar$~and (h) $\dPhibt$. In each plot the data are compared with predictions from various MC generators.
PWG+PY8 corresponds to the \POWPY sample, PWG+H7 to the \POWHER sample and MCatNLO+PY8 to the \MGNLOPY sample.
The orange band
shows the size of the statistical uncertainty and the yellow band shows the size of the total uncertainty. The lower two panels
in each subfigure display the ratios of the different predictions to the data.}
\label{fig:diffxstop}
\end{figure}
 
\begin{figure}
\centering
\subfloat[]{
\includegraphics[width=0.32\textwidth]{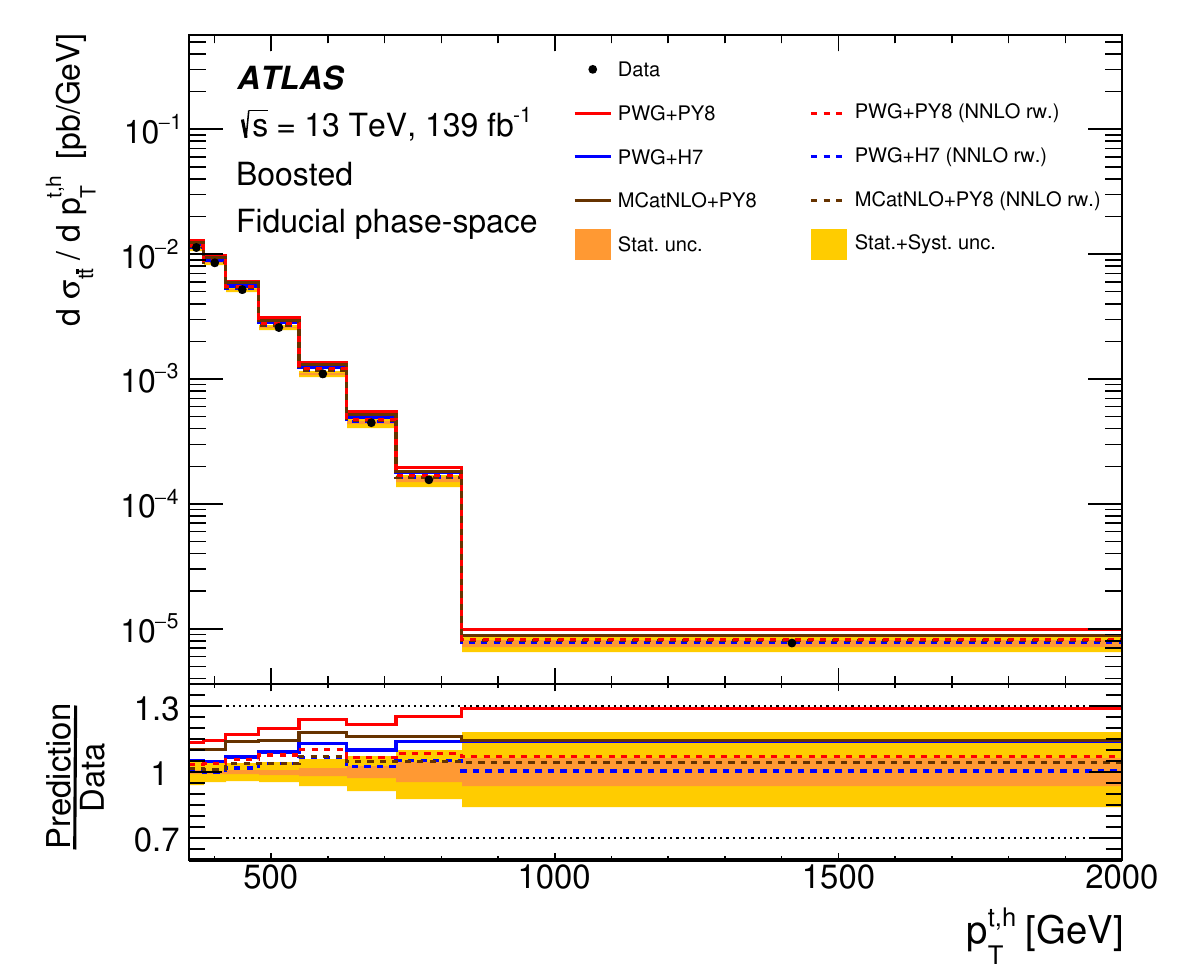}}
\subfloat[]{
\includegraphics[width=0.32\textwidth]{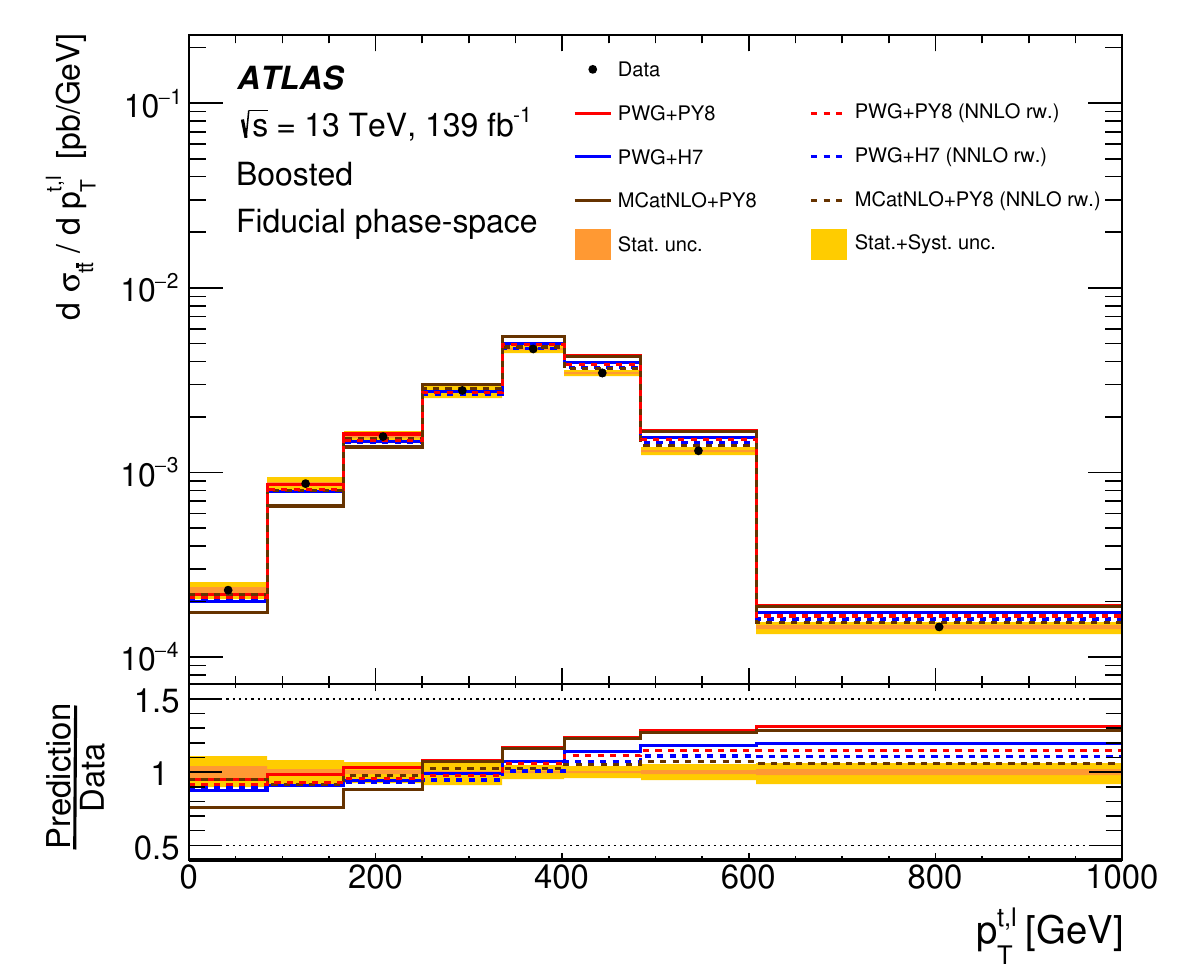}}
\subfloat[]{
\includegraphics[width=0.32\textwidth]{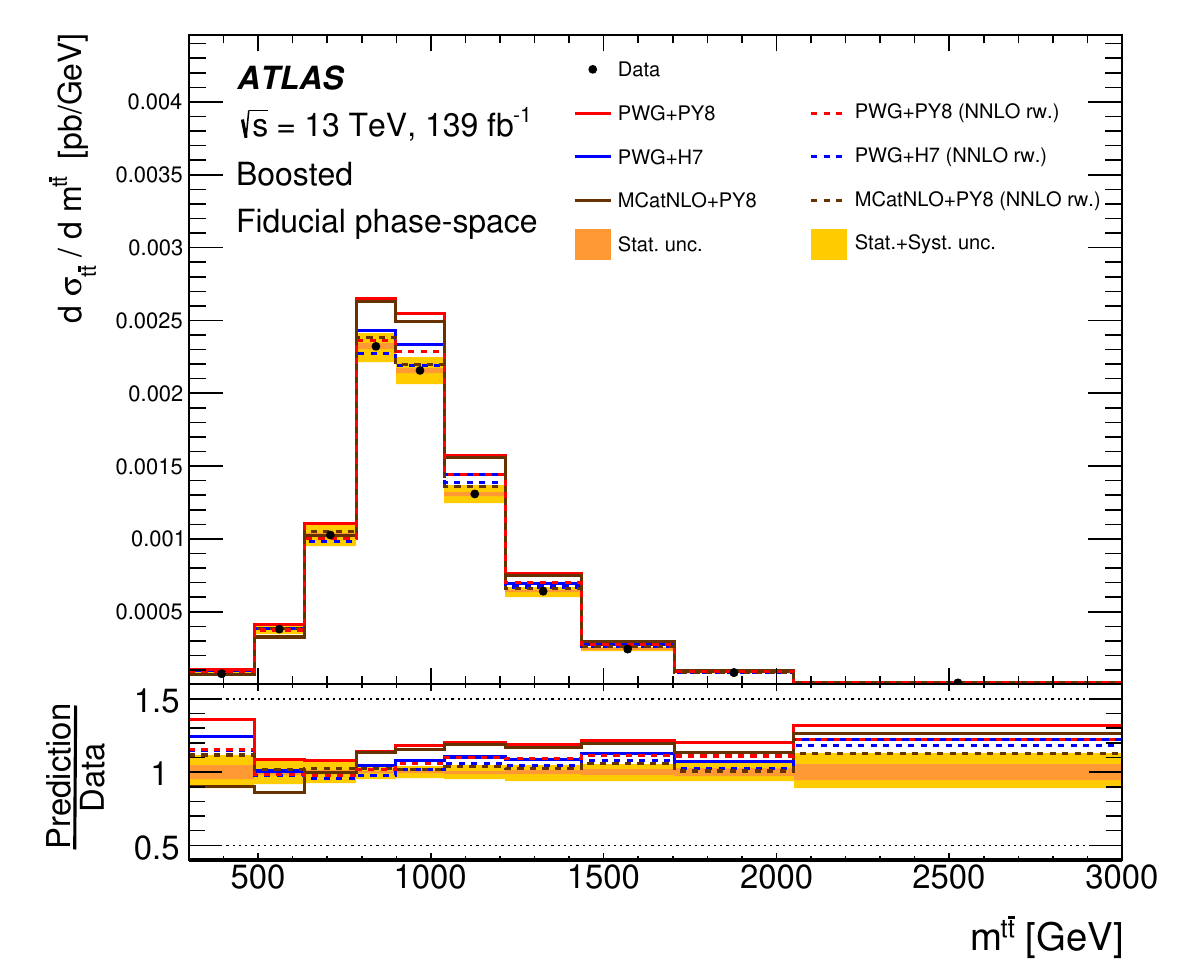}}
 
\subfloat[]{
\includegraphics[width=0.32\textwidth]{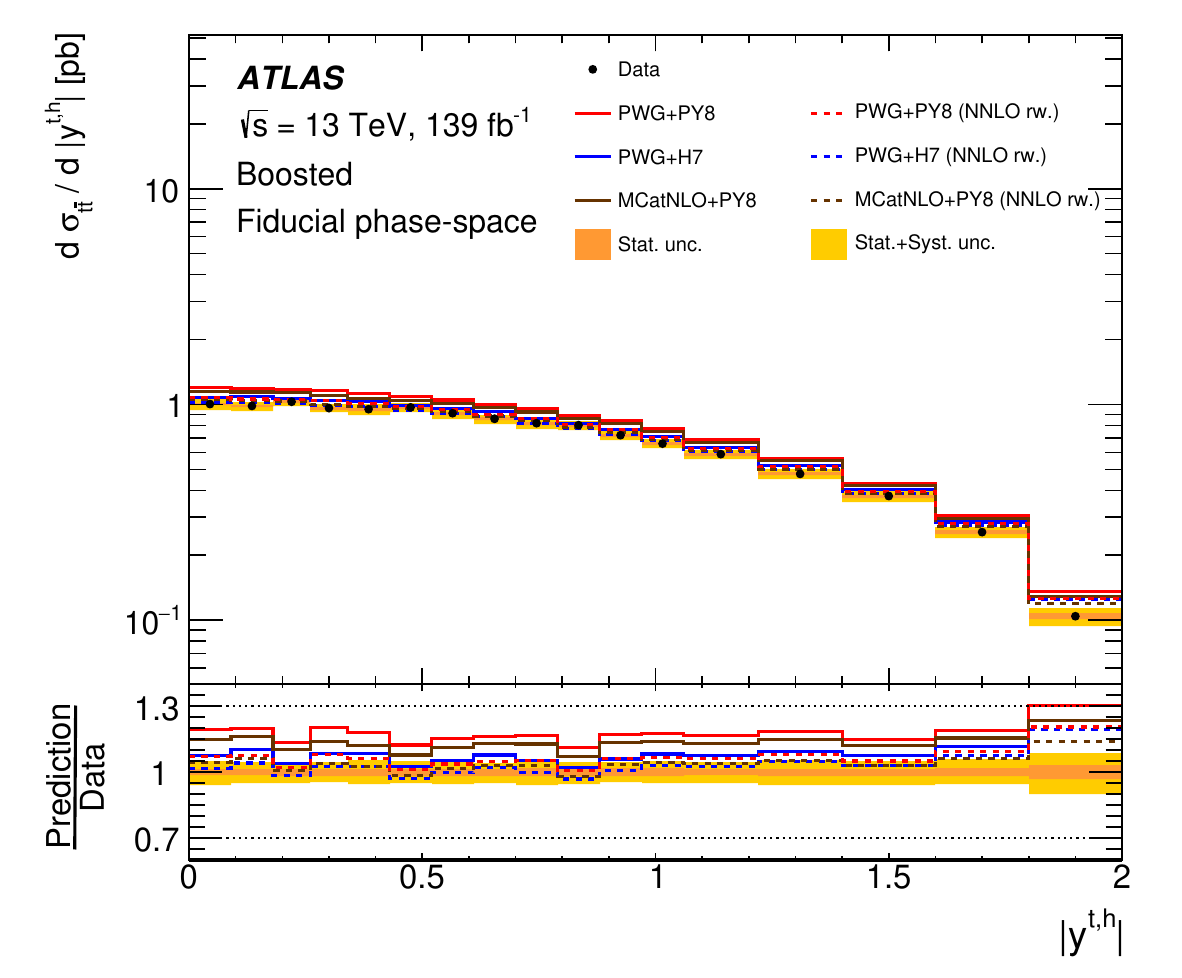}}
\subfloat[]{
\includegraphics[width=0.32\textwidth]{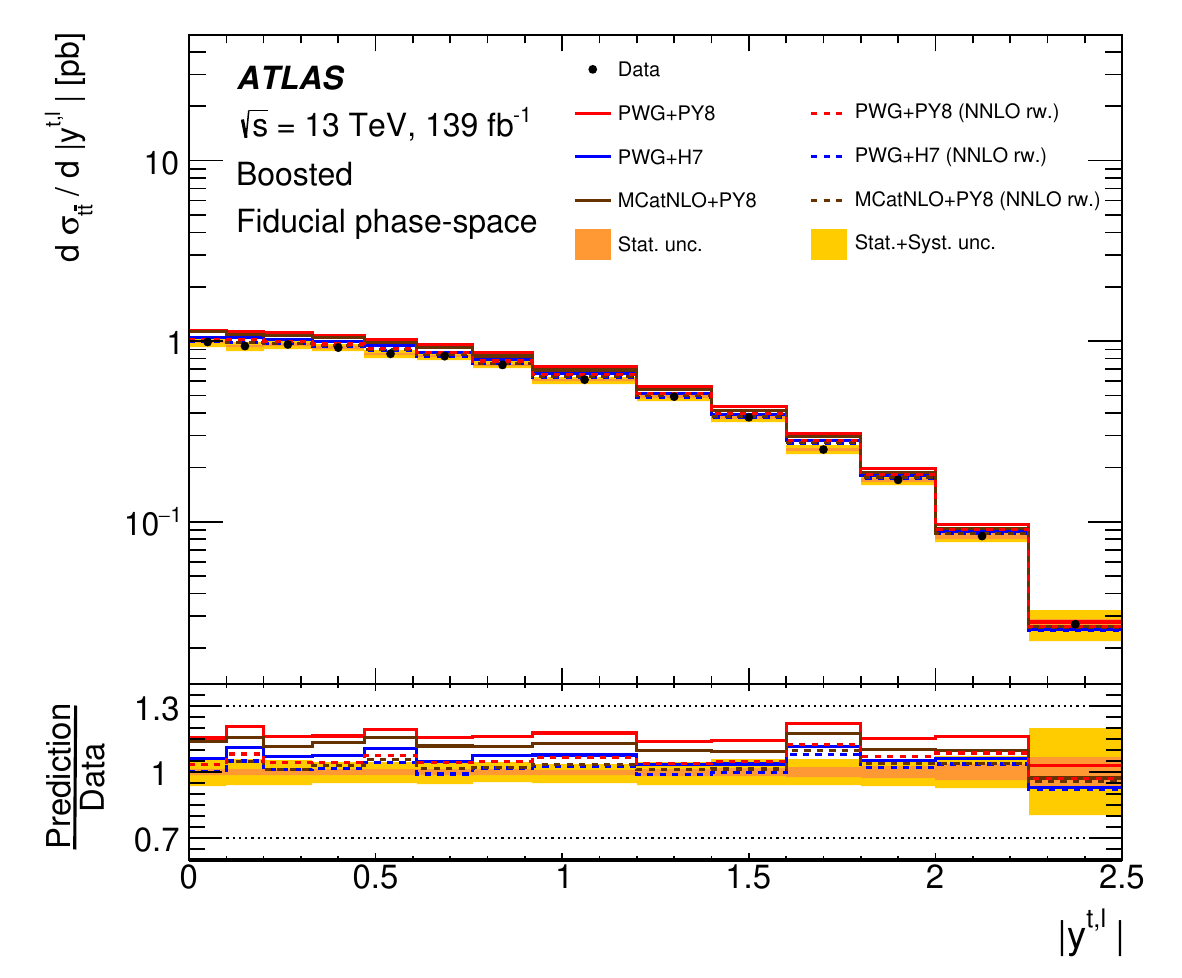}}
\subfloat[]{
\includegraphics[width=0.32\textwidth]{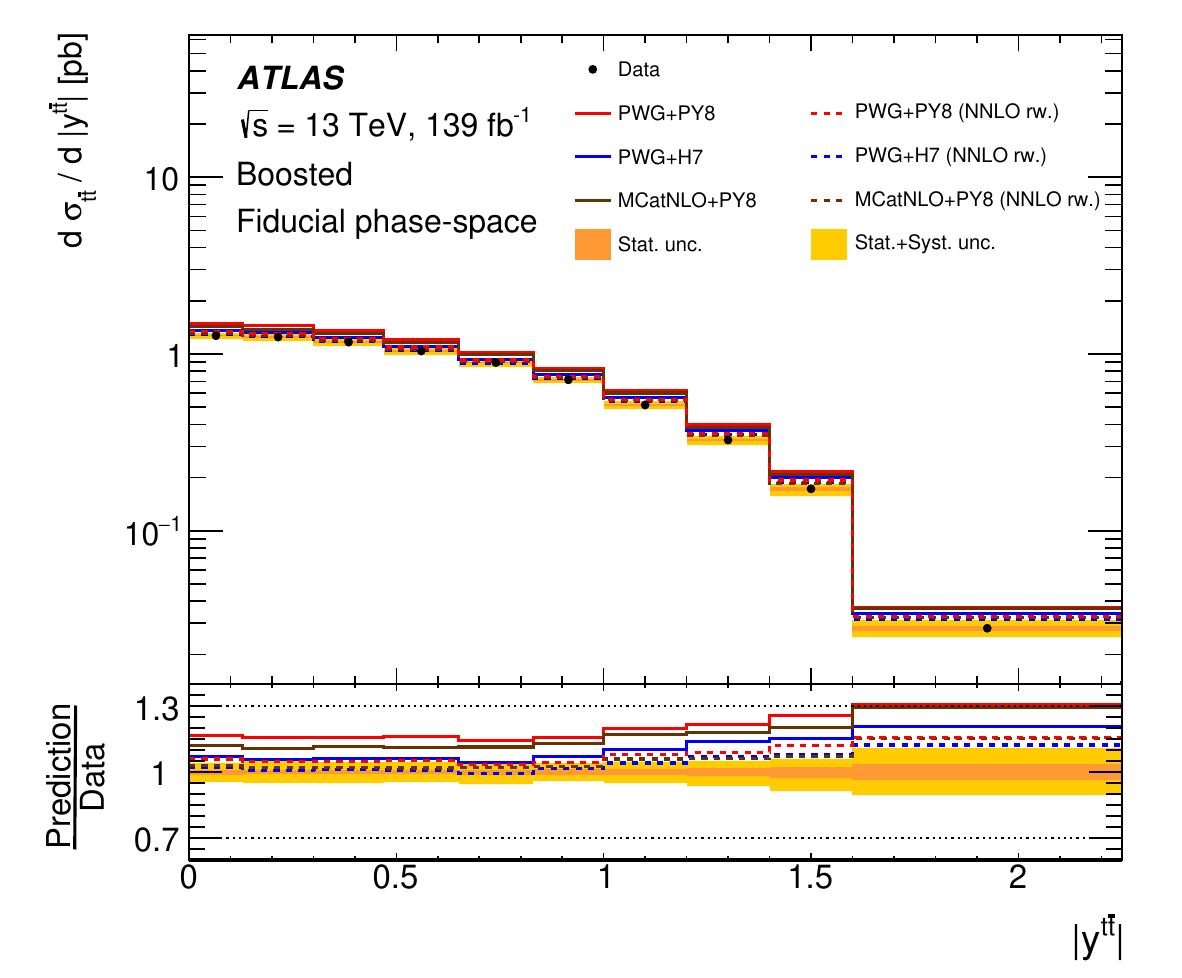}}
 
\subfloat[]{
\includegraphics[width=0.32\textwidth]{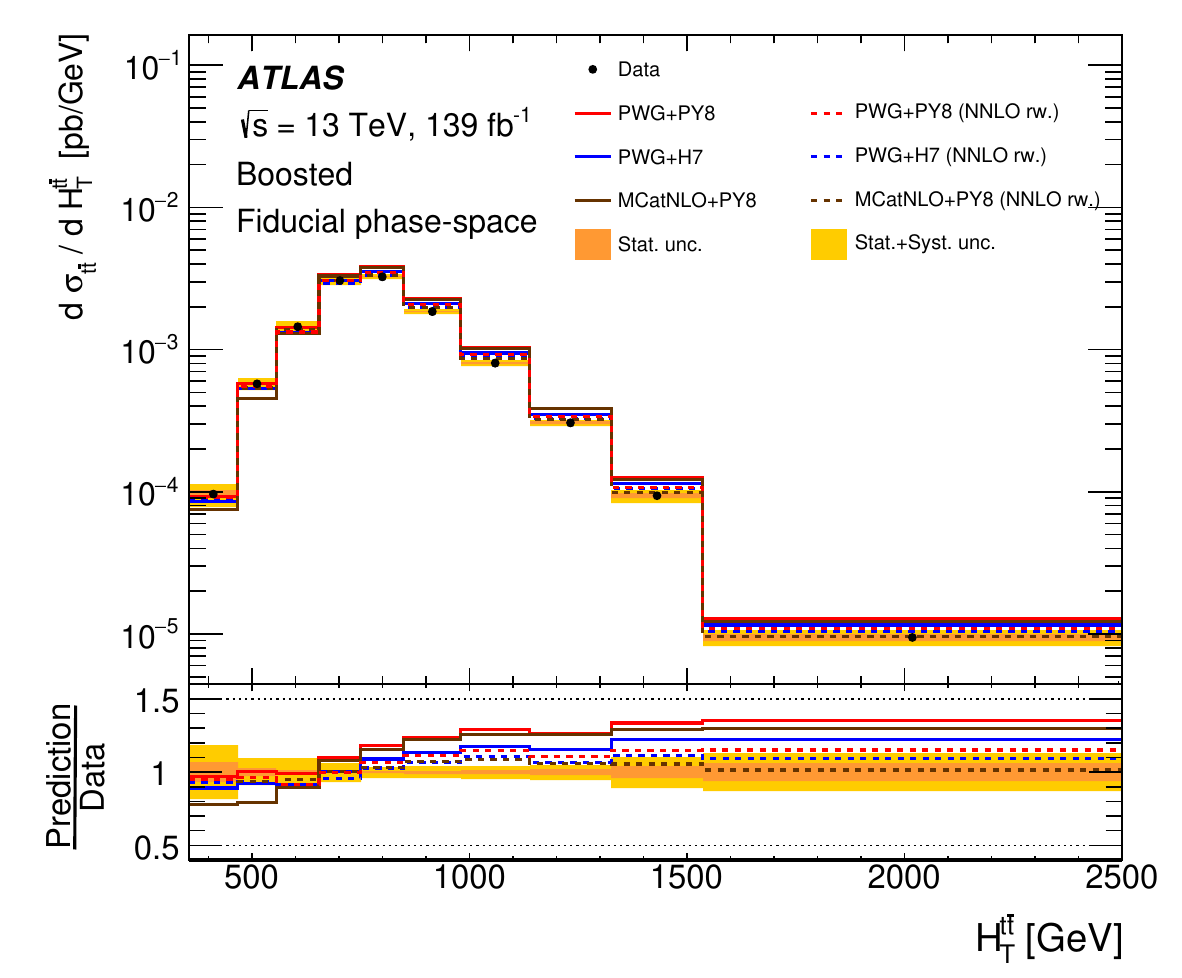}}
\subfloat[]{
\includegraphics[width=0.32\textwidth]{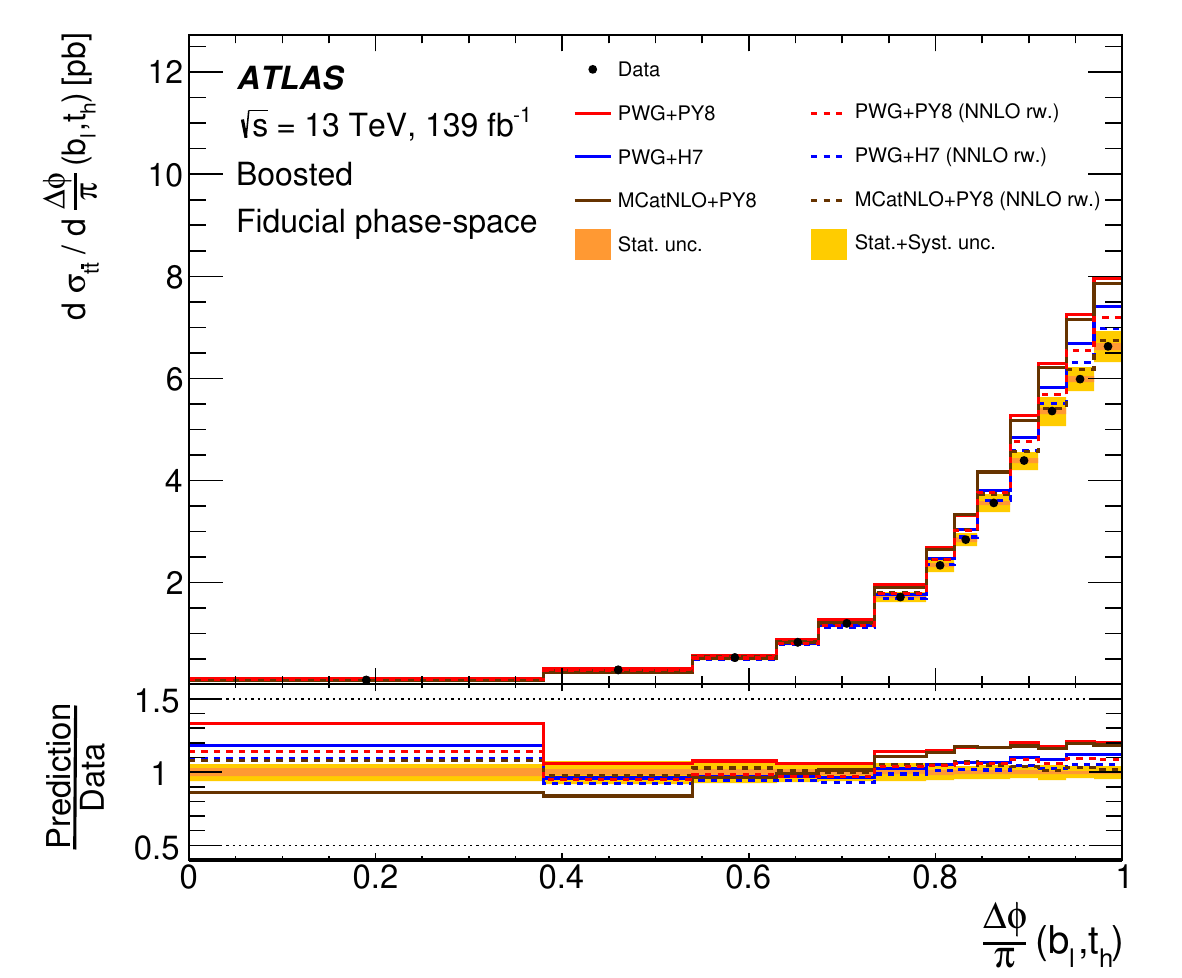}}
\caption{Differential cross-section measurements for the observables related to the kinematics of the top quarks.
The cross-section is shown
as a function of (a) $\pThad$, (b) $\pTlep$, (c) $\mttbar$, (d) $\yhad$, (e) $\ylep$,
(f) $\yttbar$, (g) $\HTttbar$~and (h) $\dPhibt$. In each plot the data are compared with predictions from the NLO generators with and without the NNLO reweighting.
PWG+PY8 corresponds to the \POWPY sample, PWG+H7 to the \POWHER sample and MCatNLO+PY8 to the \MGNLOPY sample.
The orange band
shows the size of the statistical uncertainty and the yellow band shows the size of the total uncertainty. The lower panel
in each subfigure displays the ratios of the different predictions to the data.}
\label{fig:diffxstopNNLO}
\end{figure}
 
The modelling of the additional radiation is probed by the $\pTttbar$, $\dPhitt$~and $\HTall$\ observables,
and the observables involving the additional jets (defined in Section~\ref{sec-meas-obs}). These are shown in Figures~\ref{fig:diffxsExtra1} and~\ref{fig:diffxsExtra2}.
None of the tested models provides a good description of all the measured observables. The \SHERPA\ sample provides the best description of the number
of additional jets (Figure~\ref{fig:diffxsExtra1:nj}), indicating that matrix elements for the additional jets need to be included to model this distribution accurately.
The shape of the leading additional jet's $\pT$~distribution (Figure~\ref{fig:diffxsExtra1:leadaddjet_pt}) is observed to be noticeably different between the \MGNLO\ and \POWHEG\ set-ups,
indicating that the technique for matching the matrix element and the parton shower is a relevant issue.
The difficulty of accurately modelling the extra jet radiation is illustrated in Figures~\ref{fig:diffxsExtra1:leadaddjet_pt} and~\ref{fig:diffxsExtra2:subleadaddjet_pt}
by the leading and sub-leading additional jets' $\pT$~distributions, where using the \POWPY sample with less
ISR radiation gives better agreement than the nominal sample in $\pTe{2}$\ but worse agreement in $\pTe{1}$.
The observables that involve the sub-leading additional jet are expected to have sensitivity to the parton shower in the tested MC models.
Differences between the models using \PYTHIA\ and \HERWIG\ are seen in the $\dPhiee$~and $\dPhieth{2}$~distributions, where
the \POWHER set-up is seen to have worse agreement with the data than the set-ups using \PYTHIA.
 
\begin{figure}
\centering
\subfloat[]{
\includegraphics[width=0.32\textwidth]{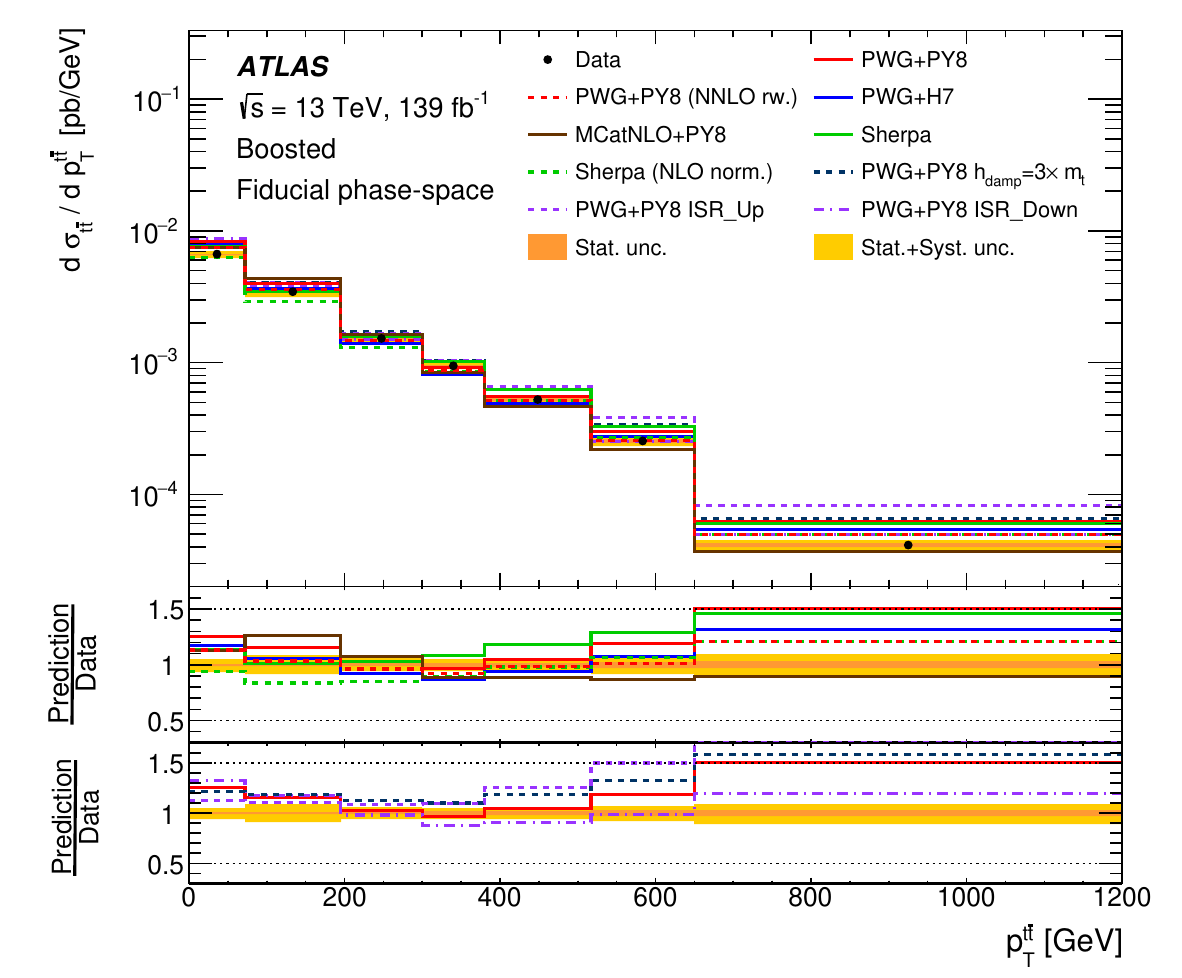}}
\subfloat[]{
\includegraphics[width=0.32\textwidth]{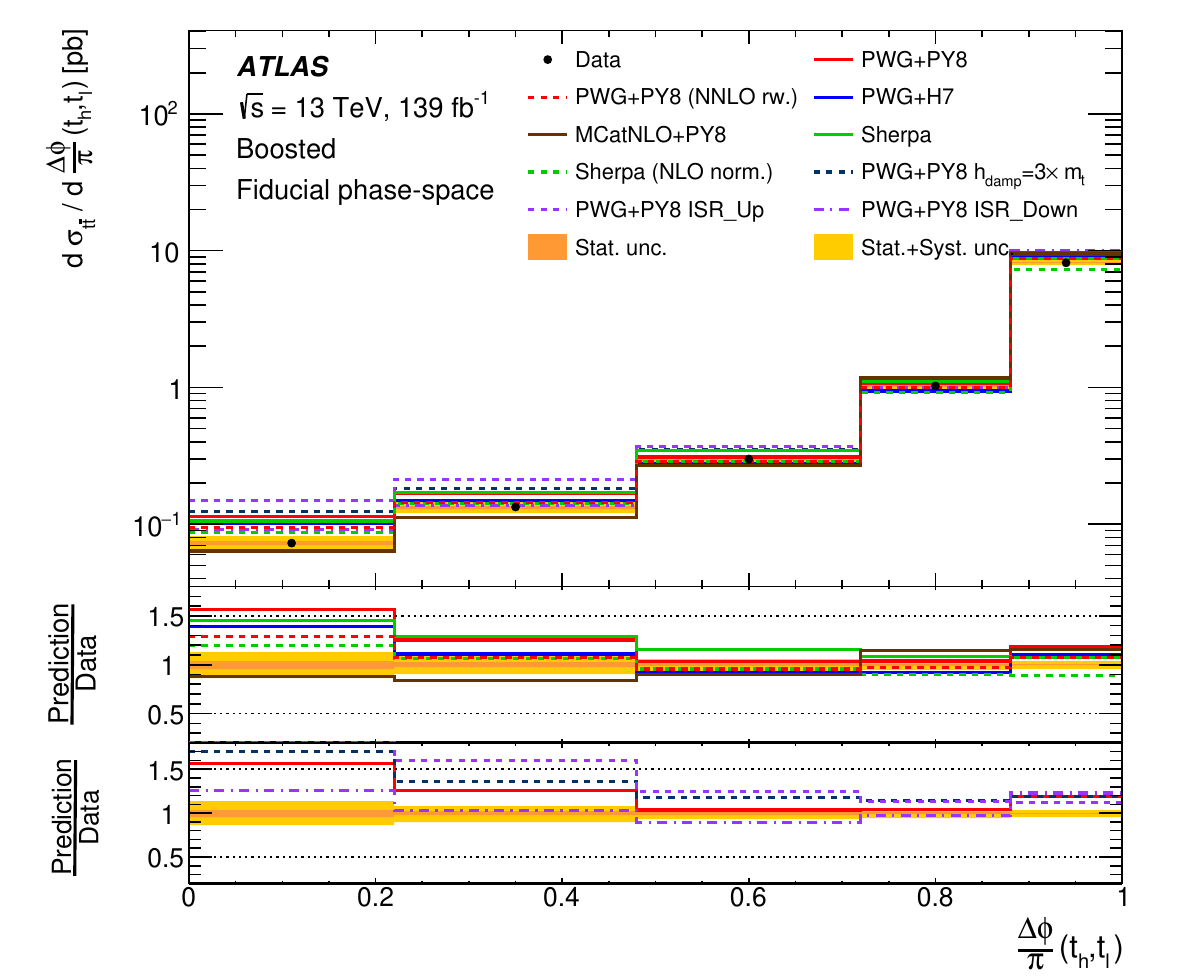}}
\subfloat[]{
\includegraphics[width=0.32\textwidth]{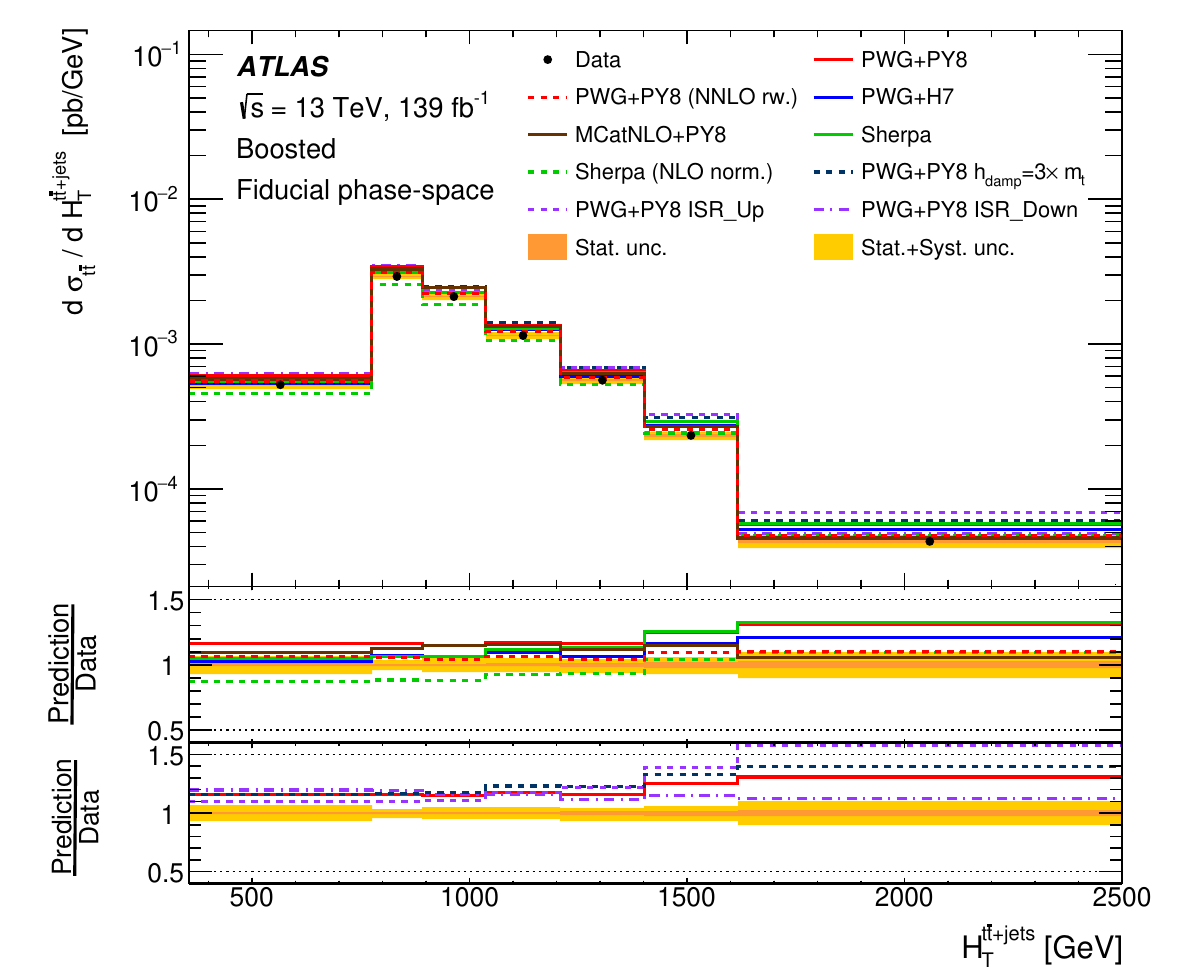}}
 
\subfloat[]{
\includegraphics[width=0.32\textwidth]{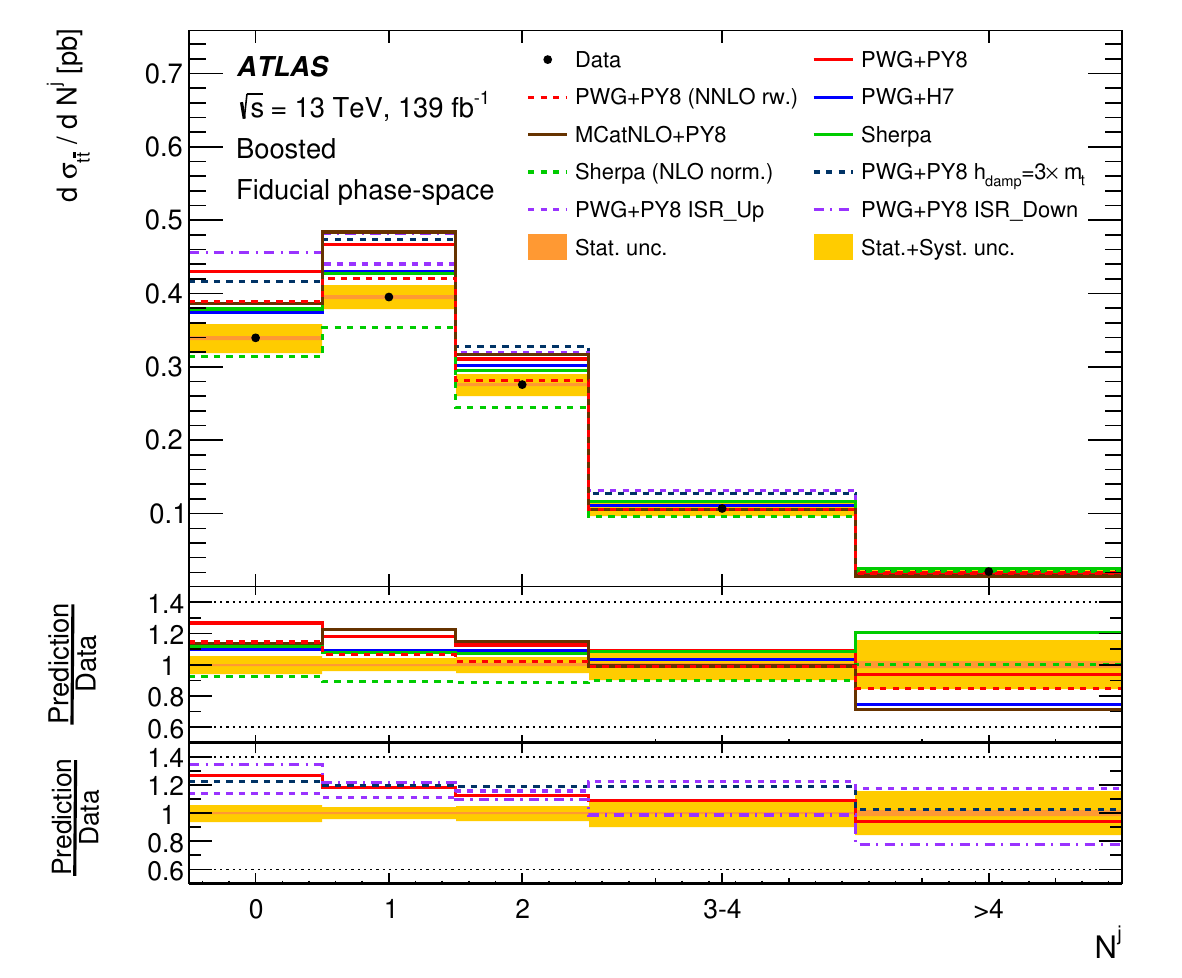}\label{fig:diffxsExtra1:nj}}
\subfloat[]{
\includegraphics[width=0.32\textwidth]{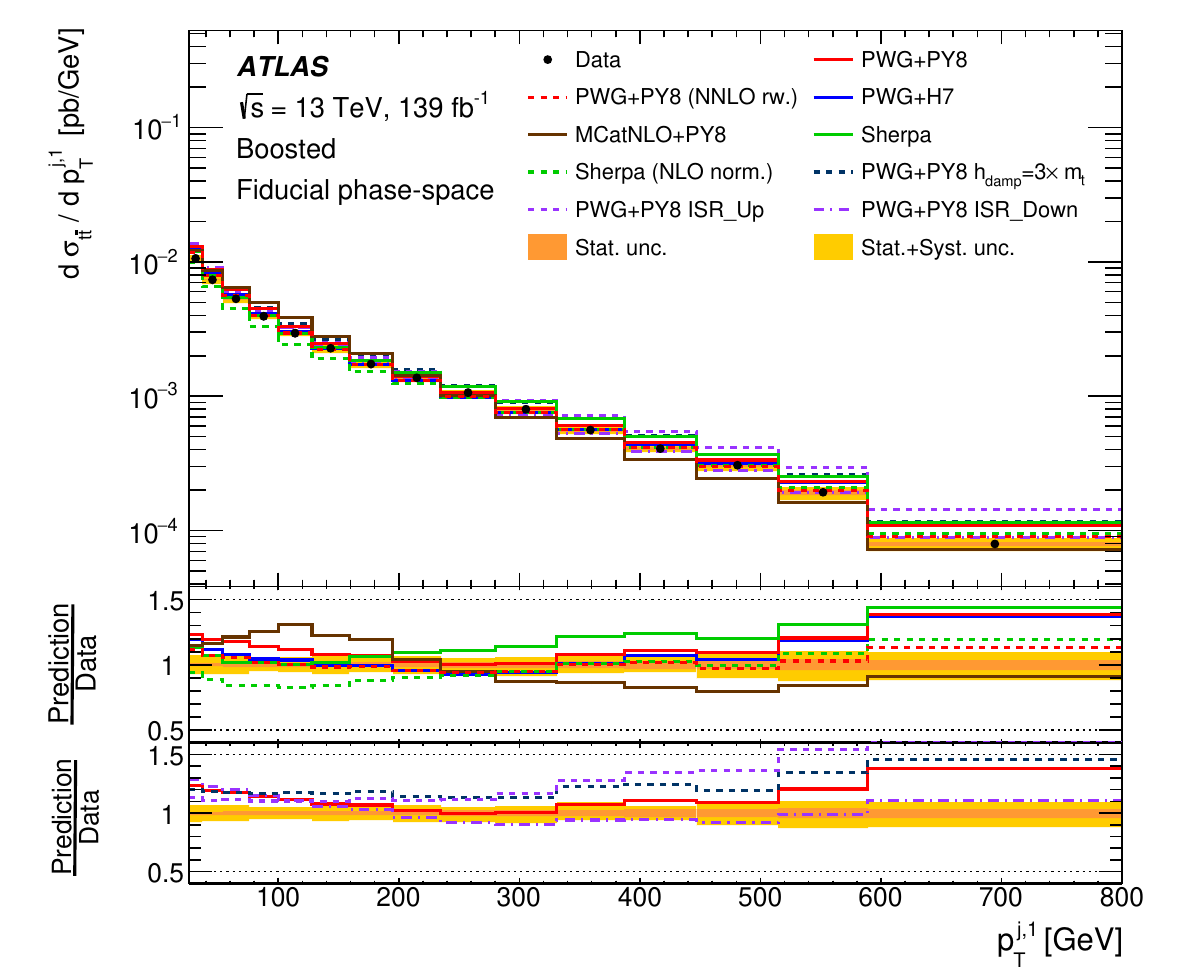}\label{fig:diffxsExtra1:leadaddjet_pt}}
\subfloat[]{
\includegraphics[width=0.32\textwidth]{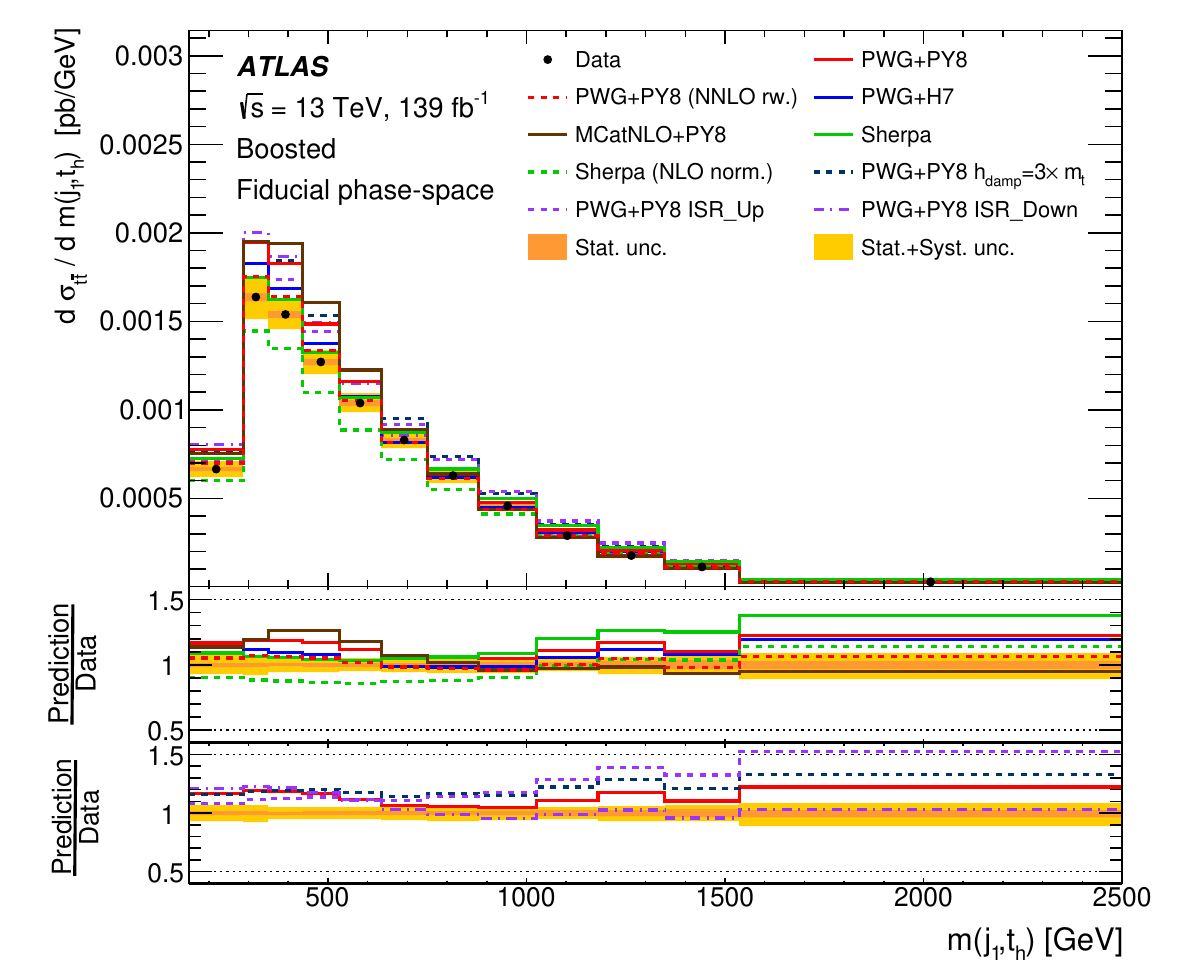}}
 
\subfloat[]{
\includegraphics[width=0.32\textwidth]{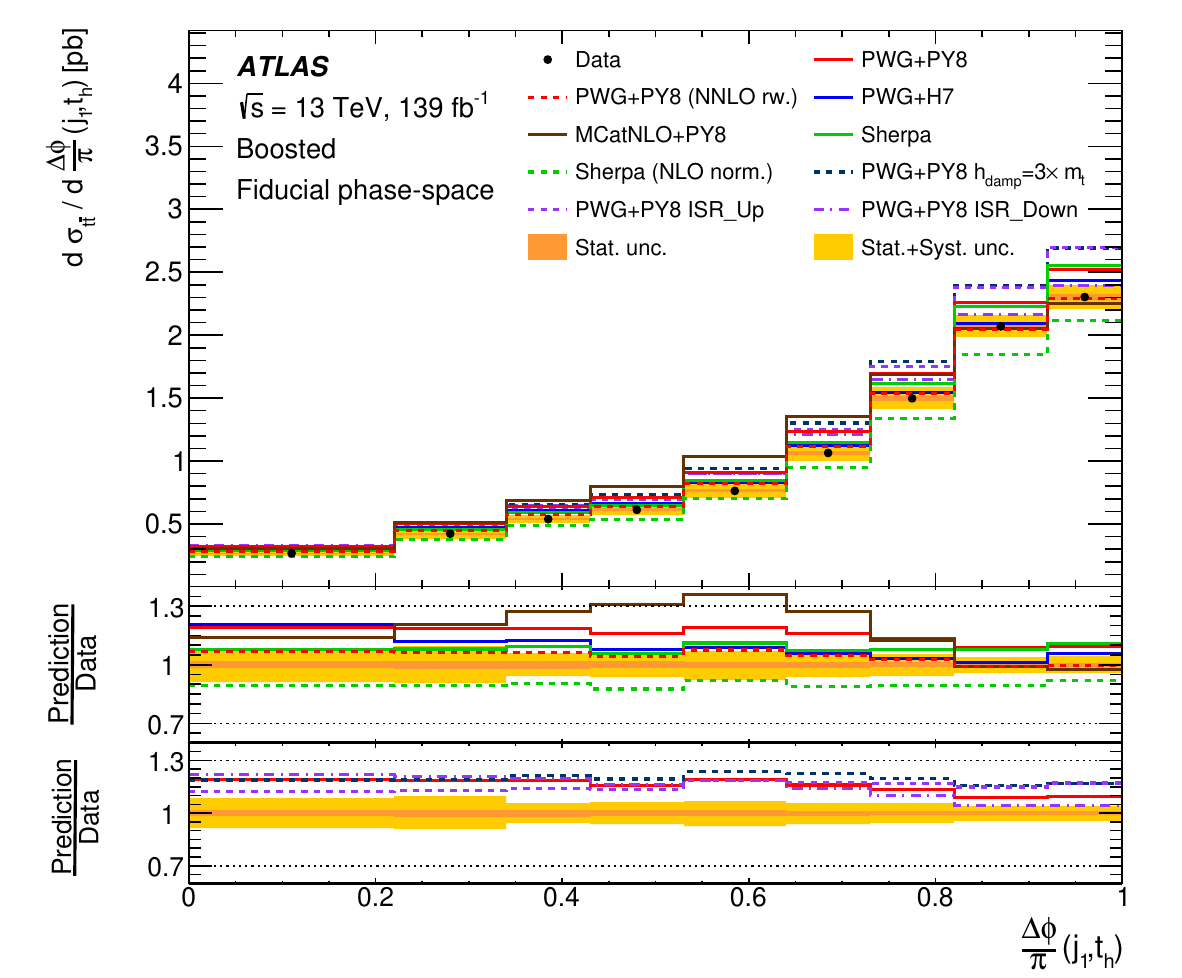}}
\caption{Differential cross-section measurements for the observables related to jets produced in association with the $\ttbar$\ system.
The cross-section is shown as a function of
(a) $\pTttbar$, (b) $\dPhitt$, (c) $\HTall$, (d) $\Ne$,
(e) $\pTe{1}$, (f) $\meth$ and (g) $\dPhieth{1}$.
In each plot the data are compared with predictions from various MC generators.
PWG+PY8 corresponds to the \POWPY sample, PWG+H7 to the \POWHER sample and MCatNLO+PY8 to the \MGNLOPY sample.
The orange band
shows the size of the statistical uncertainty and the yellow band shows the size of the total uncertainty. The lower two panels
in each subfigure display the ratios of the different predictions to the data.}
\label{fig:diffxsExtra1}
\end{figure}
 
\begin{figure}
\centering
\subfloat[]{
\includegraphics[width=0.32\textwidth]{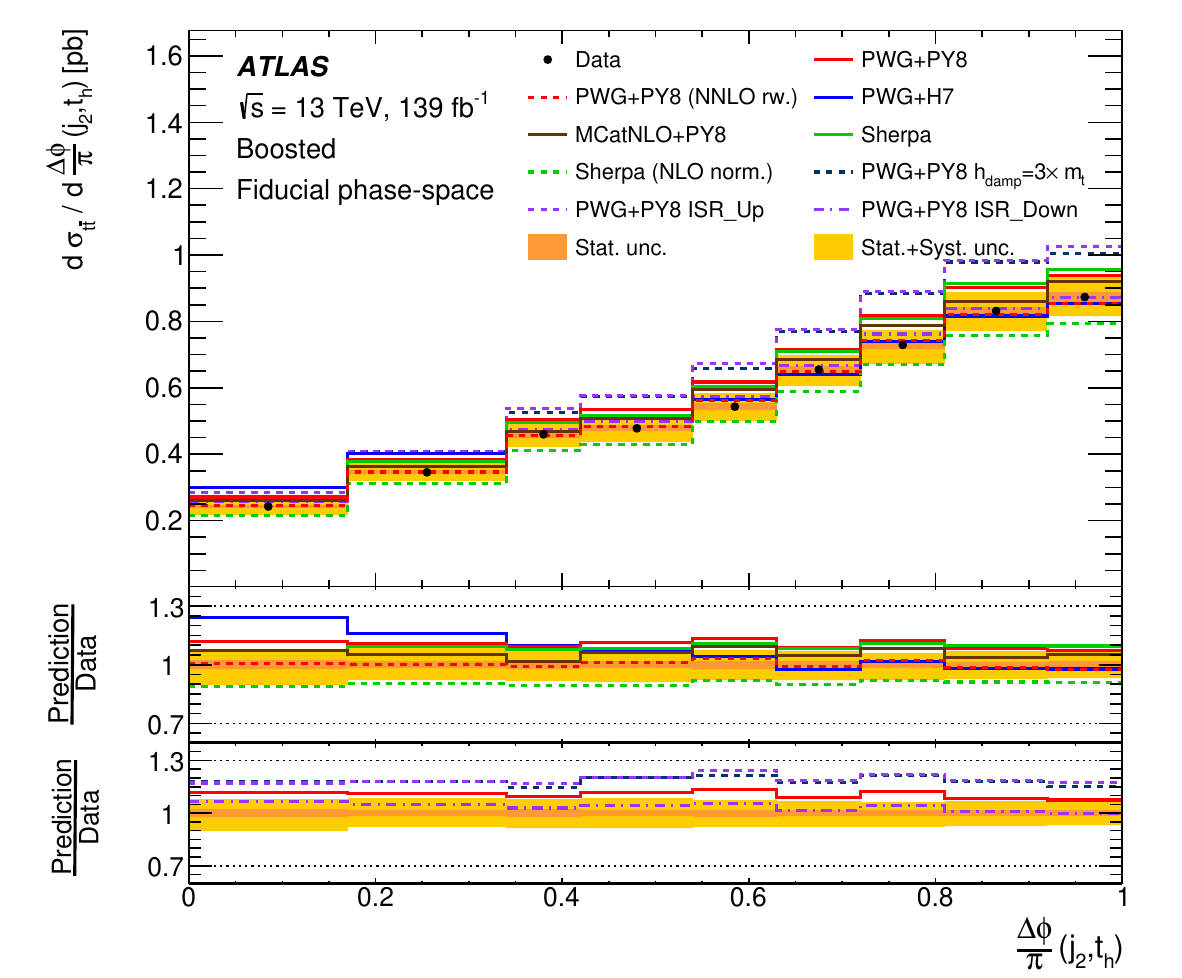}}
\subfloat[]{
\includegraphics[width=0.32\textwidth]{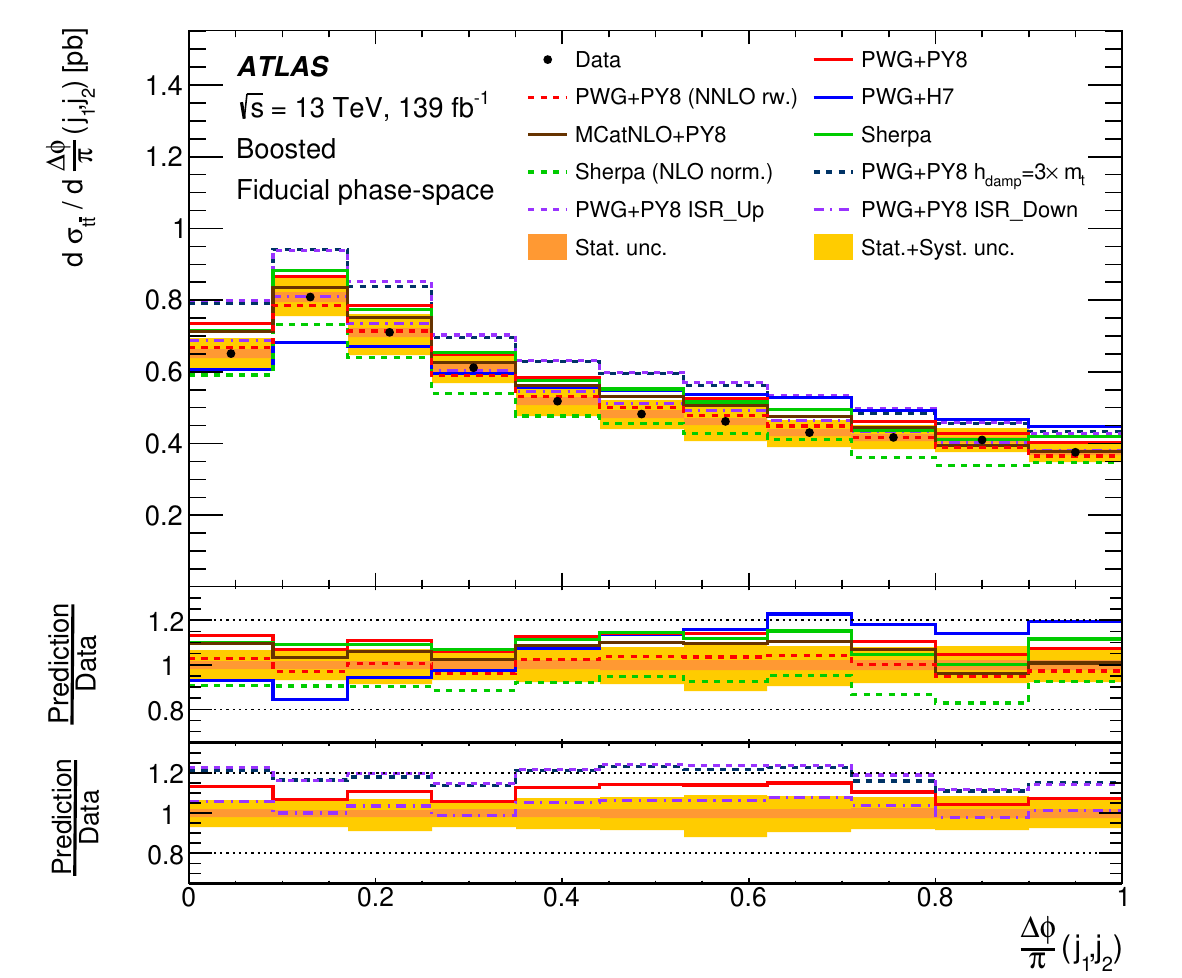}}
\subfloat[]{
\includegraphics[width=0.32\textwidth]{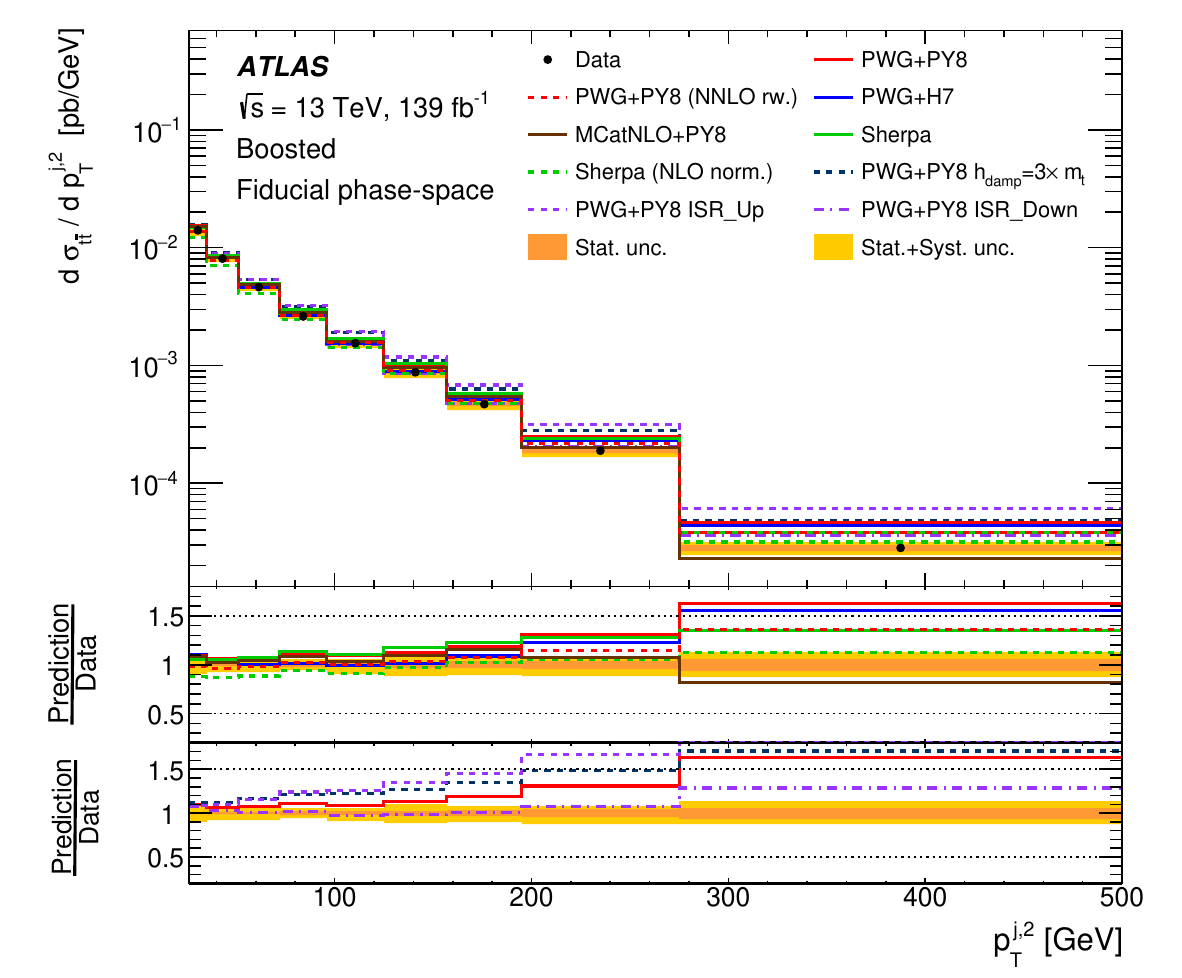}\label{fig:diffxsExtra2:subleadaddjet_pt}}
\caption{
Differential cross-section measurements for the observables probing the second leading additional jet.
The cross-section is shown as a function of
(a) $\dPhieth{2}$, (b) $\dPhiee$ and (c) $\pTe{2}$.
In each plot the data are compared with predictions from various MC generators.
PWG+PY8 corresponds to the \POWPY sample, PWG+H7 to the \POWHER sample and MCatNLO+PY8 to the \MGNLOPY sample.
The orange band
shows the size of the statistical uncertainty and the yellow band shows the size of the total uncertainty. The lower two panels
in each subfigure display the ratios of the different predictions to the data.}
\label{fig:diffxsExtra2}
\end{figure}
 
The two-dimensional cross-section measurement results are shown in Figures~\ref{fig:Results_xsection_abs_2D_leadaddjet_pt_V_naddjet}--\ref{fig:Results_xsection_abs_2D_leadadd_top_deltaPhi_V_naddjet}.
The disagreement in the leading additional jet $\pT$~is observed
to be larger for events with high $\pThad$, and the $\dPhieth{1}$~distribution is also seen to be poorly modelled at high $\pThad$ by all the tested
models.
The differences between the NLO models are often larger than the precision of the measurements, which indicates that
comparisons with generators
with higher precision would be beneficial. Higher precision could be obtained with a full NNLO plus parton shower MC model that
may become available in the future~\cite{mazzitelli2020nexttonexttoleading}.


\begin{figure}[htbp]
\centering
\subfloat[]{
\includegraphics[width=0.5\textwidth]{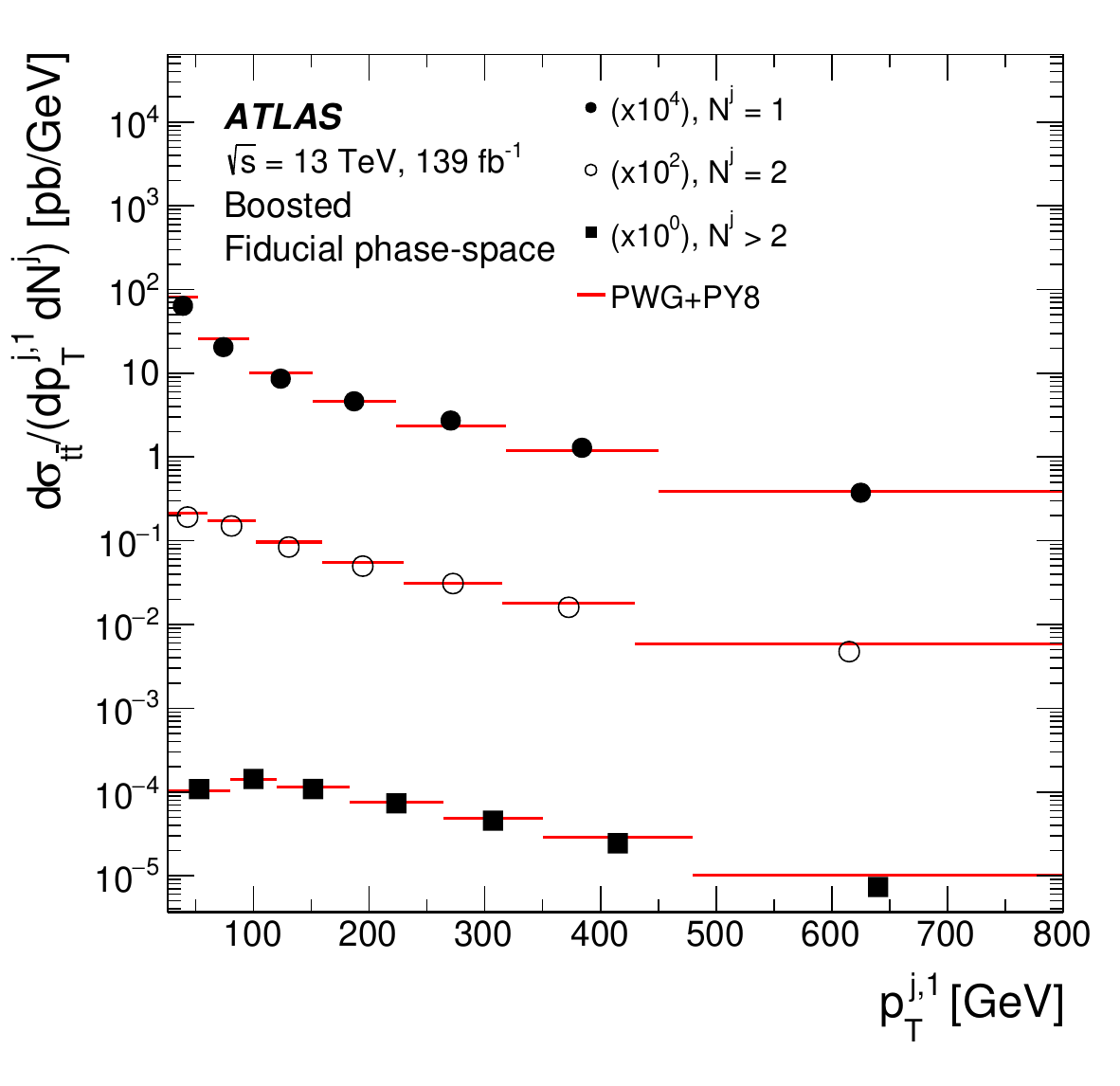}}
 
\subfloat[]{
\includegraphics[width=0.9\textwidth]{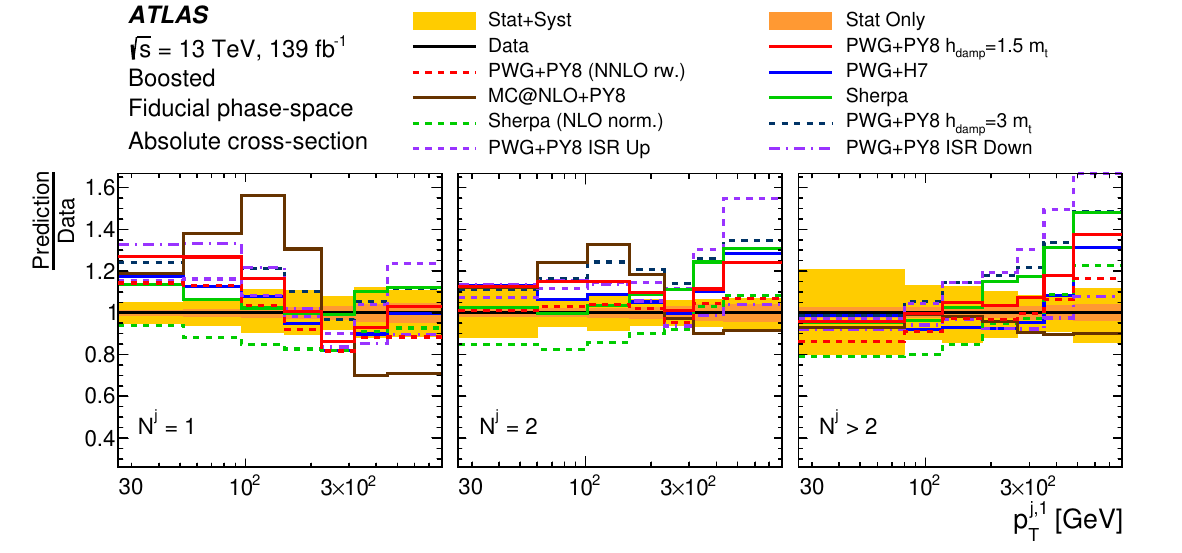}}
\caption{(a) Differential cross-section measurements as a function of the $\pt$\ of the leading additional jet in bins of additional jet multiplicity are compared with the prediction from the \POWPY MC generator. The measurement and the predictions are normalised by the factors shown in parentheses to aid visibility. (b) Ratio of the measured absolute cross-section to different NLO, and NLO reweighted to NNLO, predictions of $t\bar{t}$ signal for the same differential variables.
PWG+PY8 corresponds to the \POWPY sample, PWG+H7 to the \POWHER sample and MCatNLO+PY8 to the \MGNLOPY sample.
The yellow band represents the total uncertainty of the measured differential cross-section while the orange band shows the statistical component.}
\label{fig:Results_xsection_abs_2D_leadaddjet_pt_V_naddjet}
\end{figure}

\begin{figure}[p]
\centering
\subfloat[]{
\includegraphics[width=0.5\textwidth]{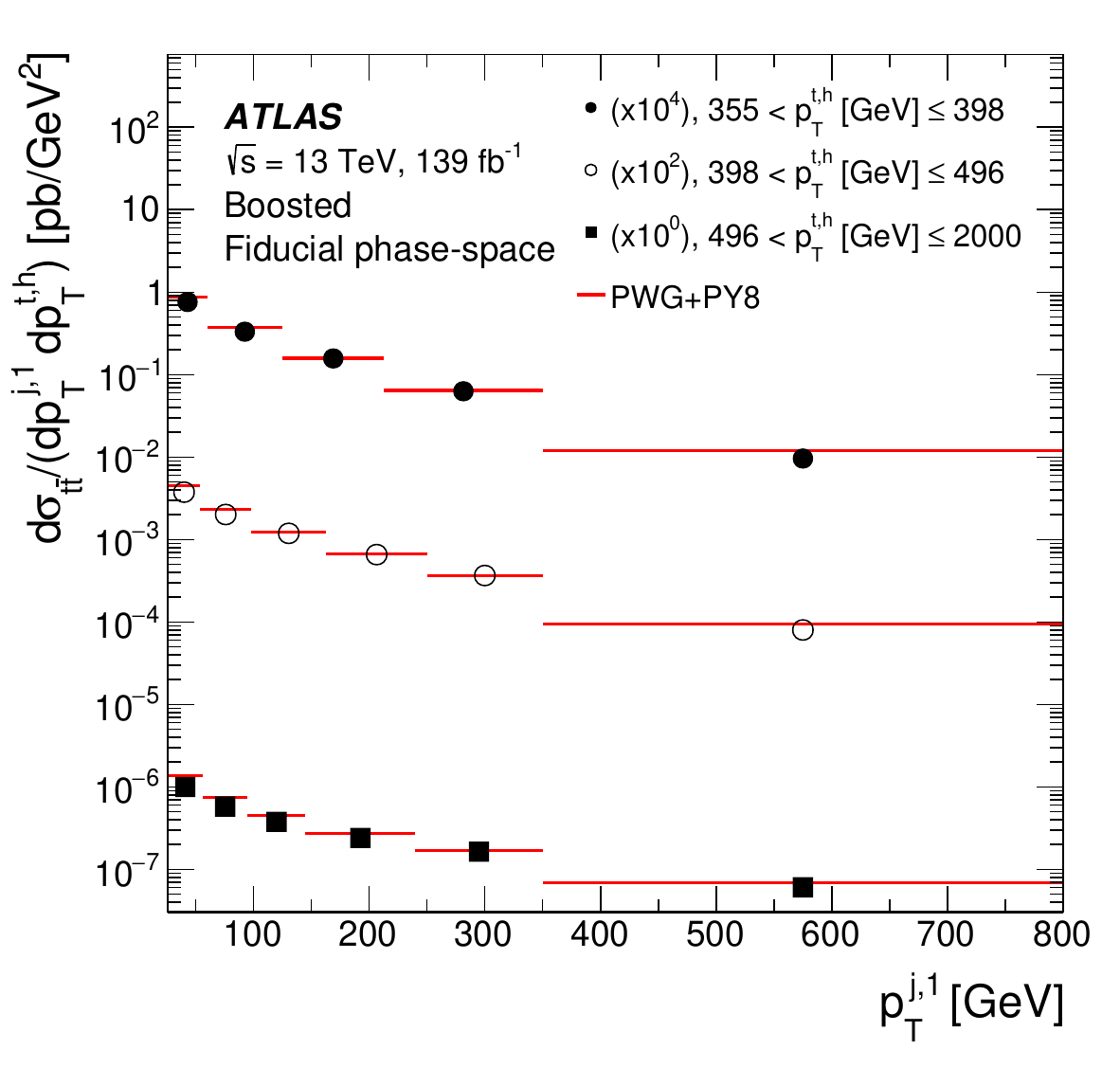}}
 
\subfloat[]{
\includegraphics[width=0.9\textwidth]{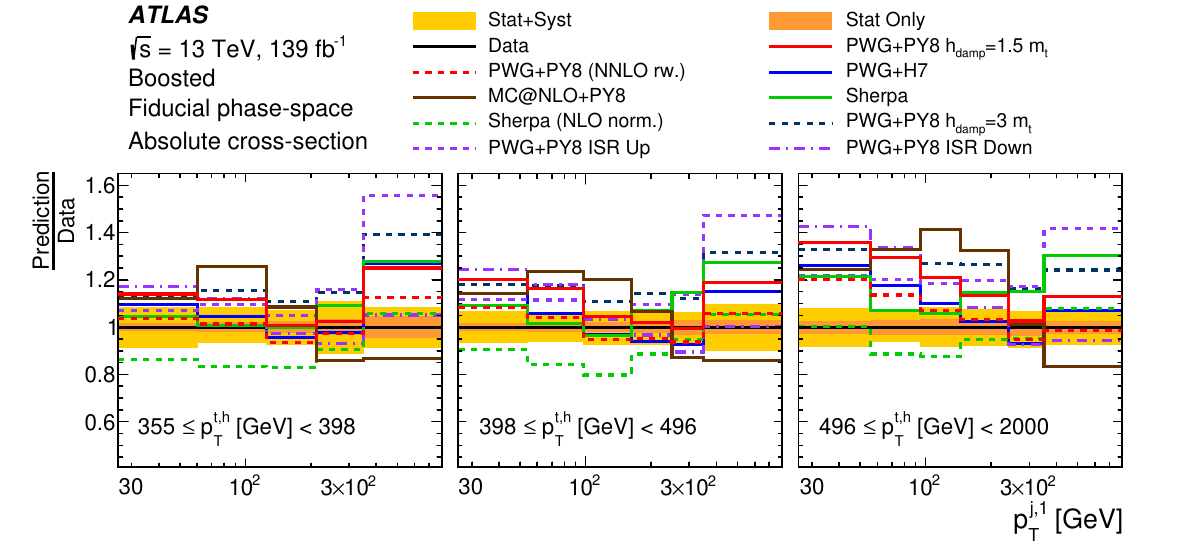}}
\caption{(a) Differential cross-section measurements as a function of the $\pt$\ of the leading additional jet in bins of $\pThad$\ are compared with the prediction from the \POWPY MC generator. The measurement and the predictions are normalised by the factors shown in parentheses to aid visibility. (b) Ratio of the measured absolute cross-section to different NLO, and NLO reweighted to NNLO, predictions of $t\bar{t}$\ signal for the same differential variables.
PWG+PY8 corresponds to the \POWPY sample, PWG+H7 to the \POWHER sample and MCatNLO+PY8 to the \MGNLOPY sample.
The yellow band represents the total uncertainty of the measured differential cross-section while the orange band shows the statistical component.}
\label{fig:Results_xsection_abs_2D_leadaddjet_pt_V_topjet_pt}
\end{figure}

\begin{figure}[p]
\centering
\subfloat[]{
\includegraphics[width=0.5\textwidth]{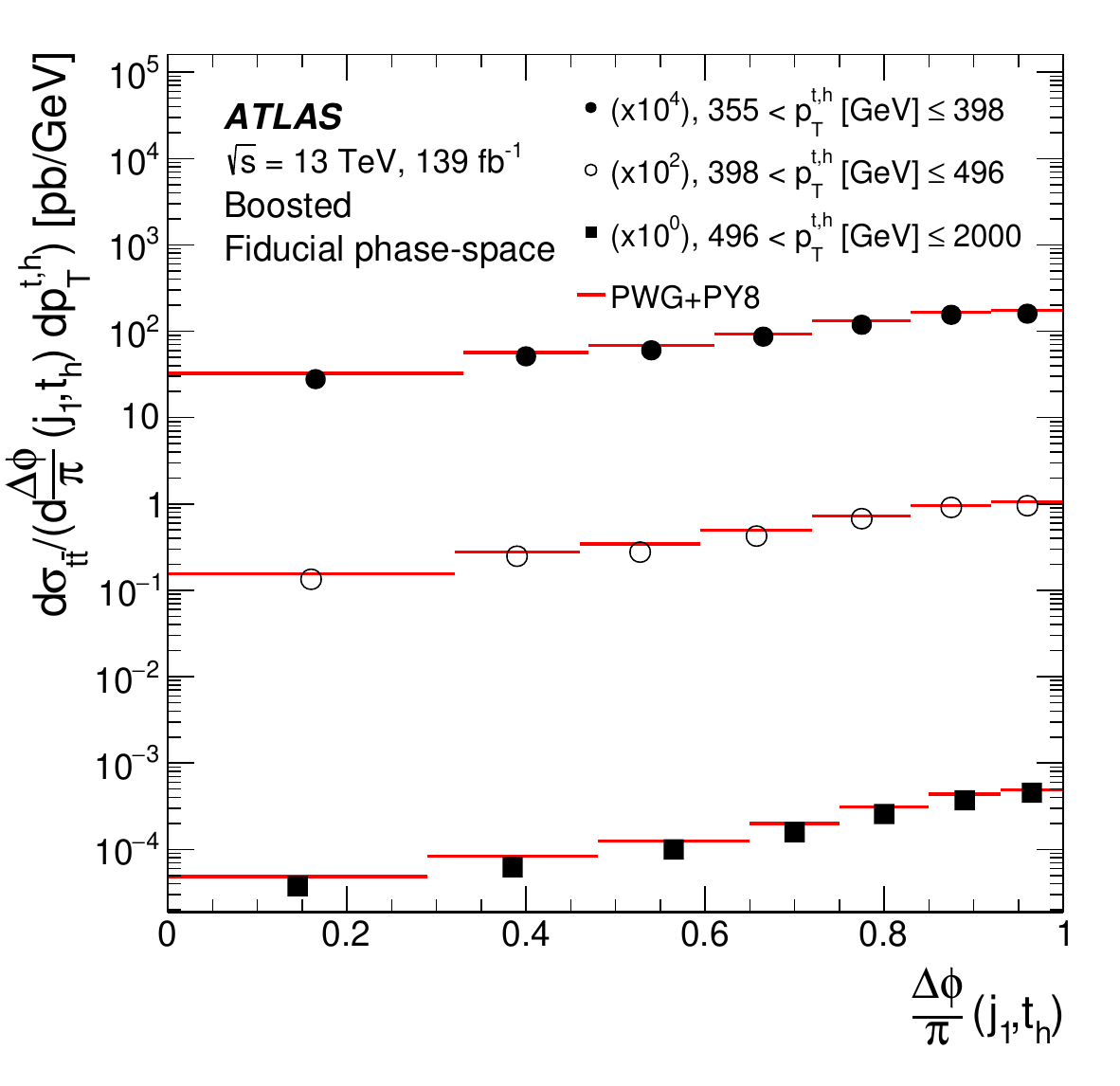}}
 
\subfloat[]{
\includegraphics[width=0.9\textwidth]{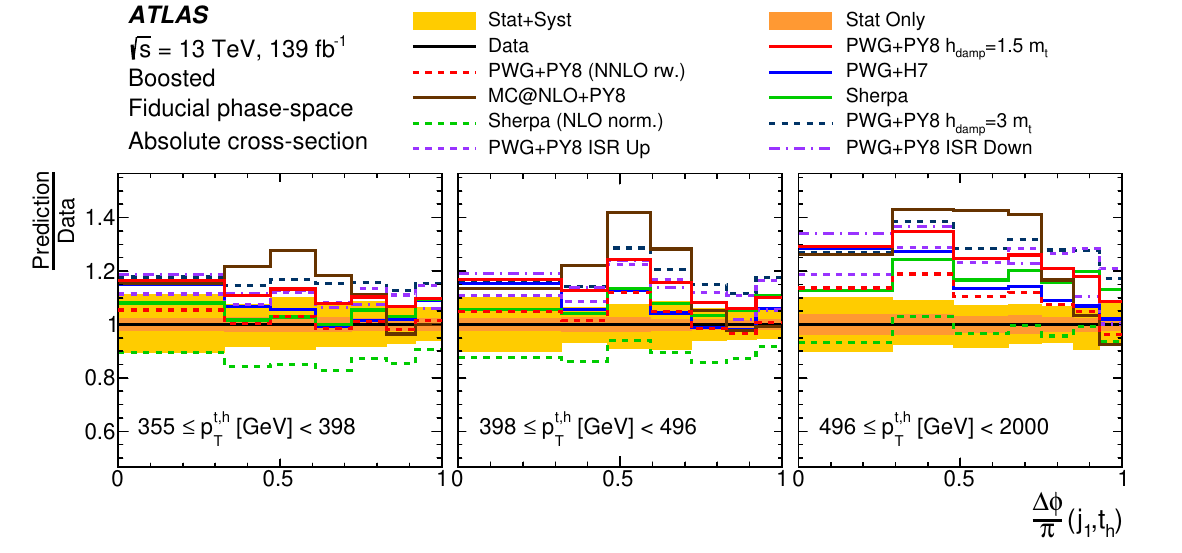}}
\caption{(a) Differential cross-section measurements as a function of the $\Delta\phi$ angle between the leading additional jet and the hadronically decaying top quark in bins of $\pThad$\ are compared with the prediction from the \POWPY MC generator. The measurement and the predictions are normalised by the factors shown in parentheses to aid visibility. (b) Ratio of the measured absolute cross-section to different NLO, and NLO reweighted to NNLO, predictions of $t\bar{t}$\ signal for the same differential variables.
PWG+PY8 corresponds to the \POWPY sample, PWG+H7 to the \POWHER sample and MCatNLO+PY8 to the \MGNLOPY sample.
The yellow band represents the total uncertainty of the measured differential cross-section while the orange band shows the statistical component.}
\label{fig:Results_xsection_abs_2D_leadadd_top_deltaPhi_V_topjet_pt}
\end{figure}

\begin{figure}[p]
\centering
\subfloat[]{
\includegraphics[width=0.5\textwidth]{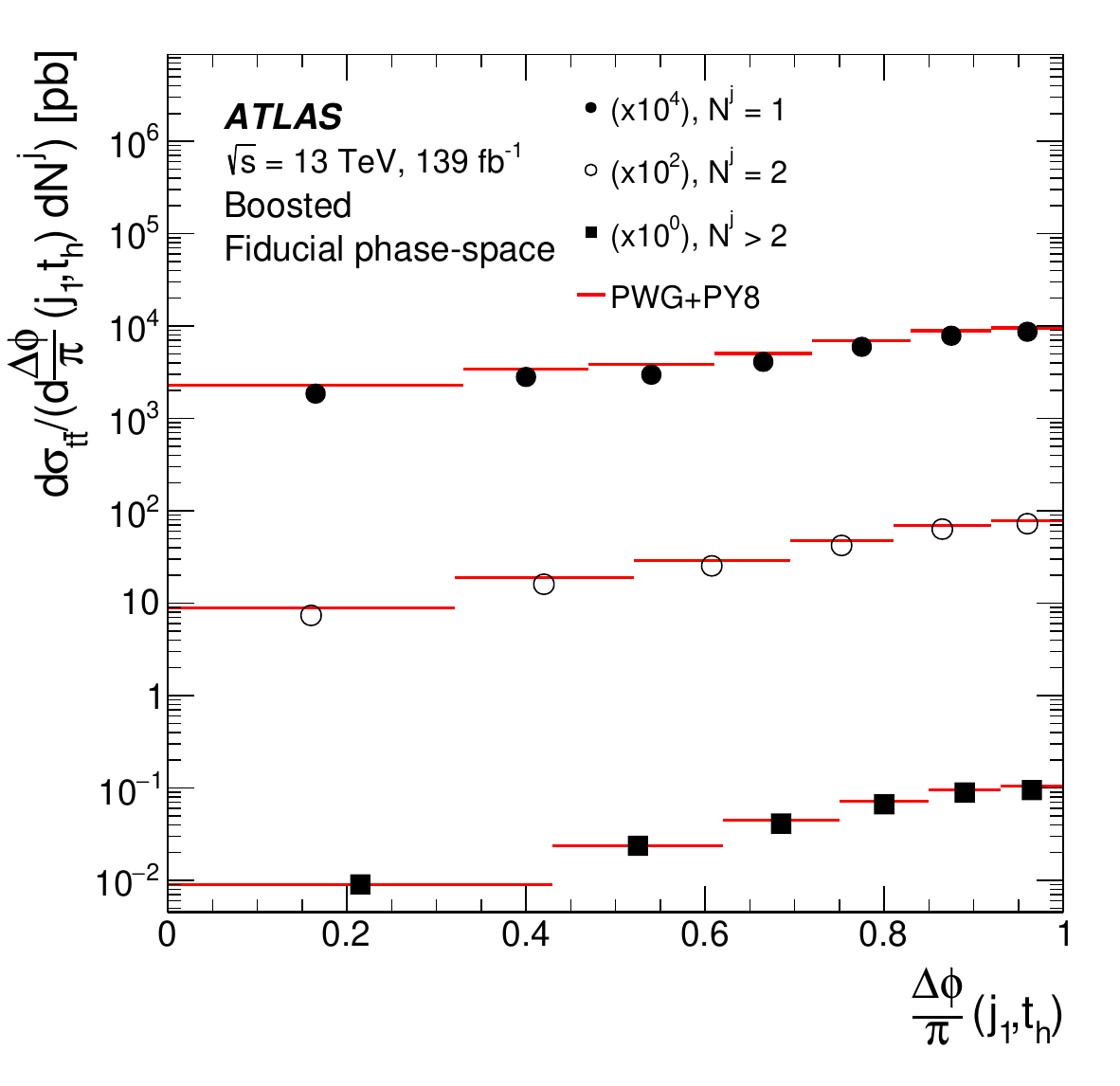}}
 
\subfloat[]{
\includegraphics[width=0.9\textwidth]{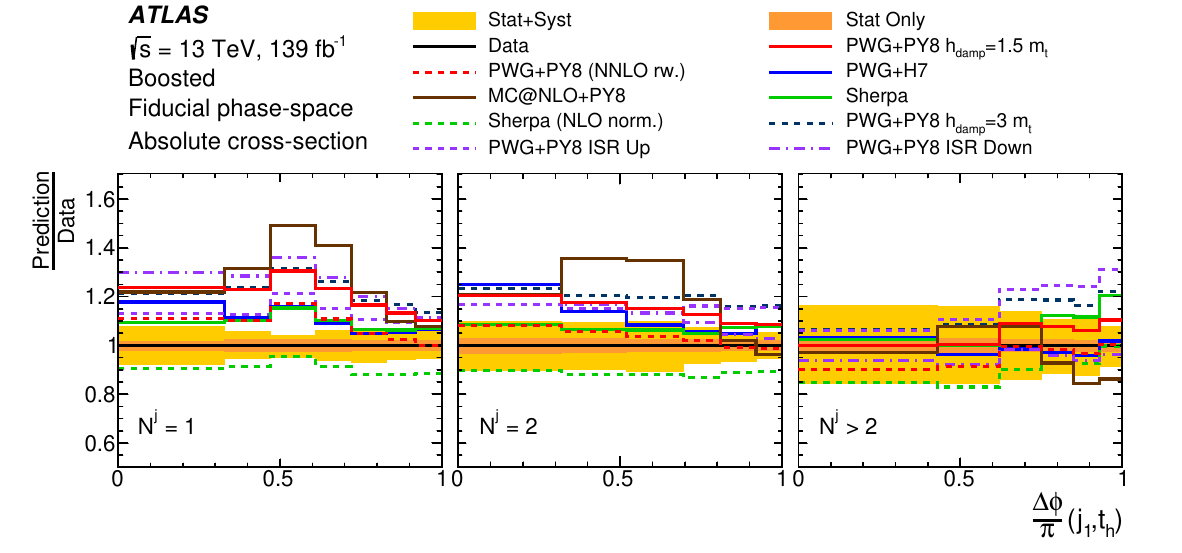}}
\caption{(a) Differential cross-section measurements as a function of the $\Delta\phi$ angle between the leading additional jet and the hadronically decaying top quark in bins of additional jet multiplicity are compared with the prediction from the \POWPY MC generator. The measurement and the predictions are normalised by the factors shown in parentheses to aid visibility. (b) Ratio of the measured absolute cross-section to different NLO, and NLO reweighted to NNLO, predictions of $t\bar{t}$\ signal for the same differential variables.
PWG+PY8 corresponds to the \POWPY sample, PWG+H7 to the \POWHER sample and MCatNLO+PY8 to the \MGNLOPY sample.
The yellow band represents the total uncertainty of the measured differential cross-section while the orange band shows the statistical component.}
\label{fig:Results_xsection_abs_2D_leadadd_top_deltaPhi_V_naddjet}
\end{figure}

\FloatBarrier
 
\clearpage
 
\section{Limits on EFT operators}
\label{sec-eft}
 
The sensitivity of the analysis to potential new physics in $\ttbar$~production is explored
by interpreting the measured $\pThad$~distribution in terms of dimension-six operators
within the effective field theory framework.
The interpretation allows two Wilson coefficients that are sensitive to $\ttbar$~production to be non-zero:
$\ctg$~and $\ctq$.
All other Wilson coefficients, for both the dimension-six and higher-dimensional operators, are assumed to be zero.
The $\Otg$~operator primarily changes the overall rate of $\ttbar$\ production
while the $\Otq$~operator results in additional $\ttbar$\ events at high energy.
Example LO Feynman diagrams for the two operators are shown in Figure~\ref{fig:eftfeyn}.
The two operators can be disentangled by fitting a differential cross-section measurement.
The $\pThad$~distribution, shown in Figure~\ref{fig:diffxstop:toppt}, is chosen as the observable, based on simulation studies that considered
the sensitivity of the variables as well as the stability of the variables when going from LO to NLO QCD.
 
\begin{figure}[h]
\centering
\subfloat[]{
\includegraphics[height=0.2\textwidth]{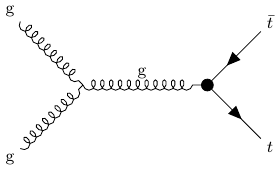}}
\hspace{5mm}
\subfloat[]{
\includegraphics[height=0.2\textwidth]{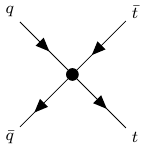}}
\caption{Example LO Feynman diagrams for EFT contributions from (a) $\ctg$\ and (b) $\ctq$. Couplings affected by EFT contributions are marked by the black dots.}
\label{fig:eftfeyn}
\end{figure}
 
The differential cross-section is parameterised as a polynomial of second degree in the two Wilson coefficients under consideration:
\begin{equation}\nonumber
\sigma^{j}(\ctg, \ctq)= p_{0}^{j} + p_{1}^{j}\cdot \ctg  + p_{2}^{j}\cdot \ctq + p_{3}^{j}\cdot (\ctg)^2 + p_{4}^{j}\cdot (\ctq)^2 + p_{5}^{j}\cdot \ctg\cdot \ctq
\label{eq:eft_model}
\end{equation}
where the index $j$~labels the bins of the differential distribution. The parameters $p_i^j$\ with $i \geq 1$~are extracted from fits to the
EFT samples with $(\ctg, \ctq)=(\pm 1 \text{~or~} 0,~\pm 1 \text{~or~} 0)$; see Section~\ref{sec-datamc} for a description of the samples.
Two separate fits are performed, one using samples
containing the full EFT contributions (i.e.\ proportional to $\Lambda^{-2}$\ and $\Lambda^{-4}$),
and the second containing only EFT contributions proportional to $\Lambda^{-2}$.
This allows the impact of the $\Lambda^{-4}$\ terms on the sensitivity of the results to be examined and the two set-ups
are referred to as the $\Lambda^{-4}$\ and $\Lambda^{-2}$\ models in the rest of this section.
 
The parameters $p_0^j$\ determine the Standard Model prediction
and are taken from the \POWPY sample after reweighting it to the NNLO prediction.
The parameterisation of the EFT effects assumes that the background estimate is independent of the tested operators, which
is reasonable given the high purity of the selected event sample.
 
The \EFTfitter~\cite{Castro:2016jjv} package is used to extract the limits on the Wilson coefficients by minimising the likelihood:
\begin{equation}\nonumber
-2 \ln p( \ctg, \ctq | \boldsymbol{m}) = \left(\boldsymbol{m} - \boldsymbol{\sigma}(\ctg, \ctq)\right)^T M^{-1} \left(\boldsymbol{m} - \boldsymbol{\sigma}(\ctg, \ctq)\right)
\label{eq:eftchi2}
\end{equation}
where $\boldsymbol{m}$~is the vector of measurements of each bin of the observed distribution and $\boldsymbol{\sigma}(\ctg, \ctq)$~is
the vector of corresponding predictions for each bin that depends on the two Wilson coefficients of interest (\ctg, \ctq).
The covariance matrix, $M$, is a sum of the matrices corresponding to the experimental and theoretical uncertainties, $M = M_\mathrm{e} + M_\mathrm{t}$.
This procedure assumes that the theoretical uncertainties in the predicted cross-section are not correlated with the theoretical uncertainties in the experimental measurement (described in Section~\ref{sec-syst-ttbarmodel}).
The covariance matrix for the experimental uncertainties corresponds exactly to the uncertainties described in Section~\ref{sec-syst}.
The theoretical uncertainties are determined using the \POWPY $\ttbar$ sample reweighted to NNLO.
The scales \mur\ and \muf\ are varied to $2 \mu$~and $\mu / 2$, with the condition $1/2 \leq \mur / \muf \leq 2$, giving seven variations.
The envelope of those variations is used to define alternative shapes for the three parton-level distributions used in the NNLO
reweighting procedure.
The difference between each alternative prediction and the nominal one gives an estimate of the uncertainty in the shape of the distributions provided by the NNLO calculations.
The uncertainty in all three distributions is included and assumed to be uncorrelated
with the uncertainty in the other distributions. The uncertainty due to the choice of PDF set is also included.
Since all these uncertainties cover only the shape differences, an additional 6$\%$ uncertainty is included
for the uncertainty in the inclusive $\ttbar$\ cross-section, covering the variations of the scales, PDFs, $\alpha_s$ and $m_t$.
The posterior probability distribution for the Wilson coefficients is extracted using the equation
of Bayes and Laplace, as implemented in the Bayesian Analysis Toolkit~\cite{Caldwell:2008fw}.
The prior probability distribution for the Wilson coefficients is taken to be uniform.
To make the dependence of the sensitivity to the Wilson coefficients on the energy scale of the new physics explicit, the results are presented for the product $C_i (\TeV / \Lambda)^2$.
This also facilitates straightforward comparisons with results where $\Lambda = 1$~\TeV.
The credible interval for each Wilson coefficient is extracted by marginalising over the other coefficient.
The fit is also performed with only one Wilson coefficient as a free parameter and the other fixed to zero; these are referred to as individual
fits.
 
The measured $\pThad$\ differential cross-section  is compared with the SM prediction and uncertainty in Figure~\ref{fig:top_pt_eft_model}.
The figure also displays the best fit for the EFT models, where the fitted Wilson coefficients are $\ctg=\ctgFitquad^{+0.16}_{-0.25}~\EFTu$, $\ctq=\ctqFitquad^{+0.40}_{-0.06}~\EFTu$ for $\Lambda^{-4}$\ and $\ctg=\ctgFit \pm 0.23~\EFTu$, $\ctq=\ctqFit \pm 0.17~\EFTu$ for $\Lambda^{-2}$.
These values agree with zero within two standard deviations, indicating there is no evidence of new physics in the data.
The fit prefers negative values for $\ctg$\ because the measured cross-section is lower than the SM prediction.
Table~\ref{tab:eftlimits} shows the expected and observed marginalised credible intervals for the nominal fit and the individual fits.
The expected and observed posterior distributions are shown in Figure~\ref{fig:top_pt_posteriors}.
The expected and observed credible intervals are asymmetric in the $\Lambda^{-4}$\ model because the linear and quadratic terms can cancel out to some extent when
the Wilson coefficients are negative.
The impact of the different bins in the distribution is investigated by repeating the fit with a subset of the measured bins.
Figure~\ref{fig:top_pt_bin_by_bin_accum} shows the evolution of the posterior distribution as bins are added to the interpretation.
With only a single bin, the operators are largely degenerate and the fit can only constrain the combination of the two.
As bins are added to the fit, the ability of the fit to distinguish between the operators is improved.
The figure also shows that the constraint on $\ctq$~is dominated by the measurements at high $\pThad$.
The observed constraints are compared in Table~\ref{tab:eftlimits} with the individual limits obtained in a recent global analysis based on multiple measurements~\cite{Ethier:2021bye}.
The limits on $\ctg$\ obtained in this paper are significantly weaker than these, but the limits on $\ctq$\ are more stringent than those obtained
in the global fit,
which, given the use of only a single dataset, indicates that the data presented here can provide important constraining power in future global
EFT fits.
 
\begin{figure}[htbp]
\centering
\subfloat[]{\includegraphics[width=0.49\textwidth]{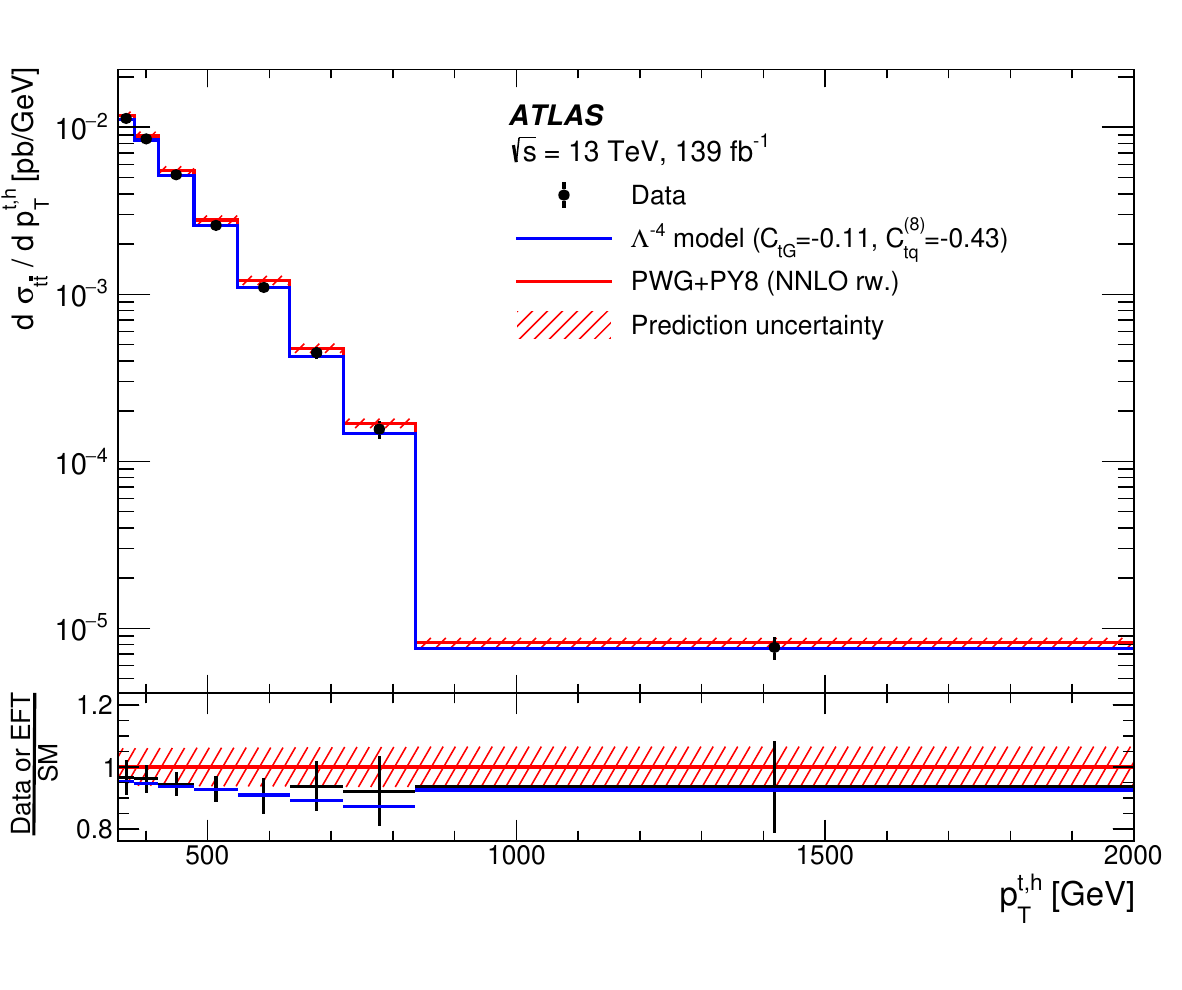}}
\subfloat[]{\includegraphics[width=0.49\textwidth]{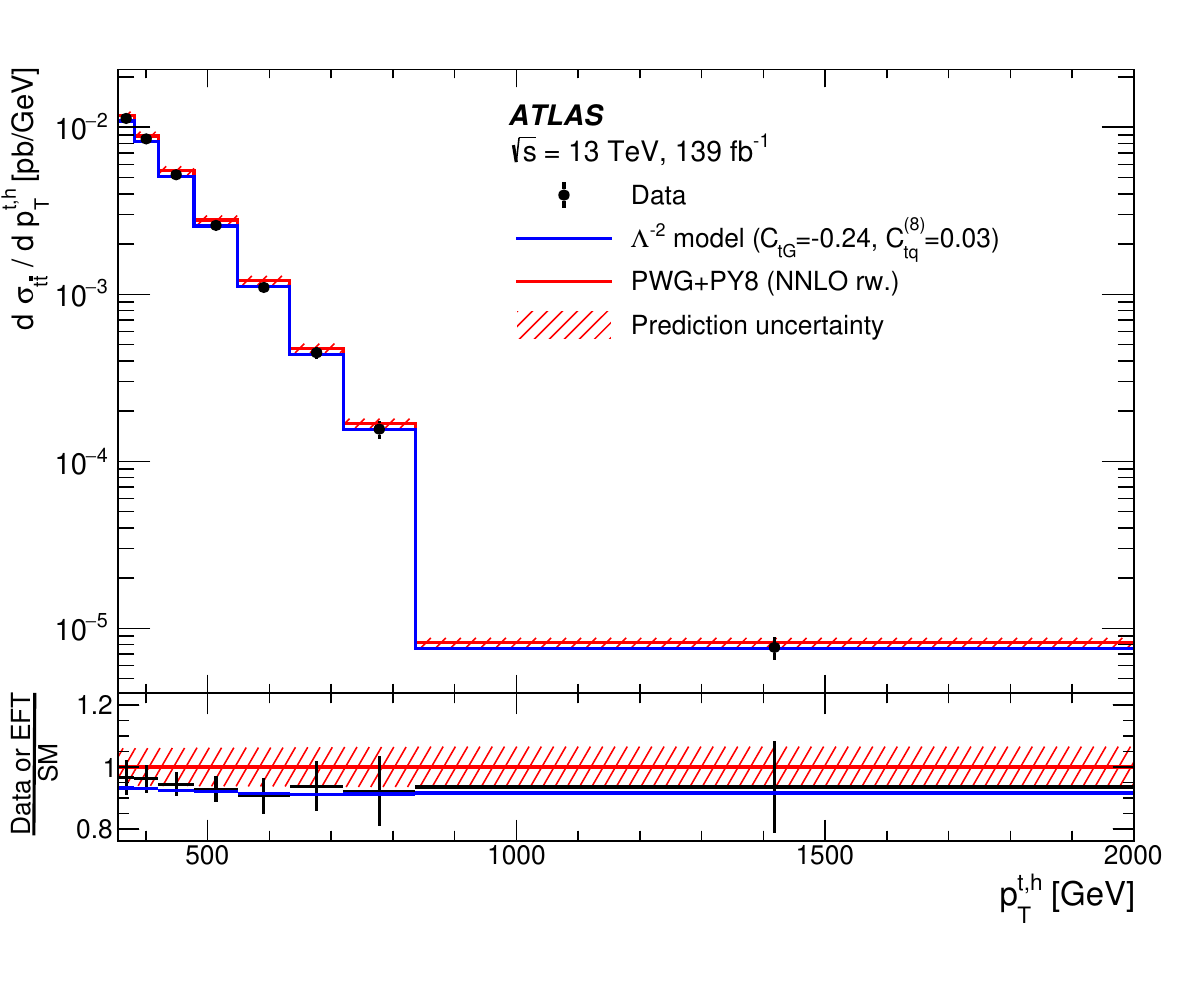}}
\caption{Differential cross-section measurement of $\pThad$\ used in the EFT interpretation. The data are compared with the SM prediction in red and the EFT model prediction in blue at the respective global modes for (a) the $\Lambda^{-4}$ model and (b) the $\Lambda^{-2}$ model.
The lower panel in each plot displays the ratio of the data or EFT model to the SM prediction. The shaded red band shows the uncertainty in the SM prediction used in the EFT fit.
}
\label{fig:top_pt_eft_model}
\end{figure}
 
\begin{figure}[htbp]
\centering
\subfloat[]{  \includegraphics[width=0.49\textwidth]{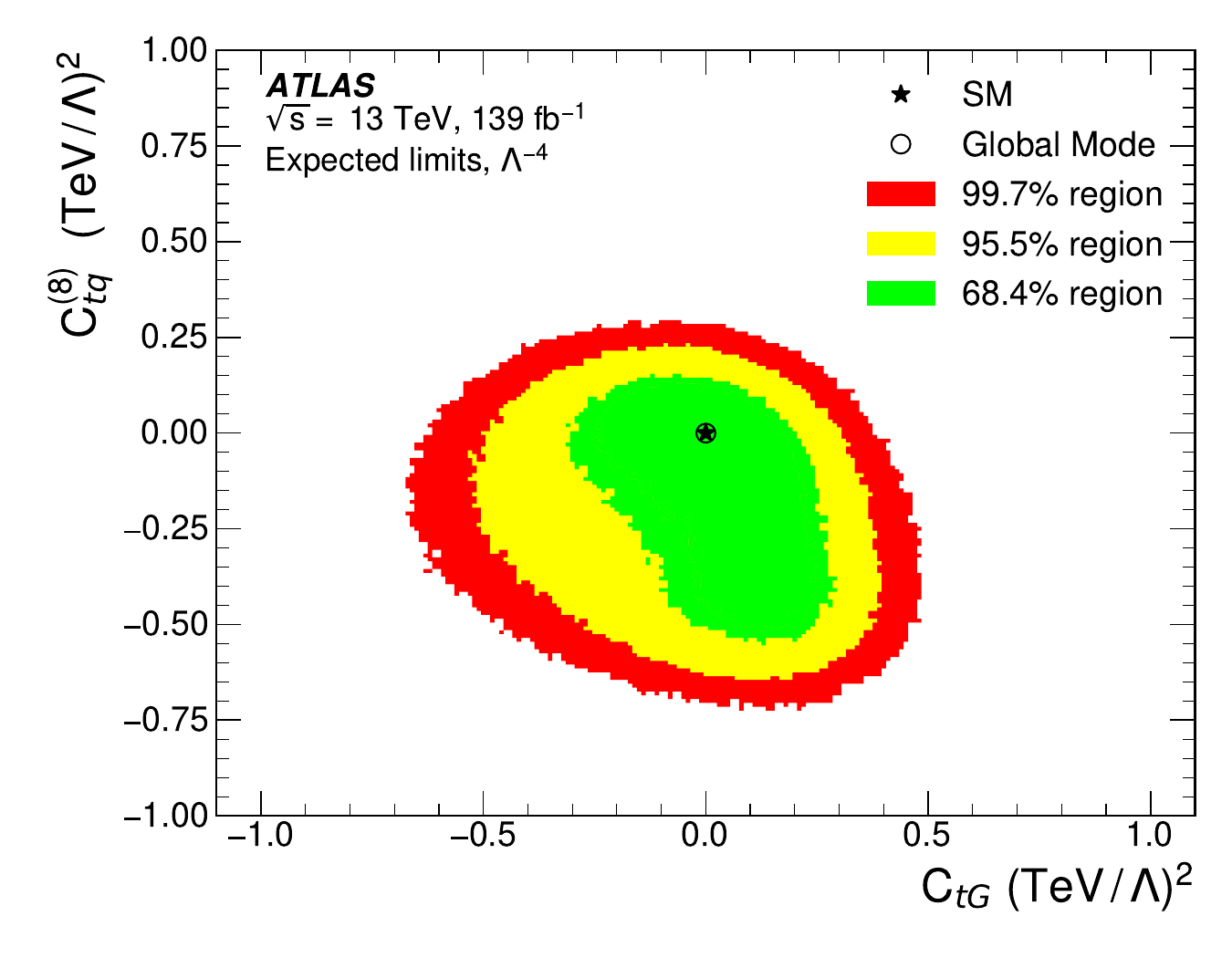}}
\subfloat[]{  \includegraphics[width=0.49\textwidth]{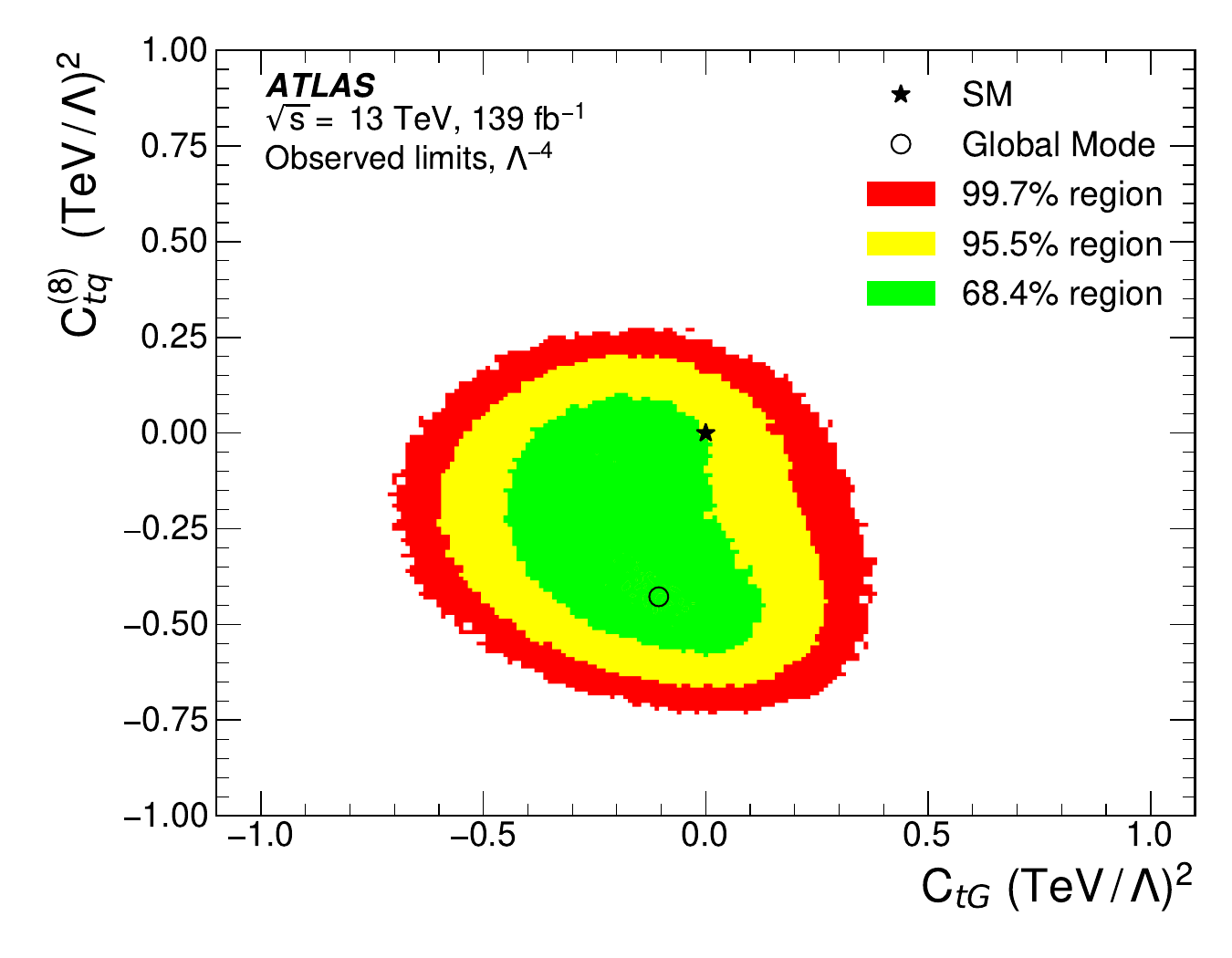}}
 
\subfloat[]{  \includegraphics[width=0.49\textwidth]{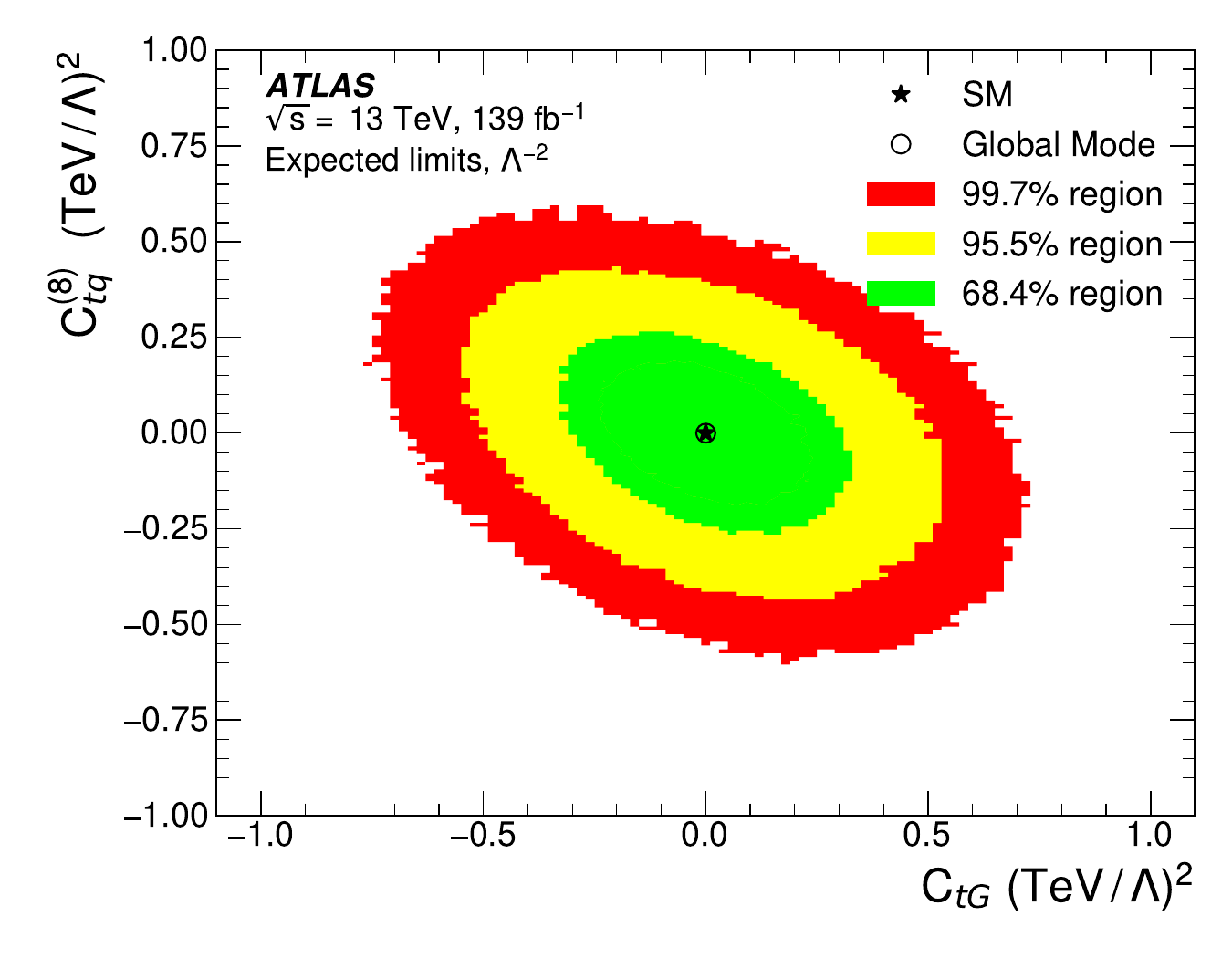}}
\subfloat[]{  \includegraphics[width=0.49\textwidth]{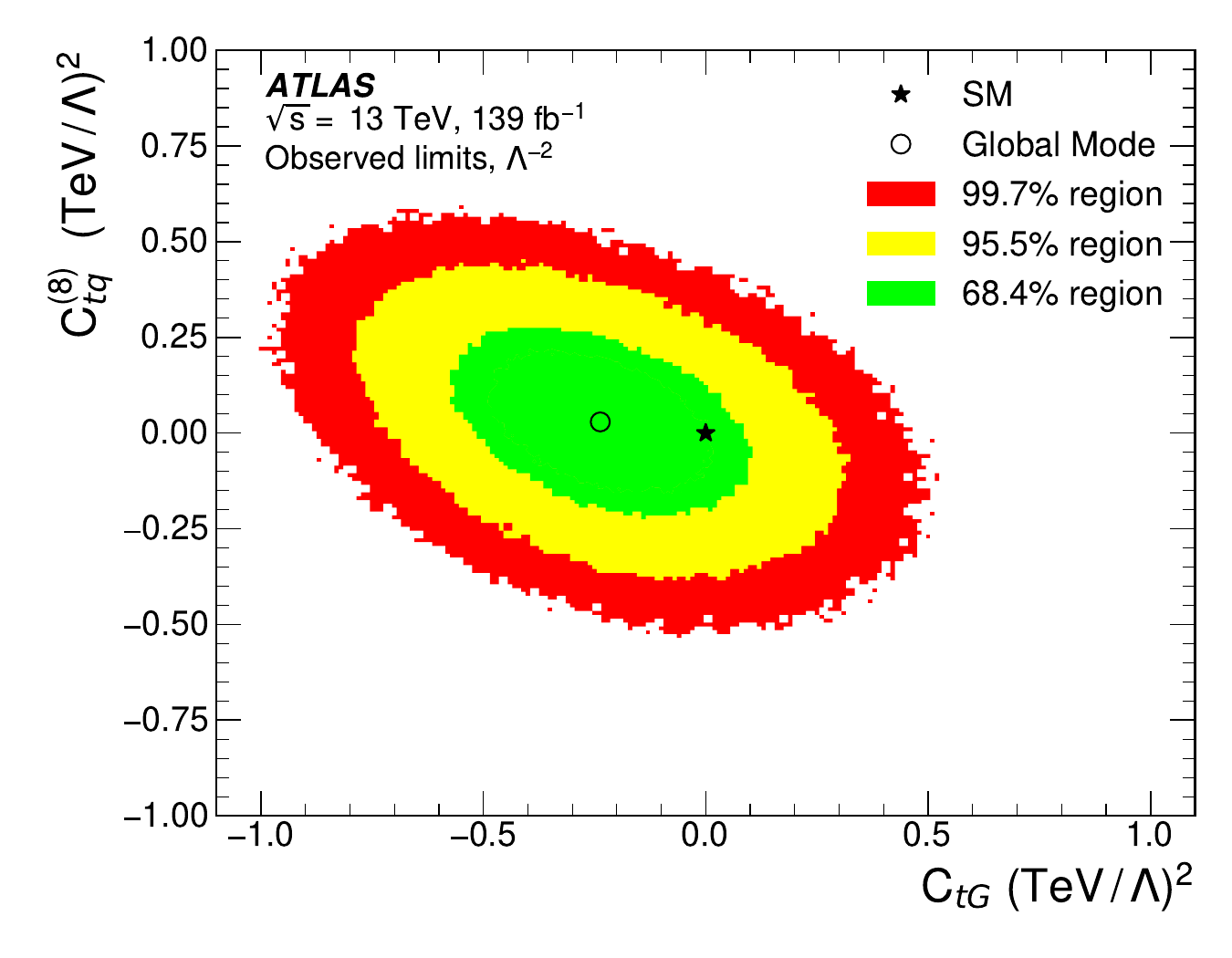}}
\caption{Two-dimensional posterior distributions for the two Wilson coefficients $\ctg$\ and $\ctq$\ obtained from fitting the $\pThad$ distribution.
(a) Shows the posterior distribution expected from the SM prediction and (b) shows the distribution obtained from the data, both for the $\Lambda^{-4}$\ model.
The same distributions are shown in (c) and (d) for the $\Lambda^{-2}$\ model.
The $\SI{68.4}{\percent}$, $\SI{95.5}{\percent}$ and $\SI{99.7}{\percent}$ regions are shown in green, yellow and red respectively.
}
\label{fig:top_pt_posteriors}
\end{figure}
 
\begin{table}[htbp]
\centering
\caption{ Expected and observed $95\%$ intervals for the Wilson coefficients ($C_i$). The marginalised results show the intervals extracted from the nominal fit where both Wilson coefficients are allowed to vary. The
individual intervals are extracted from fits where only the Wilson coefficient under study is allowed to differ from zero. The results are compared with the individual limits obtained in Ref.~\cite{Ethier:2021bye}.}
\label{tab:eftlimits}
\resizebox{\columnwidth}{!}{
\begin{tabular}{ c  | c  | c  c  | c  c  | c }
\toprule
\multirow{2}{*}{Model } & \multirow{2}{*}{$C_i~\EFTu$}  & \multicolumn{2}{c |}{Marginalised $95\%$ intervals }& \multicolumn{2}{c|}{Individual $95\%$ intervals  } & Global fit $95\%$ \\
&  & Expected & Observed  & Expected & Observed  &   limits~\cite{Ethier:2021bye} \\
\midrule
\multirow{2}{*}{$\Lambda^{-4}$} & $\ctg$ & [$\ctgDownEquad$, \ctgUpEquad]  & [$\ctgDownquad$, \ctgUpquad] & [$\ctgDownIEquad$, \ctgUpIEquad] & [$\ctgDownIquad$, \ctgUpIquad] & [0.006, 0.107] \\
& $\ctq$ &  [$\ctqDownEquad$, \ctqUpEquad]  & [$\ctqDownquad$, \ctqUpquad] & [$\ctqDownIEquad$, \ctqUpIEquad]  & [$\ctqDownIquad$, \ctqUpIquad] & [$-0.48$, 0.39] \\
\midrule
\multirow{2}{*}{$\Lambda^{-2}$} & $\ctg$ & [$\ctgDownE$, \ctgUpE]  & [$\ctgDown$, \ctgUp] & [$\ctgDownIE$, \ctgUpIE] & [$\ctgDownI$, \ctgUpI] & [0.007, 0.111] \\
& $\ctq$ &  [$\ctqDownE$, \ctqUpE]  & [$\ctqDown$, \ctqUp] & [$\ctqDownIE$, \ctqUpIE]  & [$\ctqDownI$, \ctqUpI] & [$-0.40$, 0.61] \\
\bottomrule
\end{tabular}
}
\end{table}
 
\begin{figure}[htbp]
\centering
\subfloat[]{\includegraphics[width=0.49\textwidth]{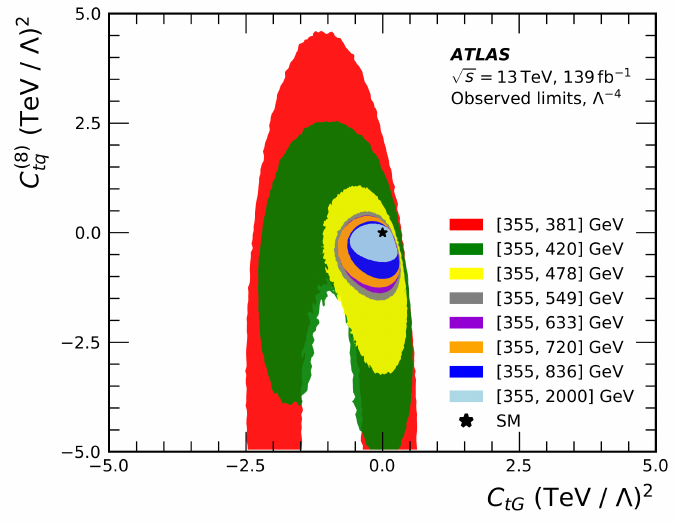}}
\subfloat[]{\includegraphics[width=0.49\textwidth]{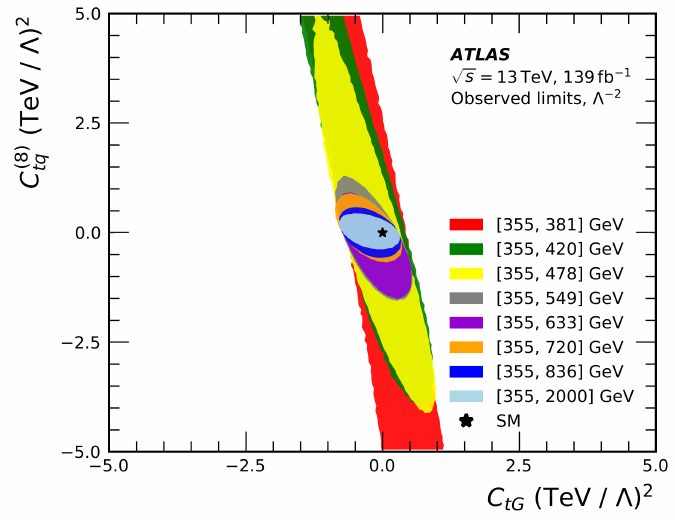}}
\caption{Evolution of the $95\%$~observed credible region when adding the bins of the measured $\pThad$\ distribution
to the interpretation one-by-one for (a) the $\Lambda^{-4}$ model and (b) the $\Lambda^{-2}$ model.
The coloured ellipses show the $95\%$\ regions obtained from fitting the $\pThad$\ range indicated in the legend.}
\label{fig:top_pt_bin_by_bin_accum}
\end{figure}

\FloatBarrier
 
\section{Conclusions}
\label{sec-conclusions}
 
This article presents precise cross-section measurements of $\ttbar$~events containing a high transverse momentum top quark that has decayed hadronically ($\pThad > 355$~\GeV).
The events were selected from the full 139~\ifb\ Run-2 ATLAS dataset of 13~\TeV\ proton--proton collisions at the LHC and the cross-sections were extracted by unfolding the reconstructed distributions.
The precision of the results is significantly better than in previous measurements. This improvement
in precision is largely driven by the introduction of a novel technique to use the invariant mass of the selected large-radius jet from the hadronically decaying top quark to reduce the
impact of jet energy scale uncertainties.
In addition, the background from $W\text{\,+\,jets}$\ and single top-quark production is reduced compared to the previous ATLAS measurement by requiring at least two $b$-tagged jets
and $m_{\ell b} < 180$~\GeV, which reduces the impact of the background uncertainties on the measurements.
The fiducial cross-section is measured to be $1.267 \pm 0.005 \mathrm{~(stat.)} \pm 0.053 \mathrm{~(syst.)}$~pb.
The measurements are compared with predictions from NLO+PS MC generators.
No single generator is able to describe all the measured variables well.
Applying parton-level reweighting to match NNLO QCD predictions gives better agreement with the data for all generators, indicating
that these corrections are relevant given the precision of the measurements.
The number of additional jets is best modelled by the \SHERPA\ generator, which provides NLO accuracy for the first additional jet, but the details of
the additional radiation are not well described by any of the MC predictions.
The sensitivity of the measurement to new physics beyond the Standard Model is illustrated by using the transverse momentum distribution of the hadronically decaying top
quark to set limits in the context of effective field theory.
No evidence of new physics is seen and the $95\%$\ credible intervals of the Wilson coefficients are
$\ctg \in [\ctgDownquad, \ctgUpquad]~\EFTu$~and $\ctq \in [\ctqDownquad, \ctqUpquad]~\EFTu$\ when
including EFT contributions proportional to $\Lambda^{-2}$ and $\Lambda^{-4}$.
The results demonstrate that the analysis can disentangle the $\Otg$~and $\Otq$\ operators,
and the stringent limits placed on $\ctq$\ demonstrate the data will be highly relevant in future global fits.

\section*{Acknowledgements}
 

We thank CERN for the very successful operation of the LHC, as well as the
support staff from our institutions without whom ATLAS could not be
operated efficiently.
 
We acknowledge the support of
ANPCyT, Argentina;
YerPhI, Armenia;
ARC, Australia;
BMWFW and FWF, Austria;
ANAS, Azerbaijan;
SSTC, Belarus;
CNPq and FAPESP, Brazil;
NSERC, NRC and CFI, Canada;
CERN;
ANID, Chile;
CAS, MOST and NSFC, China;
Minciencias, Colombia;
MEYS CR, Czech Republic;
DNRF and DNSRC, Denmark;
IN2P3-CNRS and CEA-DRF/IRFU, France;
SRNSFG, Georgia;
BMBF, HGF and MPG, Germany;
GSRI, Greece;
RGC and Hong Kong SAR, China;
ISF and Benoziyo Center, Israel;
INFN, Italy;
MEXT and JSPS, Japan;
CNRST, Morocco;
NWO, Netherlands;
RCN, Norway;
MEiN, Poland;
FCT, Portugal;
MNE/IFA, Romania;
JINR;
MES of Russia and NRC KI, Russian Federation;
MESTD, Serbia;
MSSR, Slovakia;
ARRS and MIZ\v{S}, Slovenia;
DSI/NRF, South Africa;
MICINN, Spain;
SRC and Wallenberg Foundation, Sweden;
SERI, SNSF and Cantons of Bern and Geneva, Switzerland;
MOST, Taiwan;
TAEK, Turkey;
STFC, United Kingdom;
DOE and NSF, United States of America.
In addition, individual groups and members have received support from
BCKDF, CANARIE, Compute Canada and CRC, Canada;
COST, ERC, ERDF, Horizon 2020 and Marie Sk{\l}odowska-Curie Actions, European Union;
Investissements d'Avenir Labex, Investissements d'Avenir Idex and ANR, France;
DFG and AvH Foundation, Germany;
Herakleitos, Thales and Aristeia programmes co-financed by EU-ESF and the Greek NSRF, Greece;
BSF-NSF and GIF, Israel;
Norwegian Financial Mechanism 2014-2021, Norway;
NCN and NAWA, Poland;
La Caixa Banking Foundation, CERCA Programme Generalitat de Catalunya and PROMETEO and GenT Programmes Generalitat Valenciana, Spain;
G\"{o}ran Gustafssons Stiftelse, Sweden;
The Royal Society and Leverhulme Trust, United Kingdom.
 
The crucial computing support from all WLCG partners is acknowledged gratefully, in particular from CERN, the ATLAS Tier-1 facilities at TRIUMF (Canada), NDGF (Denmark, Norway, Sweden), CC-IN2P3 (France), KIT/GridKA (Germany), INFN-CNAF (Italy), NL-T1 (Netherlands), PIC (Spain), ASGC (Taiwan), RAL (UK) and BNL (USA), the Tier-2 facilities worldwide and large non-WLCG resource providers. Major contributors of computing resources are listed in Ref.~\cite{ATL-SOFT-PUB-2021-003}.
 

\clearpage
\appendix
\part*{Appendix}
\addcontentsline{toc}{part}{Appendix}
\section{Normalised differential cross-section results}
\label{app:normalisedResults}
 
In this appendix, the differential cross-section results after normalising to the cross-section measured in the fiducial region are presented.
These results allow comparisons of the shape of the measured observables with MC predictions. Tables~\ref{table:chi2_table_NNLO_relative} and \ref{table:chi2_table_relative}
show the $\chi^2$ values for the normalised distributions for the different MC models.
Figure~\ref{figaux:diffxstop_1D_rel} shows the differential
cross-section measurements of $\pThad$,
$\pTlep$,
$\mttbar$,
$\yhad$,
$\ylep$,
$\yttbar$,
$\HTttbar$\ and
$\dPhibt$
compared with the predictions from different NLO generator set-ups. The same variables are compared in Figure~\ref{figaux:diffxstopNNLO_rel} with the MC generators with and
without the reweighting to the NNLO prediction. The variables sensitive to additional radiation are shown in Figures~\ref{figaux:diffxsExtra1_1D_rel} and~\ref{figaux:diffxsExtra2_1D_rel}.
The double-differential measurements are shown in Figures~\ref{figaux:diffxs_2D_leadaddjet_pt_V_naddjet_rel}-\ref{figaux:diffxs_2D_leadadd_top_deltaPhi_V_naddjet_rel}.
 
\tableVSpace{1.1}
\begin{table}[p]
\centering
\caption{$\chi^2$ and $p$-values quantifying the level of agreement between the relative unfolded spectra, several NLO+PS predictions and the respective NNLO reweighted spectrum. PWG+PY8 corresponds to the \POWPY sample, PWG+H7 to the \POWHER sample and MC@NLO+PY8 to the \MGNLOPY sample.}
\resizebox{\textwidth}{!}{
\begin{tabular}[t]{|c| r @{/} l r  || r @{/} l r  || r @{/} l r  || r @{/} l r  || r @{/} l r  || r @{/} l r  |} \toprule
Observable& \multicolumn{3}{c|}{\textsc{PWG+PY8}}& \multicolumn{3}{c|}{\textsc{PWG+PY8(NNLO weight)}}& \multicolumn{3}{c|}{\textsc{MC@NLO+PY8}}& \multicolumn{3}{c|}{\textsc{MC@NLO+PY8(NNLO weight)}}& \multicolumn{3}{c|}{\textsc{PWG+H7}}& \multicolumn{3}{c|}{\textsc{PWG+H7(NNLO weight)}}\\ \midrule
& \multicolumn{2}{c}{$\chi^{2}$/NDF} &  ~$p$-value& \multicolumn{2}{c}{$\chi^{2}$/NDF} &  ~$p$-value& \multicolumn{2}{c}{$\chi^{2}$/NDF} &  ~$p$-value& \multicolumn{2}{c}{$\chi^{2}$/NDF} &  ~$p$-value& \multicolumn{2}{c}{$\chi^{2}$/NDF} &  ~$p$-value& \multicolumn{2}{c}{$\chi^{2}$/NDF} &  ~$p$-value\\
\hline
$\pThad$ &{\ } 3 & 7 &0.84 &{\ } 2 & 7 &0.95 &{\ } 3 & 7 &0.85 &{\ } 3 & 7 &0.89 &{\ } 3 & 7 &0.90 &{\ } 3 & 7 &0.92\\
$\pTlep$ &{\ } 24 & 7 &<0.01 &{\ } 15 & 7 &0.03 &{\ } 76 & 7 &<0.01 &{\ } 7 & 7 &0.39 &{\ } 24 & 7 &<0.01 &{\ } 13 & 7 &0.07\\
$\pTttbar$ &{\ } 95 & 6 &<0.01 &{\ } 34 & 6 &<0.01 &{\ } 109 & 6 &<0.01 &{\ } 20 & 6 &<0.01 &{\ } 96 & 6 &<0.01 &{\ } 35 & 6 &<0.01\\
$\HTall$ &{\ } 9 & 6 &0.17 &{\ } 4 & 6 &0.71 &{\ } 5 & 6 &0.57 &{\ } 21 & 6 &<0.01 &{\ } 14 & 6 &0.03 &{\ } 11 & 6 &0.08\\
$\HTttbar$ &{\ } 39 & 9 &<0.01 &{\ } 28 & 9 &<0.01 &{\ } 63 & 9 &<0.01 &{\ } 13 & 9 &0.14 &{\ } 33 & 9 &<0.01 &{\ } 26 & 9 &<0.01\\
$\yhad$ &{\ } 21 & 16 &0.18 &{\ } 23 & 16 &0.12 &{\ } 21 & 16 &0.20 &{\ } 21 & 16 &0.17 &{\ } 23 & 16 &0.10 &{\ } 25 & 16 &0.06\\
$\ylep$ &{\ } 14 & 13 &0.40 &{\ } 13 & 13 &0.44 &{\ } 12 & 13 &0.50 &{\ } 11 & 13 &0.63 &{\ } 20 & 13 &0.09 &{\ } 18 & 13 &0.16\\
$\yttbar$ &{\ } 7 & 9 &0.61 &{\ } 6 & 9 &0.77 &{\ } 8 & 9 &0.51 &{\ } 5 & 9 &0.85 &{\ } 8 & 9 &0.51 &{\ } 6 & 9 &0.72\\
$\mttbar$ &{\ } 18 & 9 &0.03 &{\ } 19 & 9 &0.02 &{\ } 51 & 9 &<0.01 &{\ } 6 & 9 &0.74 &{\ } 20 & 9 &0.02 &{\ } 21 & 9 &0.01\\
$\pTe{1}$ &{\ } 84 & 14 &<0.01 &{\ } 35 & 14 &<0.01 &{\ } 318 & 14 &<0.01 &{\ } 168 & 14 &<0.01 &{\ } 125 & 14 &<0.01 &{\ } 65 & 14 &<0.01\\
$\pTe{2}$ &{\ } 29 & 8 &<0.01 &{\ } 18 & 8 &0.02 &{\ } 20 & 8 &<0.01 &{\ } 51 & 8 &<0.01 &{\ } 37 & 8 &<0.01 &{\ } 28 & 8 &<0.01\\
$\Ne$ &{\ } 7 & 4 &0.12 &{\ } 8 & 4 &0.08 &{\ } 48 & 4 &<0.01 &{\ } 72 & 4 &<0.01 &{\ } 47 & 4 &<0.01 &{\ } 58 & 4 &<0.01\\
$\dPhieth{1}$ &{\ } 9 & 8 &0.32 &{\ } 7 & 8 &0.51 &{\ } 113 & 8 &<0.01 &{\ } 68 & 8 &<0.01 &{\ } 37 & 8 &<0.01 &{\ } 29 & 8 &<0.01\\
$\dPhieth{2}$ &{\ } 5 & 8 &0.78 &{\ } 5 & 8 &0.80 &{\ } 6 & 8 &0.67 &{\ } 15 & 8 &0.06 &{\ } 67 & 8 &<0.01 &{\ } 61 & 8 &<0.01\\
$\dPhibt$ &{\ } 35 & 12 &<0.01 &{\ } 21 & 12 &0.06 &{\ } 92 & 12 &<0.01 &{\ } 13 & 12 &0.35 &{\ } 32 & 12 &<0.01 &{\ } 19 & 12 &0.08\\
$\dPhitt$ &{\ } 62 & 4 &<0.01 &{\ } 28 & 4 &<0.01 &{\ } 74 & 4 &<0.01 &{\ } 76 & 4 &<0.01 &{\ } 71 & 4 &<0.01 &{\ } 33 & 4 &<0.01\\
$\dPhiee$ &{\ } 16 & 10 &0.11 &{\ } 16 & 10 &0.10 &{\ } 26 & 10 &<0.01 &{\ } 51 & 10 &<0.01 &{\ } 197 & 10 &<0.01 &{\ } 186 & 10 &<0.01\\
$\meth$ &{\ } 30 & 11 &<0.01 &{\ } 19 & 11 &0.06 &{\ } 166 & 11 &<0.01 &{\ } 34 & 11 &<0.01 &{\ } 34 & 11 &<0.01 &{\ } 18 & 11 &0.07\\
$\pTe{1}$ vs $\Ne$ &{\ } 261 & 20 &<0.01 &{\ } 194 & 20 &<0.01 &{\ } 470 & 20 &<0.01 &{\ } 273 & 20 &<0.01 &{\ } 231 & 20 &<0.01 &{\ } 154 & 20 &<0.01\\
$\pTe{1}$ vs $\pThad$ &{\ } 80 & 16 &<0.01 &{\ } 50 & 16 &<0.01 &{\ } 291 & 16 &<0.01 &{\ } 127 & 16 &<0.01 &{\ } 105 & 16 &<0.01 &{\ } 71 & 16 &<0.01\\
$\dPhieth{1}$ vs $\pThad$ &{\ } 48 & 20 &<0.01 &{\ } 39 & 20 &<0.01 &{\ } 333 & 20 &<0.01 &{\ } 196 & 20 &<0.01 &{\ } 68 & 20 &<0.01 &{\ } 56 & 20 &<0.01\\
$\dPhieth{1}$ vs $\Ne$ &{\ } 68 & 18 &<0.01 &{\ } 58 & 18 &<0.01 &{\ } 404 & 18 &<0.01 &{\ } 210 & 18 &<0.01 &{\ } 73 & 18 &<0.01 &{\ } 58 & 18 &<0.01\\
\bottomrule
\end{tabular}
}
\label{table:chi2_table_NNLO_relative}
\end{table}
\tableVSpace{1.1}
\begin{table}[p]
\centering
\caption{$\chi^2$ and $p$-values quantifying the level of agreement between the relative unfolded spectra and several NLO+PS predictions. PWG+PY8 corresponds to the \POWPY sample. \textsc{SHERPA} (NLO norm.) refers to the \SHERPA sample with its default normalisation. All other samples are normalised to the inclusive NNLO+NNLL $\ttbar$\ cross-section prediction.}
\resizebox{\textwidth}{!}{
\begin{tabular}[t]{|c| r @{/} l r  || r @{/} l r  || r @{/} l r  || r @{/} l r  || r @{/} l r  || r @{/} l r  |} \toprule
Observable& \multicolumn{3}{c|}{\textsc{PWG+PY8}}& \multicolumn{3}{c|}{\textsc{PWG+PY8(ISR Down)}}& \multicolumn{3}{c|}{\textsc{PWG+PY8(ISR Up)}}& \multicolumn{3}{c|}{\textsc{PWG+PY8($h_{\mathrm{damp}}$=3.0 $m_{t}$)}}& \multicolumn{3}{c|}{\textsc{SHERPA}}& \multicolumn{3}{c|}{\textsc{SHERPA} (NLO norm.)}\\ \midrule
& \multicolumn{2}{c}{$\chi^{2}$/NDF} &  ~$p$-value& \multicolumn{2}{c}{$\chi^{2}$/NDF} &  ~$p$-value& \multicolumn{2}{c}{$\chi^{2}$/NDF} &  ~$p$-value& \multicolumn{2}{c}{$\chi^{2}$/NDF} &  ~$p$-value& \multicolumn{2}{c}{$\chi^{2}$/NDF} &  ~$p$-value& \multicolumn{2}{c}{$\chi^{2}$/NDF} &  ~$p$-value\\
\hline
\hline
$\pThad$ &{\ } 3 & 7 &0.84 &{\ } 3 & 7 &0.90 &{\ } 6 & 7 &0.57 &{\ } 5 & 7 &0.63 &{\ } 5 & 7 &0.69 &{\ } 5 & 7 &0.69\\
$\pTlep$ &{\ } 24 & 7 &<0.01 &{\ } 57 & 7 &<0.01 &{\ } 2 & 7 &0.95 &{\ } 8 & 7 &0.33 &{\ } 4 & 7 &0.73 &{\ } 4 & 7 &0.73\\
$\pTttbar$ &{\ } 95 & 6 &<0.01 &{\ } 133 & 6 &<0.01 &{\ } 258 & 6 &<0.01 &{\ } 59 & 6 &<0.01 &{\ } 57 & 6 &<0.01 &{\ } 57 & 6 &<0.01\\
$\HTall$ &{\ } 9 & 6 &0.17 &{\ } 4 & 6 &0.69 &{\ } 64 & 6 &<0.01 &{\ } 20 & 6 &<0.01 &{\ } 28 & 6 &<0.01 &{\ } 28 & 6 &<0.01\\
$\HTttbar$ &{\ } 39 & 9 &<0.01 &{\ } 57 & 9 &<0.01 &{\ } 22 & 9 &0.01 &{\ } 26 & 9 &<0.01 &{\ } 18 & 9 &0.03 &{\ } 18 & 9 &0.03\\
$\yhad$ &{\ } 21 & 16 &0.18 &{\ } 20 & 16 &0.24 &{\ } 23 & 16 &0.10 &{\ } 20 & 16 &0.20 &{\ } 15 & 16 &0.51 &{\ } 15 & 16 &0.51\\
$\ylep$ &{\ } 14 & 13 &0.40 &{\ } 16 & 13 &0.23 &{\ } 13 & 13 &0.44 &{\ } 11 & 13 &0.65 &{\ } 14 & 13 &0.37 &{\ } 14 & 13 &0.37\\
$\yttbar$ &{\ } 7 & 9 &0.61 &{\ } 9 & 9 &0.45 &{\ } 5 & 9 &0.80 &{\ } 5 & 9 &0.81 &{\ } 14 & 9 &0.13 &{\ } 14 & 9 &0.13\\
$\mttbar$ &{\ } 18 & 9 &0.03 &{\ } 32 & 9 &<0.01 &{\ } 43 & 9 &<0.01 &{\ } 16 & 9 &0.07 &{\ } 21 & 9 &0.01 &{\ } 21 & 9 &0.01\\
$\pTe{1}$ &{\ } 84 & 14 &<0.01 &{\ } 108 & 14 &<0.01 &{\ } 187 & 14 &<0.01 &{\ } 37 & 14 &<0.01 &{\ } 113 & 14 &<0.01 &{\ } 113 & 14 &<0.01\\
$\pTe{2}$ &{\ } 29 & 8 &<0.01 &{\ } 11 & 8 &0.21 &{\ } 96 & 8 &<0.01 &{\ } 32 & 8 &<0.01 &{\ } 23 & 8 &<0.01 &{\ } 23 & 8 &<0.01\\
$\Ne$ &{\ } 7 & 4 &0.12 &{\ } 16 & 4 &<0.01 &{\ } 12 & 4 &0.01 &{\ } 12 & 4 &0.02 &{\ } 13 & 4 &0.01 &{\ } 13 & 4 &0.01\\
$\dPhieth{1}$ &{\ } 9 & 8 &0.32 &{\ } 27 & 8 &<0.01 &{\ } 4 & 8 &0.86 &{\ } 4 & 8 &0.82 &{\ } 4 & 8 &0.85 &{\ } 4 & 8 &0.85\\
$\dPhieth{2}$ &{\ } 5 & 8 &0.78 &{\ } 6 & 8 &0.64 &{\ } 6 & 8 &0.69 &{\ } 6 & 8 &0.63 &{\ } 1 & 8 &1.00 &{\ } 1 & 8 &1.00\\
$\dPhibt$ &{\ } 35 & 12 &<0.01 &{\ } 48 & 12 &<0.01 &{\ } 192 & 12 &<0.01 &{\ } 46 & 12 &<0.01 &{\ } 29 & 12 &<0.01 &{\ } 29 & 12 &<0.01\\
$\dPhitt$ &{\ } 62 & 4 &<0.01 &{\ } 76 & 4 &<0.01 &{\ } 156 & 4 &<0.01 &{\ } 41 & 4 &<0.01 &{\ } 29 & 4 &<0.01 &{\ } 29 & 4 &<0.01\\
$\dPhiee$ &{\ } 16 & 10 &0.11 &{\ } 15 & 10 &0.13 &{\ } 18 & 10 &0.06 &{\ } 16 & 10 &0.11 &{\ } 19 & 10 &0.04 &{\ } 19 & 10 &0.04\\
$\meth$ &{\ } 30 & 11 &<0.01 &{\ } 88 & 11 &<0.01 &{\ } 46 & 11 &<0.01 &{\ } 11 & 11 &0.44 &{\ } 45 & 11 &<0.01 &{\ } 45 & 11 &<0.01\\
$\pTe{1}$ vs $\Ne$ &{\ } 261 & 20 &<0.01 &{\ } 379 & 20 &<0.01 &{\ } 363 & 20 &<0.01 &{\ } 161 & 20 &<0.01 &{\ } 161 & 20 &<0.01 &{\ } 161 & 20 &<0.01\\
$\pTe{1}$ vs $\pThad$ &{\ } 80 & 16 &<0.01 &{\ } 151 & 16 &<0.01 &{\ } 170 & 16 &<0.01 &{\ } 42 & 16 &<0.01 &{\ } 103 & 16 &<0.01 &{\ } 103 & 16 &<0.01\\
$\dPhieth{1}$ vs $\pThad$ &{\ } 48 & 20 &<0.01 &{\ } 82 & 20 &<0.01 &{\ } 27 & 20 &0.13 &{\ } 37 & 20 &0.01 &{\ } 30 & 20 &0.07 &{\ } 30 & 20 &0.07\\
$\dPhieth{1}$ vs $\Ne$ &{\ } 68 & 18 &<0.01 &{\ } 146 & 18 &<0.01 &{\ } 36 & 18 &<0.01 &{\ } 40 & 18 &<0.01 &{\ } 23 & 18 &0.18 &{\ } 23 & 18 &0.18\\
\bottomrule
\end{tabular}
}
\label{table:chi2_table_relative}
\end{table}

\begin{figure}[htbp]
\centering
\subfloat[]{
\includegraphics[width=0.32\textwidth]{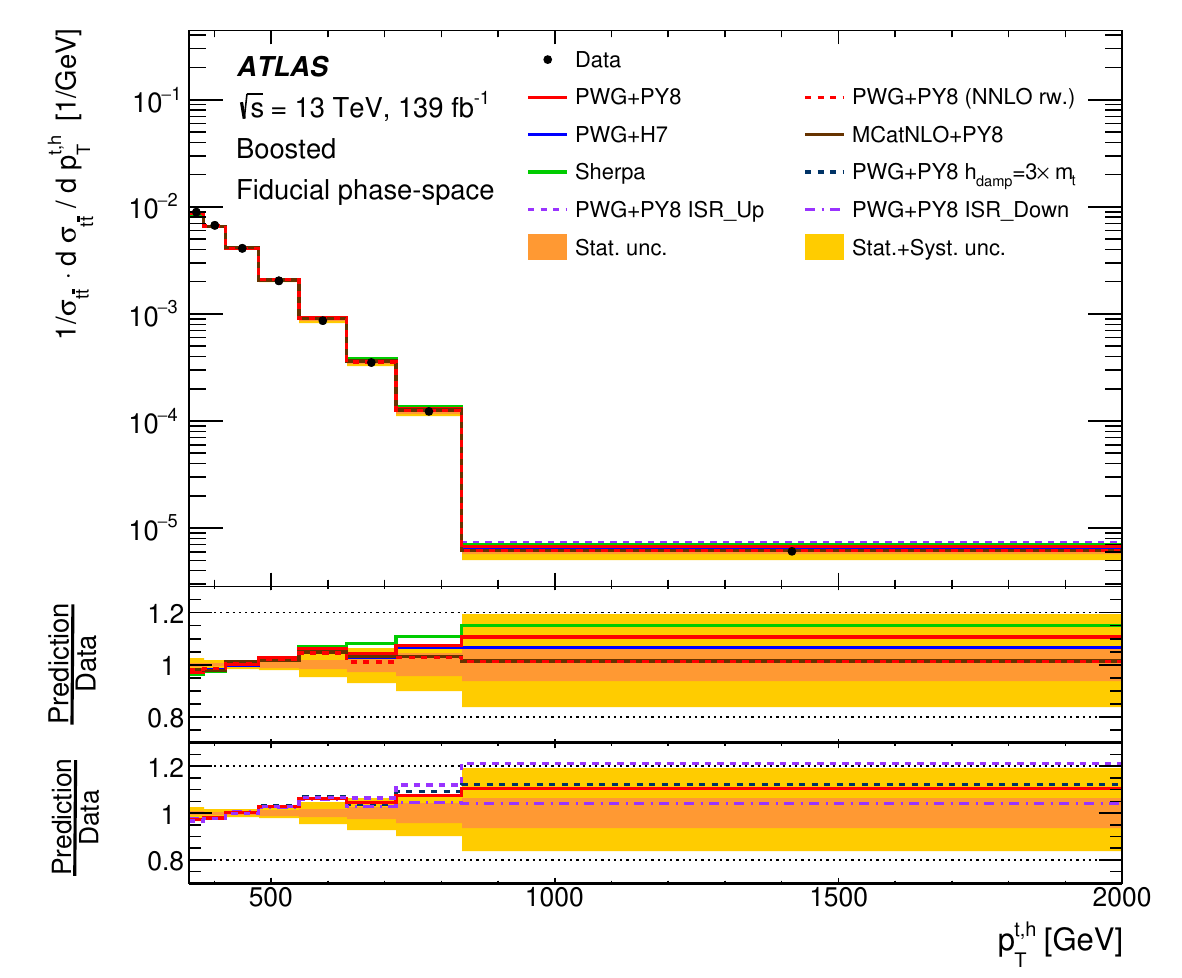}}
\subfloat[]{
\includegraphics[width=0.32\textwidth]{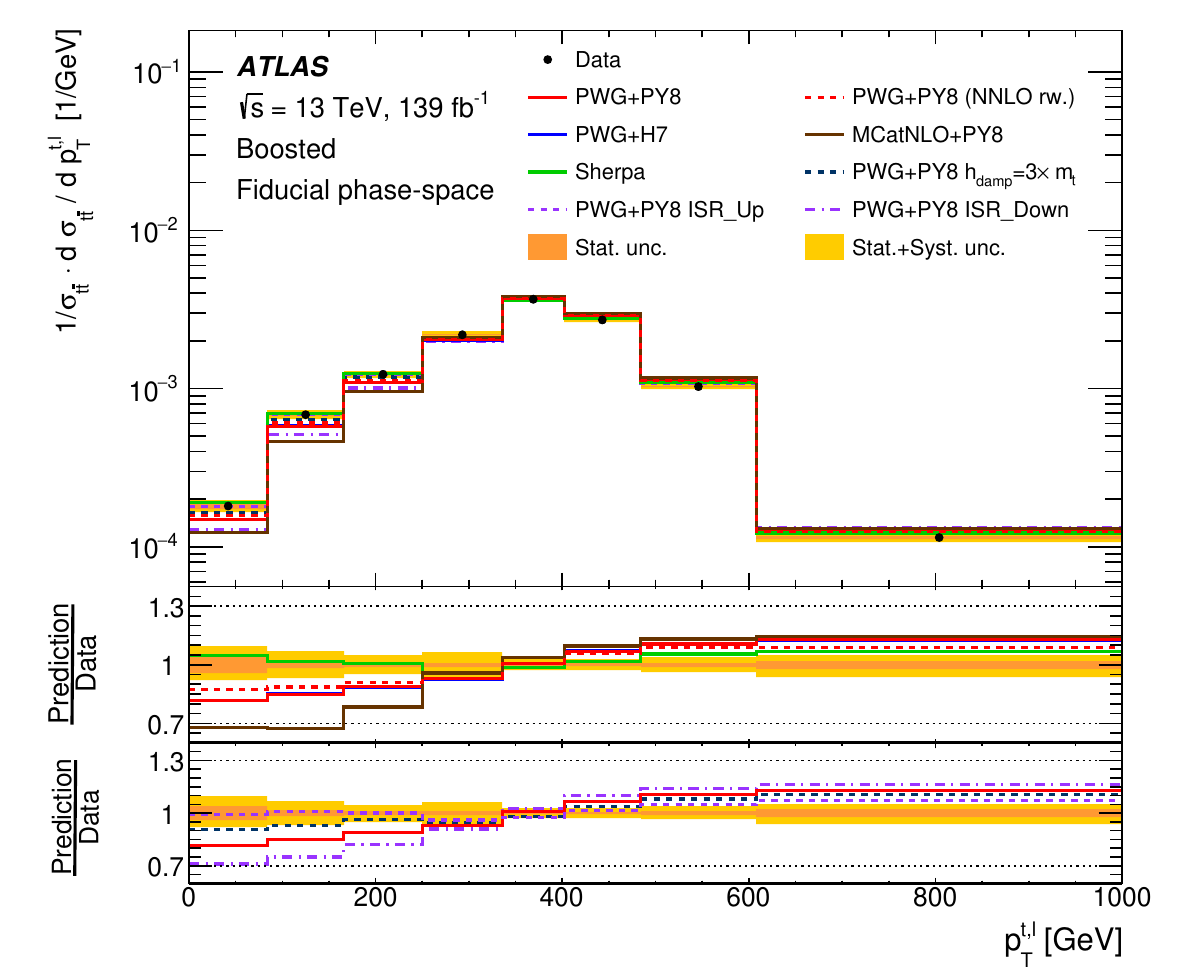}}
\subfloat[]{
\includegraphics[width=0.32\textwidth]{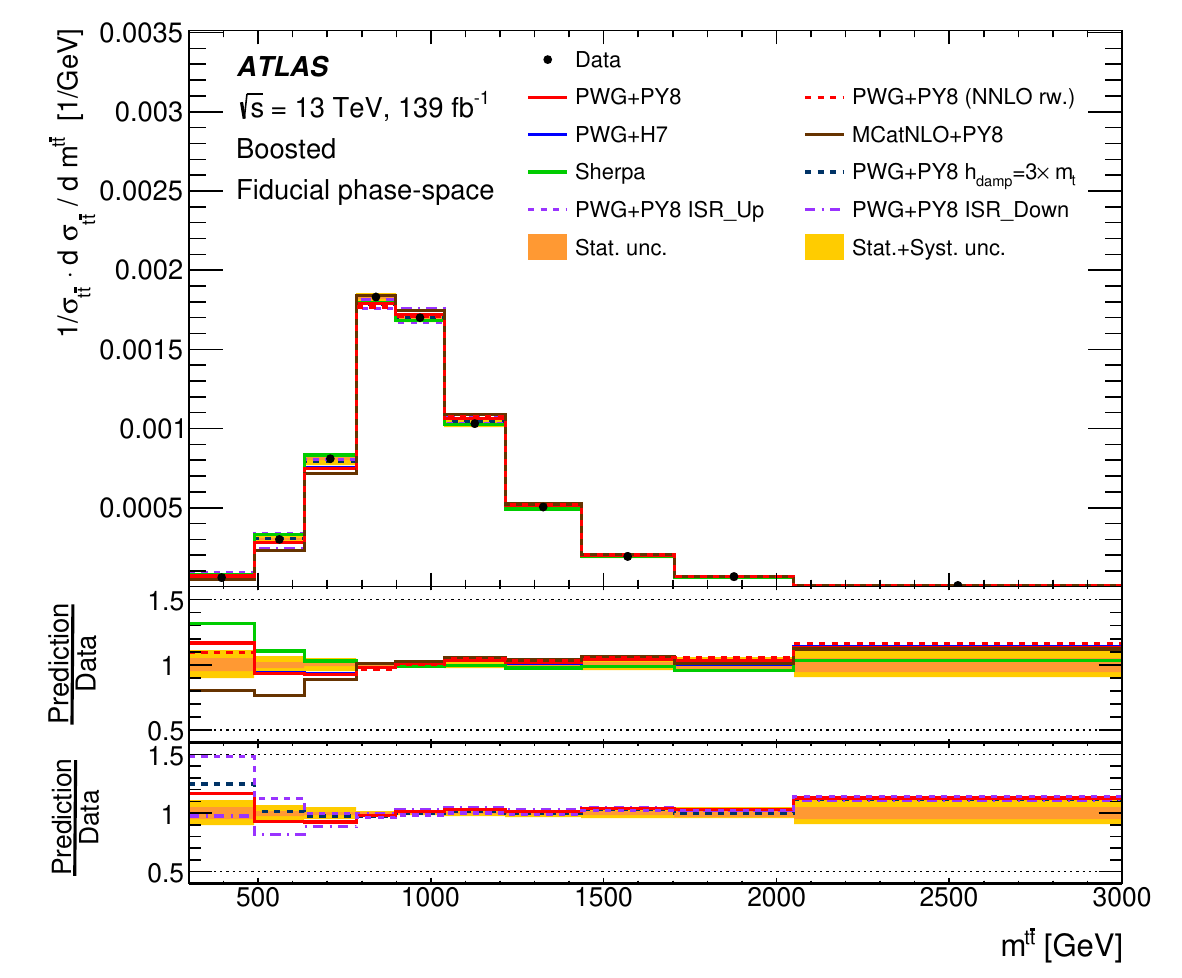}}
 
\subfloat[]{
\includegraphics[width=0.32\textwidth]{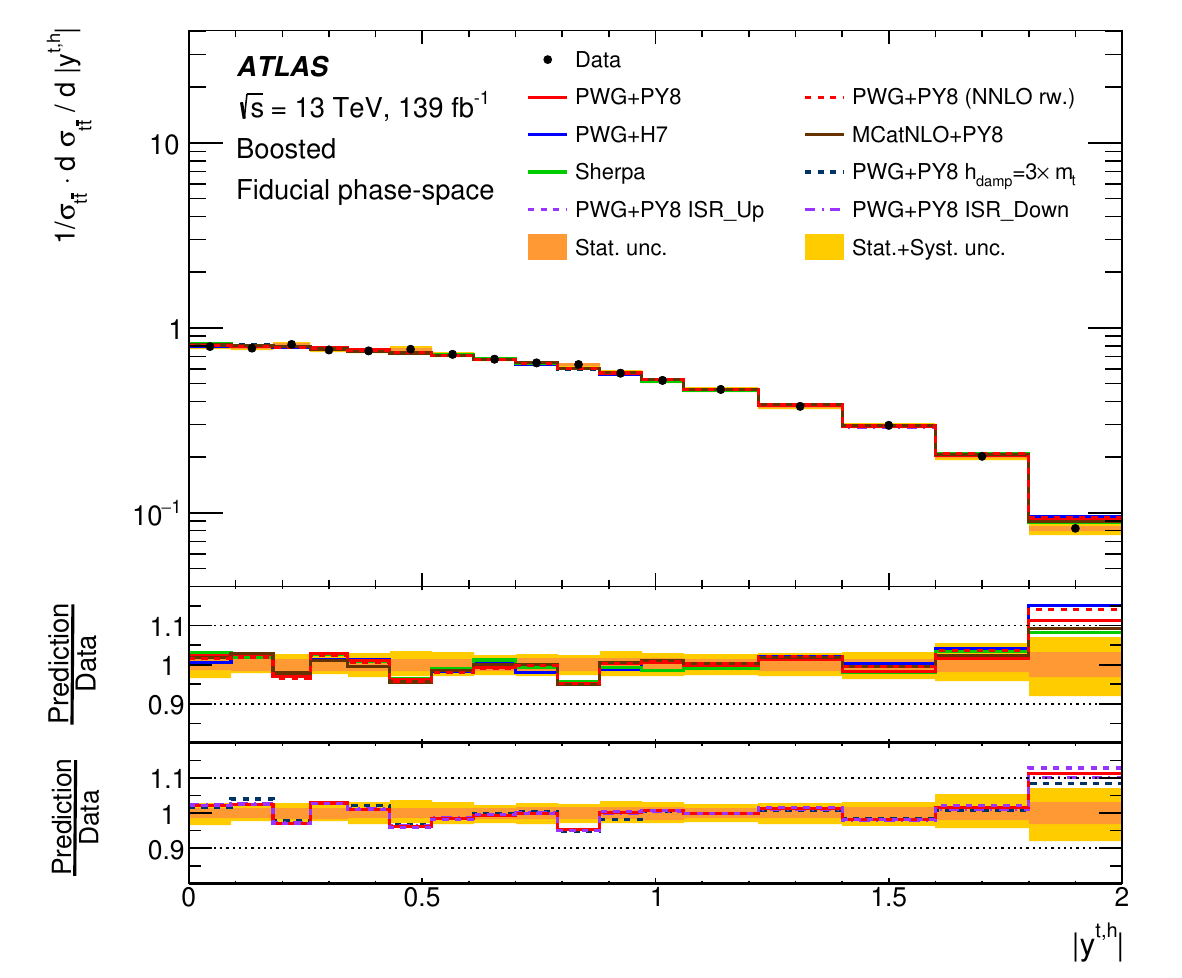}}
\subfloat[]{
\includegraphics[width=0.32\textwidth]{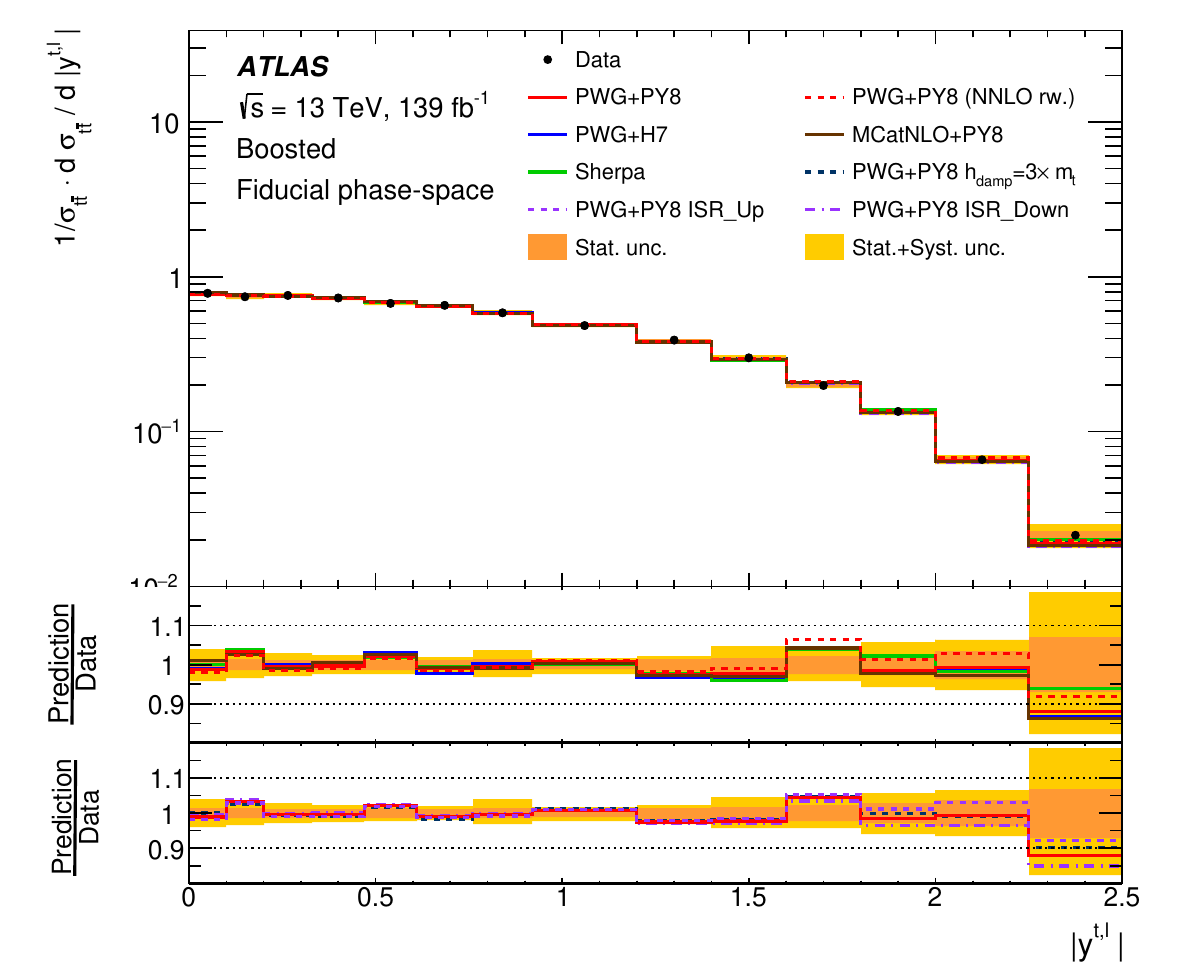}}
\subfloat[]{
\includegraphics[width=0.32\textwidth]{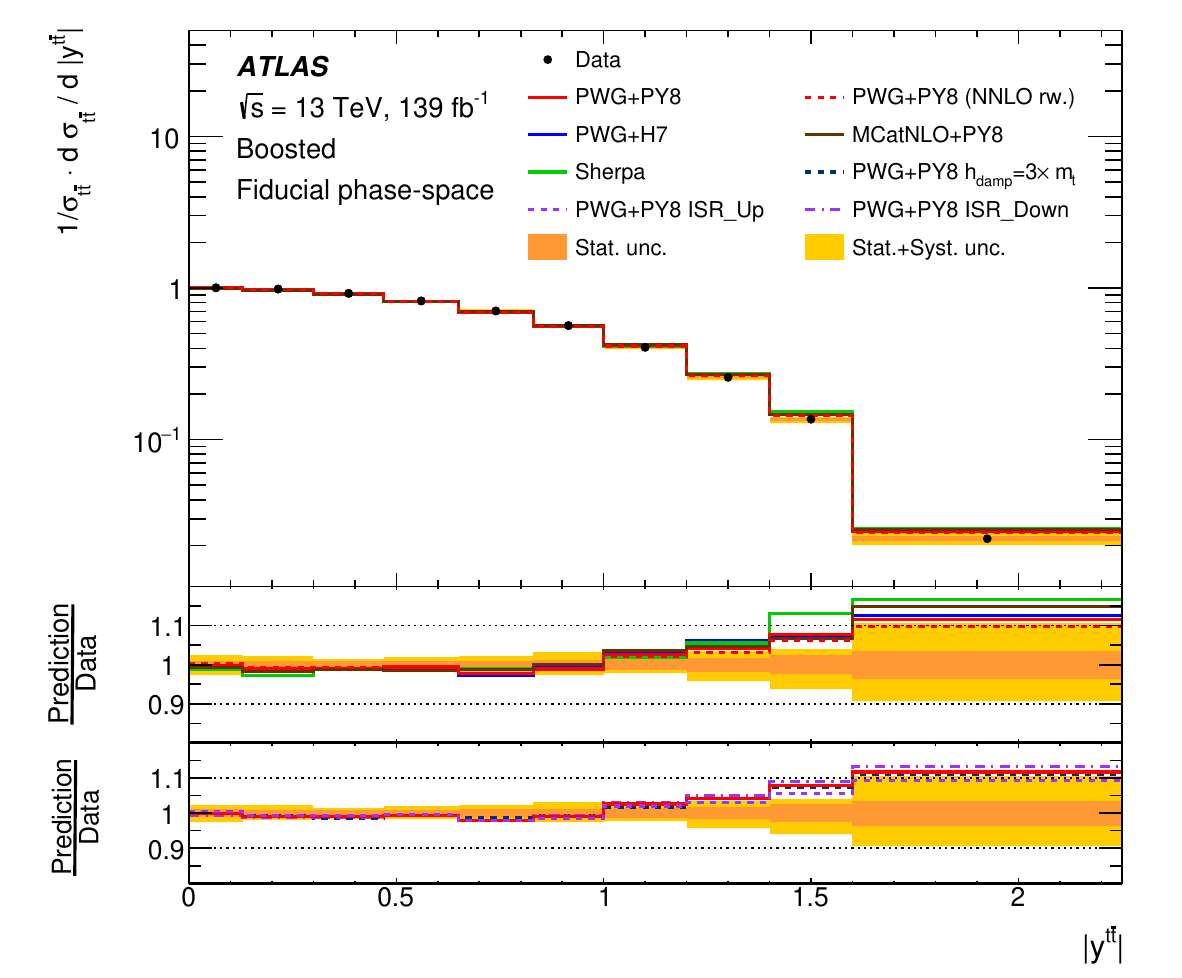}}
 
\subfloat[]{
\includegraphics[width=0.32\textwidth]{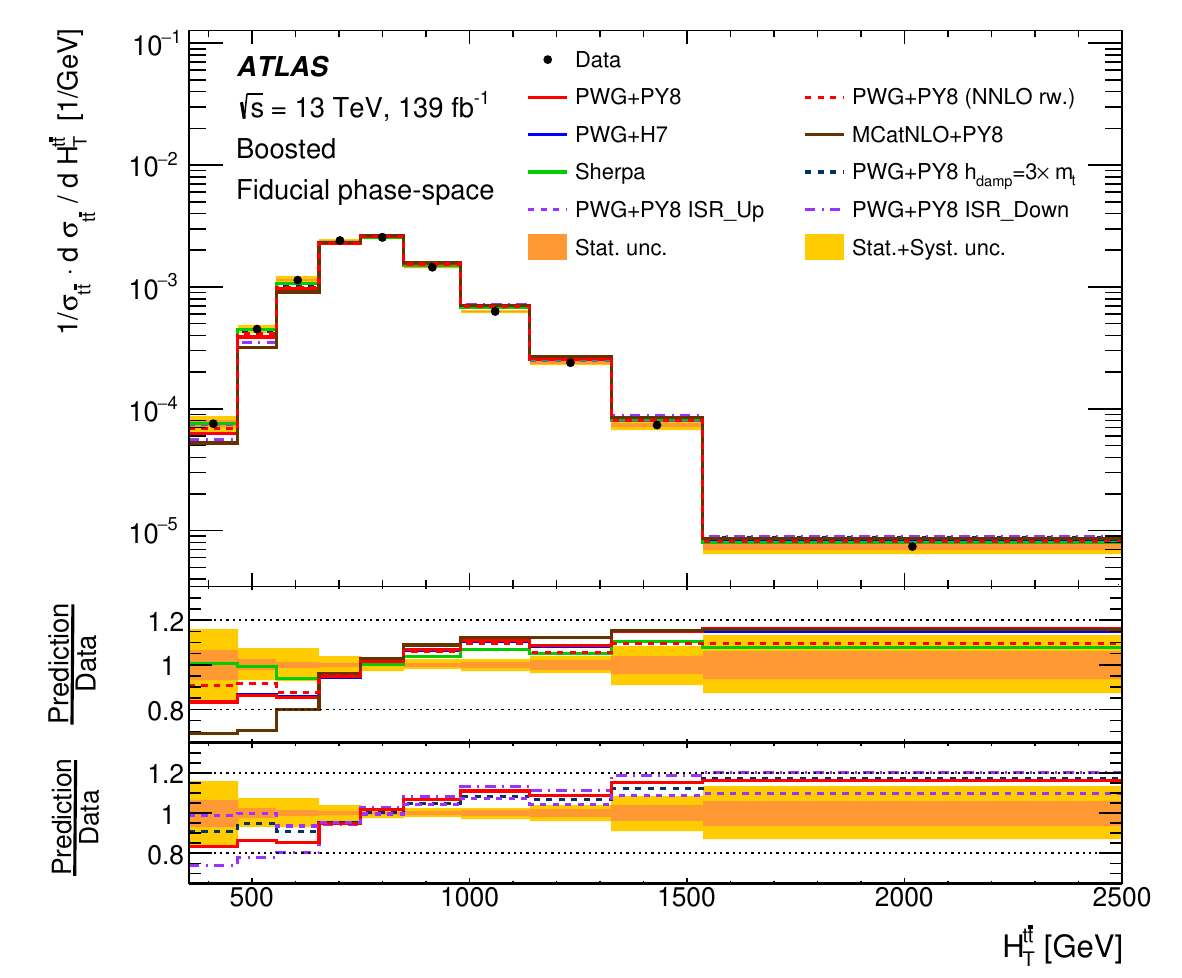}}
\subfloat[]{
\includegraphics[width=0.32\textwidth]{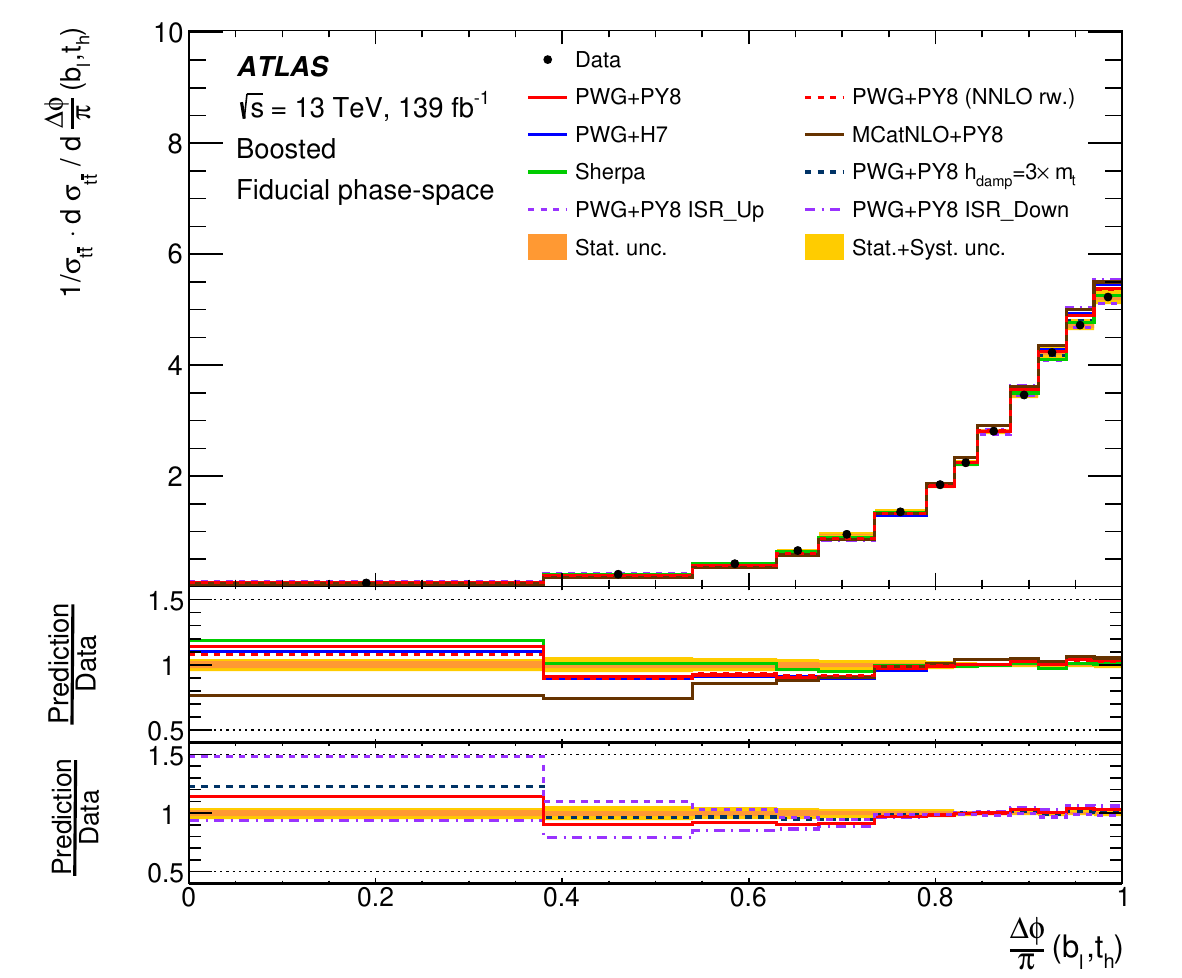}}
\caption{Normalised differential cross-section measurements for the observables related to the kinematics of the top quarks. The cross-section is shown as a function of
(a) $\pThad$,
(b) $\pTlep$,
(c) $\mttbar$,
(d) $\yhad$,
(e) $\ylep$,
(f) $\yttbar$,
(g) $\HTttbar$\ and
(h) $\dPhibt$,
In each plot the data are compared with predictions from various MC generators.
PWG+PY8 corresponds to the \POWPY sample, PWG+H7 to the \POWHER sample and MCatNLO+PY8 to the \MGNLOPY sample.
The orange band shows the size of the statistical uncertainty and the yellow band shows the size of the total uncertainty. The lower two panels
in each subfigure display the ratios of the different predictions to the data.
}
\label{figaux:diffxstop_1D_rel}
\end{figure}
 
\begin{figure}[htbp]
\centering
\subfloat[]{
\includegraphics[width=0.32\textwidth]{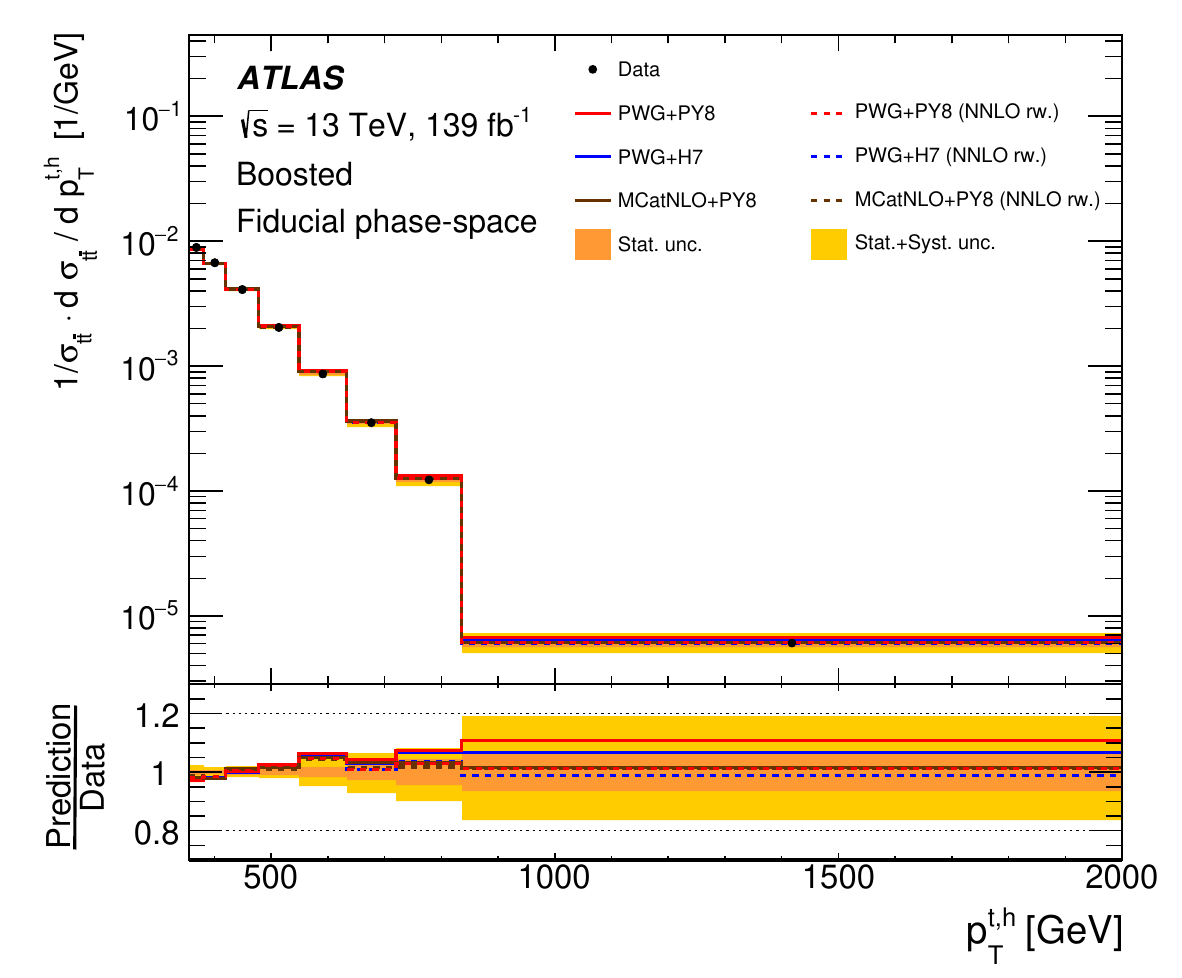}}
\subfloat[]{
\includegraphics[width=0.32\textwidth]{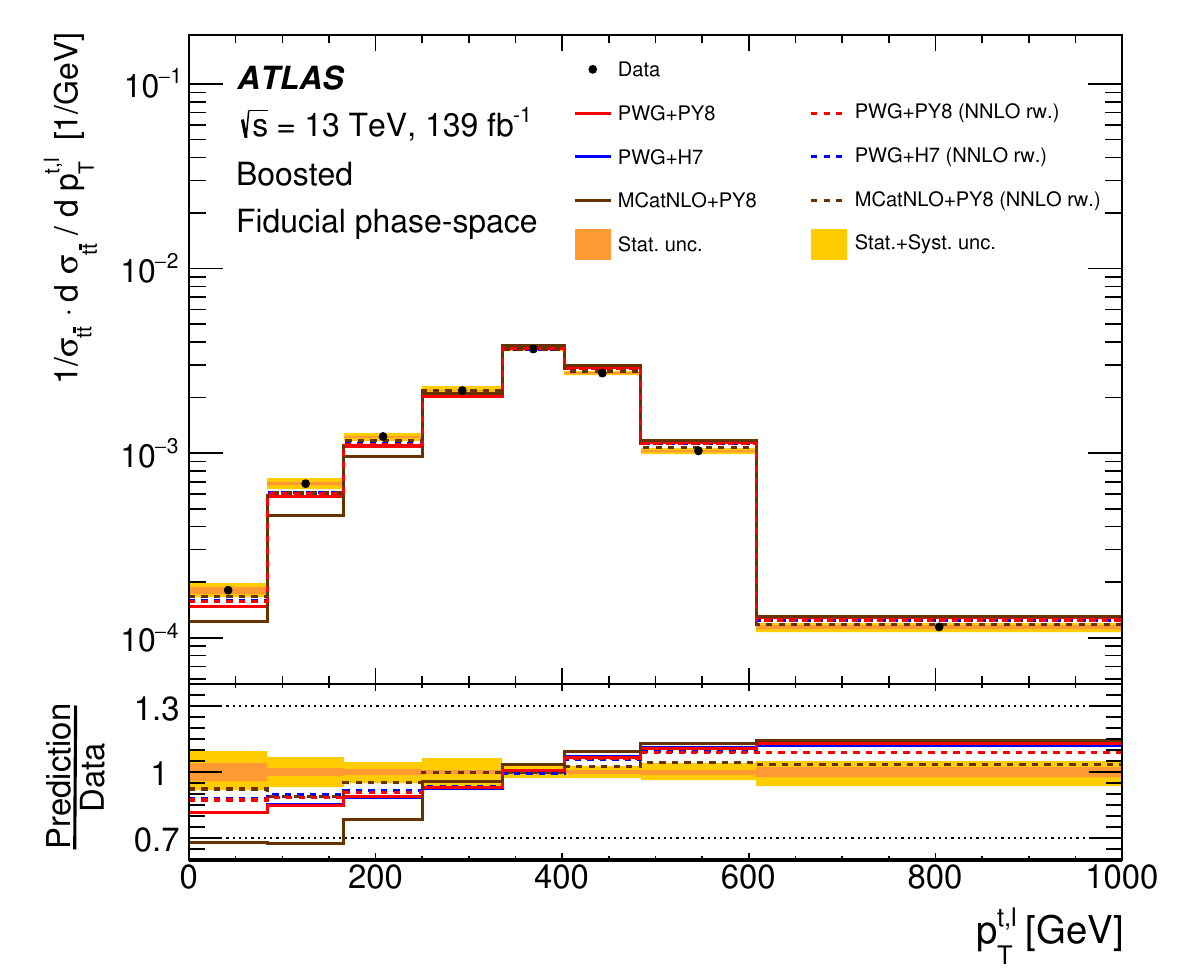}}
\subfloat[]{
\includegraphics[width=0.32\textwidth]{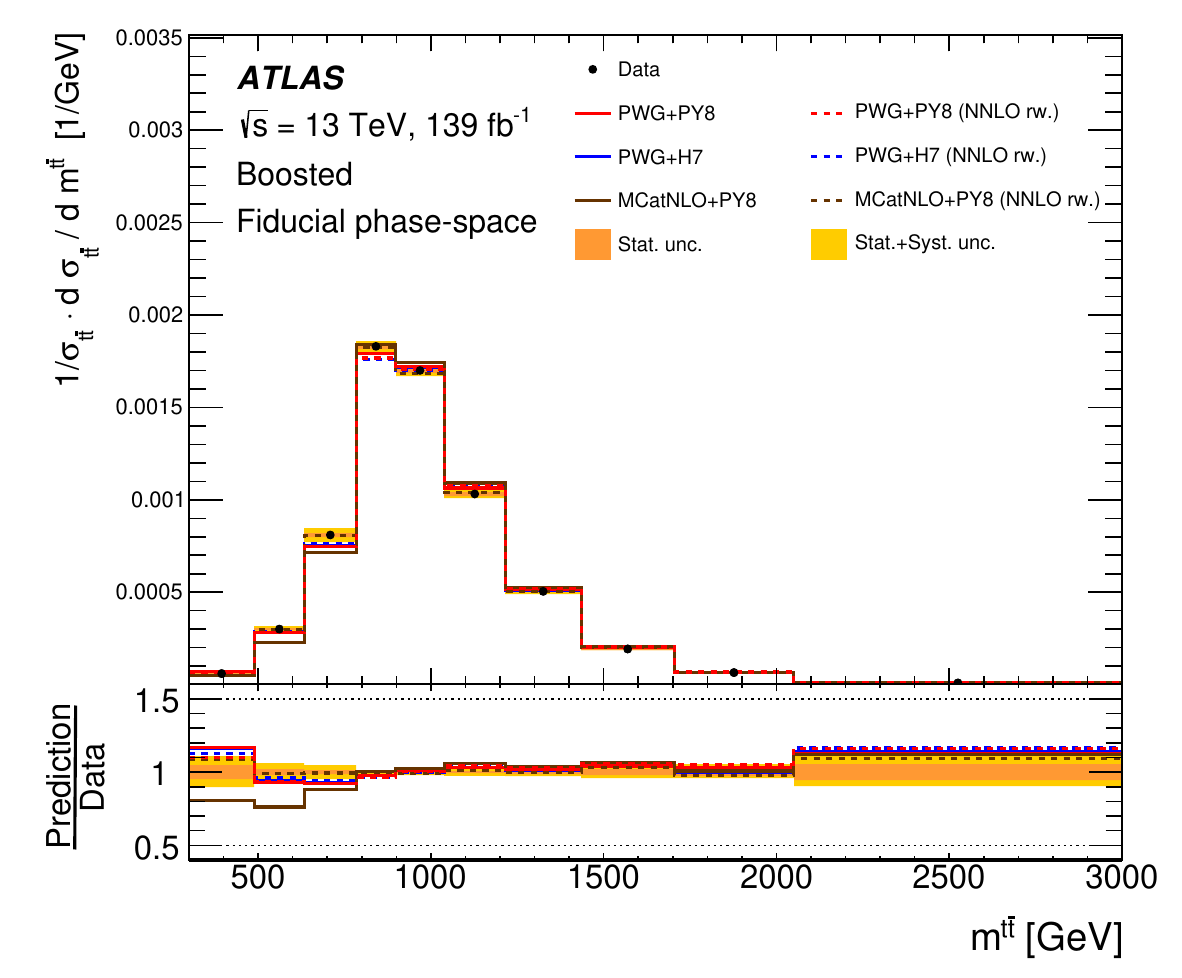}}
 
\subfloat[]{
\includegraphics[width=0.32\textwidth]{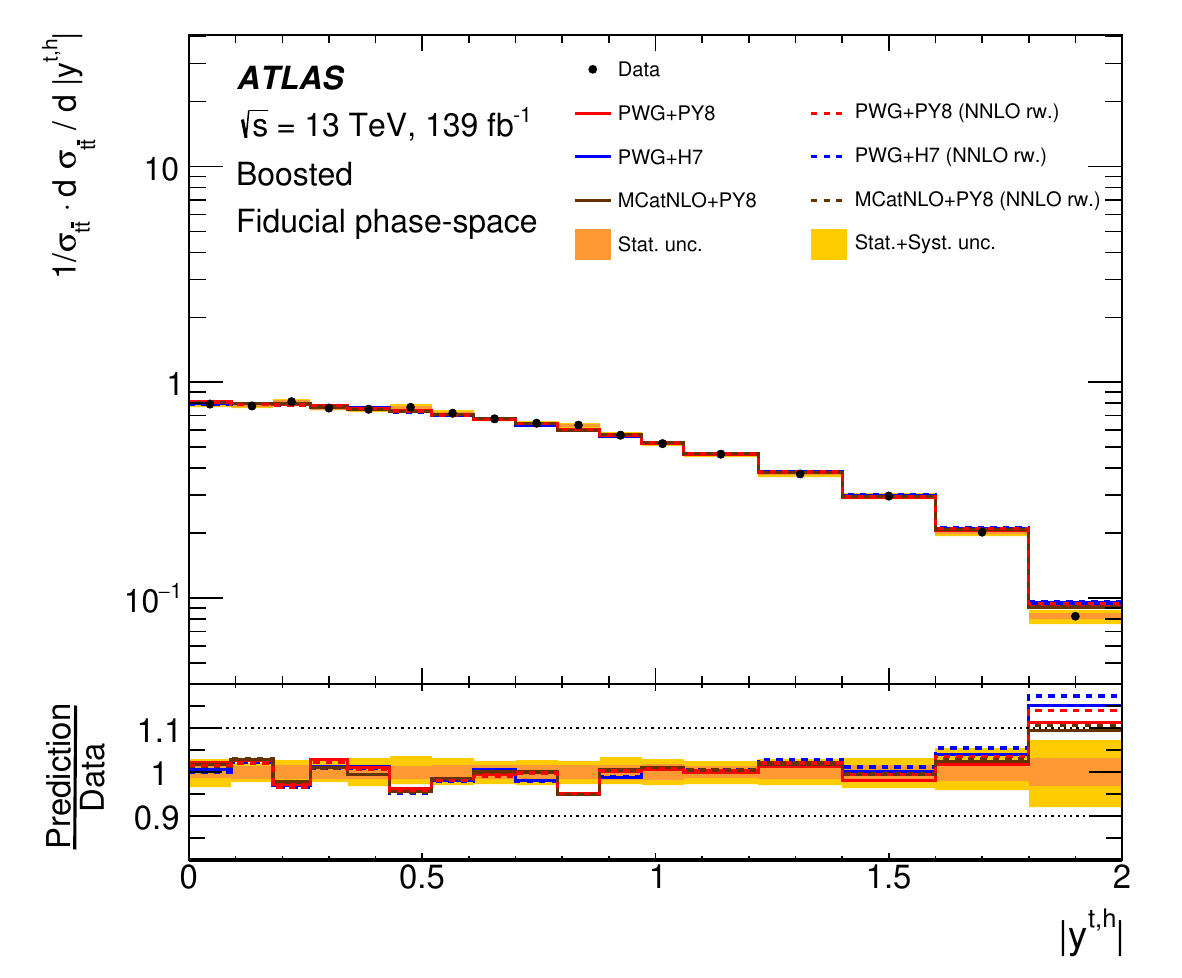}}
\subfloat[]{
\includegraphics[width=0.32\textwidth]{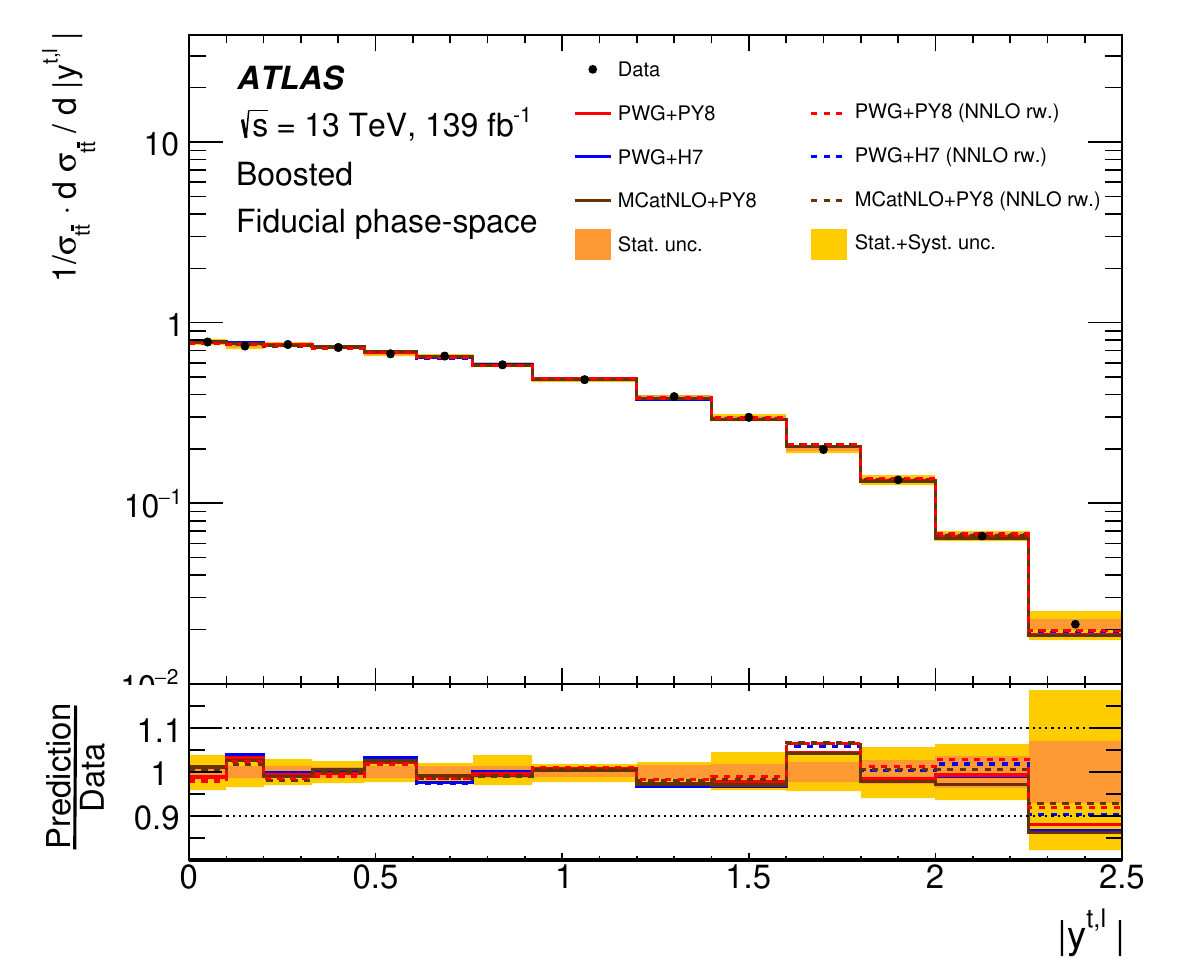}}
\subfloat[]{
\includegraphics[width=0.32\textwidth]{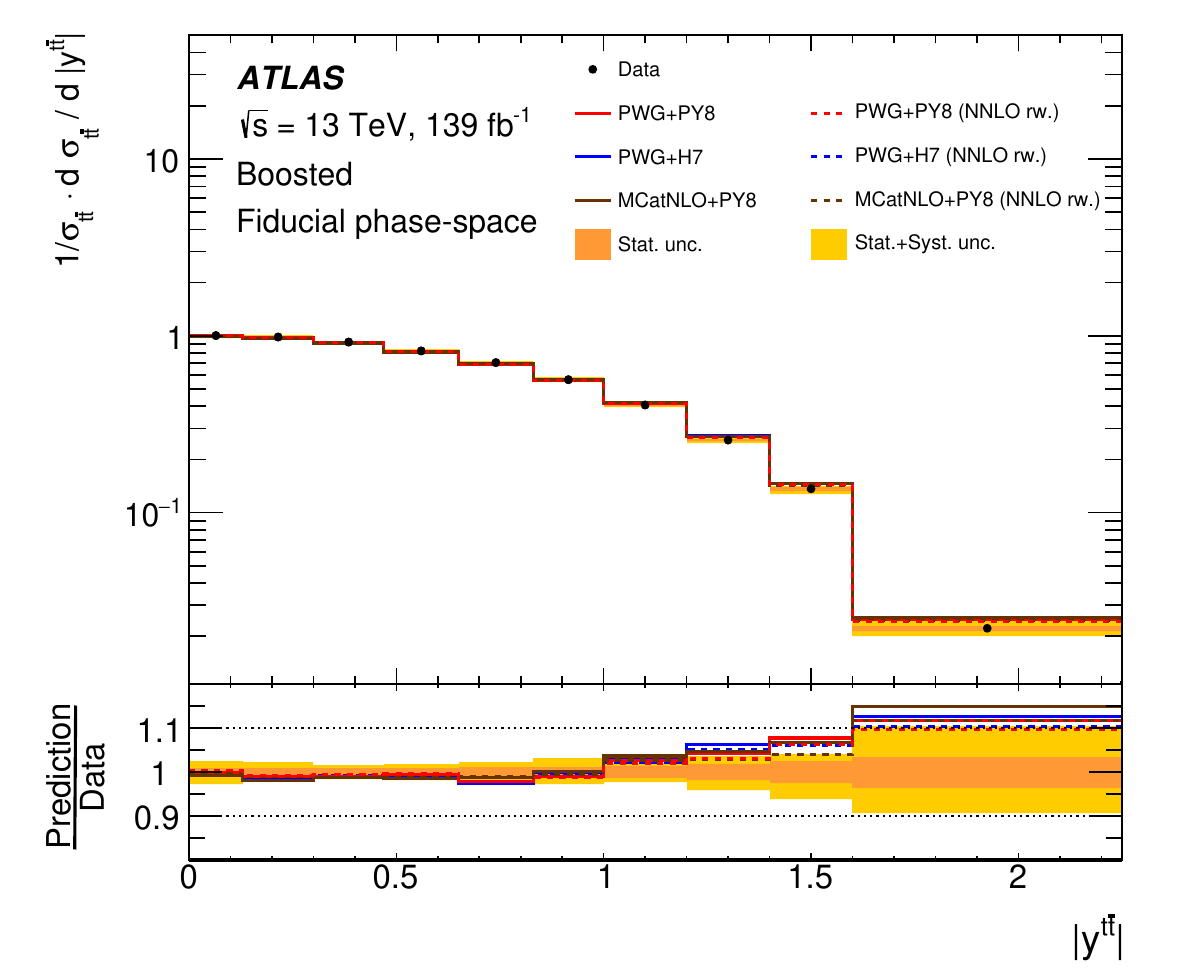}}
 
\subfloat[]{
\includegraphics[width=0.32\textwidth]{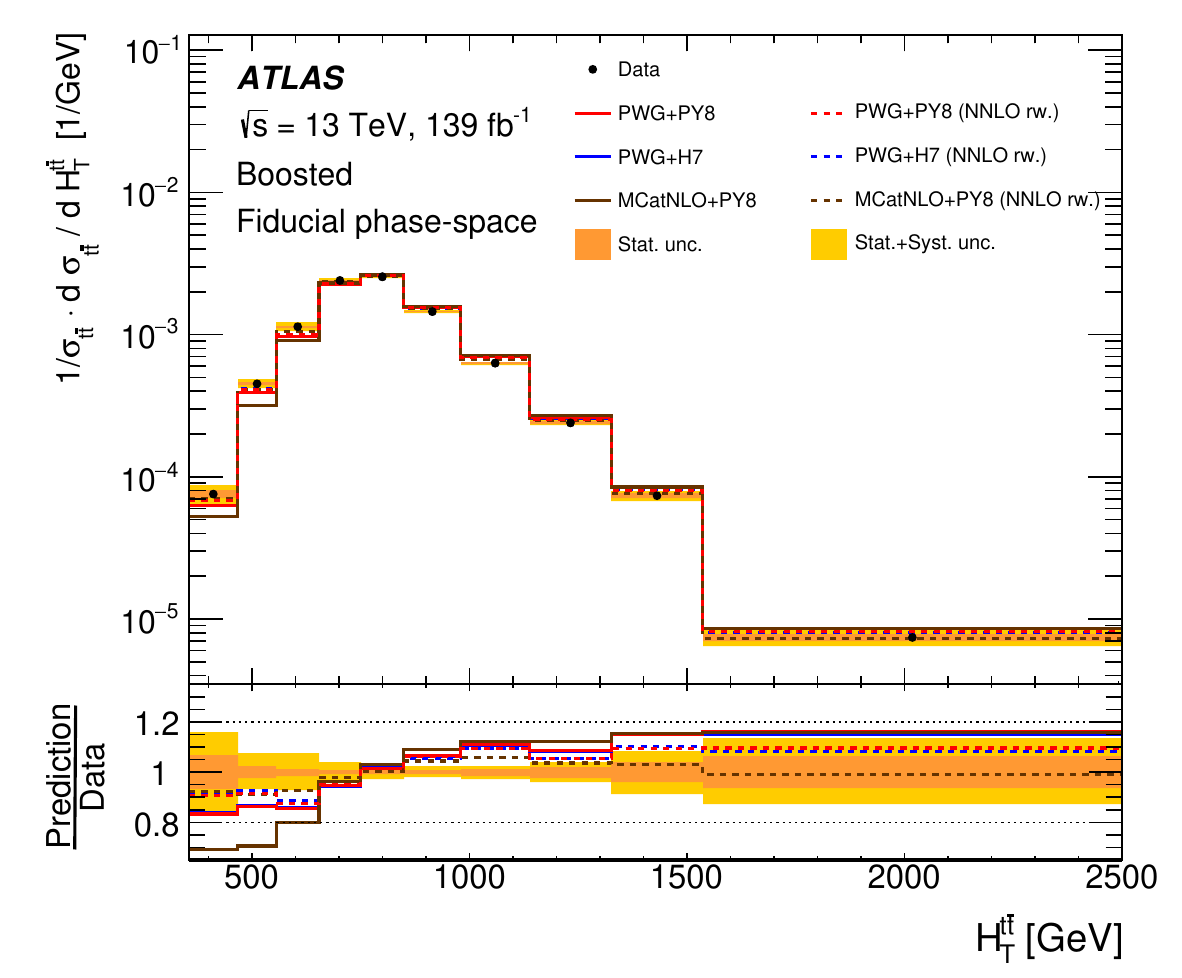}}
\subfloat[]{
\includegraphics[width=0.32\textwidth]{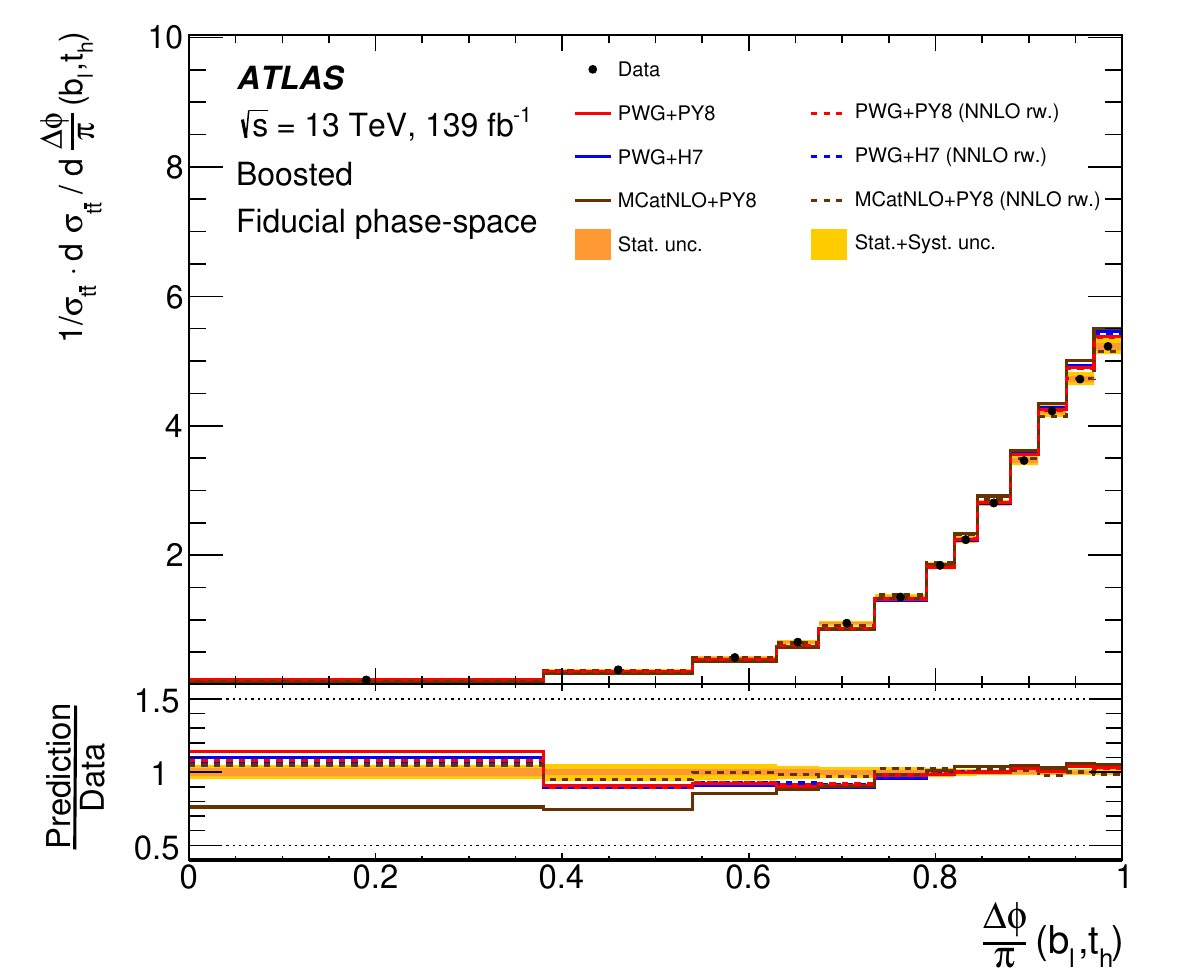}}
\caption{Normalised differential cross-section measurements for the observables related to the kinematics of the top quarks. The cross-section is shown as a function of
(a) $\pThad$,
(b) $\pTlep$,
(c) $\mttbar$,
(d) $\yhad$,
(e) $\ylep$,
(f) $\yttbar$,
(g) $\HTttbar$ ~and (h) $\dPhibt$.
In each plot the data are compared with predictions from the NLO generators with and without the NNLO reweighting.
PWG+PY8 corresponds to the \POWPY sample, PWG+H7 to the \POWHER sample and MCatNLO+PY8 to the \MGNLOPY sample.
The orange band shows the size of the statistical uncertainty and the yellow band shows the size of the total uncertainty. The lower panel
in each subfigure displays the ratios of the different predictions to the data.
}
\label{figaux:diffxstopNNLO_rel}
\end{figure}
 
\begin{figure}[htbp]
\centering
\subfloat[]{
\includegraphics[width=0.32\textwidth]{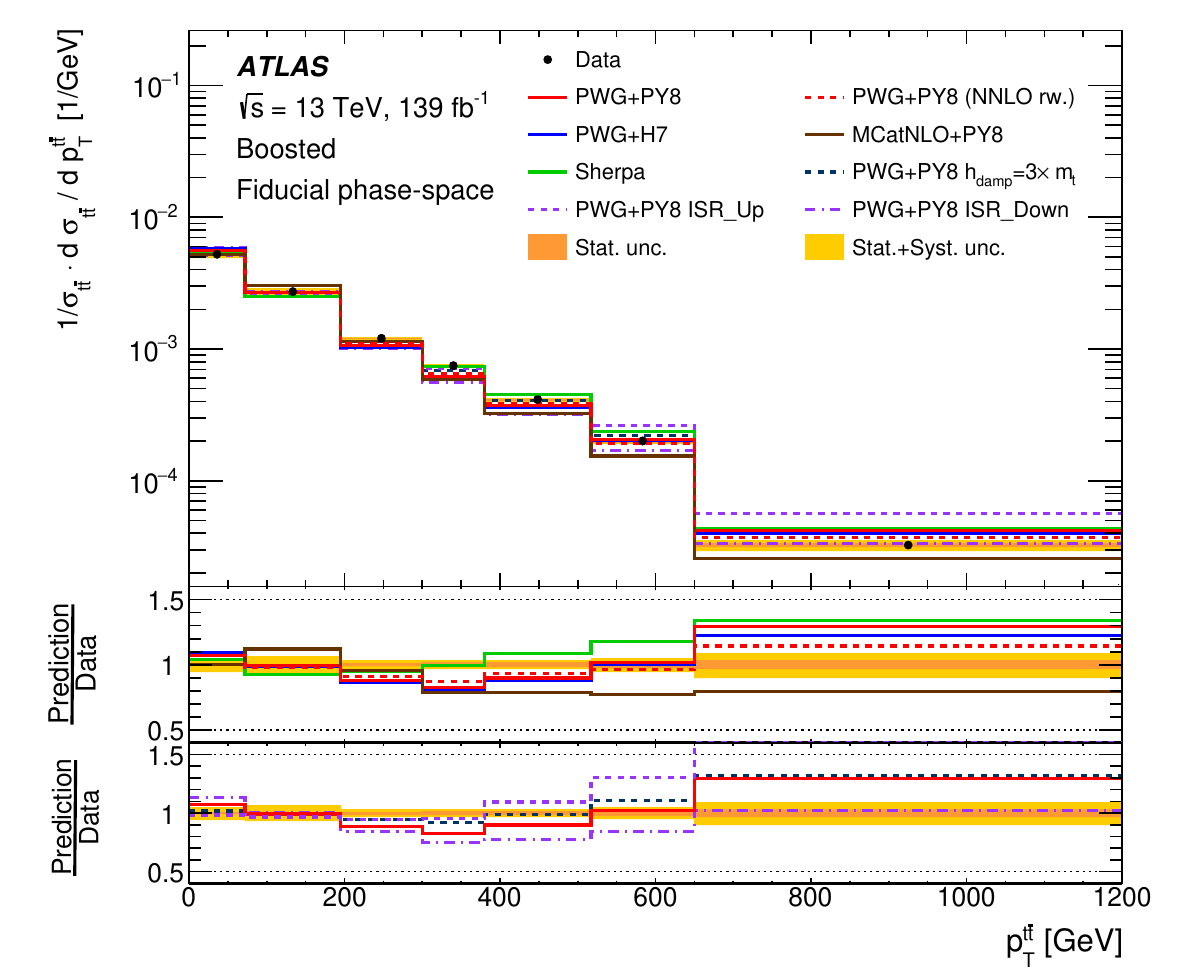}}
\subfloat[]{
\includegraphics[width=0.32\textwidth]{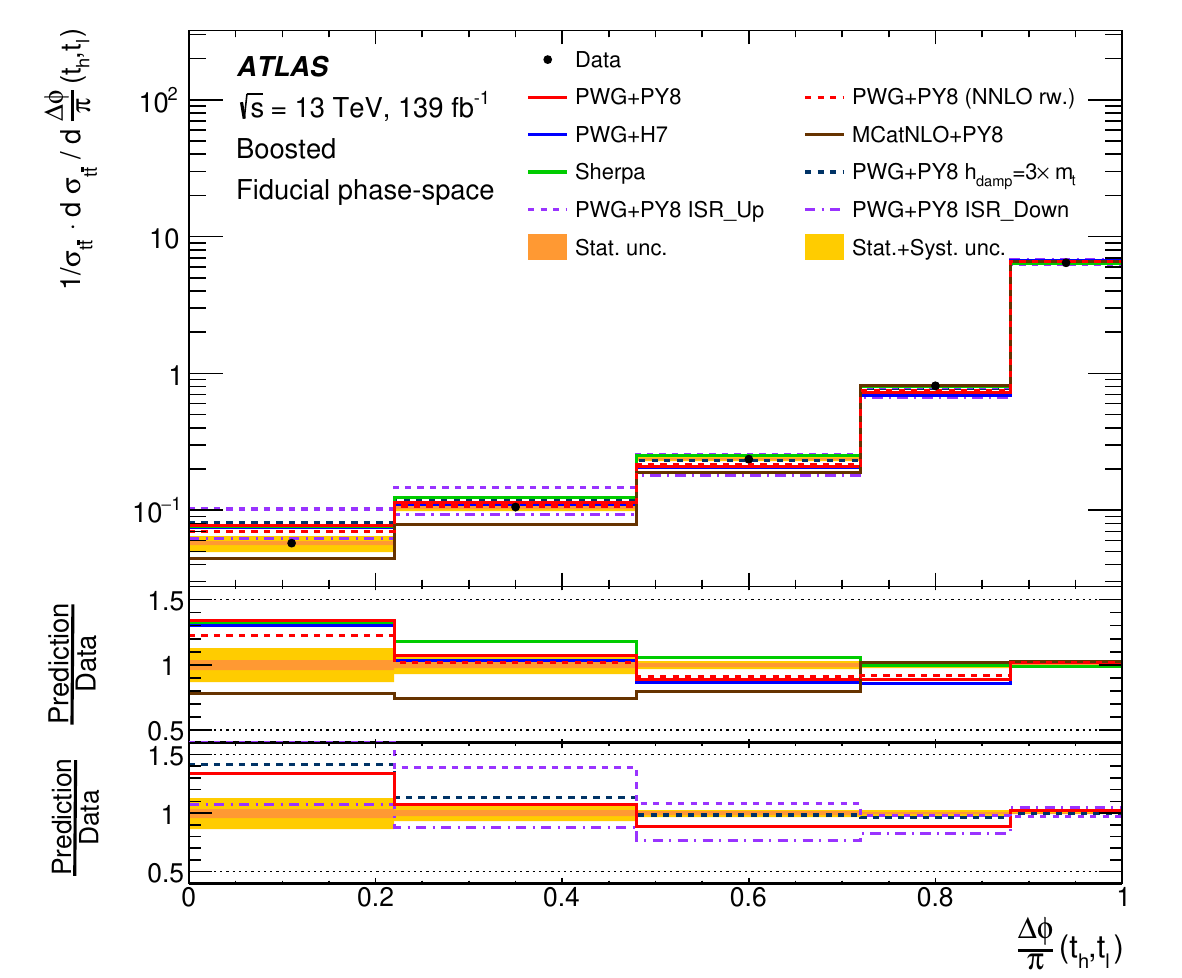}}
\subfloat[]{
\includegraphics[width=0.32\textwidth]{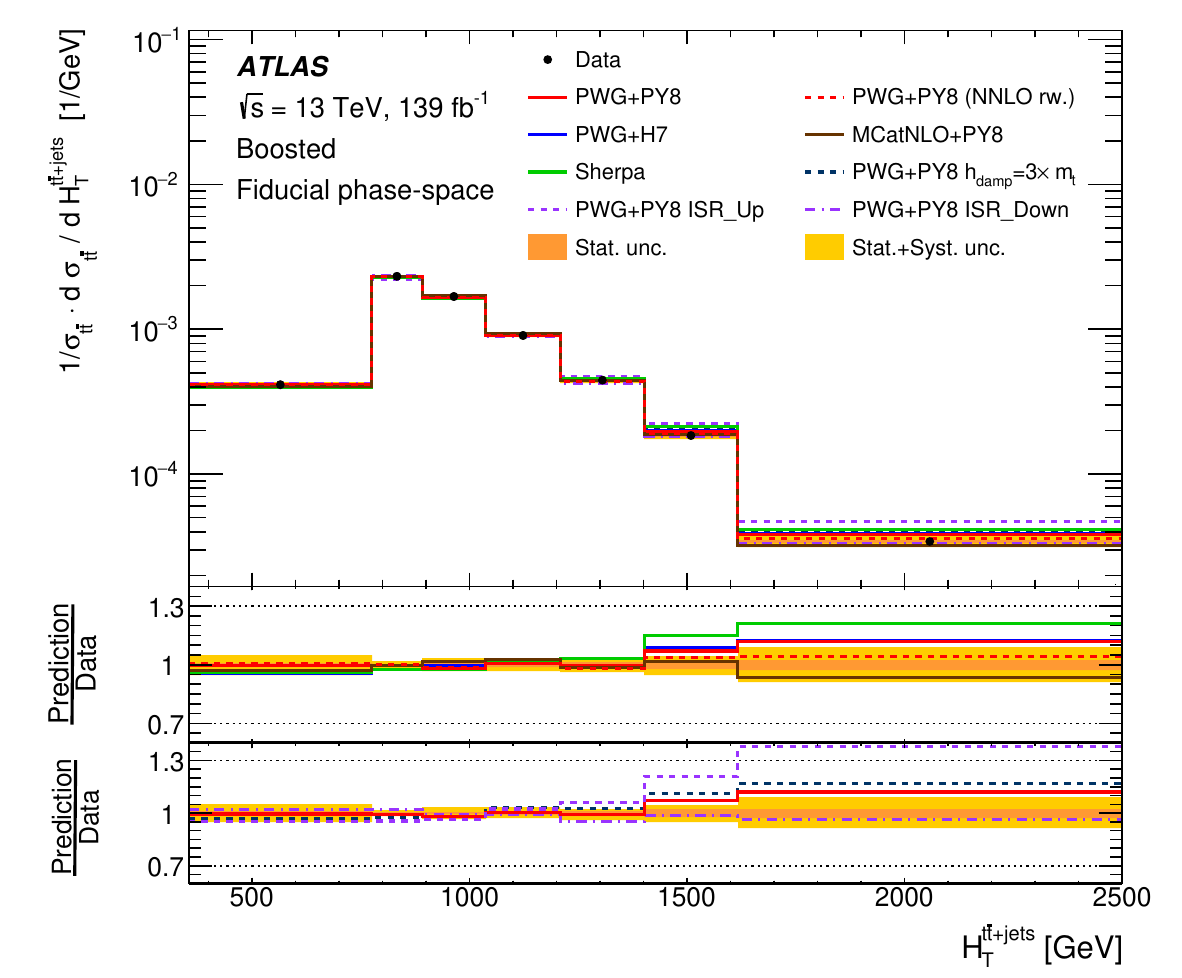}}
 
\subfloat[]{
\includegraphics[width=0.32\textwidth]{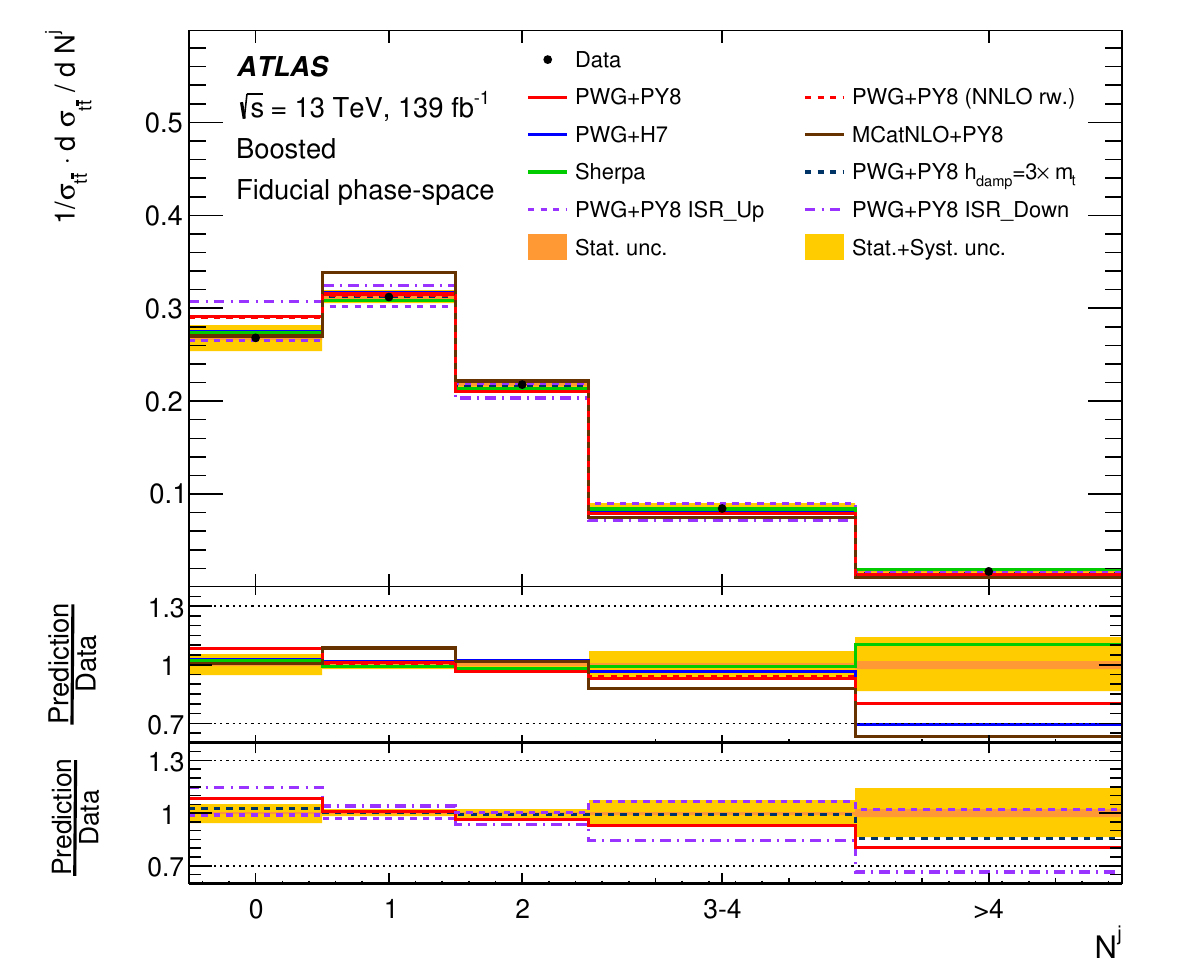}}
\subfloat[]{
\includegraphics[width=0.32\textwidth]{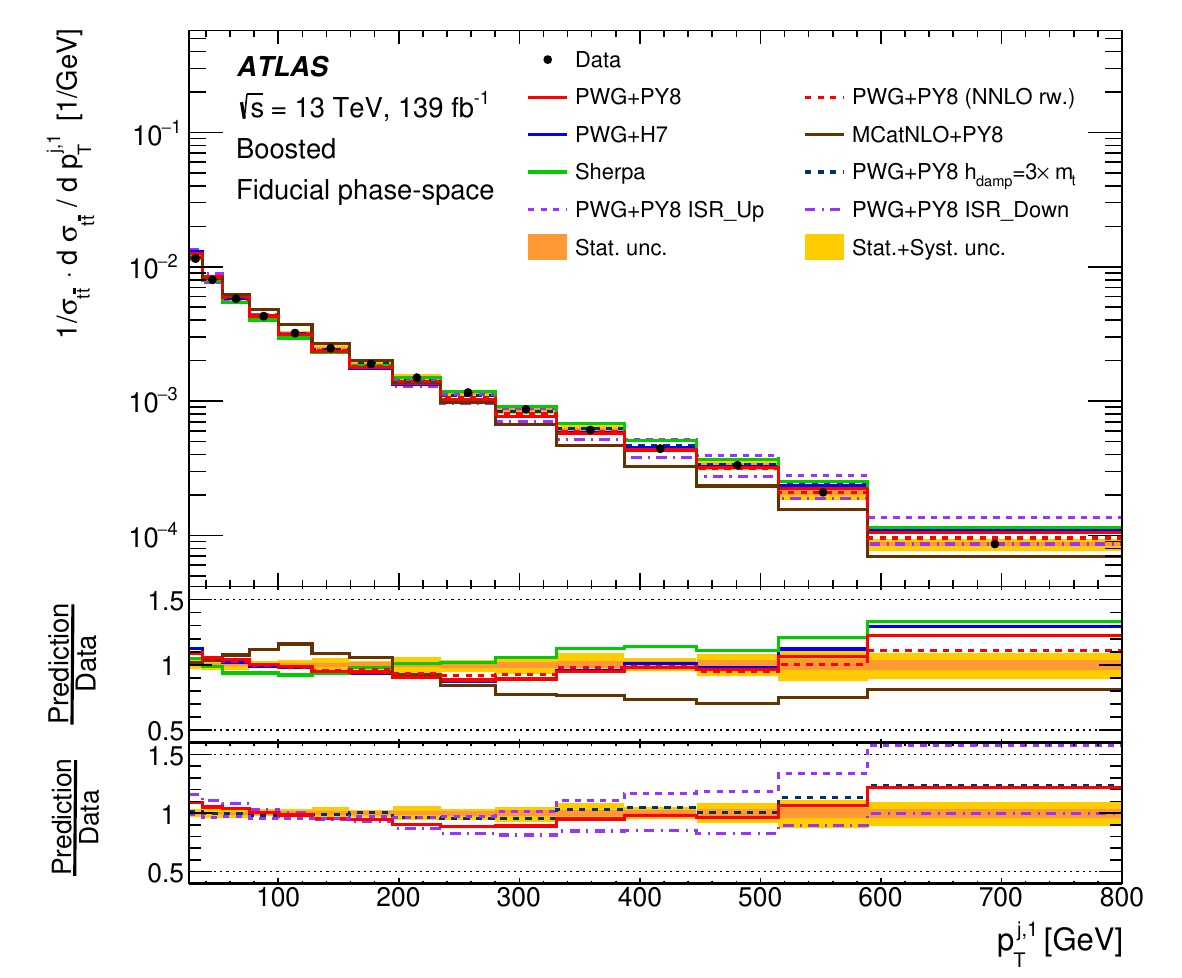}}
\subfloat[]{
\includegraphics[width=0.32\textwidth]{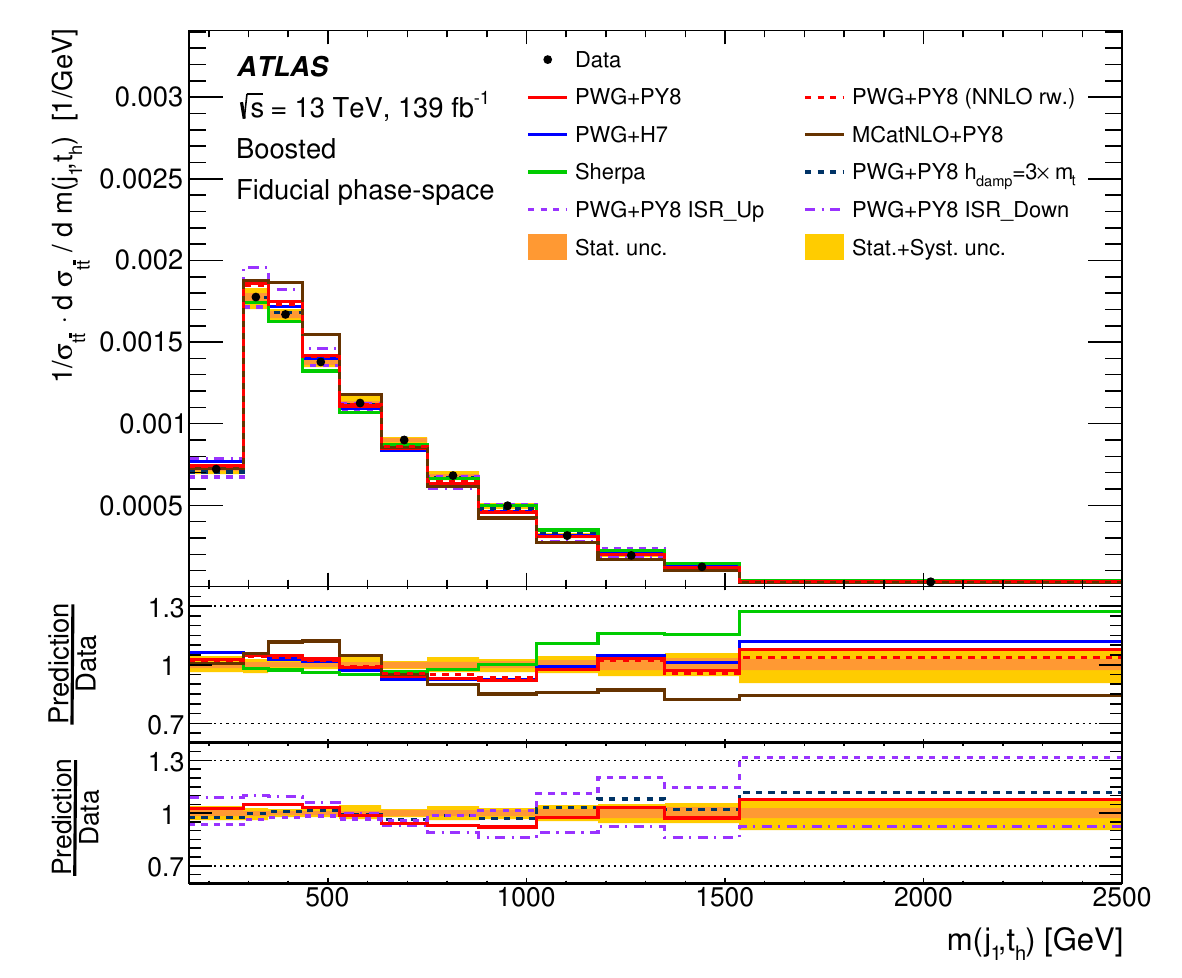}}
 
\subfloat[]{
\includegraphics[width=0.32\textwidth]{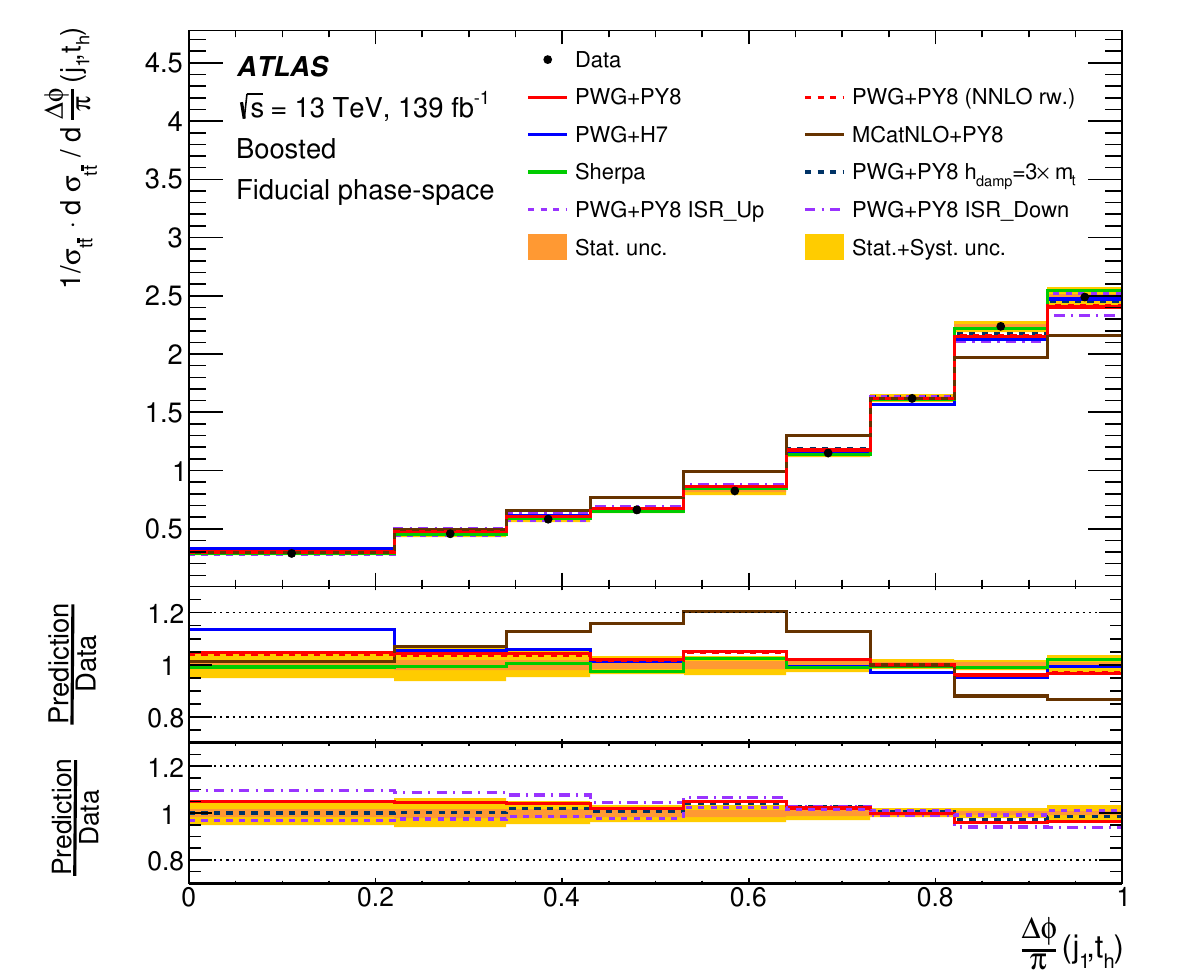}}

\caption{Normalised differential cross-section measurements for the observables related to jets produced in association with the $\ttbar$\ system. The cross-section is shown as a function of
(a) $\pTttbar$,
(b) $\dPhitt$,
(c) $\HTall$,
(d) $\Ne$,
(e) $\pTe{1}$,
(f) $\meth$\ and
(g) $\dPhieth{1}$.
In each plot the data are compared with predictions from various MC generators.
PWG+PY8 corresponds to the \POWPY sample, PWG+H7 to the \POWHER sample and MCatNLO+PY8 to the \MGNLOPY sample.
The orange band shows the size of the statistical uncertainty and the yellow band shows the size of the total uncertainty. The lower two panels
in each subfigure display the ratios of the different predictions to the data.
}
\label{figaux:diffxsExtra1_1D_rel}
\end{figure}
 
\begin{figure}[htbp]
\centering
\subfloat[]{
\includegraphics[width=0.32\textwidth]{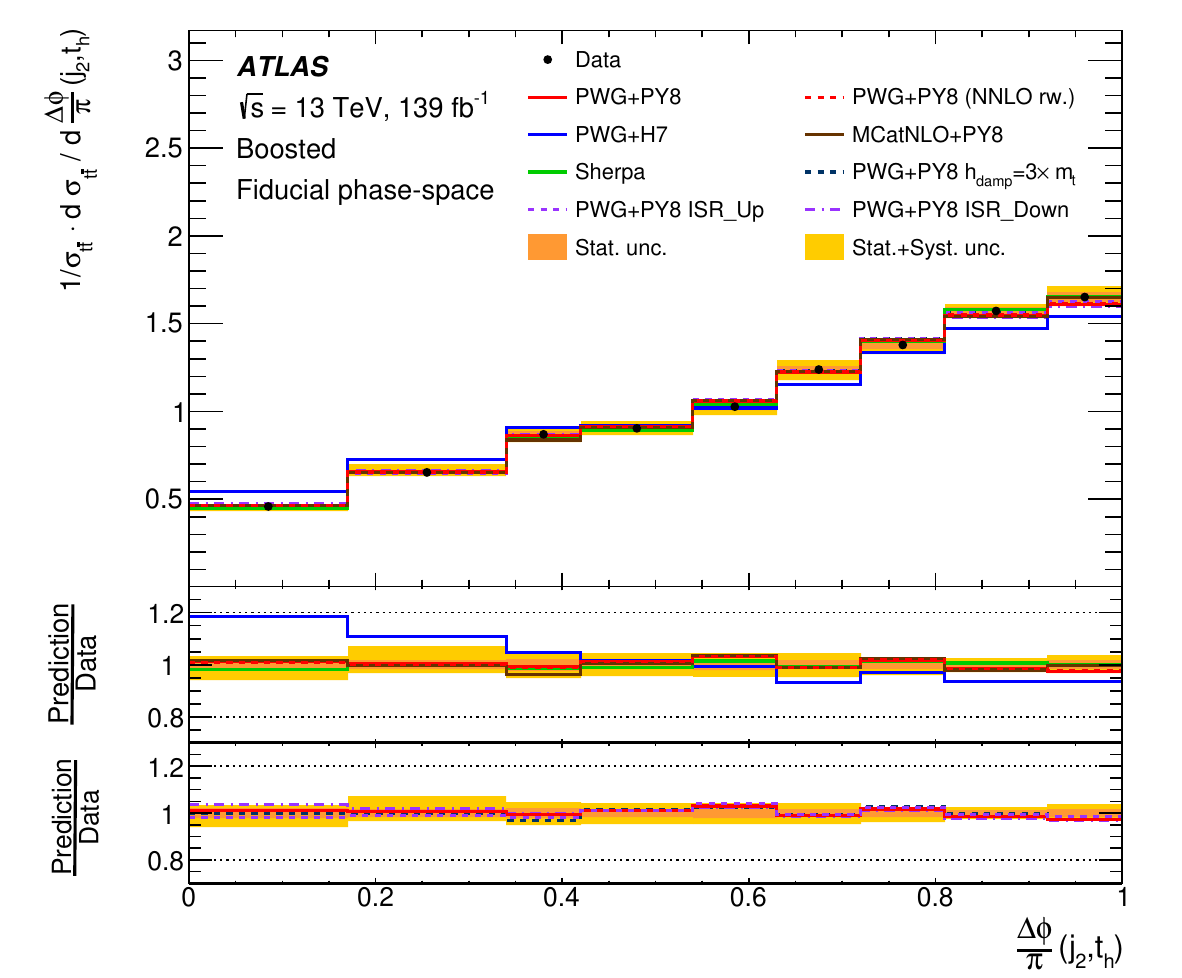}}
\subfloat[]{
\includegraphics[width=0.32\textwidth]{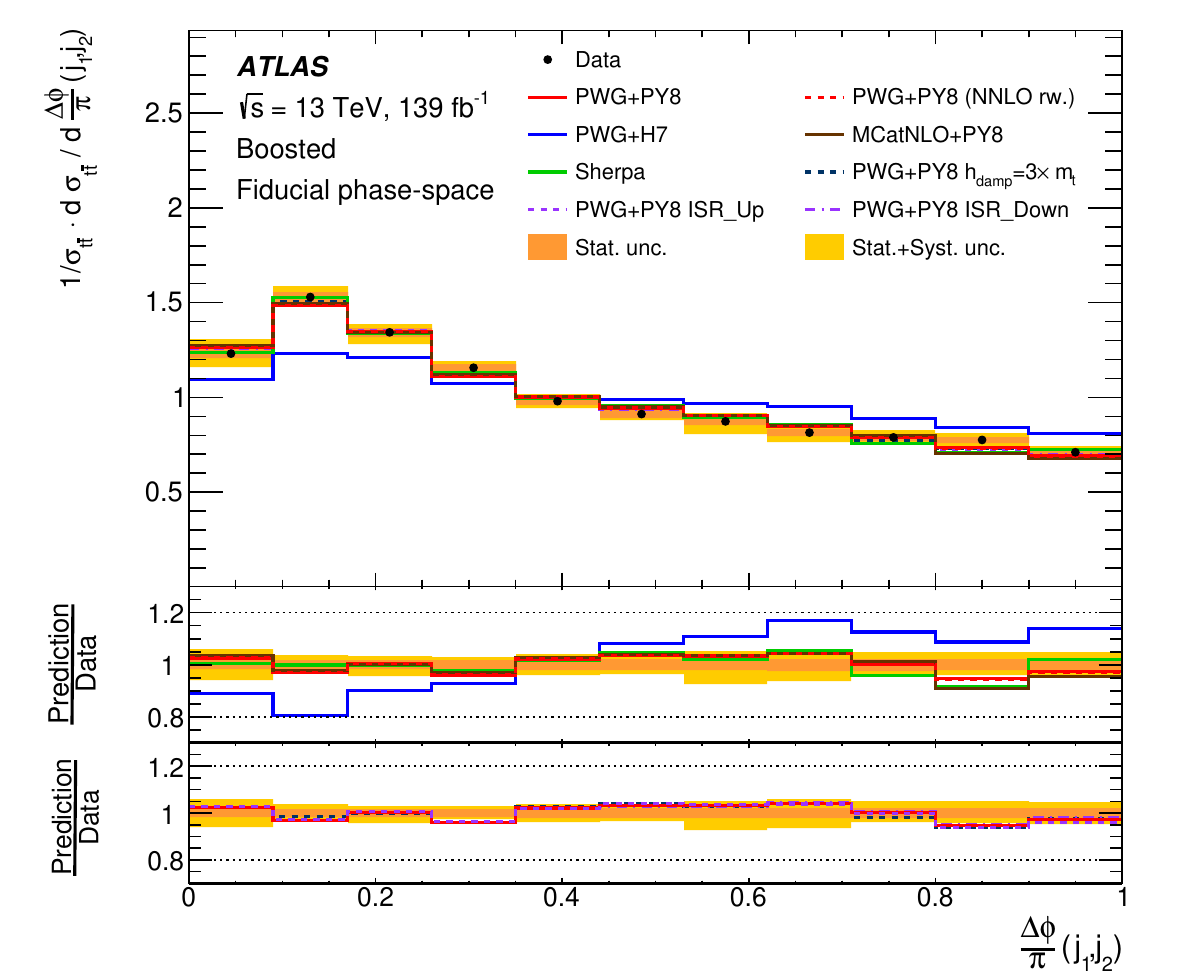}}
\subfloat[]{
\includegraphics[width=0.32\textwidth]{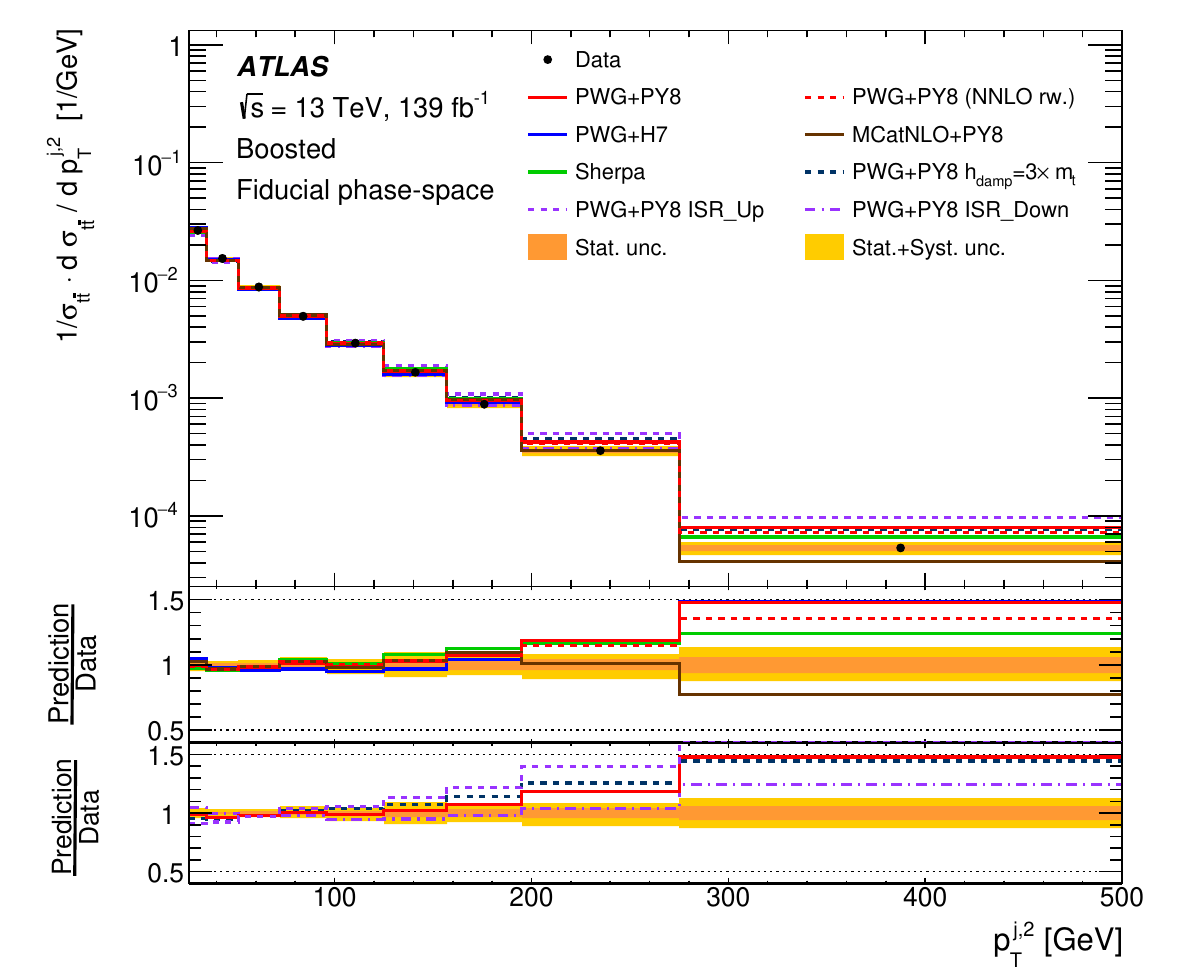}}
\caption{Normalised differential cross-section measurements for the observables related to jets produced in association with the $\ttbar$\ system. The cross-section is shown as a function of
(a) $\dPhieth{2}$,
(b) $\dPhiee$\ and
(c) $\pTe{2}$.
}
\label{figaux:diffxsExtra2_1D_rel}
\end{figure}

\begin{figure}[htbp]
\centering
\subfloat[]{
\includegraphics[width=0.5\textwidth]{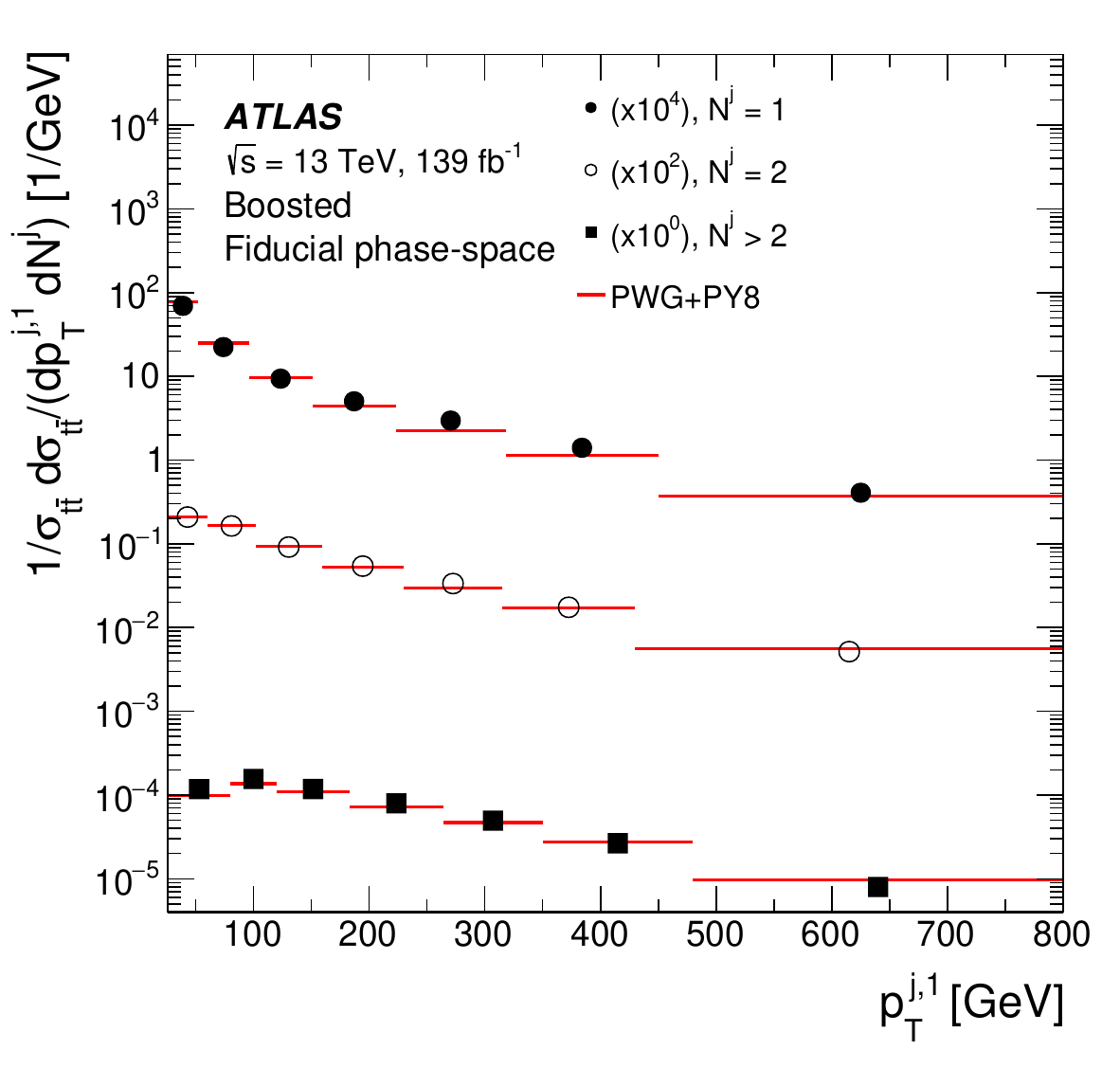}} \\
\subfloat[]{
\includegraphics[width=0.9\textwidth]{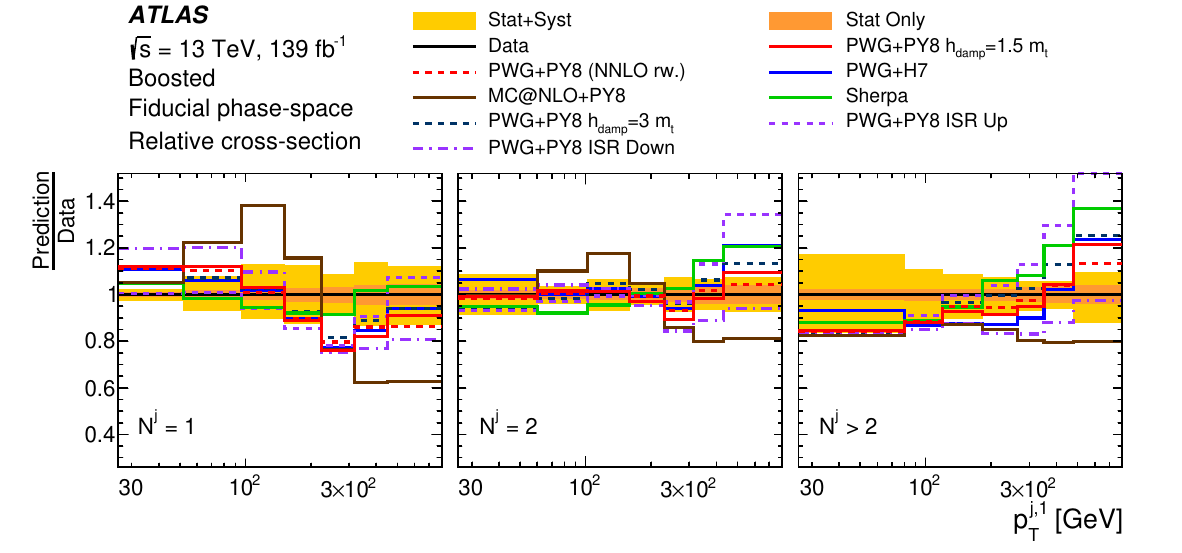}}
\caption{(a) Normalised differential cross-section measurements as a function of the $\pt$\ of the leading additional jet in bins of additional jet multiplicity are compared with the prediction from the \POWPY MC generator. The measurement and the predictions are further normalised by the factors shown in parentheses to aid visibility. (b) Ratio of the measured normalised cross-section to different NLO, and NLO reweighted to NNLO, predictions of $t\bar{t}$ signal for the same differential variables.
PWG+PY8 corresponds to the \POWPY sample, PWG+H7 to the \POWHER sample and MCatNLO+PY8 to the \MGNLOPY sample.
The yellow band represents the total uncertainty of the measured differential cross-section while the orange band shows the statistical component.
}
\label{figaux:diffxs_2D_leadaddjet_pt_V_naddjet_rel}
\end{figure}

\begin{figure}[p]
\centering
\subfloat[]{
\includegraphics[width=0.5\textwidth]{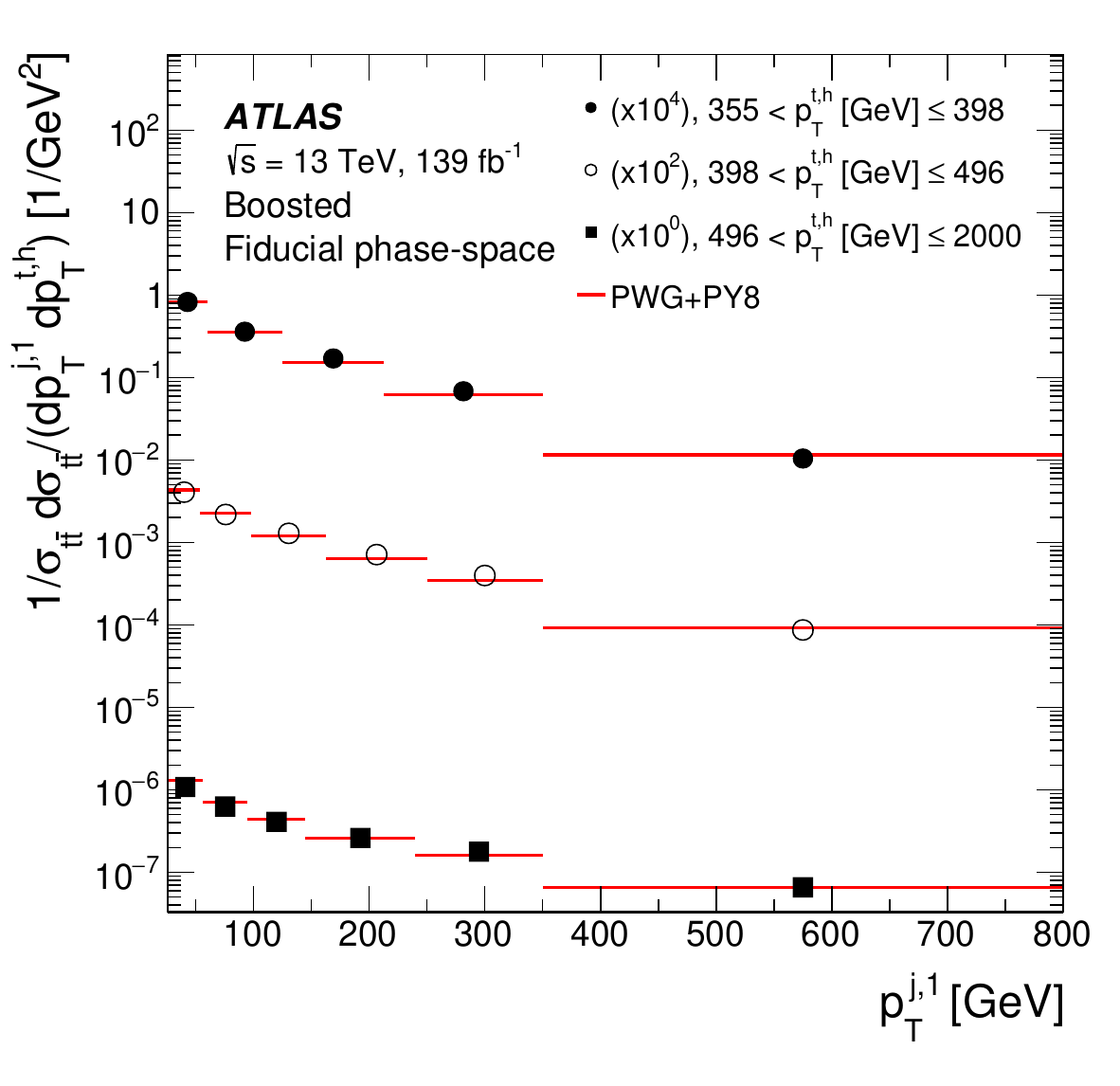}}\\
\subfloat[]{
\includegraphics[width=0.9\textwidth]{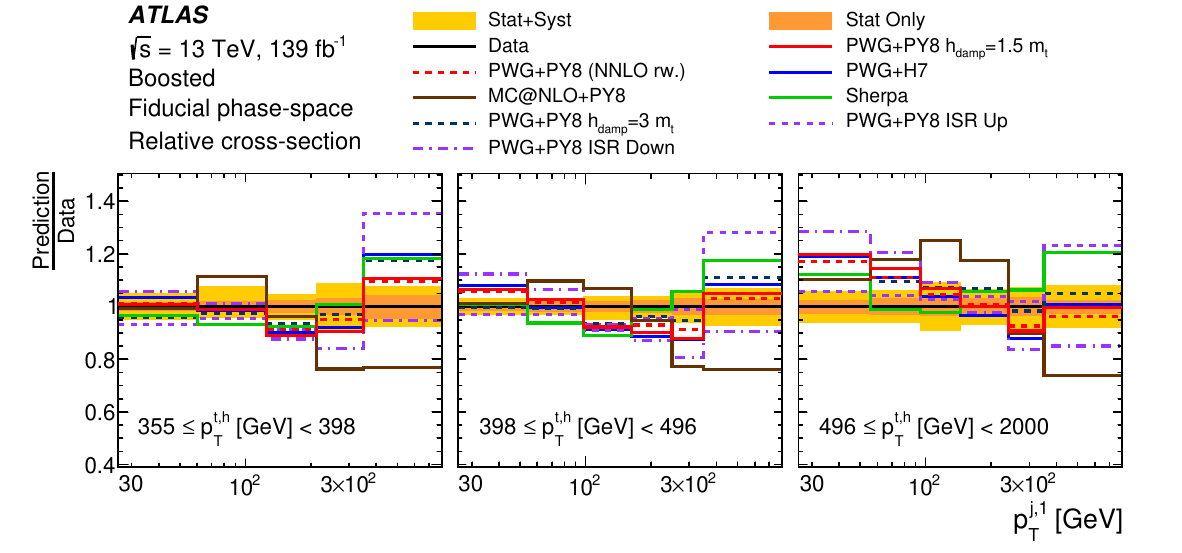}}
\caption{(a) Normalised differential cross-section measurements as a function of the $\pt$\ of the leading additional jet in bins of $\pThad$\ are compared with the prediction from the \POWPY MC generator. The measurement and the predictions are further normalised by the factors shown in parentheses to aid visibility. (b) Ratio of the measured normalised cross-section to different NLO, and NLO reweighted to NNLO, predictions of $t\bar{t}$\ signal for the same differential variables.
PWG+PY8 corresponds to the \POWPY sample, PWG+H7 to the \POWHER sample and MCatNLO+PY8 to the \MGNLOPY sample.
The yellow band represents the total uncertainty of the measured differential cross-section while the orange band shows the statistical component.}
\label{figaux:diffxs_2D_leadaddjet_pt_V_topjet_pt_rel}
\end{figure}

\begin{figure}[p]
\centering
\subfloat[]{
\includegraphics[width=0.5\textwidth]{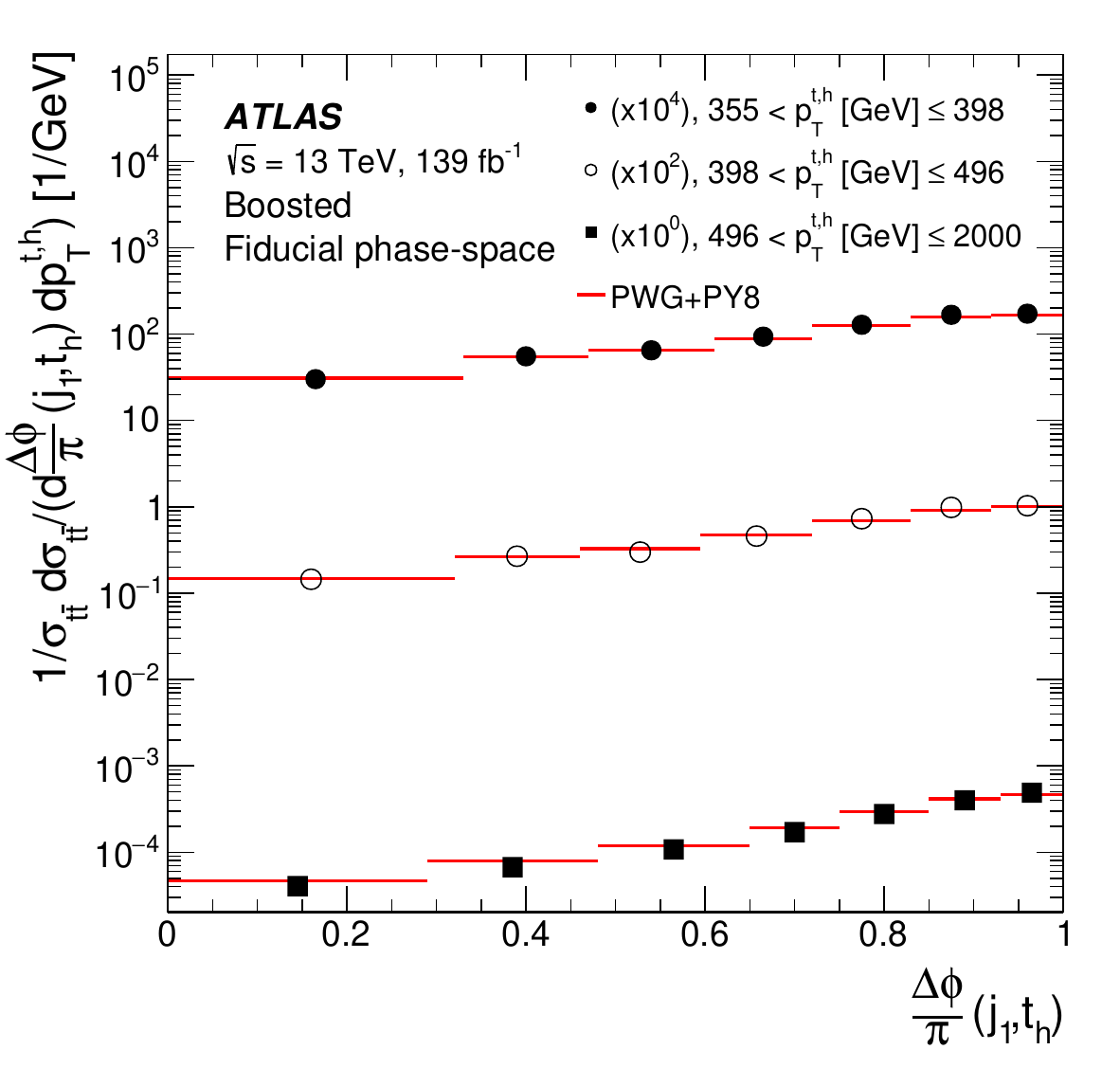}}\\
\subfloat[]{
\includegraphics[width=0.9\textwidth]{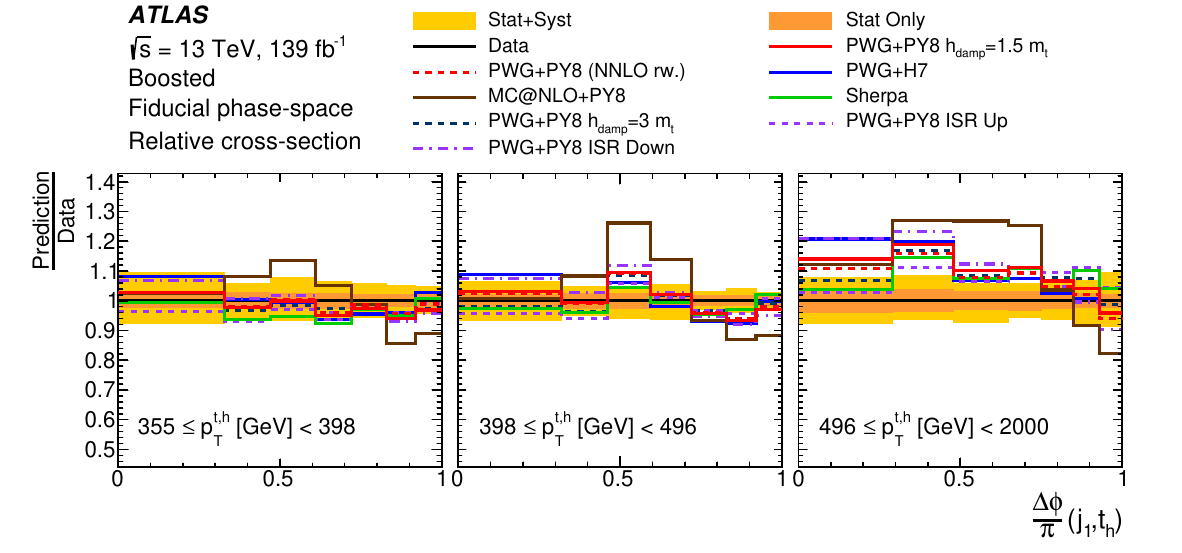}}
\caption{(a) Normalised differential cross-section measurements as a function of the $\Delta\phi$ angle between the leading additional jet and the hadronically decaying top quark in bins of $\pThad$\ are compared with the prediction from the \POWPY MC generator. The measurement and the predictions are further normalised by the factors shown in parentheses to aid visibility. (b) Ratio of the measured normalised cross-section to different NLO, and NLO reweighted to NNLO, predictions of $t\bar{t}$\ signal for the same differential variables.
PWG+PY8 corresponds to the \POWPY sample, PWG+H7 to the \POWHER sample and MCatNLO+PY8 to the \MGNLOPY sample.
The yellow band represents the total uncertainty of the measured differential cross-section while the orange band shows the statistical component.}
\label{figaux:diffxs_2D_leadadd_top_deltaPhi_V_topjet_pt_rel}
\end{figure}

\begin{figure}[p]
\centering
\subfloat[]{
\includegraphics[width=0.5\textwidth]{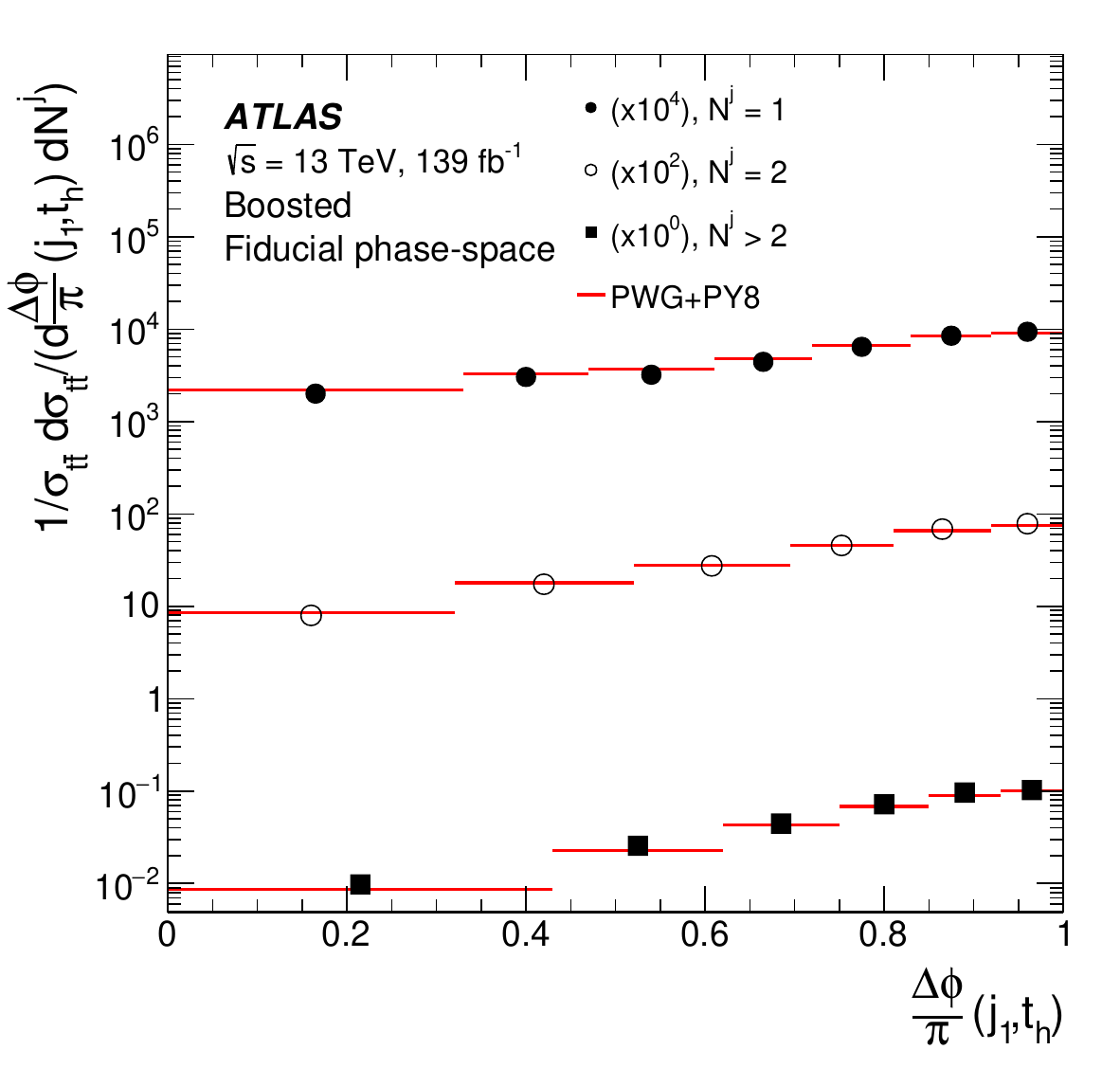}}\\
\subfloat[]{
\includegraphics[width=0.9\textwidth]{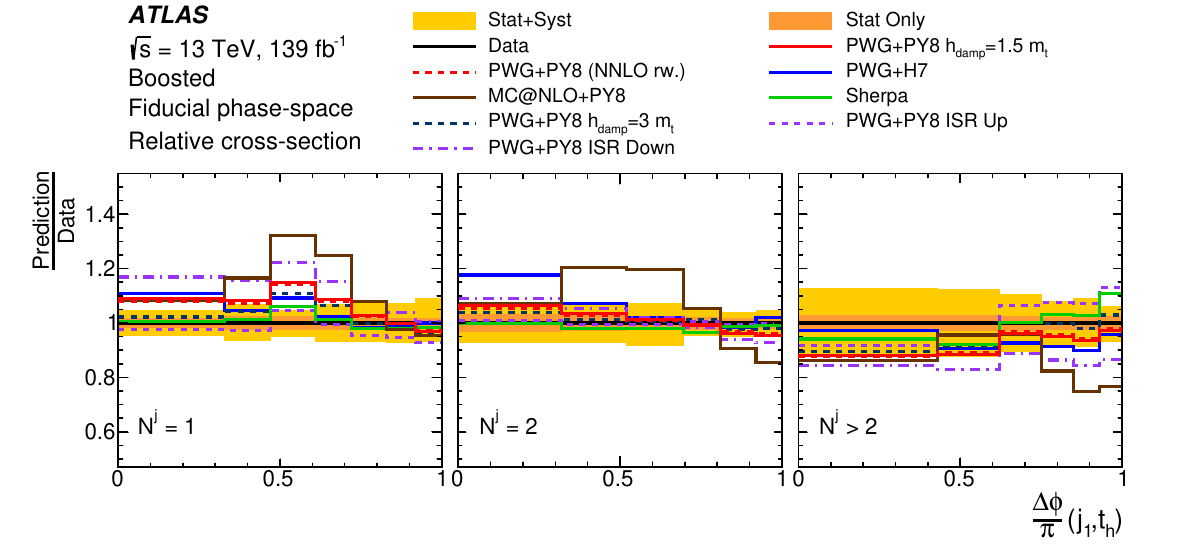}}
\caption{(a) Normalised differential cross-section measurements as a function of the $\Delta\phi$ angle between the leading additional jet and the hadronically decaying top quark in bins of additional jet multiplicity are compared with the prediction from the \POWPY MC generator. The measurement and the predictions are further normalised by the factors shown in parentheses to aid visibility. (b) Ratio of the measured normalised cross-section to different NLO, and NLO reweighted to NNLO, predictions of $t\bar{t}$\ signal for the same differential variables.
PWG+PY8 corresponds to the \POWPY sample, PWG+H7 to the \POWHER sample and MCatNLO+PY8 to the \MGNLOPY sample.
The yellow band represents the total uncertainty of the measured differential cross-section while the orange band shows the statistical component.}
\label{figaux:diffxs_2D_leadadd_top_deltaPhi_V_naddjet_rel}
\end{figure}
 
\FloatBarrier

\printbibliography
 
\clearpage
 
\begin{flushleft}
\hypersetup{urlcolor=black}
{\Large The ATLAS Collaboration}

\bigskip

\AtlasOrcid[0000-0002-6665-4934]{G.~Aad}$^\textrm{\scriptsize 100}$,    
\AtlasOrcid[0000-0002-5888-2734]{B.~Abbott}$^\textrm{\scriptsize 126}$,    
\AtlasOrcid[0000-0002-7248-3203]{D.C.~Abbott}$^\textrm{\scriptsize 101}$,    
\AtlasOrcid[0000-0002-2788-3822]{A.~Abed~Abud}$^\textrm{\scriptsize 36}$,    
\AtlasOrcid[0000-0002-1002-1652]{K.~Abeling}$^\textrm{\scriptsize 53}$,    
\AtlasOrcid[0000-0002-2987-4006]{D.K.~Abhayasinghe}$^\textrm{\scriptsize 93}$,    
\AtlasOrcid[0000-0002-8496-9294]{S.H.~Abidi}$^\textrm{\scriptsize 29}$,    
\AtlasOrcid[0000-0002-9987-2292]{A.~Aboulhorma}$^\textrm{\scriptsize 35e}$,    
\AtlasOrcid[0000-0001-5329-6640]{H.~Abramowicz}$^\textrm{\scriptsize 158}$,    
\AtlasOrcid[0000-0002-1599-2896]{H.~Abreu}$^\textrm{\scriptsize 157}$
% ...
\bigskip
\\

% ...
$^{29}$Physics Department, Brookhaven National Laboratory, Upton NY; United States of America.\\
% ...
$^{35}$$^{(a)}$Facult\'e des Sciences Ain Chock, R\'eseau Universitaire de Physique des Hautes Energies - Universit\'e Hassan II, Casablanca;$^{(b)}$Facult\'{e} des Sciences, Universit\'{e} Ibn-Tofail, K\'{e}nitra;$^{(c)}$Facult\'e des Sciences Semlalia, Universit\'e Cadi Ayyad, LPHEA-Marrakech;$^{(d)}$LPMR, Facult\'e des Sciences, Universit\'e Mohamed Premier, Oujda;$^{(e)}$Facult\'e des sciences, Universit\'e Mohammed V, Rabat;$^{(f)}$Mohammed VI Polytechnic University, Ben Guerir; Morocco.\\
$^{36}$CERN, Geneva; Switzerland.\\
% ...
$^{53}$II. Physikalisches Institut, Georg-August-Universit\"{a}t G\"ottingen, G\"ottingen; Germany.\\
% ...
$^{93}$Department of Physics, Royal Holloway University of London, Egham; United Kingdom.\\
% ...
$^{100}$CPPM, Aix-Marseille Universit\'e, CNRS/IN2P3, Marseille; France.\\
$^{101}$Department of Physics, University of Massachusetts, Amherst MA; United States of America.\\
% ...
$^{126}$Homer L. Dodge Department of Physics and Astronomy, University of Oklahoma, Norman OK; United States of America.\\
% ...
$^{157}$Department of Physics, Technion, Israel Institute of Technology, Haifa; Israel.\\
$^{158}$Raymond and Beverly Sackler School of Physics and Astronomy, Tel Aviv University, Tel Aviv; Israel.\\
% ...
$^{a}$ Also at Borough of Manhattan Community College, City University of New York, New York NY; United States of America.\\
$^{b}$ Also at Bruno Kessler Foundation, Trento; Italy.\\
$^{c}$ Also at Center for High Energy Physics, Peking University; China.\\
$^{d}$ Also at Centro Studi e Ricerche Enrico Fermi; Italy.\\
$^{e}$ Also at CERN, Geneva; Switzerland.\\
$^{f}$ Also at D\'epartement de Physique Nucl\'eaire et Corpusculaire, Universit\'e de Gen\`eve, Gen\`eve; Switzerland.\\
% ...

\end{flushleft}


\end{document}